\newif\ifarxivmode

\newif\ifQRtradeoffsvarietyplotinclude
\newif\ifanalysisofstochnetworksinclude
\newif\ifcompletedetailsofdiffinclusionconvproofinclude
\newif\ifsemicontinuityproofsinclude

\QRtradeoffsvarietyplotincludetrue
\arxivmodetrue

\documentclass[11pt,onecolumn]{IEEEtran}

\usepackage{amsmath}
\usepackage{amssymb}
\usepackage{amsthm}
\usepackage{graphicx}
\usepackage{epsfig}
\usepackage{graphics}
\usepackage{array}
\usepackage{pdfsync}
\usepackage{color}
\usepackage{textcomp} 
\usepackage{accents}
\usepackage{float}
\usepackage{bm} 
\definecolor{mycitecolor}{rgb}{.55,0,0}
\definecolor{mylinkcolor}{rgb}{0,0,.75}
\definecolor{myurlcolor}{rgb}{0,.45,.0}

	\newenvironment{assumption}{\vspace{1ex}\noindent{\bf Assumption}\hspace{0.0cm}}
	{\hfill\vspace{1ex}}	
		\newenvironment{assumptions}{\vspace{1ex}\noindent{\bf Assumptions}\hspace{0.0cm}}
	{\hfill\vspace{0ex}}	

\newtheorem{theorem}{Theorem}

\newtheorem{corollary}{Corollary}
\newtheorem{lemma}{Lemma}

\theoremstyle{definition}
\newtheorem{definition}{Definition}

\newcommand{\proj}{\widetilde}
\newcommand{\exn}{{\mathbb E}}
\newcommand{\defeq}{{\mathrel{\mathop:}=}}
\newcommand{\expn}{{\mathbb E}}

\newcommand*{\bdt}[1]{%
  \accentset{\mbox{\large\bfseries .}}{#1}}

\newcommand{\iiN}{i\in{\cal N}}

\newcommand{\ciC}{c\in{\cal C}}

\usepackage{setspace}
\title{NOVA: QoE-driven Optimization of \\DASH-based Video Delivery in Networks\footnote{This is an extended version of (9-page) conference version available at http://wncg.org/publications/dl.php?file=VJGdV\_conf\_13.pdf}
\thanks{This research was supported in part by Intel and Cisco under the VAWN program, and by the NSF under Grant CNS-0917067. 
We thank Zheng Lu, Xiaoqing Zhu, Chao Chen and Sarabjot Singh for helpful discussions.}}
\begin{document}
\author{Vinay Joseph and Gustavo de Veciana,
\\Department of Electrical and Computer Engineering, 
\ The University of Texas at Austin
}
\maketitle
\begin{abstract}
We consider the problem of optimizing video delivery for a network 
supporting video clients streaming stored video.
Specifically, we consider the problem of jointly optimizing 
network resource allocation and video quality adaptation.
Our objective is to fairly maximize video clients' Quality of Experience (QoE) 
realizing tradeoffs among the mean quality, temporal variability in quality, and fairness, 
incorporating user preferences on rebuffering and cost of video delivery. 
We present a \emph{simple asymptotically optimal online} algorithm, NOVA, to solve the problem.
NOVA is \emph{asynchronous}, and using minimal communication, \emph{distributes} the tasks 
of resource allocation to network controller, 
and quality adaptation to respective video clients. 
Video quality adaptation in NOVA is also optimal for standalone video clients,
and is well suited for use with DASH framework.
Further, we extend NOVA for use with more general QoE models, networks shared with other traffic loads 
and networks using fixed/legacy resource allocation.
\end{abstract}

\section{Introduction}
\label{introduction_section}

There has been tremendous growth in video traffic in the past decade. 
Current trends (see \cite{cisco_trends}) suggest that mobile video traffic will more than double 
each year till 2015, with two-thirds of mobile data traffic being video by 2015.
It is unlikely that wireless infrastructure can keep up with such growth.
Even brute force densification (using HetNets etc.) would not resolve the problem since variability in throughput 
would likely worsen due to increased throughput sensitivity to the dynamic number of users sharing an access point and/or dynamic interference.
Given these challenges, optimizing video delivery to make the best use of available network resources is one of the critical networking problems today.

The main focus of this paper is to develop solutions for optimizing the delivery of stored video, i.e., video stored in video servers, 
streamed by video clients that can \emph{adapt} their video quality.
Our solution is designed to achieve better Quality of Experience (QoE) 
while taking important video client preferences like rebuffering and data cost into account.
Further, it is suited for operation in settings that present video clients with 
heterogeneous preferences, channels and video content.

In this paper, we view the video delivery optimization problem for a network as a problem of 
\emph{fairly} maximizing the video clients' QoE subject to network constraints.
Here, QoE is a proxy for `video client satisfaction'.
A comprehensive solution to this problem requires two components- an allocation component and an adaptation component.
The allocation component decides how the resources (e.g., bandwidth, power etc) in the network are allocated to the video clients.
The adaptation component decides how the video clients adapt their video quality (or video compression rate) in response to the allocated resources, the nature of the video etc.
In this paper, 
we develop a distributed algorithm Network Optimization for Video Adaptation (NOVA) to jointly optimize the two components.
The adaptation component itself has strong optimality guarantees, and can be used in standalone video clients
and, in particular, the adaptation in NOVA can be used with video clients using the DASH (Dynamic Adaptive Streaming over HTTP) framework.
Under the DASH framework, video associated with each video client is stored at the respective video server (at the content provider),
and is a concatenation of several short duration videos called segments which for example could be a GOP (Group Of Pictures). 
Various `representations' of a segment are obtained by compressing it to different sizes by 
changing various parameters associated with the segment like quantization, resolution, frame rate etc,
and typically high quality representations of a segment are larger in size.
Video clients can \emph{adapt} their video quality across segments, i.e.,
can pick different representations for different segments.
The choice of representations can be based on several factors such as the state of playback buffer, 
current channel capacity, features of video content being downloaded etc. 
For instance, the video client can request representations of smaller size to \emph{adapt} to poor channel conditions.

We identify the following four key factors determining the QoE of a video client:
(a) average quality, (b) temporal variability in quality,
(c) fraction of time spent rebuffering, and
(d) cost to the video client and video content provider.
Our main focus is on solving the video delivery optimization problem OPT-BASIC given below which takes these key factors into account:
\begin{eqnarray} \label{opt_basic_obj}
 &&\hspace{-.5cm}\max \quad \sum_{\iiN}  U^E_i  \left(  \mbox{Mean Quality}_i  - \mbox{Quality Variability}_i \right)
\\&&\hspace{-1cm} \mbox{subject to} \ \mbox{ Rebuffering$_i$, Cost$_i$, and Network constraints,} \label{opt_basic_network_cap_constraint}
\end{eqnarray}
where ${\cal N}$ is the set of video clients supported by the network and $U^E_i$ is a `nice' concave function chosen 
in accordance with the fairness desired in the network.
Network constraint captures time varying constraints on network resource allocation allowing us to model
wide range variability in resource availability found in real networks.

Next, we discuss the four key factors mentioned above.
We measure mean quality for a video session as the average across segments of Short Term Quality (STQ)
associated with the downloaded representations of (short duration, e.g., 1 second) segments.
STQ of a downloaded segment should ideally capture the 
viewer's \emph{subjective} evaluation of the quality of the downloaded representation of the segment,
although in practice, this subjective metric will be measured approximately using \emph{objective} video quality assessment metrics (see \cite{Seshadrinathan_survey} for a survey) 
like PSNR, SSIM, MSSSIM etc (see \cite{ssim_reference,ssim_for_video}).

While the benefit of high mean quality is clear, 
the detrimental impact of temporal variability on QoE (see \cite{evaluation_of_temporal_variation_of_video_quality,kim_reduce_quality_variability,video_tutorial}),
and fundamental tradeoff between average quality and temporal variability in quality is often ignored.
Indeed \cite{evaluation_of_temporal_variation_of_video_quality} even suggests that 
temporal variability in quality can result in a QoE that is \emph{worse} than that of a constant quality 
video with {\em lower} average quality.
Two prominent sources for such variability (\cite{joseph_gustavo_varaiance_aware_video_tech_report}) 
are time varying capacity and the time varying nature of video content.
Time varying capacity is especially relevant when considering wireless networks 
where such variations can be caused by
fast fading (on faster time scales, e.g.,  ms) and slow fading due to shadowing, dynamic interference, mobility,
and changing loads (on slower time scales, e.g. secs).
The second, is the time varying nature of the dependence of a segment's STQ
on parameters like compression rate.
Perhaps the key contributor to such change is 
the video content itself, for instance,
segments of same size and same duration could have very different STQ, for e.g.,
consider two such segments where the first segment is of an action scene (where there
is a lot of changing visual content) and the second segment is of a slower scene (where things stay the same).

Rebuffering is the event when playback buffer of a video client empties, and video playback stalls.
Rebuffering events have a significant impact on QoE.
Indeed \cite{http_qoe} points out that the total time spent rebuffering and the frequency of rebuffering events during a video session
can significantly reduce video QoE.
In our approach, we impose constraints on the fraction of the total time spent rebuffering,
and suggest simple ideas to reduce the frequency of rebuffering events.
We also provide flexibility to the video client in setting these constraints according to their preferences.
For instance, a video client who is willing to tolerate rebuffering in return for higher mean quality 
(for e.g., to watch a movie in HD over a poor network) can set these constraints accordingly.
Such constraints driven by video client preferences will often be content and device dependent, and capture important tradeoffs for the video client.

Client preferences concerning the cost of video delivery are also significant, and are important when 
viewers wish to manage their wireless data costs.
Note that content providers may also pay Content Distribution Network operators for the delivery of video data.
Thus, if the cost of data delivery is high, higher QoE often comes at higher cost, 
and the video client/content provider may want to tradeoff QoE versus delivery cost.
In our framework, we allow each video client/content provider to set a constraint on the average cost per unit video duration
which in turn reflects the desired tradeoff.

\subsection{Main contributions}
\label{The_main_result_and_an_outline_of_its_proof}
This paper presents a general optimization framework for stored video delivery optimization
that factors heterogeneity in client preferences, QoE models, capacity and video content.
Further, we develop a {\em simple online} algorithm 
NOVA (Network Optimization for Video Adaptation) to solve this 
multiuser joint resource allocation and quality adaptation problem.
The algorithm has been both rigorously analyzed and validated through extensive simulations.
NOVA's novelty lies in realizing a comprehensive set of features that meet the challenges of developing next-gen video transport protocols.
Key features of NOVA, discussed in more detail in Section \ref{section_NOVA}, are listed below:
\begin{enumerate}
\item \emph{Strong optimality}: guaranteeing
that NOVA performs as well as optimal offline scheme which is omniscient, i.e., knows everything about the evolution of channel and
video ahead of time. 
\item NOVA carries out `cross-layer' joint optimization of  resource allocation and quality adaptation.
	\item NOVA is a \emph{simple} and \emph{online} algorithm.
	\item NOVA is a \emph{distributed} algorithm where network controller carries out resource allocation and
	 video clients carry out their own quality adaptation.
	\item NOVA is an \emph{asynchronous} algorithm well suited for DASH-based video clients 
	where the network controller and video clients operate `at their own pace'.
	The value of this asynchrony (and consequential technical challenges) are discussed in the next subsection. 
	\item NOVA requires almost no statistical information about the network.
	\item \emph{Suited for current networks:} The resource allocation in NOVA requires just a simple modification of legacy schedulers.
	\item \emph{Optimal Adaptation:} 
	Quality adaptation proposed in NOVA is independently optimal and can even be used with a standalone video client, 
	and this optimality is `insensitive' to network resource allocation.	
\end{enumerate}

\subsection{Related work}

The problem of video delivery optimization in wireless networks has been studied in many works,
for instance, see
\cite{Fu10,mchiang_video_paper,hao_xzhu_icc,neely_paper_small_cell_networks,
joseph_gustavo_varaiance_aware_video_tech_report,vjsbmr_stochastic,Blaszczyszyn_paper,khan_cross_layer_optimization}
which utilize extensions of Network Utility Maximization (NUM) framework (see \cite{NUM_reference}).
The main focus of \cite{Fu10} and \cite{mchiang_video_paper} is real-time interactive video which present the challenge
of meeting strict delivery deadlines.
Papers \cite{hao_xzhu_icc} and \cite{neely_paper_small_cell_networks} study video delivery optimization in wireless networks 
considering simpler QoE models,
and do not explicitly incorporate rebuffering (nor cost) into their respective optimization frameworks, and instead control rebuffering through network congestion control.
Using static QoE models, \cite{vjsbmr_stochastic} and \cite{Blaszczyszyn_paper} study the resource allocation component for video delivery accounting for user dynamics.
A major weakness of the aforementioned papers 
is the limited nature of the associated QoE models (that are essentially just the mean quality)
and their lack of flexibility in managing/incorporating user preferences related to rebuffering and cost.

While \cite{joseph_gustavo_varaiance_aware_video_tech_report} presents a novel algorithm for realizing mean-variability tradeoffs for video delivery
(see \cite{VJGdVAA_TAC2012} for genearalizations),
the model involves a strong assumption of synchrony- the download of a segment of each video client starts at the beginning of a (network) slot and finishes by the end of the slot.
This assumption on synchrony precludes any explicit control over rebuffering
as it limits the ability of a video client to get ahead (by downloading more segments) during periods when channel is good and/or network is underloaded.
Relaxed/different versions of this assumption can be found in the theoretical frameworks used in many previous papers 
(e.g., decision making in \cite{khan_cross_layer_optimization,hao_xzhu_icc,neely_paper_small_cell_networks} is synchronous)
as it facilitates an easier extension of tools from classical NUM framework.
However, this assumption of synchrony is not ideal for DASH-based video clients in a wireless network that operate `at their own pace'- 
downloading variable sized segments (with variable download times) one after the other.
In this paper, we drop the assumption of synchrony
which allows us to exploit opportunism across video clients' state of playback buffer (channels and features of video content like quality rate tradeoffs),
and base our adaptation decision concerning a segment on network state information relevant to the download period of the segment.
We also tackle the consequent novel technical challenges related to distributed asynchronous algorithms operating in a stochastic setting.
Further, the rebuffering constraint in our asynchronous setting effectively induces a new type of constraint involving averages measured over two time scales.

\subsection{Notation}
Here, we describe some of the notation used in this paper.
We shall consider a network shared by a set ${\cal N}$ of video clients (or other entities) 
where $N\defeq|{\cal N}|$ denotes the number of video clients in the system. 
We use bold letters to denote vectors, e.g., $\mathbf{a}=\left(a_i\right)_{i\in{\cal N}}$.
Given a collection of $T$ objects $\left(a(t)\right)_{1\le t \le T}$ or a sequence $\left(a(t)\right)_{t\in \mathbb{N}}$, 
we let $\left(a\right)_{1:T}$  denote the finite length sequence $\left(a(t)\right)_{1\le t \le T}$
(in the space associated with the objects of the sequence).  
For example, consider a sequence $\left(\mathbf{a}(t)\right)_{t\in \mathbb{N}}$ where each element is a vector.
Then $\left(\mathbf{a}\right)_{1:T}$ denotes the $T$ length sequence containing the first $T$ vectors
of the sequence $\left(\mathbf{a}\right)_{1:T}$, and $\left(a_i\right)_{1:T}$ denotes the sequence containing the $i$th component of the first $T$ vectors.
Let $\mathbb{N}$, $\mathbb{R}$ and $\mathbb{R}_+$ denote the sets of positive integers,
real numbers and nonnegative real numbers respectively.
For any function $U$ on $\mathbb{R}$, let $U'$ denote its derivative.
For any positive integer $M$, $\mathbf{a},\mathbf{b}\in{\mathbb{R}}^M$ and set ${\cal A}\subset {\mathbb{R}}^M$, let
$$
d_M \left(\mathbf{a},\mathbf{b}\right)
\defeq \sqrt{\sum_{i=1}^{M}  \left(a_{i} - b_{i}\right)^2  }, 
\quad d_M \left(\mathbf{b},{\cal A}\right)\defeq \inf_{\mathbf{a}\in{\cal A}} d_M \left(\mathbf{a},\mathbf{b}\right).
$$

\subsection{Organization of the paper}
Section \ref{section_System_model} introduces the system model and assumptions. 
We formulate the problem OPT-BASIC as an offline optimization problem in Section \ref{section_Optimal_finite_horizon_policy}.
In section \ref{section_NOVA}, we present NOVA to essentially solve the offline optimization problem, and discuss its optimality properties.
Section \ref{section_proof_optimality_NOVA} is devoted to the proof of optimality of NOVA.
We discuss few useful extensions of NOVA in Section \ref{nova_qoe_ch_extensions},
and conclude the paper in Section \ref{section_conclusion}.

\section{System model}
\label{section_System_model}

We consider a network serving video to a fixed
set of video clients ${\cal N}$ where $| {\cal N}| =N.$ 
We consider a slotted network system where 
resources are allocated for the
duration of a slot $\tau_{slot}$, and the slots are indexed by $k\in \left\{0,1,2...\right\}$. 

We assume that resource allocation is subject to time varying constraints.
In each slot $k$, a network controller (e.g., base station in a cellular network) allocates 
$\mathbf{r}_k= \left(r_{i,k}\right)_{\iiN}\in \mathbb{R}_+^N$ bits (or $\mathbf{r}_k/\tau_{slot}$ bits per second) to the video clients such that
$c_k \left( \mathbf{r}_k \right) \le 0$, where $c_k$ is a real valued function modeling 
the current constraints on network resource allocation.
This function could be determined by various parameters like video clients' SNR, interference etc.
In the sequel, we refer to these functions as allocation constraints.
Let $C_k$ denote the random variable corresponding to the allocation constraint in slot $k$ (and $c_k$ is a realization of it).
We make the following assumptions on these allocation constraints: 
\\\line(1,0){514}
\\\begin{assumptions} \textbf{C.1-C.3 (Time varying allocation constraints)}
\\\textbf{C.1} \ $\left(C_k\right)_{k\in{\mathbb N}}$ is a \emph{stationary ergodic} process
of functions selected from a set ${\cal C}$.
\\\textbf{C.2} \ ${\cal C}$ is a (arbitrarily large) finite set of real valued functions on $\mathbb{R}_+^N$,
such that each function $c\in{\cal C}$ is \emph{convex} and continuously differentiable on an open set containing
$\left[0,r_{\max}\right]^N$ with $c \left(\mathbf{0} \right)\le 0$ and 
\begin{eqnarray}\min_{\mathbf{r}\in\left[0,r_{\max}\right]^N} c \left(\mathbf{r}\right)< 0.
\label{there_exists_strictly_feasible_point_in_interior}
\end{eqnarray}
\\\textbf{C.3} \ The feasible region for each allocation constraint is \emph{bounded}: 
there is a constant $0<r_{\max}<\infty$ such that for any $c\in{\cal C}$ and $\mathbf{r}\in\mathbb{R}_+^N$ satisfying $c \left(\mathbf{r}\right)\le 0$, we have $r_i\le r_{\max}$ for each $i\in{\cal N}$.
\\\line(1,0){514}
\end{assumptions}

As indicated in Assumption C.1, we model the evolution of the allocation constraints as a stationary ergodic process.
Hence, time averages associated with the allocation constraints will converge to their respective 
statistical averages, and the distribution of the random vector 
$\left(C_{k_1 + s},C_{k_2 + s},...,C_{k_n +s }\right)$
for any choice of indices $k_1,...,k_n$
does not depend on the shift $s$, thus the marginal distribution of $C_k$ does not depend on time. 
We denote the marginal distribution of this process by $\left(\pi(c)\right)_{c\in{\cal C}}$.
Without loss of generality, we assume that $\pi^{\cal C}(c)>0$ for each $c\in{\cal C}$.
This model (along with the generalization in Subsection \ref{more_general_channel_models}) 
captures a fairly general class of allocation constraints, including, for example, time-varying
capacity constraints associated with bandwidth allocation in wireless networks.

We express
the network constraints in \eqref{opt_basic_network_cap_constraint} of OPT-BASIC 
as the requirement $c_k\left(\mathbf{r}_k\right)\le 0$ on resource allocation $\mathbf{r}_k$
in each slot $k$.
We impose an additional requirement on the resource allocation algorithm 
to ensure that the resource allocation to each video client $\iiN$ in each slot should be at least $r_{i,\min}$
where $r_{i,\min}$ is a small positive constant.
This technical requirement can be relaxed as long as we ensure that each video client 
can be guaranteed a strictly positive amount of resource allocation over a fixed (large) number of slots.

Next, we discuss our video quality adaptation model which is compatible with that proposed in DASH.
The video associated with each video client $\iiN$ is stored at the respective video server (at the content provider),
and is a concatenation of 
segments.
Representations of a segment are obtained by compressing it to different sizes by 
changing various parameters associated with it like quantization, resolution, frame rate etc.
Video clients adapt their quality across segments by selecting different representations for different segments,
and these choices can be based on a variety of factors such as the state of playback buffer, 
current channel capacity, features of video content being downloaded etc. 

The STQ of a (downloaded) segment,
measured using objective video quality assessment metrics like PSNR, SSIM, MSSSIM etc,
typically increases with the compression rate of the corresponding downloaded representation.
Here the compression rate is the ratio of the size of the segment's representation to the duration of the segment.
Note that the size and (hence the) compression rate \emph{also} depend on the size of overheads due to metadata (like identifiers, sequence numbers etc) 
associated with the generation of the data-unit (e.g., file) associated with the segment's representation.
In the sequel, we interchangeably use the terms quality and STQ.
We abstract the relationship between 
the compression rate and quality of a segment 
using a convex increasing function referred to as QR (Quality Rate) tradeoff.
QR tradeoffs maps quality $q$ to the compression rate $f_s(q)$ (measured in bits per second).
Note that for each segment and given compression rate, we are implicitly restricting our attention
to the representation with highest quality and ignoring less efficient representations.
QR tradeoffs can be segment dependent and vary depending on the nature of the segment's video content.
For instance, a segment associated with a slow scene (where things stay the same)
will typically have a `steeper' QR tradeoff
when compared to that of an action scene (where there is a lot of changing visual content). 
For stored video these functions might be obtained offline.
For video streaming of live events, live broadcast of TV channels etc, computationally efficient 
video quality assessment metrics can be used to obtain the QR tradeoffs.

Let $f_{i,s}$ denote a realization of QR tradeoff associated with
the $s$th segment downloaded by video client $i$.
Also, let $F_{i,s}$ denote the random variable corresponding to the QR tradeoff associated with the $s$th segment of video client $i$.
Next, let $l_{i,s}$ denote a realization of length (or duration in seconds)  of the $s$th segment downloaded by video client $i$,
and let $L_{i,s}$ denote the corresponding random variable.
Thus, to obtain a quality $q$ for the $s$th segment, the size of the segment that has to be downloaded by video client $i$ is given by $l_{i,s}f_{i,s} \left(q\right)$.
For each video client $\iiN$, we make the following assumptions on the QR tradeoffs  and segment lengths associated with it:
\\\line(1,0){514}
\\\begin{assumptions} \textbf{QRL.1-QRL.3 on QR tradeoffs and segment lengths}
\\\textbf{QRL.1} $\left(F_{i,s},L_{i,s}\right)_{s\ge 0}$ is a \emph{stationary ergodic} process 
taking values in a set ${\cal FL}_i \subset {\cal F}_i \times {\cal L}_i$.
\\\textbf{QRL.2} ${\cal F}_i$ is a finite set consisting of differentiable \emph{increasing convex} functions defined on an open set containing $[0,q_{\max}]$
such that $\min_{\left\{f_i\in{\cal F}_i\right\}} f_i \left(0\right) >0 $ and $\max_{\left\{f_i\in{\cal F}_i\right\}} \left(f_i\right)'\left(q_{\max}\right)$ is finite.
\\\textbf{QRL.3} ${\cal L}_i$ is a finite set of positive real numbers.
\\\line(1,0){514}
\end{assumptions}
As indicated in Assumption QRL.1, we model the evolution of QR tradeoffs and segment lengths of each video client $\iiN$ as a stationary ergodic process.
Let 
$\left(\pi^{{\cal F},{\cal L}}_i\left( f_i,l_i  \right)\right)_{\left(f_{i},l_i\right)\in{\cal F L}_i} $
denote the associated marginal distribution.
Without loss of generality, we assume that 
$\pi^{{\cal F},{\cal L}}_i\left( f_i,l_i  \right)>0$
for each $\left(f_{i},l_i\right)\in{\cal F L}_i$.
Let
$f_{\min}\defeq\min_{\left\{i,\in{\cal N},f_i\in{\cal F}_i\right\}} f_i \left(0\right)$
which is strictly positive from QRL.2, and this gives a lower bound on segment compression rates.
Even at zero quality, there is usually overhead information associated with a representation of a segment which causes $f_{\min}$ to be positive.
The constant $q_{\max}$ represents the maximum quality that can achieved in the given network setting. 
Let $f_{\max}\defeq\max_{\left\{i,\in{\cal N},f_i\in{\cal F}_i\right\}} f_i \left(q_{\max}\right) $
denote an upper bound on segment compression rates.
From assumption QRL.3, it follows that $l_{\min}\defeq \min_{\left\{i\in{\cal N},l_i\in{\cal L}_i\right\}} l_i$ is strictly positive, and
$l_{\max}\defeq\max_{\left\{i\in{\cal N},l_i\in{\cal L}_i\right\}} l_i $ is finite although it can be arbitrarily large.

Each video client downloads the segments of its video sequentially, and we index the segments using variables like $s,s_i$ etc taking values in $\left\{0,1,2,...\right\}$.
Let $q_{i,s}$ denote the quality (i.e., STQ) associated with the segment $s$ downloaded by video client $i$.

Next, we discuss our model for QoE.
Our QoE model requires that the rebuffering constraint referred to in \eqref{opt_basic_network_cap_constraint} (discussed in more detail below)
are met and under this condition,
QoE of video client $\iiN$ depends only on the quality $\left(q_i\right)_{1:S}$ seen over segments by the video client $\iiN$.
Thus, our QoE model  maps quality seen by a video client $\iiN$ over $S$ segments, i.e. $\left(q_i\right)_{1:S}$, to a QoE metric (under the assumption that the rebuffering constraint of video client $i$ is met).
While accurate QoE models are typically very complex, we use a simple model motivated by the discussion in Section \ref{introduction_section} 
and the model proposed in \cite{evaluation_of_temporal_variation_of_video_quality}.
Let $m^S_i \left(q_i\right)$ and $\mbox{Var}^S_i \left(q_i\right)$ denote (length weighted) mean quality 
and temporal variance in quality respectively associated with the first $S$ segments downloaded by the video client $i$, i.e.,
\begin{eqnarray}
\label{mean_expression}m^S_i \left(q_i\right)&\defeq& \frac{\sum_{s=1}^{S} l_{i,s} q_{i,s}}{ \sum_{s=1}^{S} l_{i,s}  } ,
\\\label{variance_expression}
\mbox{Var}^S_i \left(q_i\right) &\defeq& \frac{ \sum_{s=1}^{S} l_{i,s} \left( q_{i,s}- m^S_i \left(q_i\right) \right)^2 } { \sum_{s=1}^{S} l_{i,s}  }.
\end{eqnarray}
Note that the arguments of $m^S_i$ and $\mbox{Var}^S_i$ are actually $S-$length sequences $\left(q_i\right)_{1:S}$ (i.e., $\left(q_{i,s}\right)_{1\le s\le S}$)
although we are using a shorthand for simplicity.
We model the QoE of video client $i$ for these $S$ segments as
\begin{eqnarray}\label{our_gen_QoE_model}
e^S_i \left(q_i\right)=
   m^S_i \left(q_i\right)	- U^V_i \left(\mbox{Var}^S \left(q_i\right)\right),
\end{eqnarray}
where 
$\left(U^V_i(.)\right)_{\iiN}$  are `nice' convex functions satisfying assumption U.V given below:
\\\line(1,0){514}
\\\begin{assumption} \textbf{U.V: (Variability penalty)}
\\\textbf{U.V} For each $\iiN$, $U^V_i$ is a continuously differentiable increasing convex function with Lipschitz continuous derivatives
defined on an open set containing $[0,q^2_{\max}]$ satisfying $\left(U^V_i\right)'(0)>0$.
\\\line(1,0){514}\end{assumption}
Thus, we could choose $U^V_i(v)=\eta_i v$ or $U^V_i(v)=\eta_i v^2$,
where $\eta_i>0$ and scales the penalty for temporal variability in quality.
Note that our approach can be extended to more general QoE models, and we discuss this in 
Section \ref{more_general_QoE_models}.

Now that we have developed our QoE model, we express the objective function \eqref{opt_basic_obj} in OPT-BASIC as
\begin{eqnarray}\label{OPTBASIC_objective}
\phi_S\left(\left(\mathbf{q}\right)_{1:S}\right)
\defeq \sum_{i\in {\cal N}}  U^E_i  \left(	 e^S_i \left(q_i\right)\right) 	
\end{eqnarray}
where $e^S_i \left(q_i\right) $ is defined in \eqref{our_gen_QoE_model},
$\left(U^E_i(.)\right)_{\iiN}$  are `nice' concave functions satisfying assumption U.E described below.
Let $e_{\min,i}\defeq -U^V_i(q^2_{\max})$ and $e_{\max,i}\defeq q_{\max}-U^V_i(0)$.
\\\line(1,0){514}
\\\begin{assumption} \textbf{U.E: (Fairness in QoE)}
\\\textbf{U.E} For each $\iiN$, we assume that $U^E_i$ is a continuously differentiable increasing concave function with Lipschitz continuous derivatives
defined on an open set containing $[e_{\min,i},e_{\max,i}]$ satisfying $\left(U^E_i\right)'(e_{\max,i})>0$.
\\\line(1,0){514}
\end{assumption}
For each $i\in{\cal N}$, although $U^E_i$ has to be defined over an open set containing $\left[e_{\min,i},e_{\max,i}\right]$, 
only the definition of the function over $\left[ -U^V_i(0),e_{\max,i}\right]$ affects the optimization.
This is because we can achieve this value of QoE for each video client by just picking representation corresponding to zero quality 
for each segment.
Thus, for example, we can choose any function from the following class of strictly concave 
increasing functions parametrized by $\alpha\in (0,\infty)$ (\cite{mo_walrand}) 
\begin{eqnarray} 
U_{\alpha}(e) = \begin{cases} \log \left(e\right) &  \mbox{ if } \alpha=1,
\\ \left(1-\alpha\right)^{-1}e^{1-\alpha} &  \mbox{ otherwise,}
\end{cases}
\end{eqnarray}
and can satisfy the above conditions by making minor modifications to the function.
For instance, we can use the following modification $U^{E,log}$ of the $\log$ function for any (small) $\delta>0$:
$U^{E,log}(e)=\log\left( e - e_{\min,i}+\delta \right),$ $e\in\left[e_{\min,i},e_{\max,i}\right]$.
The above class of functions are commonly used to enforce fairness specifically to achieve allocations that are $\alpha-$fair (see \cite{NUM_reference}). 
A larger $\alpha$ corresponds to a more fair allocation which eventually converges to max-min fair allocation
as $\alpha$ goes to infinity.

Next, we consider the rebuffering related constraint considered in \eqref{opt_basic_network_cap_constraint} of OPT-BASIC.
Let $\kappa>0$ and let $K_S= \left\lceil \kappa S\right\rceil$.
We obtain a good estimate for the fraction of time spent rebuffering by a video client under an additional 
assumption on resource allocation that for each video client $i$,
$ \frac{1}{K_S} \sum_{k=1}^{K_S} r_{i,k}$ converges (for almost all sample paths),
and hence provides an asymptotically accurate estimate for time-average resource allocation to video client $i$
as $S$ goes to infinity.
Note that this condition is satisfied by alpha-fair resource allocation policies like proportionally fair allocation, max-min fair allocation etc.
Next, note that the cumulative size of the first $S$ segments is given by
$\sum_{s=1}^{S} l_{i,s} f_{i,s} \left(q_{i,s}\right)$.
Thus, a good estimate (for large  $S$) for the time required by video client $i$ to download the first $S$ segments is 
$$  \frac{ \sum_{s=1}^{S} l_{i,s} f_{i,s} \left(q_{i,s}\right) }{ \frac{1}{\tau_{slot}K_S} \sum_{k=1}^{K_S} r_{i,k}  }$$
which is the ratio of the cumulative size of $S$ segments to the per slot allocation estimate.
In the above observation, we are implicitly assuming that the network always has video data to send to the video client.
Now, we show that the following expression is an asymptotically (as $S$ goes to infinity) accurate estimate for the 
percentage of time that video client $i$ is rebuffering while watching the $S$ segments:
\begin{eqnarray*}
\beta_{i,S}\left(\left(q_i\right)_{1:S},\left(r_i\right)_{1:K_S}\right)
\defeq \frac{ \frac{\sum_{s=1}^{S} l_{i,s} f_{i,s} \left(q_{i,s}\right)}
{ \frac{1}{\tau_{slot}K_S} \sum_{k=1}^{K_S} r_{i,k}  }}
{\sum_{s=1}^{S} l_{i,s}}
- 1.
\end{eqnarray*}
Note that the first term in the right hand side is the ratio of the estimate for time required for download of the first $S$ segments 
to the total duration $\sum_{s=1}^{S} l_{i,s}$ associated with the $S$ segments.
For video client $i$, let $T^{reb}_i(t)$ denote the fraction of time spent rebuffering till time $t\ge 0$ (measured in seconds),
and let $T^{dow}_i(S)$ denote the time required to download $S$ segments.
Then, we have
\begin{eqnarray}
T^{reb}_i(t) = \int_{0}^{t} I \left( T^{dow}_i(S^{seg}_i(u)) > u + T^{reb}_i(u) \right) du
\label{rebuf_equation}
\end{eqnarray}
where $I(.)$ is the indicator function, and 
$
S^{seg}_i(t) = \min \left\{S: \sum_{s=1}^S l_{i,s} \ge t\right\}
$
denotes the number of segments corresponding to video duration of $t$.
Rearranging \eqref{rebuf_equation}, we have
\begin{eqnarray}
T^{reb}_i(t) = \int_{0}^{t} I \left( I^{reb}_i(u) >0 \right) du
\label{rebuf_equation_rewrite}
\end{eqnarray}
where
\begin{eqnarray}
I^{reb}_i(t)=
\left(\frac{T^{dow}_i(S^{seg}_i(t))}{t} -1\right) - \frac{ T^{reb}_i(t)}{t}.
\label{indicator_rebuf_equation}
\end{eqnarray}
Using \eqref{rebuf_equation_rewrite} and \eqref{indicator_rebuf_equation},
we can show that when $I^{reb}_i(t)<0$, ($T^{reb}_i(t)$ is non-increasing and hence) $I^{reb}_i(t)$ is non-decreasing 
and strictly increasing over a large enough window of time (of duration greater than $l_{\max}$)
due to presence of the term $T^{dow}_i(S^{seg}_i(t)$ in \eqref{indicator_rebuf_equation}.
Using this observation along with the fact that $I^{reb}_i(t) \le 0$ 
(since $T^{reb}_i(t) \ge T^{dow}_i(S^{seg}_i(t)) - t  $ for any $t$),
we can conclude that
$$\lim_{t\rightarrow \infty}I^{reb}_i(t)=0.$$
Using the above observation (and set $t=\sum_{s=1}^S l_{i,s}$ in \eqref{indicator_rebuf_equation}) along with the convergence of $ \frac{1}{K_S} \sum_{k=1}^{K_S} r_{i,k}$,
we can show that that $\beta_{i,S}\left(\left(q_i\right)_{1:S},\left(r_i\right)_{1:K_S}\right)$ 
is an asymptotically accurate estimate for the 
percentage of time that video client $i$ is rebuffering while watching $S$ segments.

Note that $\beta_{i,S}\left(\left(q_i\right)_{1:S},\left(r_i\right)_{1:K_S}\right)$
can also take negative values which happens when segments are being downloaded at rate higher than the rate at which 
they are viewed.
We express the rebuffering constraint in OPT-BASIC as 
$$
\beta_{i,S}\left(\left(q_i\right)_{1:S},\left(r_i\right)_{1:K_S}\right) \le \overline{\beta}_i, \ \forall \ {i\in {\cal N}},
$$
where each video client $i$ specifies an upper bound $\overline{\beta}_i>-1$
on the percentage of time spent rebuffering.
Though setting $\overline{\beta}_i=0$ ensures that there is only an asymptotically negligible amount of rebuffering,
we can enforce more stringent constraints on rebuffering by setting $\overline{\beta}_i$ to negative values.

Next, we consider the cost constraint considered in \eqref{opt_basic_network_cap_constraint} of OPT-BASIC.
The average compression rate associated with the first $S$ segments of video client $\iiN$ is 
$\frac{\sum_{s=1}^{S} l_{i,s} f_{i,s} \left(q_{i,s}\right)} { \sum_{s=1}^{S} l_{i,s} }$.
Let $p^d_i$ denote the cost per unit of data (measured in dollar per bit) that video client $\iiN$
(or the video content provider associated with the video client) has to pay.
Then, the average amount of money per unit video duration the video client (/content provider) pays is
$$ 
p_{i,S}\left( \left(q_i\right)_{1:S} \right) 
\defeq
 p^d_i \frac{\sum_{s=1}^{S} l_{i,s} f_{i,s} \left(q_{i,s}\right)} { \sum_{s=1}^{S} l_{i,s} }.
 $$
We express the cost constraint in OPT-BASIC as 
$$
p_{i,S}\left( \left(q_i\right)_{1:S} \right) \le \overline{p}_i, \ \forall \ {i\in {\cal N}},
$$
where each video client $i$ (or the video content provider associated with the video client) sets an upper bound $\overline{p}_i> 0$
on the amount of money per unit video duration.

Rest of the paper is devoted to the derivation and analysis of an algorithm for solving OPT-BASIC that
carries out \emph{jointly} optimal quality adaptation (i.e., picks optimal $\left(q_i\right)_{1:S}$ for each video client $\iiN$) and
resource allocation (i.e., picks optimal $\left(\mathbf{r}\right)_{1:K_S}$).

\section{Offline optimization formulation}
\label{section_Optimal_finite_horizon_policy}
In this section, we formulate the problem OPT-BASIC of joint optimization of quality adaptation and resource allocation
as an offline optimization problem.
In the offline setting we assume $\left(c_k\right)_{k}$ and
$\left(l_{i,s},f_{i,s}\right)_s $, 
i.e., the realization of the processes $\left(C_k\right)_{k}$ and $\left(L_{i,s},F_{i,s}\right)_s $,
for each video client $i\in{\cal N}$ are known.

Based on the discussion in Section \ref{section_System_model}, we rewrite OPT-BASIC as the optimization problem OPT$(S)$ given below:
\begin{eqnarray} 
\label{OPTS_obj}
\nonumber & &\hspace{-1.8 cm}\max_{\left(\mathbf{q}\right)_{1:S},\left(\mathbf{r}\right)_{1:K_S}} 	
 \phi_S\left(\left(\mathbf{q}\right)_{1:S}\right)
\\\nonumber \mbox{ subject to } \hspace{-.6cm} & & 
 0\le q_{i,s} \le q_{\max} \  \forall \ s\in \left\{1,...,S\right\}, \forall \ {i\in {\cal N}},
\label{OPTS_q_non_neg_and_upper_bound}
\\\nonumber  & & r_{i,k} \ge r_{i,\min}, \  \forall \ k\in \left\{1,...,K_S\right\}, \forall \ {i\in {\cal N}},
\label{OPTS_r_non_neg}
\\\nonumber& & c_k\left(\mathbf{r}_k\right)\le 0, \ \  
\forall \ k\in \left\{1,...,K_S\right\},\label{OPTS_feasibility}
\\ && \beta_{i,S}\left(\left(q_i\right)_{1:S},\left(r_i\right)_{1:K_S}\right)  \le \overline{\beta}_i , \forall \ {i\in {\cal N}},
\label{OPTS_bound_rebuf}
\\\nonumber &&p_{i,S}\left( \left(q_i\right)_{1:S} \right) \le \overline{p}_i, \forall \ {i\in {\cal N}}.
\label{OPTS_cost_bound}
\end{eqnarray}
Although the objective function of OPT$(S)$ does not depend 
directly on the allocated resources $\left(\mathbf{r}\right)_{1:K_S}$, 
the constraint \eqref{OPTS_bound_rebuf} ties the quality adaptation of video clients (and hence the objective function) 
to their respective network resource allocation since
the constraint \eqref{OPTS_bound_rebuf} for video client $i\in{\cal N}$ is equivalent to 
$$ 
\frac{1}{\left(1 + \overline{\beta}_i \right)} 
\frac{\sum_{s=1}^{S} l_{i,s} f_{i,s} \left(q_{i,s}\right)}
{  \sum_{s=1}^{S} l_{i,s} }
\le \frac{1}{\tau_{slot} K_S} \sum_{k=1}^{K_S} r_{i,k}.
$$

We need the following assumption to ensure strict feasibility which will be used in later sections.
\\\begin{assumption}\textbf{-SF} (Strict Feasibility):
 For each $c\in {\cal C}$, $c\left( \left(r_{i,\min}\right)_{\iiN} \right)< 0$, and 
 for each $\iiN$, 
$\max_{\left\{f_i\in{\cal F}_i\right\}} \frac{\tau_{slot} f_i \left(0\right)}{ r_{i,min}} < 1$, and
$p^d_i \max_{\left\{f_i\in{\cal F}_i\right\}} f_i \left(0\right) < \overline{p}_i $.
\end{assumption}

This assumption\footnote{The assumption requires a uniform upper bound on the size of the segments at zero quality which is used in Lemma \ref{parameters_are_bounded}.
We conjecture that this per segment requirement can be replaced with a milder averaged version.} requires that the resource allocation $\left(r_{i,\min}\right)_{\iiN}$ 
is strictly feasible for any $c\in {\cal C}$, and
that the maximum size of segments at zero quality is not too large.

Let $\phi^{opt}_{S}$ denote the optimal value of objective function of OPT$(S)$.
We would solve 
the optimization problem OPT$(S)$ directly if it were possible.
However this is impossible in practice
since we need to know $\left(c_k\right)_{k}$ and $\left(f_{i,s}\right)_s $ ahead of time.
Further, a direct approach is also computationally prohibitive as the optimization is over $O(NS)$ variables. 
Thus, from a practical point of view, the main challenge is to overcome these two hurdles and obtain a \emph{simple} and \emph{online} algorithm 
that performs as well as $\phi^{opt}_{S}$ asymptotically.
We present our solution to this challenge in the next section.

\section{An online algorithm for jointly optimizing resource allocation and quality adaptation}
\label{section_NOVA}
In this section, we present our algorithm Network Optimization for Video Adaptation (NOVA), and discuss its asymptotic optimality.
The algorithm NOVA comprises three components:
\begin{enumerate}
	\item \emph{Allocate:} Network resource allocation is done at the beginning of each slot $k$ 
	by solving an optimization problem RNOVA$\left(\mathbf{b}_k,c_k\right)$ 
	which depends on the parameter $\mathbf{b}_k$ (described later in the section) and current allocation constraint $c_k$.
\item \emph{Adapt:} When a video client $i\in{\cal N}$ finishes download of $s_i$th segment,
select the quality/representation for the next segment by solving an optimization problem 
 QNOVA$_i(\bm{\theta}_{i,s_i}, f_{i,s_i+1})$
 which depends on a parameter $\bm{\theta}_{i,s_i}$ (described later in the section),
 and the QR tradeoff $f_{i,s_i+1}$ of the next segment.
\item \emph{Learn:} parameters $\left(m_{i,s_i},\mu_{i,s_i},v_{i,s_i},b_{i,k},d_{i,s_i},\lambda_{i,s_i}\right)_{\iiN}$ 
used in the optimization problems
RNOVA$\left(\mathbf{b}_k,c_k\right)$ and QNOVA$_i(\bm{\theta}_{i,s_i}, f_{i,s_i+1})$.
Here $s_i$ is the current segment index of video client $i$ and $k$ is current slot index.
The parameters $m_{i,s_i}$ and $\mu_{i,s_i}$ track mean quality, $v_{i,s_i}$ tracks variance in quality,
and $\lambda_{i,s_i}$ tracks the mean segment duration of video client $i\in{\cal N}$.
The parameters $b_{i,k}$ and $d_{i,s_i}$ serve as indicators of risk of violation of rebuffering constraints \eqref{OPTS_bound_rebuf} and cost constraints \eqref{OPTS_cost_bound} respectively  of video client $i\in{\cal N}$,
and larger the parameter, larger the risk.
\end{enumerate}

We start by describing the two optimization problems RNOVA$\left(\mathbf{b},c\right)$ 
and QNOVA$_i(\bm{\theta}_i,f_i)$ associated with NOVA before presenting the algorithm.
We can control the response of NOVA to the indicators $b_{i,k}$ and $d_{i,s_i}$ using functions discussed next.
For each $i\in{\cal N}$, let $h^B_i(.)$ and $h^D_i(.)$ be non-negative valued Lipschitz continuous functions that are strictly increasing 
over $\mathbb{R}_+$, and are such that $\lim_{b\rightarrow \infty}h^B_i(b) =\infty$ and $\lim_{d\rightarrow \infty}h^D_i(d) =\infty$.
Also, let $h^B_i(b_i)=0$  for all $b_i \le \underline{b}$ and $h^D_i(d_i)=0$  for all $d_i \le \underline{d}$
for  some constants $\underline{b}$ and $\underline{d}$ typically set as zero or small negative numbers.
Simple examples of functions satisfying these conditions are
$\max(b,0)$, $\max(b^2,0)$ etc.

Let $\mathbf{b}\in \mathbb{R}^N$ and $\ciC$.
The optimization problem RNOVA$\left(\mathbf{b},c\right)$ 
associated with network resource allocation is given below:
	\begin{eqnarray} \label{defn_phiR}
  \max_{\mathbf{r}}	&& \phi^R\left(\mathbf{r},\mathbf{b}\right)\defeq\sum_{i\in {\cal N}}   h^B_i\left(b_i\right)r_i
\label{RNOVA_objective}
\\\mbox{ subject to }& & c\left( \mathbf{r} \right)\le 0,
\label{RNOVA_channel_constraints}
\\& & r_i \ge r_{i,\min} \ \forall \ i\in{\cal N}.
\label{RNOVA_nonneg_constraints}
\end{eqnarray}
Let ${\cal R}^*\left(\mathbf{b},c\right)$ denote the set of optimal solutions to RNOVA$\left(\mathbf{b},c\right)$.
Note that the objective function \eqref{defn_phiR} gives more weight to video clients higher value of $b_i$, i.e., higher risk of violation of rebuffering constraints.

The optimization problem QNOVA$_i(\bm{\theta}_i,f_i)$ associated with quality adaptation performed by video clients.
Let 
$0\le m_i,\mu_i\le q_{\max},\ 0\le v_i\le q^2_{\max},\ b_i,d_i\in\mathbb{R}$, $\bm{\theta}_i=(m_i,\mu_i,v_i,b_i,d_i)$
and $f_i\in{\cal F}_i$.
For $i\in{\cal N}$, the optimization problem 
QNOVA$_i(\bm{\theta}_i,f_i)$ is given below:
	\begin{eqnarray} 
 \nonumber \max_{q_i} &&  \phi^Q\left(q_i,\bm{\theta}_i,f_i\right)
\label{QNOVA_objective}
\\\mbox{ subject to }& & q_i \ge 0,
\label{QRNOVA_nonneg_constraints}
\\& & q_i \le q_{\max},
\label{QRNOVA_bounded_above_constraints}
\end{eqnarray}
where
\begin{eqnarray}\label{defn_phiQ}
\phi^Q\left(q_i,\bm{\theta}_i,f_i\right) 
=\left(U^E_i\right)^{'}\left( \mu_i - U^V_i \left(v_i\right)\right) \left( q_i
-  \left(U^V_i\right)^{'} \left(v_i\right) \left(q_i - m_i\right)^2\right)
- \frac{h^B_i\left(b_i\right)}{\left(1+\overline{\beta}_i\right)} f_{i} \left(q_i\right)
 - \frac{p^d_i h^D_i\left(d_i\right)}{\overline{p}_i } f_{i} \left(q_i\right).
\end{eqnarray}
We can obtain an intuitive understanding of the objective function \eqref{defn_phiQ} of the above optimization problem by noting that the term $\left(q_i - m_i\right)^2$ ensures that an optimal solution to  QNOVA$_i(\bm{\theta}_i,f_i)$
is not too far away from $m_i$ (current estimate of mean quality), and thus avoids high variance in quality.
Also, the terms $\frac{h^B_i\left(b_i\right)}{\left(1+\overline{\beta}_i\right)} f_{i} \left(q_i\right)$
and $\frac{p^d_i h^D_i\left(d_i\right)}{\overline{p}_i } f_{i} \left(q_i\right)$ in \eqref{defn_phiQ}
penalize quality choices leading to large segment sizes
when $b_i$ or $d_i$ are high, and thus help NOVA to respond to increased risk of violation of rebuffering constraints and cost constraints.
The optimization problem QNOVA$_i(\bm{\theta}_i,f_i)$ is convex and has a unique solution due to the strict concavity of the objective function.
Let $q^*_i \left(\bm{\theta}_i,f_i\right)$  
denote the solution to QNOVA$_i(\bm{\theta}_i,f_i)$.

For each $\iiN$, define the set ${\cal H}^{(i)}$ as follows:
\begin{eqnarray}\label{defn_H_i}
{\cal H}^{(i)}= 
\left\{ \left(m_i,\mu_i,v_i,b_i,d_i,\lambda_i\right):
\ 0\le m_i,\mu_i\le q_{\max},
\ 0\le v_i\le q^2_{\max},\ b_i\ge \underline{b},\ d_i\ge \underline{d},\ l_{\min}\le  \lambda_i \le l_{\max}  \right\}.
\end{eqnarray}
Let $s_i$ be an indexing variable keeping track of the segment video client $i$ is currently downloading.
We also use auxiliary variables 
$b_{Q,i,s}$ 
and $b_{R,i,k}$ for each video client $i$ to keep track of the parameter $b_{i,.}$ in NOVA.
The algorithm NOVA is given below.
\\\line(1,0){514}\vspace{-.4cm}
\begin{center}
NOVA\vspace{-.7cm}
\end{center}
\line(1,0){514}
\begin{flushleft}
\vspace{-.3cm}\textbf{NOVA.0}: Initialization: Let 
$\left(m_{i,0},\mu_{i,0},v_{i,0},b_{i,0},d_{i,0},\lambda_{i,0}\right)
\in {\cal H}^{(i)}$ and $s_i=0$ for each $i\in{\cal N}$, and $\epsilon>0$.
\end{flushleft}
In each slot $k\ge 0$, carry out the following steps:
\\ \textbf{ALLOCATE}: At the beginning of slot $k$, let $b_{R,i,k}= b_{i,k}$ for each $\iiN$,
and
allocate resources 
according to any element of the set ${\cal R}^*\left(\mathbf{b}_k,c_k\right)$
(of optimal solutions to RNOVA$\left(\mathbf{b}_k,c_k\right)$ )
and update $\mathbf{b}_k$ as follows:
\begin{eqnarray}\label{b_allocate_update}
  b_{i,k+1} &=&  b_{i,k} + \epsilon \left( \frac{\tau_{slot}}{ \left(1+ \overline{\beta}_i\right)  } \right).
\end{eqnarray}
 \textbf{ADAPT}: In slot $k$, if any video client $i\in{\cal N}$ finishes download of $s_i$ th segment,
 let $b_{Q,i,s_i+1}= b_{i,k+1}$, $\bm{\theta}_{i,s_i}=(m_{i,s_i},\mu_{i,s_i},v_{i,s_i},b_{Q,i,s_i+1},d_{i,s_i})$.
For segment $s_i+1$ of video client $i$, select representation with quality $q^*_i(\bm{\theta}_{i,s_i}, f_{i,s_i+1})$
(i.e., optimal solution to QNOVA$_i(\bm{\theta}_{i,s_i},f_{i,s_i+1})$), denoted as $q^*_{i,s_i+1}$ for brevity,
and update parameters $m_{i,s_i+1}$, $\mu_{i,s_i+1}$, $v_{i,s_i+1}$, $b_{i,k+1}$, $d_{i,s_i+1}$ and $s_i$ as follows:
\begin{eqnarray}
\label{m_update_NOVA}  m_{i,s_i+1} &=& m_{i,s_i} 
  + \epsilon \left(U^E_i\right)^{'}\left( \mu_i - U^V_i \left(v_i\right)\right)\left(U^V_i\right)^{'} \left(v_i\right)
  \left( \frac{l_{i,s_i+1}}{\lambda_{i,s_i}} q^*_{i,s_i+1}- m_{i,s_i}   \right),
\\\label{mu_update_NOVA}  \mu_{i,s_i+1} &=& \mu_{i,s_i} 
  + \epsilon  \left( \frac{l_{i,s_i+1}}{\lambda_{i,s_i}} q^*_{i,s_i+1}- \mu_{i,s_i}   \right),
\\\label{v_update_NOVA}
v_{i,s_i+1} &=& v_{i,s_i} +  \epsilon \left( \frac{l_{i,s_i+1}}{\lambda_{i,s_i}}\left(q^*_{i,s_i+1}-m_{i,s_i} \right)^2  - v_{i,s_i}   \right),
\\\label{b_adapt_update}
b_{i,k+1} &=&\left[   b_{i,k+1}
- \epsilon \left(l_{i,s_i+1}\right)\right]_{\underline{b}},
\\\label{d_adapt_update}
d_{i,s_i+1} &=&  \left[  d_{i,s_i} + \epsilon 
\left(p^d_i \frac{ l_{i,s_i+1} f_{i,s_i+1}\left(  q^*_{i,s_i+1}\right)  }{ \overline{p}_i } - \lambda_{i,s_i}  \right)\right]_{\underline{d}},
\\\label{lambda_update_NOVA}
\lambda_{i,s_i + 1} &=& \lambda_{i,s_i} + \epsilon \left( l_{i,s_i+1} -  \lambda_{i,s_i}  \right),
\\\nonumber s_i &=& s_i+1.
\end{eqnarray}
\line(1,0){514}

Here, $\left[ x \right]_{y}=\max(x,y)$ for $x,y\in\mathbb{R}$.
The variable
$b_{Q,i,s}$ stores the value of $b_{i,.}$ used in choosing quality for the $s$th segment of video client $i$,
and $b_{R,i,k}$ stores the value of $\left(b_{i,k}\right)_{\iiN}$ used in the resource allocation in slot $k$.
These are just auxiliary variables, and do not affect the evolution of the algorithm (unlike  $b_{i,k}$
which affects the algorithm).
Further, to ensure that the video clients start downloading video segments from the beginning,
we assume that all the video clients have already downloaded 0th segment.

Under NOVA, allocation is done at the beginning of each slot
whereas adaptation is asynchronous, i.e., adaptation related decisions about a segment is made by a video client
only at the completion of download of previous segment.
The update equation \eqref{lambda_update_NOVA} associated with the parameter $\lambda_{i,s_i}$
is similar to update rules used for tracking EWMA (Exponentially Weighted Moving Averages),
and ensures that $\lambda_{i,s_i}$ tracks the mean segment duration of video client $i$.
The update rules \eqref{m_update_NOVA}-\eqref{v_update_NOVA} are similar,
and ensure that $m_{i,s_i}$ and $\mu_{i,s_i}$ track mean quality, while $v_{i,s_i}$ tracks variance in quality.
Both $m_{i,s_i}$ and $\mu_{i,s_i}$ track mean quality giving different weights to the current quality,
and we later generalize the update rule \eqref{mu_update_NOVA} so that $\mu_{i,s_i}$ tracks parameters associated with more general QoE metrics.
The weights $\frac{l_{i,s_i+1}}{\lambda_{i,s_i}}$ used in the update rules ensure that the duration of the segment is appropriately factored.
Next, we consider the evolution of the parameter the operator $b_{i,k}$ which is updated in both \eqref{b_allocate_update}
and \eqref{b_adapt_update} ignoring $\left[ .\right]_{\underline{b}}$ and setting initialization to zero.
\eqref{b_allocate_update} ensures that $b_{i,k}$ is increased by fixed amount $\epsilon \left( \frac{\tau_{slot}}{ \left(1+ \overline{\beta}_i\right)  } \right)$
at the beginning of each slot.
\eqref{b_adapt_update} ensures that when a video client completes the download of a segment, $b_{i,k}$ is reduced by $\epsilon$ times 
the duration of the next segment.
Hence, at some time $t$ seconds (or $k=t/\tau_{slot}$ slots) after starting the video,  
\begin{eqnarray*}
\frac{b_{i,k}-b_{i,0}}{\epsilon}&\approx&    \frac{t}{ \left(1+ \overline{\beta}_i\right)  }- L^D_i(t),
\end{eqnarray*}
where $L^D_i(t)$ is the duration of video downloaded till time $t$.
This sheds light on the role of  $b_{i,k}$ 
as an indicator of risk of violation of rebuffering constraint in \eqref{OPTS_bound_rebuf} for video client $i$.
In particular, we see that for $\overline{\beta}_i=0$ and small enough $\underline{b}$,
$(b_{i,k}-b_{i,0})/\epsilon$ is equal to $(t- L^D_i(t))$ which is equal to \emph{negative} of the duration of video content in playback buffer (if there is any).
Similarly, we can argue that
$d_{i,s_i}$ serves as an indicator of risk of violation of cost constraint \eqref{OPTS_cost_bound} for video client $i$.
Depending on the problem under consideration, we can drop some of the parameters from NOVA.
For instance, if $U^V_i$ is a linear function, we need not track $v_{i,s_i}$.
Or, if a video client does not have a cost constraint, we need not track $d_{i,s_i}$.

It is interesting to note that the quality adaptation proposed in NOVA does not directly use any information about the allocation constraints.
Neither does the resource allocation directly use any information about QR tradeoffs of the video clients.
Yet, the joint resource allocation and quality adaptation under NOVA has strong optimality properties (which are presented later in this section).
This is mainly due to the fact that 
the variables $\left(b_{i,k}\right)_{\iiN}$ carry almost all the information about the video clients' quality adaptation
that is required by the network controller
to carry out optimal resource allocation,
and the variable $b_{i,k}$ carries almost all the information 
that the quality adaptation at video client $i$ needs to know about the resource allocation (to the client).
For e.g., consider a video client $i$ in the network that has very few unwatched segments in the playback buffer,
i.e., the video client is about to experience rebuffering.
We see that the update rules for $b_{i,k}$ (and a large enough initialization) ensure that $b_{i,k}$ will be large in this scenario,
and this forces the video client and the network controller to make the right moves,
i.e., this forces the video client to switch to low quality representations (while taking current QR tradeoffs into account),
and forces the network controller to give higher priority to this video client in the resource allocation 
(while taking the current allocation constraints also into account).

Next, we discuss some important features of NOVA that make it attractive from a practical point of view.
\begin{itemize}
\item NOVA carries out `cross-layer' joint optimization of  resource allocation and quality adaptation,
with \emph{strong optimality guarantees} (given in Theorem \ref{main_optimality_theorem}).
\item NOVA is an \emph{online} algorithm as it only uses \emph{current} information, i.e., 
NOVA only needs the allocation constraint $c_k$ for slot $k$,
and for quality adaptation of segment $s_i+1$ of video client $i$, it only requires the QR tradeoff $f_{i,s_i+1}$ 
for the optimization and $l_{i,s_i+1}$ for updates associated with that segment.
\item NOVA is a \emph{simple} algorithm since
RNOVA$\left(\mathbf{b},c\right)$ is convex optimization problem in $N$ variables.
Further, if allocation constraints are linear, RNOVA$\left(\mathbf{b},c\right)$ is just a linear program which often has enough structure to allow for very efficient evaluation of the solutions.
Also, note that QNOVA$_i(\bm{\theta}_i,f_i)$ is just a scalar convex optimization problem.
	\item The asynchronous nature of NOVA ensures that the video clients can work at their own pace and the adaptation prescribed in NOVA is entirely \emph{client driven} requiring no assistance from the network controller,
and is thus well suited for DASH framework.
\item NOVA can be implemented in a \emph{distributed} manner with minimal signaling since quality
adaptation is client driven and for the resource allocation, the network controller
need only know $\mathbf{b}_k$
which are indicators of risk of violation of rebuffering constraints associated with the video clients
(illustrated in Fig.~\ref{distributed_implementation_fig}).
To ensure that the network controller knows the current value of $\mathbf{b}_k$,
each video client can send a signal to the base station indicating the latest value of $b_{i,k}$
at the end of each segment download which usually occurs at a low frequency (typically once a second).
On receiving this signal from video client $\iiN$, the network controller can then update $b_{i,k}$ until the next signal from video client $i$
using the simple update rule in \eqref{b_allocate_update} that requires only constant increments.

\begin{figure}[ht]
	\centering
	\ifarxivmode
\includegraphics[scale=.41]{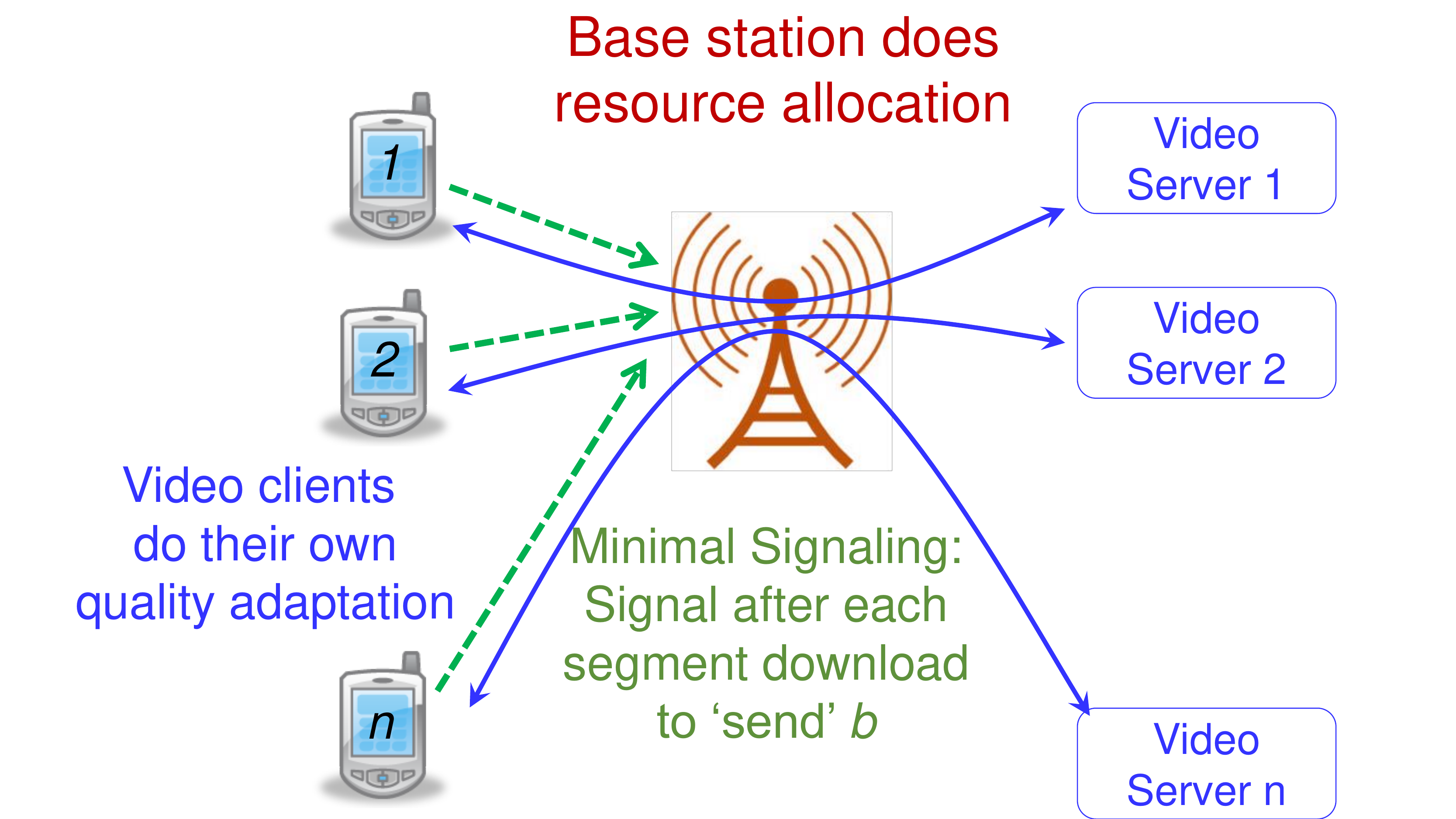}
\else
\includegraphics[scale=.41]{figs/distributed_implementation_fig.pdf}
\fi
	\caption{Distributed Implementation of NOVA}
	\label{distributed_implementation_fig}
\end{figure}

\item Optimization algorithm for resource allocation, RNOVA$\left(\mathbf{b},c\right)$
requires only a simple modification of legacy schedulers like proportionally fair schedulers (see \cite{kushner_prop_fair_conv}).
In fact, the optimization problem associated with proportionally fair schedulers
is almost the same as RNOVA$\left(\mathbf{b},c\right)$ except that it uses a function of current estimate of average throughput instead of $\mathbf{b}_k$.
\end{itemize}

The preceding discussion of NOVA suggests it is intuitively doing the right things.
The discussion in the rest of this section and Section \ref{section_proof_optimality_NOVA} is a rigorous analysis of NOVA aimed at 
establishing the strong optimality result for NOVA given in Theorem \ref{main_optimality_theorem}.
Proving Theorem \ref{main_optimality_theorem} requires other intermediary results.
We have devoted Section \ref{section_proof_optimality_NOVA} to these results and give a proof of Theorem \ref{main_optimality_theorem} at the end of that section.
\begin{theorem}\label{main_optimality_theorem}
Suppose $\left(m_{i,.},\mu_{i,.},v_{i,.},b_{i,.},d_{i,.},\lambda_{i,.}\right)$ evolve according to the update rules in NOVA.
\\(a) \emph{Feasibility:} NOVA asymptotically satisfies the constraints on rebuffering and cost, i.e., for each $\iiN$
\begin{eqnarray}
\label{NOVA_satifies_rebuf_constraints}
\mbox{limsup}_{S\rightarrow \infty}\beta_{i,S}\left(\left(q^*_i\right)_{1:S},\left(r^*_i\right)_{1:K_S}\right)  &\le& \overline{\beta}_i,
\\\label{NOVA_satifies_cost_constraints}
\mbox{limsup}_{S\rightarrow \infty}p_{i,S}\left( \left(q^*_i\right)_{1:S} \right) &\le& \overline{p}_i.
\end{eqnarray}
\\(b) \emph{Optimality:} Let $S_{\epsilon}=\frac{ S}{\epsilon}$. Then, 
\begin{eqnarray*}
\lim_{S\rightarrow \infty} \lim_{\epsilon \rightarrow 0}
 \left( 
 \phi_{S_{\epsilon}}\left(\left( \left( q^*_i \left(\bm{\theta}_{i,s} , f_{i,s} \right)\right)_{\iiN} \right)_{1\le s \le S_{\epsilon}}  \right)
-\phi^{opt}_{S_{\epsilon}}\right)
\end{eqnarray*}
goes to zero in probability.
\end{theorem}
The above result tells us that the difference in performance 
of the \emph{online} algorithm NOVA (i.e., $\phi_{S_{\epsilon}}\left(\left( \mathbf{q}^* \right)_{1:S_{\epsilon}}\right)$ ) and that of the optimal \emph{offline} scheme 
goes to zero for long enough videos and small enough $\epsilon$.
Recall that $\phi^{opt}_{S_{\epsilon}}$ is the optimal value of the OPT$(S_{\epsilon})$, i.e., the performance of the optimal omniscient offline scheme
which knows all the allocation constraints $\left(c_k\right)_{k}$ and QR tradeoffs and segment lengths $\left(f_{i,s},l_{i,s}\right)_s $ ahead of time.
Note that although choosing small $\epsilon$ is beneficial for long videos, it can significantly affect the performance (initial transient and tracking ability) of NOVA
for short videos.

In the rest of this section, we discuss some useful properties of NOVA that will be used in Section \ref{section_proof_optimality_NOVA}.
We start with optimality conditions associated with solutions to RNOVA$\left(\mathbf{b},c\right)$ and QNOVA$_i(\bm{\theta}_i,f_i)$.
The optimization problem RNOVA$\left(\mathbf{b},c\right)$ is convex, and using Assumption-SF, 
we can show that it satisfies Slater's condition (see \cite{boyd_convex} for reference).
Thus, KKT conditions are necessary (and sufficient) for optimality.
Hence,
if $\mathbf{r}^*\left(\mathbf{b},c\right)$ is an optimal solution to RNOVA$\left(\mathbf{b},c\right)$,
there exist constants $\chi^*(c)$ and  $\left(\omega^*_i(c)\right)_{i\in{\cal N}}$ such that
for each $i\in{\cal N}$,
	\begin{eqnarray} 
  h^B_i\left(b_i\right) &=&\chi^*(c) c^{'}_i \left( \mathbf{r}^*\left(\mathbf{b},c\right) \right) + \omega^*_i(c),
\label{rvbar_gradient_zero_condtion}
\\ \chi^*(c) c\left( \mathbf{r}^*\left(\mathbf{b},c\right) \right) &=&0,
\label{rvbar_complimentary_slackness_for_capacity_constraint}
\\ \omega^*_i(c) \left(r^*_i\left(\mathbf{b},c\right) - r_{i,\min}\right) &=&0,
\label{rvbar_complimentary_slackness_for_resource_allocation_lower_bound} 
	\end{eqnarray} 
The optimization problem QNOVA$_i(\bm{\theta}_i,f_i)$ is also convex and satisfies Slater's condition (since the constraints are all linear),
and thus, KKT conditions are necessary (and sufficient) for optimality.
Thus, there exist constants $\gamma_i(\bm{\theta}_i,f_i)$
and $\overline{\gamma}_i(\bm{\theta}_i,f_i)$ such that
\begin{eqnarray} 
\nonumber\left(U^E_i\right)^{'}\left( \mu_i - U^V_i \left(v_i\right)\right) 
 \left( 1-  2\left(U^V_i\right)^{'} \left(v_i\right) \left(q^*_i \left(\bm{\theta}_i,f_i\right) - m_i\right)\right)
 + \gamma_i \left(\bm{\theta}_i,f_i\right) 
 \hspace{-0cm}&& 
 \\ \hspace{-1cm} - \overline{\gamma}_i \left(\bm{\theta}_i,f_i\right) - \frac{h^B_i\left(b_i\right)}{\left(1+\overline{\beta}_i\right)} 
 \left(f_{i}\right)^{'} \left(q^*_i \left(\bm{\theta}_i,f_i\right)\right) 
- p^d_i \frac{h^D_i\left(d_i\right)}{\overline{p}_i }  
\left(f_{i}\right)^{'} \left(q^*_i \left(\bm{\theta}_i\right),f_i\right)&=& 0,
\label{KKT_QNOVA_gradient_zero}
\\  \hspace{0cm} \gamma_i \left(\bm{\theta}_i,f_i\right) q^*_i \left(\bm{\theta}_i,f_i\right)&=& 0,
\label{KKT_QRNOVA_nonneg_constraints}
\\ \hspace{0cm} \overline{\gamma}_i \left(\bm{\theta}_i,f_i\right)\left(q^*_i \left(\bm{\theta}_i,f_i\right) - q_{\max}\right)&=&0.
\label{KKTQRNOVA_bounded_above_constraints}
\end{eqnarray}
	
	The next result states that the parameters in NOVA stay in a compact set.
\begin{lemma}\label{parameters_are_bounded}
For any initialization $\left(m_{i,0},\mu_{i,0},v_{i,0},b_{i,0},d_{i,0},\lambda_{i,0}\right)_{\iiN}
\in \prod_{\iiN} {\cal H}^{(i)}$, 
the parameters 
evolving according to NOVA satisfy the following:
for each $\iiN$, $s\ge 1$ and $k\ge 1$, we have $0\le m_{i,s},\mu_{i,s}\le q_{\max},
\ 0\le v_{i,s}\le q^2_{\max}$,
and $\ l_{\min}\le  \lambda_{i,s} \le l_{\max}$.
Further, $\underline{b}\le b_{i,k}\le \overline{b}$, $\underline{d}\le d_{i,s}\le \overline{d}$
for some finite constants $\overline{b}$ and $\overline{d}$ and for all $k$ and $s$ large enough.
\end{lemma}
\begin{IEEEproof}
It is easy to establish the result for the parameters
$m_{i,s},\mu_{i,s},v_{i,s}$ and $\lambda_{i,s}$ using the initialization of these parameters in NOVA 
and the boundedness of the quantities involved 
in the respective update rules.
For instance, we can use \eqref{m_update_NOVA}, \eqref{mu_update_NOVA}  and the fact that
 $0\le q^*_{i,s_i+1} \le q_{\max}$ 
 to obtain the result for $m_{i,.}$ and $\mu_{i,.}$.
 
Next, we show that there exists a finite $\overline{b}$ such that
$\underline{b}\le b_{i,k}\le \overline{b}$ for all $k$ large enough.
The lower bound is easy to show and holds for all $k$.
We establish the upper bound by showing the following property 
regarding the optimal solution to QNOVA$_i(\bm{\theta}_i,f_i)$:
for each $\iiN$, there is a finite constant $\overline{b}_i$ such that
$$
\max_{\left\{f_i\in{\cal F}_i\right\}} \frac{f_i \left( q^*_i \left(\bm{\theta}_i,f_i\right)\right)}{\left(1 + \overline{\beta}_i \right)} - \frac{r_{i,min}}{\tau_{slot}} 
\le  0.5 \left(\max_{\left\{f_i\in{\cal F}_i\right\}} \frac{f_i \left(0\right)}{\left(1 + \overline{\beta}_i \right)} - \frac{r_{i,min}}{\tau_{slot}} \right)
$$
for any $\bm{\theta}_i=(m_i,\mu_i,v_i,b_i,d_i)$ satisfying
$b_i\ge\overline{b}_i, \ 0\le m_i,\mu_i\le q_{\max},\ 0\le v_i\le q^2_{\max},\ d_i\ge \underline{d}$.
Note that $\max_{\left\{f_i\in{\cal F}_i\right\}} \frac{f_i \left(0\right)}{\left(1 + \overline{\beta}_i \right)} - \frac{r_{i,min}}{\tau_{slot}}<0$ from
Assumption-SF.
Hence, if this property holds, 
we can conclude that for large enough $b_{i,k}$, i.e.  $b_{i,k}\ge \overline{b} \defeq \max_{\iiN}\max \left( \overline{b}_i\right)$, 
the time required to download a segment is strictly less than $\left(1 + \overline{\beta}_i \right)$ times the duration 
of video associated with the segment.
Thus, $b_{i,k}$ is strictly decreasing when it is greater than $\overline{b}$.
Hence for any initialization $b_{i,k}\ge \underline{b}$,
we can show that $b_{i,k}\ge \overline{b}$ for large enough $k$.

We establish the above property next.
Using \eqref{KKT_QNOVA_gradient_zero} and the fact that
$f_i$ are convex increasing functions, we have
\begin{eqnarray*}
 \left(U^E_i\right)^{'}\left( \mu_i - U^V_i \left(v_i\right)\right) 
 \left( 1-  2\left(U^V_i\right)^{'} \left(v_i\right) q^*_i \left(\bm{\theta}_i,f_i\right)  \right)\hspace{0cm}
\ge \frac{h^B_i\left(b_i\right)}{\left(1+\overline{\beta}_i\right) } 
 \left(f_{i}\right)^{'} \left( q^*_i \left(\bm{\theta}_i,f_i\right) \right)
   -  \gamma_i \left(\bm{\theta}_i,f_i\right).
\end{eqnarray*}
Let 
\begin{eqnarray*}
\eta_1&=&\max_{\iiN} \max_{e_i\in [e_{\min,i},e_{\max,i}]}\left(1+\overline{\beta}_i\right) \left(U^E_i\right)^{'}\left( e_i \right),
\\\eta_2&=&\min_{\iiN} \min_{e_i\in [e_{\min,i},e_{\max,i}],v_i\in [0,q^2_{\max}]} 
2\left(1+\overline{\beta}_i\right) \left(U^E_i\right)^{'}\left( e_i \right) \left(U^V_i\right)^{'}\left( v_i \right).
\end{eqnarray*}
Recall that
$e_{\min,i}=-U^V_i(q^2_{\max})$ and $e_{\max,i}=q_{\max}-U^V_i(0)$.
Since, 
$\left(U^E_i\right)^{'}\left( . \right)$ and $\left(U^V_i\right)^{'}\left( . \right)$
are continuous,
$\left(U^V_i\right)'(0)>0$ and $\left(U^E_i\right)'(e_{\max,i})>0$,
$\eta_1$ is finite, and $\eta_2>0$.
 Hence, for any $\iiN$ and $\bm{\theta}_i$ 
 \begin{eqnarray*}
 \eta_1-  \eta_2 q^*_i \left(\bm{\theta}_i,f_i\right)  
 \ge h^B_i\left(b_i\right)
 \left(f_{i}\right)^{'} \left( q^*_i \left(\bm{\theta}_i,f_i\right) \right)
 -  \gamma_i \left(\bm{\theta}_i,f_i\right),
\end{eqnarray*}
Using the above inequality, and using the facts that $ \left(f_{i}\right)^{'} \left( q \right)>0$ for each $q>0$
and $\lim_{b\rightarrow \infty}h^B_i(b) =\infty$,
we can show that 
$\lim_{b\rightarrow \infty} q^*_i \left(\bm{\theta}_i,f_i\right) =0$.
Also, from Assumption-SF, $\max_{\left\{f_i\in{\cal F}_i\right\}} \frac{f_i \left(0\right)}{\left(1 + \overline{\beta}_i \right)} - \frac{r_{i,min}}{\tau_{slot}}<0$.
Now, (using continuity of the functions in ${\cal F}_i$ and finiteness of $\left|{\cal F}_i\right|$) we can conclude that there is some finite constant $\overline{b}_i$ such that
$\max_{\left\{f_i\in{\cal F}_i\right\}} \frac{f_i \left( q^*_i \left(\bm{\theta}_i,f_i\right)\right)}{\left(1 + \overline{\beta}_i \right)} - \frac{r_{i,min}}{\tau_{slot}} 
\le  0.5 \left(\max_{\left\{f_i\in{\cal F}_i\right\}} \frac{f_i \left(0\right)}{\left(1 + \overline{\beta}_i \right)} - \frac{r_{i,min}}{\tau_{slot}} \right)$
when $b_i\ge\overline{b}_i$.

The proof for $d_{i,s}$ can be completed using an approach similar to that given for $b_{i,k}$.
\end{IEEEproof}

For the next two results, let $\bm{\theta}_i=(m_i,\mu_i,v_i,b_i,d_i)$ where
$0\le m_i,\mu_i\le q_{\max},\ 0\le v_i\le q^2_{\max}$ and $b_i,d_i\in\mathbb{R}$.
The next result provides smoothness properties for the optimal solutions of RNOVA$\left(\mathbf{b},c\right)$ and QNOVA$_i(\bm{\theta}_i,f_i)$.
\begin{lemma}\label{continuity_of_solutions_to_QNOVA_and_RNOVA}
(a) For each $\iiN$ and $f_i\in {\cal F}_i$,
$q^*_i \left(\bm{\theta}_i,f_i\right) $ is a continuous function of $\bm{\theta}_i$.
\\(b) For each $c\in{\cal C}$, ${\cal R}^*\left(\mathbf{b},c\right)$ 
is a convex and compact set.
Further, ${\cal R}^*\left(\mathbf{b},c\right)$ is an upper semi-continuous set valued map of $\mathbf{b}$.
\\(c) For each $\ciC$ and $\mathbf{r}^*\left(\mathbf{b},c\right)\in {\cal R}^*\left(\mathbf{b},c\right)$, $\phi^R\left(\mathbf{r}^*\left(\mathbf{b},c\right),\mathbf{b}\right)$ is a continuous function of $\mathbf{b}$.
\end{lemma}
\begin{IEEEproof}
Part (a) follows from Theorem 2.2 in \cite{fiacco_sensitivity_analysis}
which provides sufficient conditions for verifying continuity of the optimal solution $q^*_i \left(\bm{\theta}_i,f_i\right) $.
We can verify that the conditions given in Theorem 2.2 are satisfied since
 $\phi^Q\left(q_i,\bm{\theta}_i,f_i\right)$ is continuous in $\left(q_i,\bm{\theta}_i\right)$,
QNOVA$_i(\bm{\theta}_i,f_i)$ has a unique solution,
and since the set of feasible solutions is a compact set.

Part (b) follows from Theorem 2.4 in \cite{fiacco_sensitivity_analysis}
which provides sufficient conditions for verifying continuity of the set of optimal solutions
${\cal R}^*\left(\mathbf{b},c\right)$ of RNOVA$\left(\mathbf{b},c\right)$..

Part (c) follows from Theorem 2.1 in \cite{fiacco_sensitivity_analysis}
which provides sufficient conditions for verifying continuity of the optimal value
$\phi^R\left(\mathbf{r}^*\left(\mathbf{b},c\right),\mathbf{b}\right)$ of RNOVA$\left(\mathbf{b},c\right)$.
\end{IEEEproof}

In the next result, we discuss concavity and differentiability properties of the 
optimal value of QNOVA$_i(\bm{\theta}_i,f_i)$, i.e., 
$\phi^Q\left(q^*_i \left(\bm{\theta}_i,f_i\right),\bm{\theta}_i,f_i\right)$.
\begin{lemma}\label{concavity_and_derivative_of_obj_of_QNOVA}
The following statements hold for each $\iiN$ and $f_i\in{\cal F}_i$.
\\(a) The optimal value of QNOVA$_i(\bm{\theta}_i,f_i)$, i.e., 
$\phi^Q\left(q^*_i \left(\bm{\theta}_i,f_i\right),\bm{\theta}_i,f_i\right)$,
is a strictly concave function of $m_i$
where $\bm{\theta}_i=(m_i,\mu_i,v_i,b_i,d_i)$.
\\(b) The partial derivative of $\phi^Q\left(q^*_i \left(\bm{\theta}_i,f_i\right),\bm{\theta}_i,f_i\right)$
with respect of $m_i$ is given by:
	\begin{eqnarray} 
 \frac{\partial \phi^Q\left(q^*_i \left(\bm{\theta}_i,f_i\right),\bm{\theta}_i,f_i\right)} {\partial m_i}
 =2\left(U^E_i\right)^{'}\left( \mu_i - U^V_i \left(v_i\right)\right)   \left(U^V_i\right)^{'} \left(v_i\right) 
 \left(q^*_i \left(\bm{\theta}_i,f_i\right) - m_i\right).
 \end{eqnarray} 
 \\(c) Let $\bm{\theta}^{(m)}_i=(m,\mu_i,v_i,b_i,d_i)$, i.e., 
 $\bm{\theta}_i$ with the first component set to $m$. 
 If $m \neq m_i$, the optimal value 
 of QNOVA$_i(\bm{\theta}^{(m)}_i,f_i)$ satisfies
	\begin{eqnarray} 
	 &&\phi^Q\left(q^*_i \left(\bm{\theta}^{(m)}_i,f_i\right),\bm{\theta}^{(m)}_i,f_i\right)< \phi^Q\left(q^*_i \left(\bm{\theta}_i,f_i\right),\bm{\theta}_i,f_i\right)
	 \\\nonumber && \hspace{2.5cm}
 +2 \left(m - m_i\right)  \left(U^E_i\right)^{'}\left( \mu_i - U^V_i \left(v_i\right)\right)   \left(U^V_i\right)^{'} \left(v_i\right) \left(q^*_i \left(\bm{\theta}_i,f_i\right) - m_i\right).
  \end{eqnarray}  
\end{lemma}
\begin{IEEEproof}
Part (a) follows from Proposition 2.8 from \cite{fiacco_convexity}
which provides sufficient conditions for verifying strict concavity of the optimal value of an optimization problem
with respect to parameters associated with the problem.

Part (b) follows from Theorem 4.1 related to sensitivity analysis of optimal value function given in \cite{shapiro_sensitivity_analysis_paper},
and the remark following the theorem.
Part (c) follows from strict concavity in part (a) and using the expression for the partial derivative in part (b).
\end{IEEEproof}

	\section{Performance evaluation of NOVA via simulation}
\label{section_simulations}

In this section, we carry out an evaluation of NOVA using Matlab simulations to compare the performance of a wireless network operating under NOVA
vs one using Proportionally Fair (PF) network resource allocation (see \cite{kushner_prop_fair_conv}) 
and quality adaptation based on Rate Matching (RM). 
We discuss PF and RM in detail below. 
The main objective of this section is to use the simulation results to answer questions like:
\begin{itemize}
	\item What are the typical gains under NOVA? In particular, we are interested in the following questions:
\begin{itemize}
	\item What are the typical capacity gains (defined later)? 
	\item How does NOVA perform in terms of rebuffering?
	\item How do price constraints affect the gains?
	\item Does NOVA penalize mean quality by too much to reduce variability in quality?
	\item Is NOVA fair?
\end{itemize}
	\item What is the loss in the performance of NOVA (and for that matter, any good adaptation algorithm utilizing information about
	QR tradeoffs) in the absence of
	accurate QR tradeoffs? In particular, we are interested in the following questions:
\begin{itemize}
	\item What is the reduction in performance if we have access only to partially accurate QR tradeoffs, 
	e.g., those based on less sophisticated video quality assessment metrics like PSNR or
	if we just know `averaged' QR tradeoffs?
	\item What is the reduction in performance if we do not have any information about QR tradeoffs?
\end{itemize}
\end{itemize}

\subsection{Simulation setting}

We consider a wireless network with $\tau_{slot}=10$ msecs,
and with allocation constraints of the form 
$c_k\left( \mathbf{r}_k \right) = \sum_{\iiN} \frac{r_{i,k}}{p_{i,k}}-1$ in each slot $k$,
where $p_{i,k}$ denotes the peak resource allocation for video client $i$ in slot $k$,
i.e., if we only allocate resources to video client $i$ in slot $k$, then $r_{i,k}=p_{i,k}$ is the maximum resource allocation to the video client 
that does not violate the allocation constraint in the slot.
To obtain traces of peak resource allocation for the video clients, we generated 300 sequences of length 150000 each using Markov Chain Monte Carlo method,
in such a way that the values in consecutive slots are positively correlated (the positive correlation reflects the correlation of the wireless channel in adjacent slots)
and the marginal distribution of the stationary process is that of an appropriately scaled version of the sequence is equal to a distribution 
which is representative of capacities seen by a randomly placed wireless user with 
single antenna equalizer in an HSDPA system with 50\% load (and thus associated interference) from its neighbors
\footnote{This data was provided by a service provider and is based on a simulation framework for such a system.}.

Unless mentioned otherwise, in our simulations, we consider settings with heterogeneous channels: 
we uniformly and at random pick a sequence for each video client from the 300 sequences,
scale the sequence by a uniformly distributed random number in the range $[0.5,1.5]$, and use the scaled sequence
as the peak resource allocation seen by the video clients over 15000 slots.
Thus, a video client with random scaling close to 0.5  
sees the worst wireless channel on average,
whereas one with random scaling close to 1.5 sees the best.
We also present simulation results for a setting with homogeneous channels later in the section.
In the setting with homogeneous channels, we uniformly at random pick a sequence for each video client from the 300 sequences 
and just use the sequence (without any additional scaling)
as the peak resource allocation seen by the video clients.

Under PF, network resource allocation in slot $k$ is an optimal solution to 
	\begin{eqnarray} \label{PF_obj}
  \max_{\mathbf{r}}	 
  \left\{  \sum_{i\in {\cal N}} \frac{r_i}{\rho_{i,k}}: \ c_k\left( \mathbf{r} \right)\le 0,\ r_i \ge r_{i,\min} \ \forall \ i\in{\cal N}\right\}
	\end{eqnarray}
	where the parameters $\left(\rho_{i,k}\right)_{\iiN}$ track the mean resource allocation to the video clients, and are updated using \eqref{rho_NOVA_update}
	with $\epsilon$ set to 0.01
	(this is a good choice since $\rho_{i,k}$ is getting updated at a high rate of once every $\tau_{slots}=10$ msecs).

	In our simulations, we consider video clients downloading different parts of three open source movies
	Oceania (about 55 mins long), Route 66 (about 100 mins long) and Valkaama (about 90 mins long).
	The movies Oceania and Valkaama are compressed at rates 0.1, 0.2, 0.3, 0.6, 0.9 and 1.5 Mbps,
	with segments of duration 1 second each (hence, each segment is available in six representations).
	The movie Route 66 is compressed at rates 0.1, 0.2, 0.3, 0.6 and 0.9 Mbps,
	with segments of duration 1 second each (hence, each segment is available in five representations).
	Unless mentioned otherwise, in each simulation, video clients pick a movie and starting segment (index) for the movie at random, and start downloading the rest of the movie from that segment onwards. A video client on reaching the last segment of a movie continues viewing the movie from the first segment.
	We measure STQ of a representation using a proxy for DMOS (Differential Mean Opinion Score) score (see \cite{wirelessVQA}
	for a discussion on DMOS) associated and mapping: STQ=100-DMOS.
	We chose this mapping as it roughly maps the proxy DMOS scores to the range $[0,100]$,
	and an increase in STQ (unlike that for DMOS) corresponds to an improvement in quality.
	The proxy DMOS score for a representation is 
	obtained from the value of the video quality assessment metric MSSSIM-Y (see \cite{MSSSIM}) associated with the representation,
	by using the following mapping (obtained from the model used in \cite{rajiv_dmos_from_msssim_map_ref}): $\mbox{DMOS}= 13.6056\times\log(1+(1-\mbox{MSSSIM-Y})/0.0006)$.
	The MSSIM-Y value of each segment was obtained as the average MSSIM-Y of the constituent frames obtained using the code given in \cite{LIVE_software_Releases}.
	To summarize, the QR tradeoffs used in our simulations map STQ values (obtained essentially from MS-SSIM metric) 
	for five/six representations (each one second long)
	to its associated compression rate. Recall that this compression rate also accounts for
	the size of overheads due to metadata.
	The diversity of QR tradeoffs associated with these movies is illustrated in Figs.~\ref{variety_route}-\ref{variety_valkaama}.

\ifQRtradeoffsvarietyplotinclude
	\begin{figure}[!ht]
	\centering
	\ifarxivmode	
	\includegraphics[scale=.5]{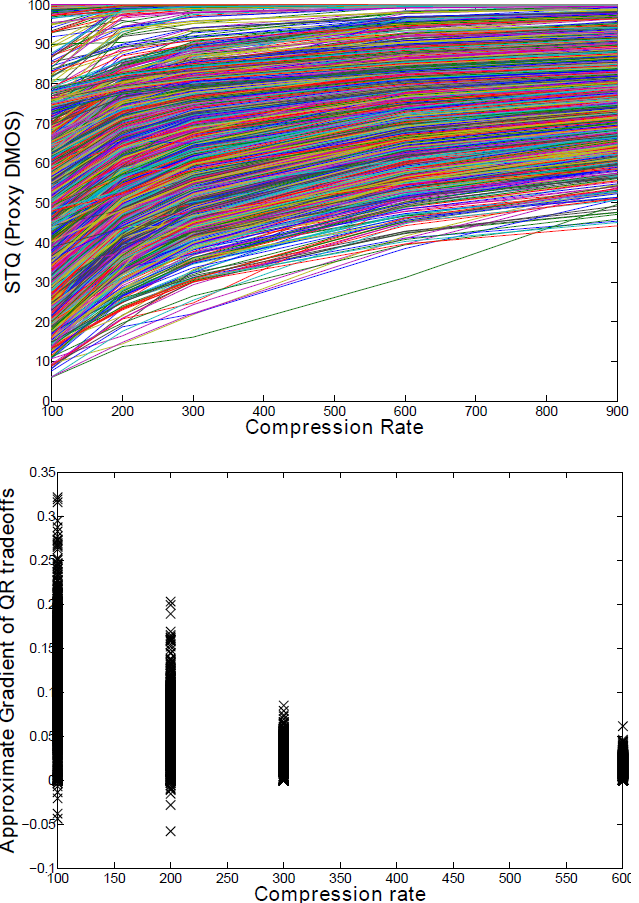}
	\else	\includegraphics[scale=.5]{../../../papers/video_varaiability_buffer_and_cost_aware/simulations/thesis_simlns/plot_QR_variety/route_lowdef.png}
	\fi
	\caption{Diversity in QR tradeoffs of movie Route 66.}
	\label{variety_route}
\end{figure}

\begin{figure}[!ht]
	\centering
	\ifarxivmode	\includegraphics[scale=.5]{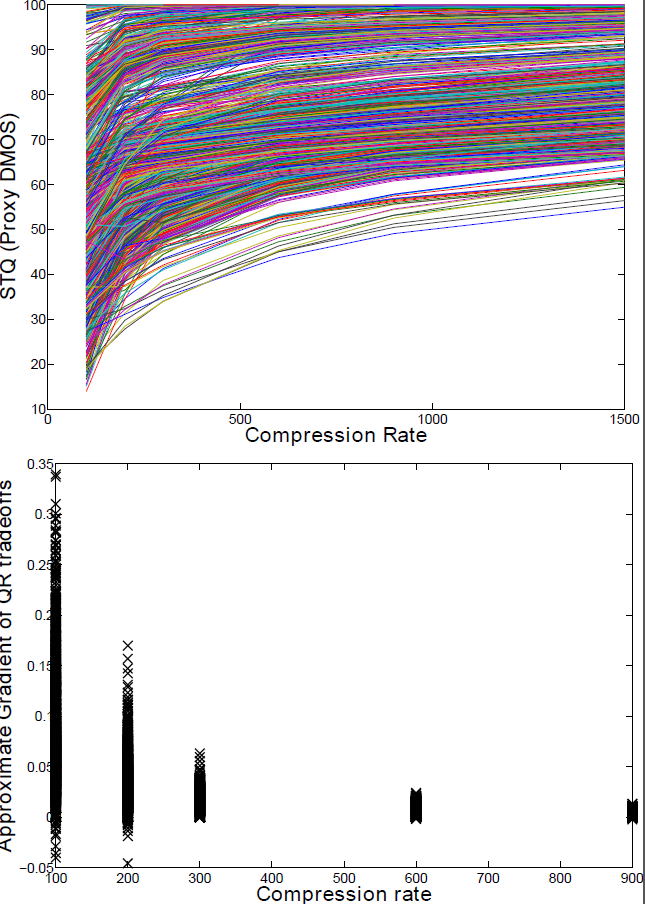}
	\else	\includegraphics[scale=.5]{../../../papers/video_varaiability_buffer_and_cost_aware/simulations/thesis_simlns/plot_QR_variety/oceania_lowdef.png}
	\fi
	\caption{Diversity in QR tradeoffs of movie Oceania.}
	\label{variety_oceania}
\end{figure}

\begin{figure}[!ht]
	\centering
	\ifarxivmode	
	\includegraphics[scale=.5]{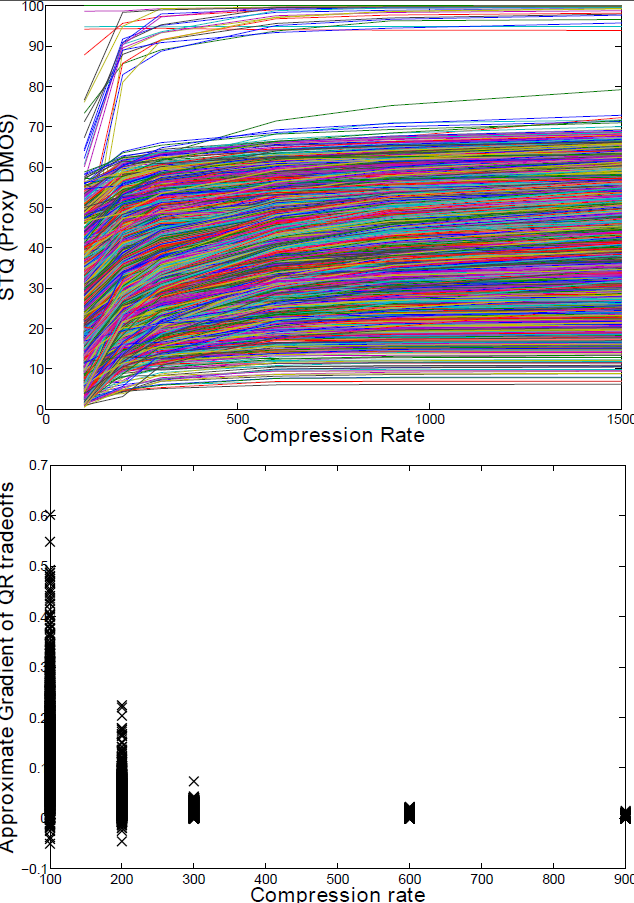}
	\else	\includegraphics[scale=.5]{../../../papers/video_varaiability_buffer_and_cost_aware/simulations/thesis_simlns/plot_QR_variety/valkaama_lowdef.png}
	\fi
	\caption{Diversity in QR tradeoffs of movie Valkaama.}
	\label{variety_valkaama}
\end{figure}

\else \fi

	Till now, we assumed that we have a continuous set of quality choices for each segment.
	However, in practice, video segments are only available in a \emph{finite} number of representations,
	and this is also the case with QR tradeoffs we obtained for the three movies.
	Thus for our simulations, we modify the optimization problem QNOVA$_i(\bm{\theta}_i,f_i)$,
	used in quality adaptation of NOVA, by
imposing an additional restriction that the quality for segment $s$ of video client $i$ 
is picked from the finite set
${\cal Q}_{i,s}$ of available quality choices associated with the segment.
We discuss this modification in more detail in Subsection \ref{NOVA_with_finite_number_of_representations}.

	In quality adaptation based on RM (Rate Matching), 
	the video client essentially tries to `match' the compression rate of the selected representation
	to (current estimate of) mean resource allocation in bits per second,
	and further modifies the selection to respond to the state of the playback buffer.
	This is basic feature in many compression rate adaptation algorithms,
	for instance, see \cite{Akhshabi_paper} where (following their terminology) we see that `requested bitrate' (i.e., size of the representation) 
	stays close to the
	`average throughput' (i.e., $\rho_{i,k}$ in our setting) in Microsoft Smooth Streaming player and Netflix player.
	For each video client $\iiN$, the variables $I^{cautious}_i(k)$ and $I^{aggressive}_i(k)$
	are used to enable RM to respond to low and high playback buffer respectively.
	The variable $I^{cautious}_i(k)$ is set to one if in slot $k$, the playback buffer has video content 
	of duration less than 10 seconds and is set to zero if  the playback buffer has video content 
	of duration greater than 15 seconds.
	The variable $I^{aggressive}_i(k)$ is set to one if in slot $k$, the playback buffer has video content 
	of duration greater than 30 seconds and is set to zero if  the playback buffer has video content 
	of duration less than 25 seconds.
	The quality adaptation in RM works as follows:
	if any video client $i\in{\cal N}$ finishes download of $(s-1)$ th segment in slot $k$,
	we first find the representation 
	with quality equal to
	$\mbox{argmax}_{q_i} \left\{q_i\in{\cal Q}_{i,s}:\ f_{i,s}(q_i) \le 0.99 \rho_{i,k}, \ p^d_if_{i,s}(q_i) \le \overline{p}_i \right\} $,
	and let $M^{RM,0}_{i,s}$ denote the index of this representation.
	We denote the index of the representation picked by RM for segment $s$ of video client $i$ by $M^{RM}_{i,s}$
	which is given by
	\begin{eqnarray*}
	M^{RM}_{i,s} = 
	\begin{cases} \max \left(\min \left(M^{RM,0}_{i,s} + I^{aggressive}_i(k)-I^{cautious}_i(k),M^{RM}_{i,s,max}\right),1\right), 
	\\1, \mbox{ if playback buffer has video content 
	duration less than 5 secs,}
	\end{cases}
	\end{eqnarray*}
	where $\overline{M}^{RM}_{i,s}$ is the number of representations available for segment $s$ of video client $i$.
	Hence, RM picks a lower representation if $I^{cautious}_i(k)=1$ (i.e., when playback buffer is low),
	picks a higher representation if $I^{aggressive}_i(k)=1$  (i.e., when playback buffer is high),
	and picks the lowest representation when the risk of rebuffering is high.
	Thus, RM meets the price constraint by ensuring that it is met for each segment.
	While considering price constraints in our simulations,	
	we let $p^d_i=0.01$ dollars per bit,
	and explicitly indicate the price constraints by referring to RM as RM$(\overline{p}_i)$ when we have cost constraint $\overline{p}_i$ for video client $i$.
	
.

	 For our simulations of NOVA, we let $\epsilon=0.05$ and $r_{i,min}=0.001$ bits.
	 We set $U^E_i(e)= e_i$ and $U^V_i(v)=0.05 v_i$ for each $\iiN$,
	 and hence we only have to track the parameters $\left(m_{i,.},b_{i,.},d_{i,.}\right)$
	 in this implementation of NOVA.
	 We let $\overline{\beta}_i=0$ for each $\iiN$,
	 and consider settings with two types of price constraints: in the first setting there are no price constraints,
	 and in the second, each user $\iiN$ has a price constraint of $\overline{p}_i=3$ dollars per second.
	 While evaluating the rebuffering time in the simulation results, we allow for a startup delay of 3 secs
	 (which does not count towards rebuffering time).	
	  For each $\iiN$, we chose $h^D_i(d_i)=10 d_i$ and
		\begin{eqnarray}\label{hb_used_for_simulations}
		h^B_i(b_i)= h_{i,0} \left( \frac{b_i}{0.05} +  \max \left( \frac{b_i-20}{0.05},0\right)^2\right),
		\end{eqnarray}
	 with $h_{i,0}=0.005$ (see Subsection \ref{choiceof_epsilon_etc} for a discussion about this choice of $h^B_i(.)$).
	 In all our simulations, we use the following initialization of NOVA parameters for each $\iiN$: $m_{i,0}=25$,
	 $b_{i,0}=\frac{40}{0.05}$ and $d_{i,0}=1$.
	 Note that these initializations are used in all simulations ranging from lightly loaded (e.g. $N=12$) to heavily loaded networks (e.g. $N=33$),
	 and for users seeing very good wireless channels to very bad wireless channels.
	 Given the challenge of operating in these diverse settings, 
	 we enable NOVA to quickly `learn' the setting 
	 by starting with larger values of $\epsilon$ for a few slots and segments initially 
	 and we keep reducing it until it reaches 0.1.
	 
	 Each point in the plots discussed below is obtained by running the associated algorithm in 50 times
	where each simulation is run until all the users have downloaded a video of duration at least 10 minutes (i.e., 600 segments).
	Each point corresponds to a fixed number $N$ of video clients in the network,
	and we vary $N$ over the set $\left\{12,\ 15,\ 18,\ 21,\  24,\ 27, \ 30,\ 33\right\}$.
	We refer to the combination of PF resource allocation and RM quality adaptation as PF-RM.
	To study the effectiveness of the quality adaptation in NOVA,
	we also study the performance of PF-QNOVA obtained by using PF resource allocation
	and the quality adaptation in NOVA.
	We refer to the modification of NOVA, PF-QNOVA and PF-RM with price constraint of 3 dollars per bit
	using the phrases NOVA(3), PF-QNOVA(3) and PF-RM(3) respectively.
	While implementing NOVA(3) and PF-QNOVA(3) with price constraint of 3 dollars per bit,
	we used a more stringent price constraint of $0.95\times 3$ to ensure that the constraint is met for short videos 
	(note that Theorem \ref{main_optimality_theorem} guarantees that the constraint will be met for long enough videos
	without any additional tightening of the constraint).

	\subsection{Simulation results}
	 

\begin{figure}[ht]
	\centering
	\ifarxivmode	\includegraphics[scale=.42]{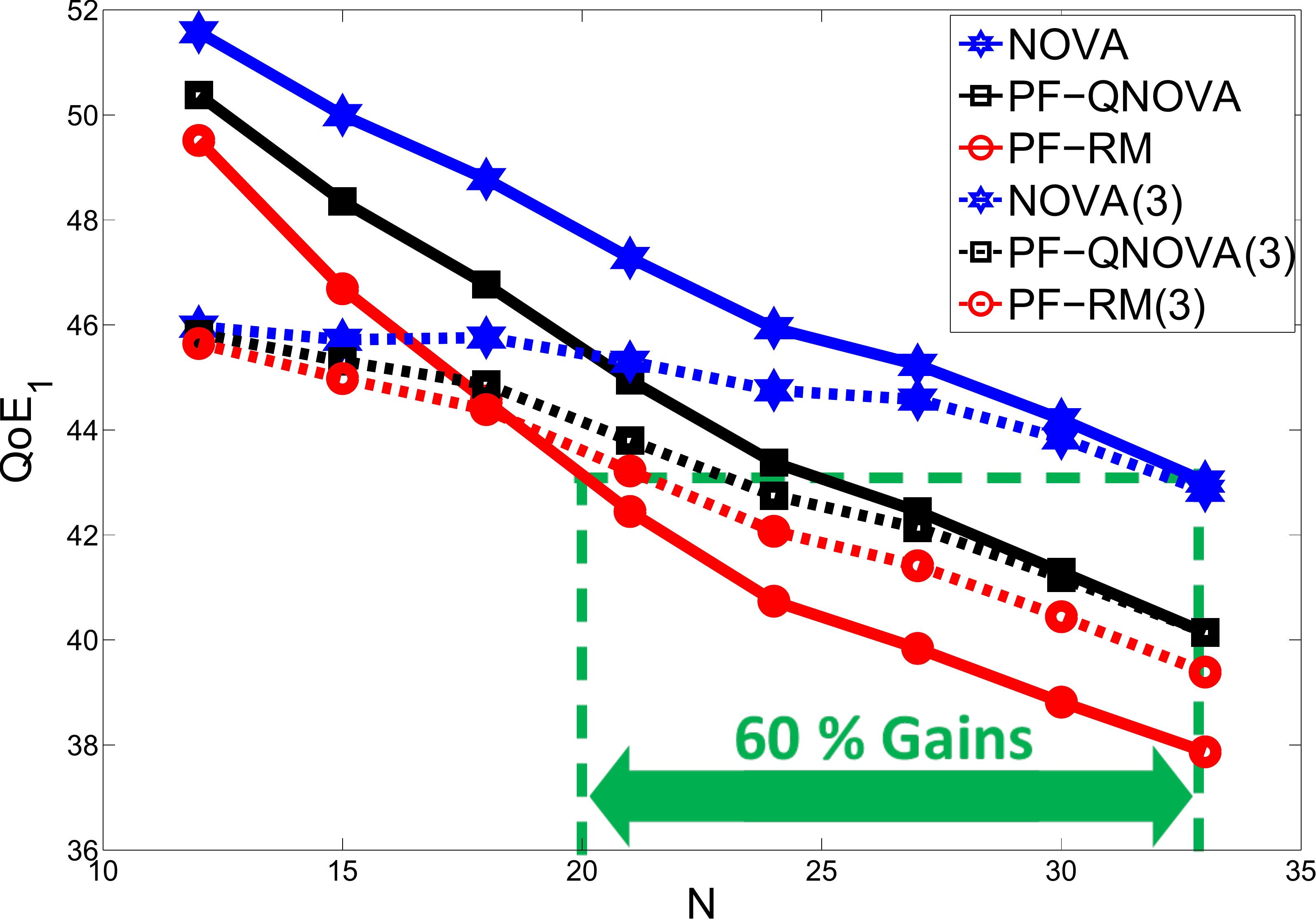}
	\else	\includegraphics[scale=.42]{../../../papers/video_varaiability_buffer_and_cost_aware/simulations/thesis_simlns/main_het_network/qoe1.pdf}
	\fi
	\caption{QoE$_1$ gains from NOVA.}
	\label{qoe1_gains}
\end{figure}

In Fig.~\ref{qoe1_gains}, we compare the QoE of the video clients under different algorithms,
where we measure QoE using the metric QoE$_1$ which is the average across simulation runs of 
\begin{eqnarray*}
\frac{1}{\left|{\cal  N}\right|} \sum_{\iiN} 
  \left( m^{600}_i \left(q_i\right)	- \sqrt {\mbox{Var}^{600}_i \left(q_i\right)}\right),
\end{eqnarray*}
where $m^{600}_i \left(q_i\right)	- \sqrt {\mbox{Var}^{600}_i \left(q_i\right)}$ 
is the metric proposed in \cite{evaluation_of_temporal_variation_of_video_quality}
with the scaling constant for $ \sqrt {\mbox{Var}^{600}_i \left(q_i\right)}$ set to unity
(and $m^{600}_i \left(q_i\right)$ and $\mbox{Var}^{600}_i \left(q_i\right)$ are defined in 
\eqref{mean_expression} and \eqref{variance_expression}).

On comparing QoE$_1$ using Fig.~\ref{qoe1_gains}, we see that NOVA performs much better than PF-RM and PF-QNOVA,
and in fact provides `network capacity gains' of about 60\% over PF-RM,
i.e., 
given a requirement on (user) average QoE$_1$, we can support about 60\% more video clients by using NOVA than that under PF-RM.
For instance, if we consider the horizontal dashed line in Fig.~\ref{qoe1_gains} that corresponds to an average QoE$_1$
requirement of about 43, we see that PF-RM can only support 20 video clients while meeting this requirement
whereas NOVA can support almost 33 video clients.
Under price constraint (of 3 dollars per second) also, we see that NOVA(3) provides network capacity gains of about 60\% over PF-RM(3).

The gain from the adaptation component of NOVA is also visible in Fig.~\ref{qoe1_gains},
where
we see that PF-QNOVA provides network capacity gains of about 25\% 
over PF-RM respectively.

\begin{figure}[ht]
	\centering
	\ifarxivmode	\includegraphics[scale=.42]{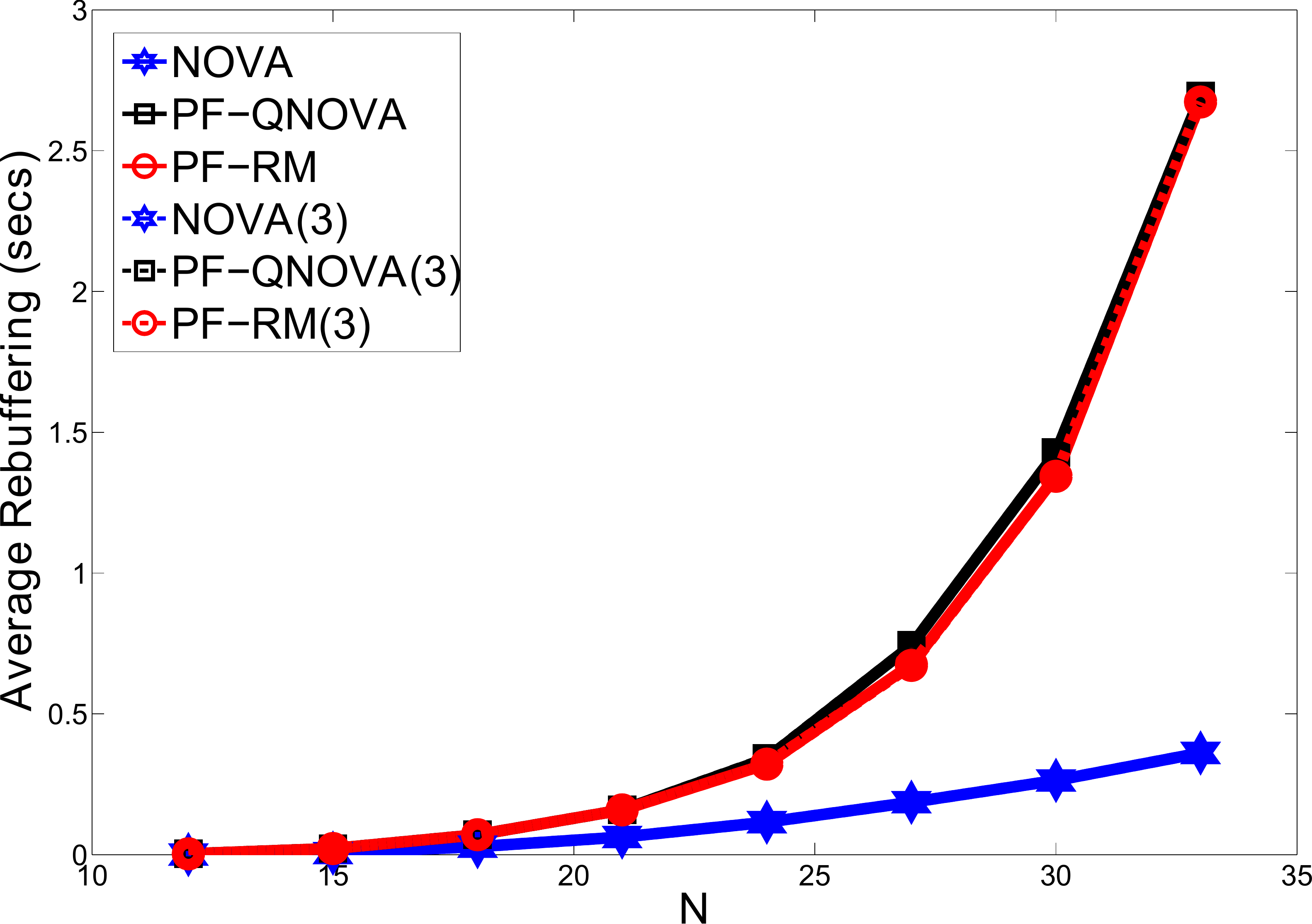}
	\else	\includegraphics[scale=.42]{../../../papers/video_varaiability_buffer_and_cost_aware/simulations/thesis_simlns/main_het_network/rebuf.pdf}
	\fi
	\caption{Reduction in rebuffering using NOVA.}
	\label{rebuf_reduction}
\end{figure}

The results in Fig.~\ref{rebuf_reduction} depict the significant reduction in the amount of time spent rebuffering under NOVA and NOVA(3).
Using Figs \ref{qoe1_gains}-\ref{rebuf_reduction}, we see that NOVA outperforms 
PF-RM in both the metric QoE$_1$ and the amount of time spent rebuffering which cover some of the most important factors affecting users' QoE 
(see the discussion in Section \ref{introduction_section}).

\begin{figure}[ht]
	\centering
	\ifarxivmode
\includegraphics[scale=.42]{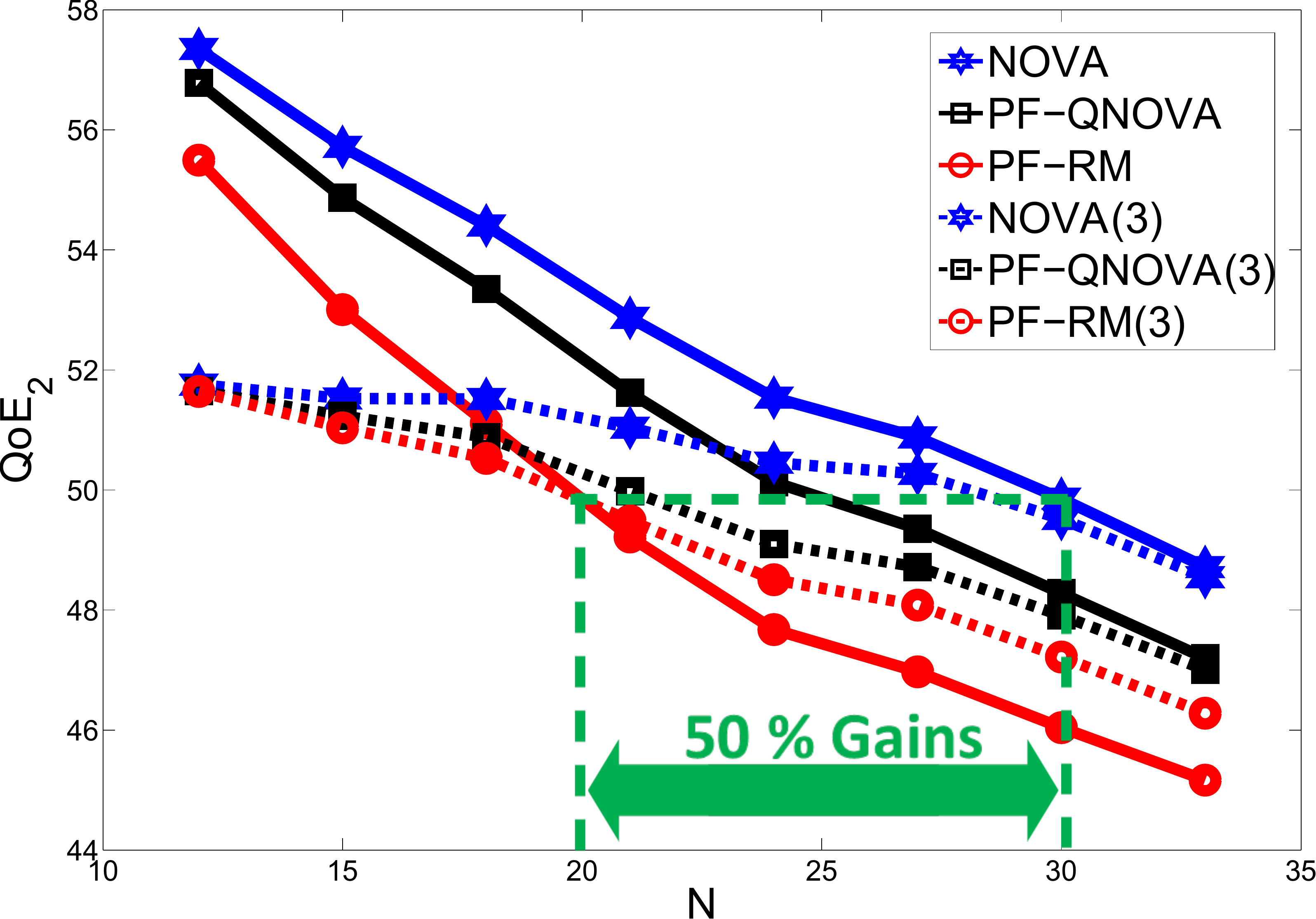}
\else
\includegraphics[scale=.42]{../../../papers/video_varaiability_buffer_and_cost_aware/simulations/thesis_simlns/main_het_network/qoe2.pdf}
\fi
	\caption{QoE$_2$ gains from NOVA.}
	\label{qoe2_gains}
\end{figure}

In Fig.~\ref{qoe2_gains}, we compare the performance of different algorithms
using another metric QoE$_2$ which is the average across simulation runs of 
\begin{eqnarray*}
\frac{1}{\left|{\cal  N}\right|}
  \sum_{\iiN}\left( m^{600}_i \left(q_i\right)	- \sqrt {\mbox{MSD}^{600}_i \left(q_i\right)}\right),
\end{eqnarray*}
where 
$$
\mbox{MSD}^{600}_i \left(q_i\right) \defeq \frac{1}{600} \sum_{s=1}^{600} \left( q_{i,s+1}- q_{i,s} \right)^2 .
$$
Note that this metric is similar to the metric QoE$_1$,
except that it penalizes short term variability in quality (i.e., variability across consecutive segments).
From Fig.~\ref{qoe2_gains}, we see that NOVA provides gains similar to those in the case of the metric QoE$_1$ (in Fig.~\ref{qoe1_gains}),
By comparing QoE$_2$, we see that NOVA and NOVA(3) provide network capacity gains of about 50\% over PF-RM and PF-RM(3)
respectively.

\begin{figure}[ht]
	\centering
	\ifarxivmode
\includegraphics[scale=.42]{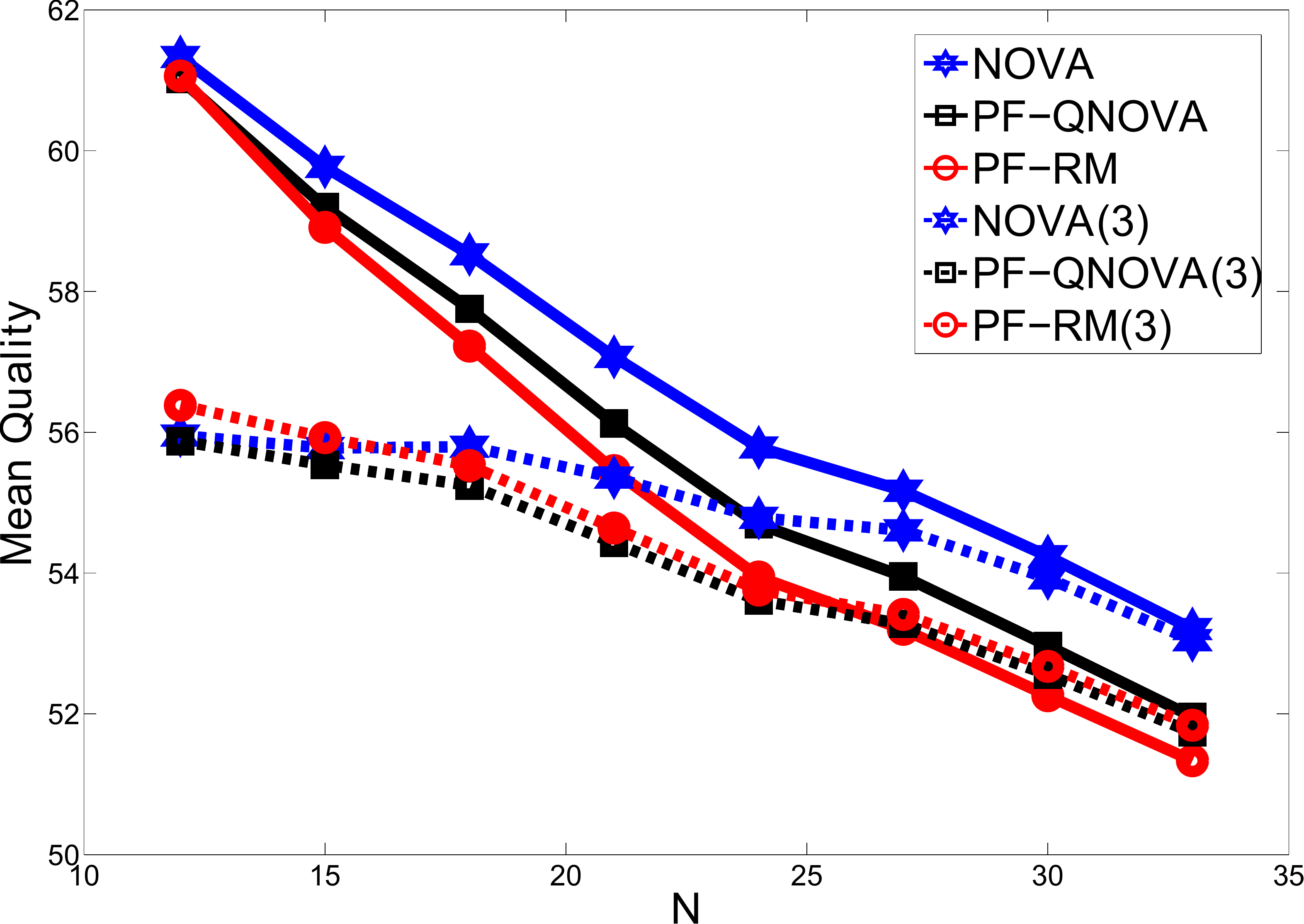}
\else
\includegraphics[scale=.42]{../../../papers/video_varaiability_buffer_and_cost_aware/simulations/thesis_simlns/main_het_network/mean_quality.pdf}
\fi
	\caption{Mean quality gains from NOVA.}
	\label{mean_quality_gains}
\end{figure}

The results in Fig.~\ref{mean_quality_gains} show that the improvement in QoE$_1$ and QoE$_2$ under NOVA 
does not come at the cost of significant reduction in mean quality.
In fact, the results suggest that NOVA has better mean quality (in addition to lower variability in quality)
in all but lightly loaded networks (i.e., $N=12$).
Also note that we can further increase the mean quality under NOVA (at the cost higher variability in quality)
if we scale down the functions $\left(U^V_i\right)_{\iiN}$.

\begin{figure}[ht]
	\centering
		\ifarxivmode
\includegraphics[scale=.42]{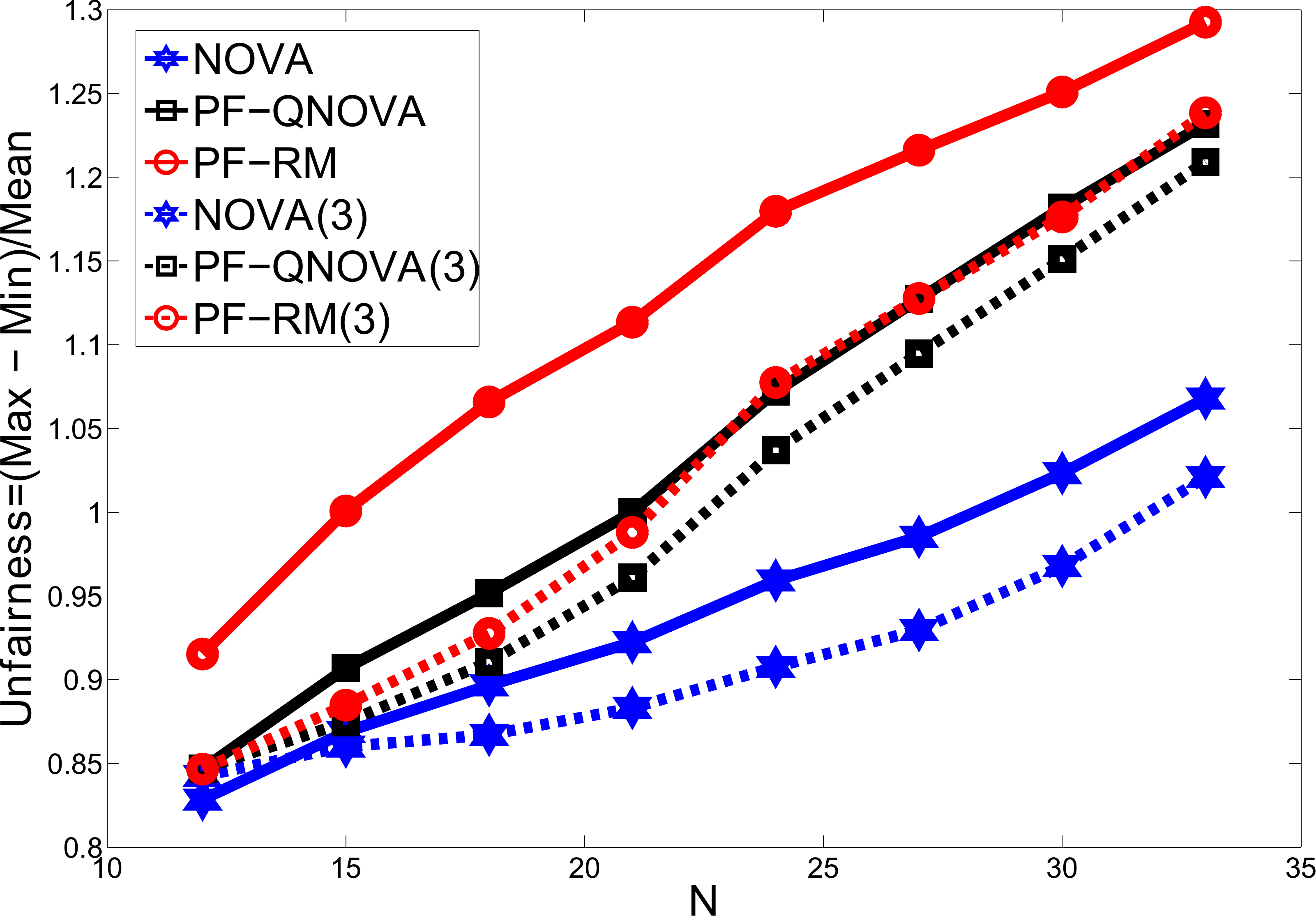}
\else
\includegraphics[scale=.42]{../../../papers/video_varaiability_buffer_and_cost_aware/simulations/thesis_simlns/main_het_network/unfairness.pdf}
\fi
	\caption{Fairness gains using NOVA.}
	\label{fairness_gains}
\end{figure}

\begin{figure}[ht]
	\centering
		\ifarxivmode
\includegraphics[scale=.42]{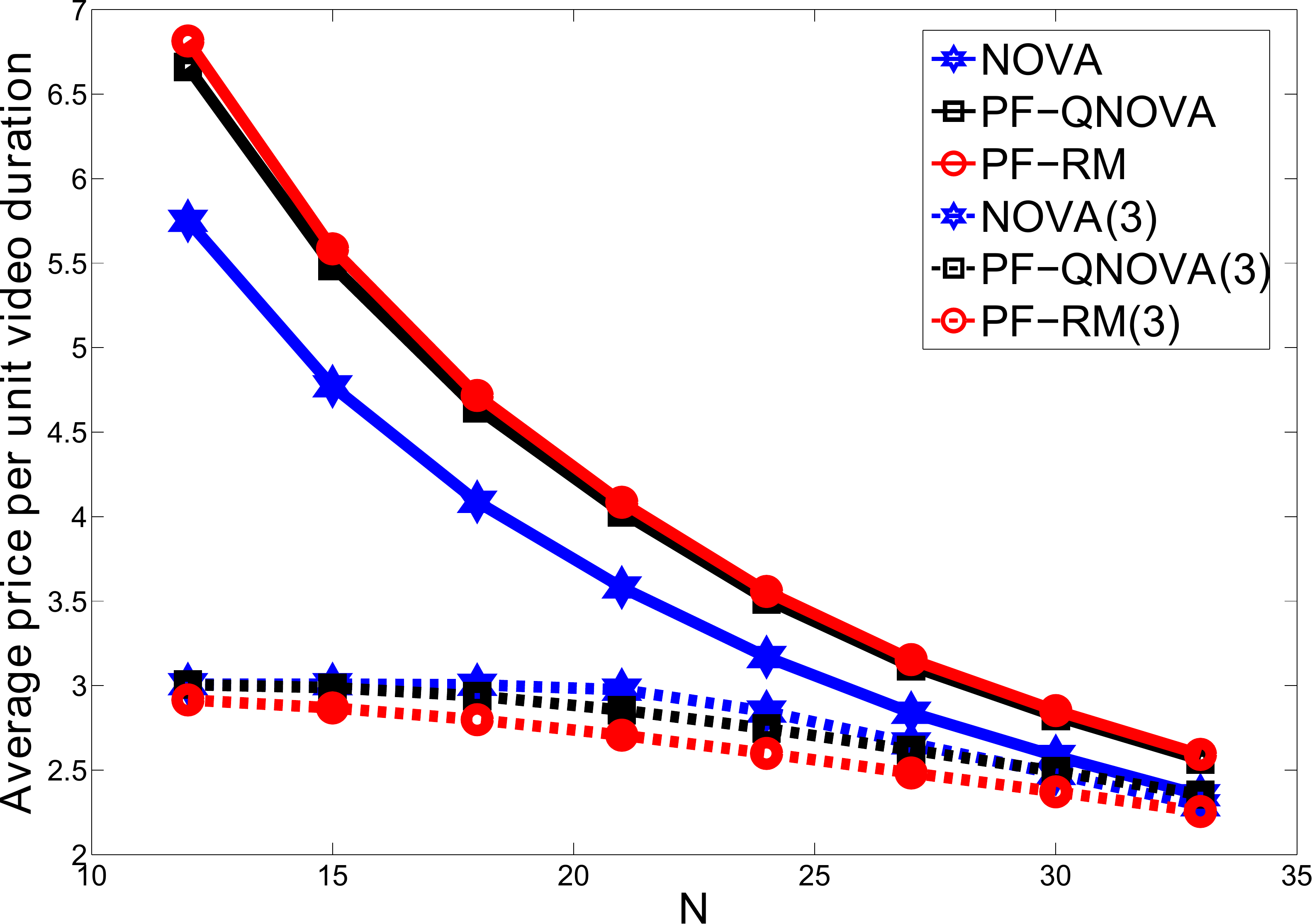}
\else
\includegraphics[scale=.42]{../../../papers/video_varaiability_buffer_and_cost_aware/simulations/thesis_simlns/main_het_network/price.pdf}
\fi
	\caption{NOVA meets cost constraints.}
	\label{cost_constraints_met}
\end{figure}

The results in Fig.~\ref{fairness_gains} indicate that, when compared to PF-RM, NOVA is more fair in QoE$_1$ delivered to the video clients.
Here, we measure fairness as (the average across simulations of) the ratio 
of the difference between maximum and minimum of QoE$_1$ across users to the mean (across users of) QoE$_1$.
Although we chose $\left(U^E_i\right)_{\iiN}$ to be linear functions,
the fairness associated with NOVA in these results can be attributed to the concavity of inverse of QR tradeoffs (i.e., convexity of QR tradeoffs)
and the structure of the objective function (see \eqref{defn_phiQ}) of the optimization problem associated with quality adaptation.
Further, from Fig.~\ref{cost_constraints_met}, we see that NOVA(3) meets cost constraints (of 3 dollars per second).

\begin{figure}[ht]
	\centering
		\ifarxivmode
\includegraphics[scale=.29]{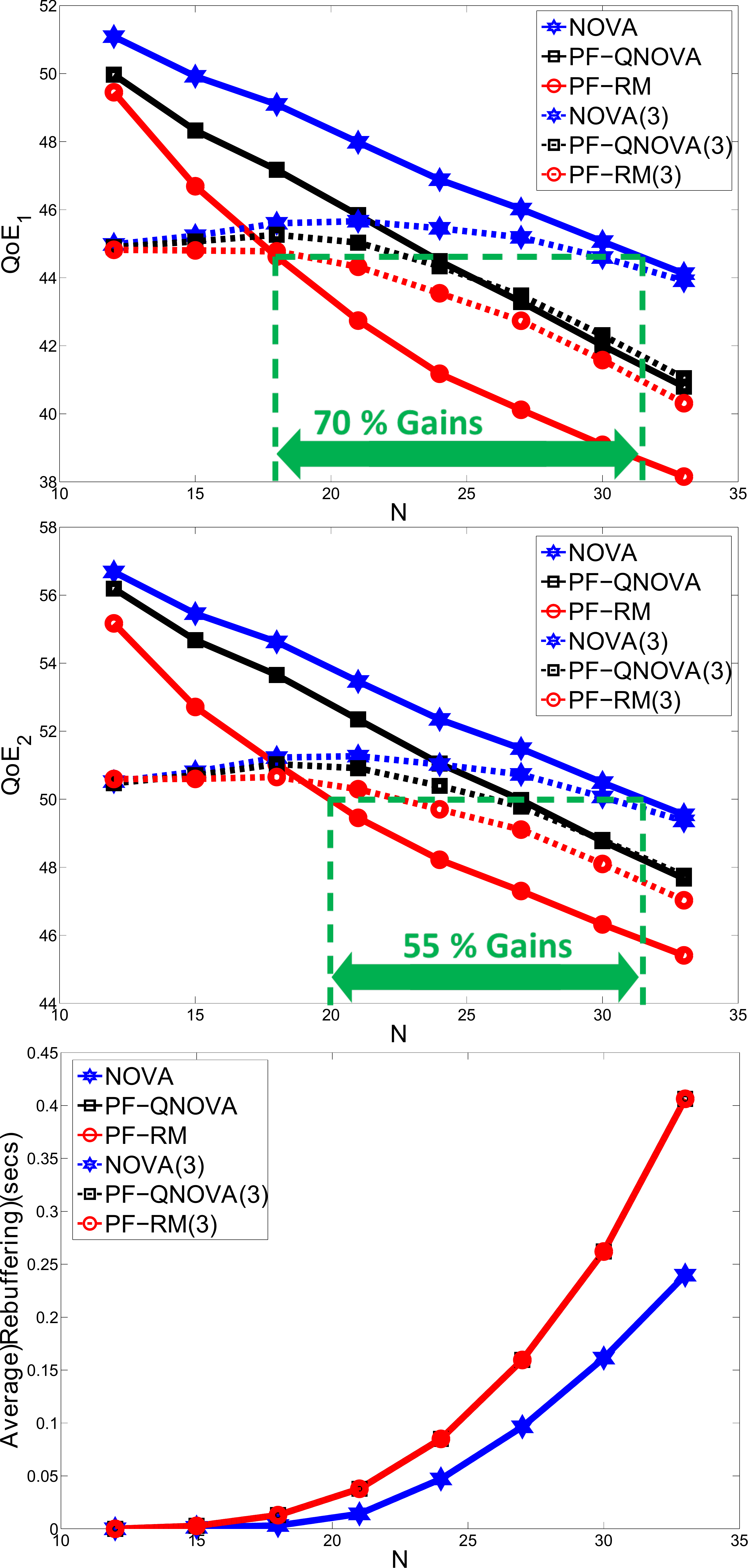}
\else
\includegraphics[scale=.29]{../../../papers/video_varaiability_buffer_and_cost_aware/simulations/thesis_simlns/main_hom_network/combined.pdf}
\fi
	\caption{Performance gains using NOVA: Homogeneous channels.}
	\label{perf_gains_homog}
\end{figure}

We have depicted the results obtained using simulations for the setting with homogeneous channels (see the discussion about the setting in the beginning of this section)
in Fig.~\ref{perf_gains_homog}.
We see that the performance gains under homogeneous channels are slightly higher than those in the case of heteregeneous ones.

\begin{figure}[ht]
	\centering
	\ifarxivmode
\includegraphics[scale=.42]{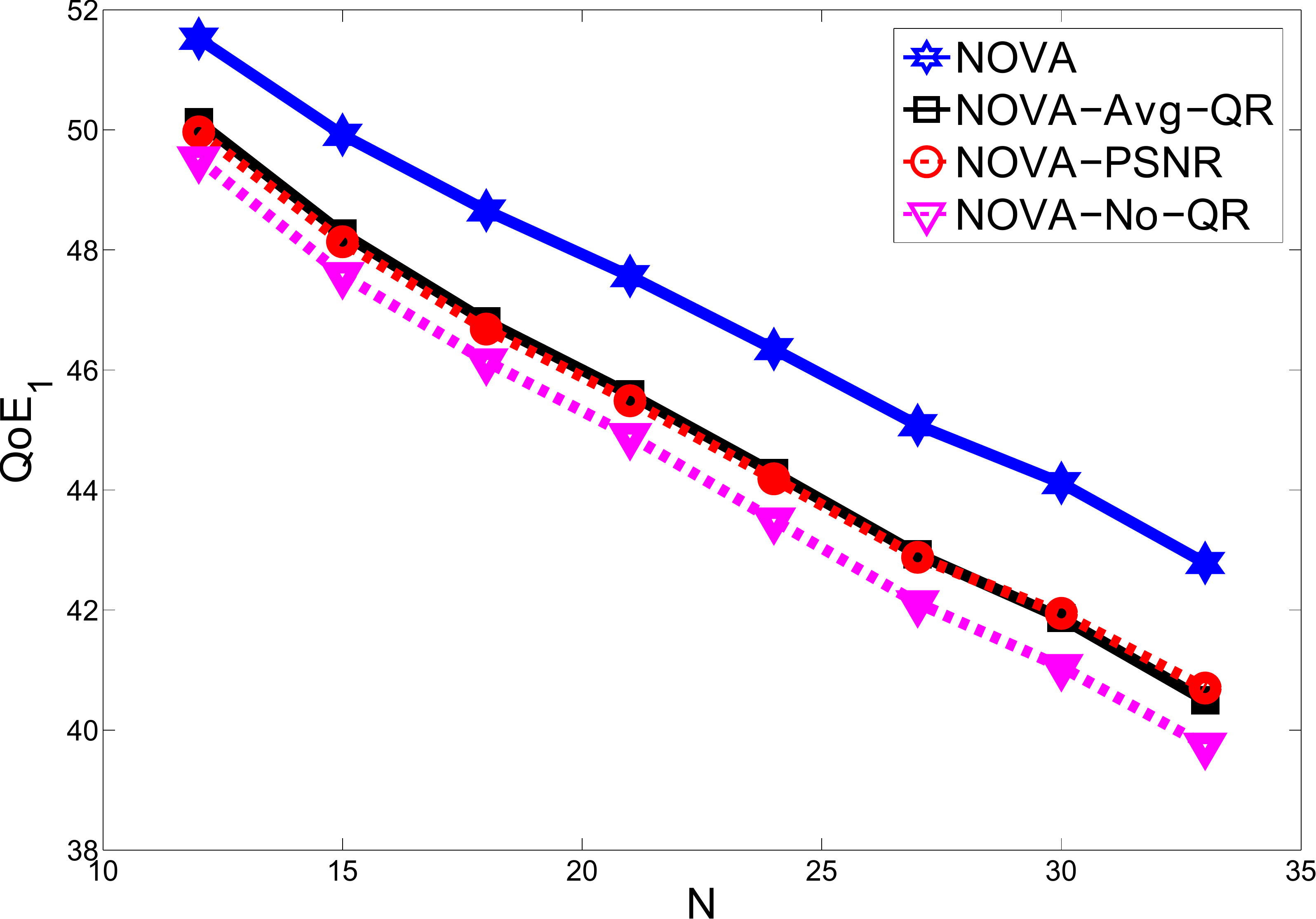}
\else
\includegraphics[scale=.42]{../../../papers/video_varaiability_buffer_and_cost_aware/simulations/thesis_simlns/knowledge_value_network/qoe1_gains.pdf}
\fi
	\caption{Value of knowing QR tradeoffs.}
	\label{value_of_knowing_QR}
\end{figure}

To assess the value of knowing accurate QR tradeoffs, we carried out simulations for NOVA 
with STQ based on less sophisticated video quality assessment metrics.
In particular, we carried out simulations where we used the same QR tradeoff for all segments and
this QR tradeoff was equal to the average of QR tradeoffs of all segments
of the movie being viewed by the video client.
Thus, instead of using segment level QR tradeoff information, we are using an approximation based on long term features of the videos
being viewed by the video clients.
The results associated with this setting is depicted using the curve NOVA-Avg-QR in Fig.~\ref{value_of_knowing_QR}.
We also carried out simulations using NOVA with STQ equal to PSNR, and the results associated with this setting are depicted using the curve NOVA-PSNR in Fig.~\ref{value_of_knowing_QR}.
We picked $h_{i,0}$ in \eqref{hb_used_for_simulations} as $0.0025$ for the simulations with STQ equal to PSNR.
In a setting, where we do not have any information about the QR tradeoffs, we could use crude metrics like $10\log(\mbox{Representation Size})$,
and the results associated with this setting are depicted using the curve NOVA-No-QR in Fig.~\ref{value_of_knowing_QR}.
Although NOVA-No-QR  has (expectedly) the worst performance, 
we see that there is a significant reduction in performance (i.e., QoE$_1$) when we do not have accurate segment level QR tradeoff information,
and further these reductions are not too different for NOVA-Avg-QR or NOVA-PSNR when compared to that for NOVA-No-QR.

\begin{figure}[ht]
	\centering
		\ifarxivmode
\includegraphics[scale=.29]{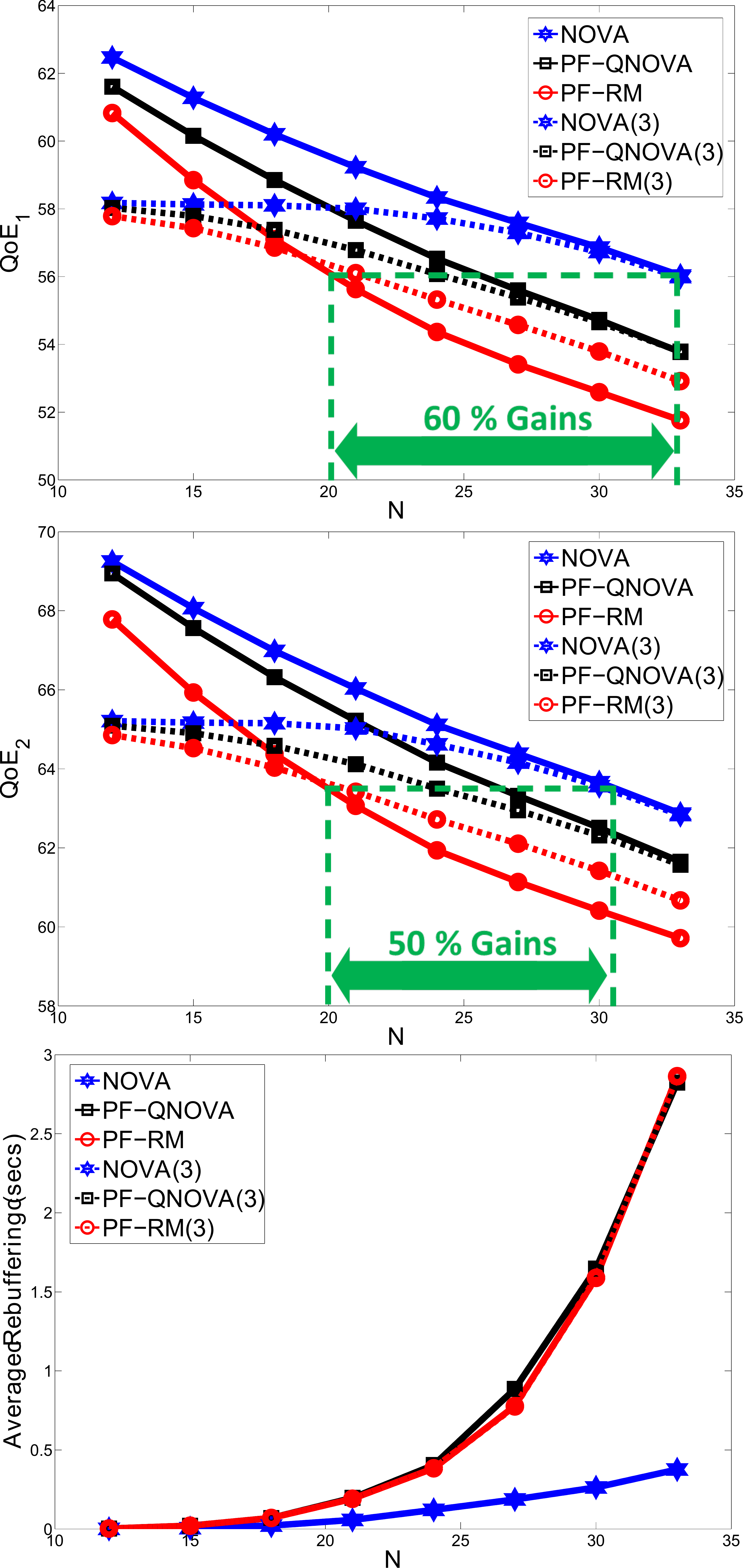}
\else
\includegraphics[scale=.29]{../../../papers/video_varaiability_buffer_and_cost_aware/simulations/thesis_simlns/main_het_network_O/combinedO.pdf}
\fi
	\caption{Performance gains using NOVA: Streaming movie Oceania.}
	\label{perf_gains_oceania}
\end{figure}

\begin{figure}[ht]
	\centering
		\ifarxivmode
\includegraphics[scale=.29]{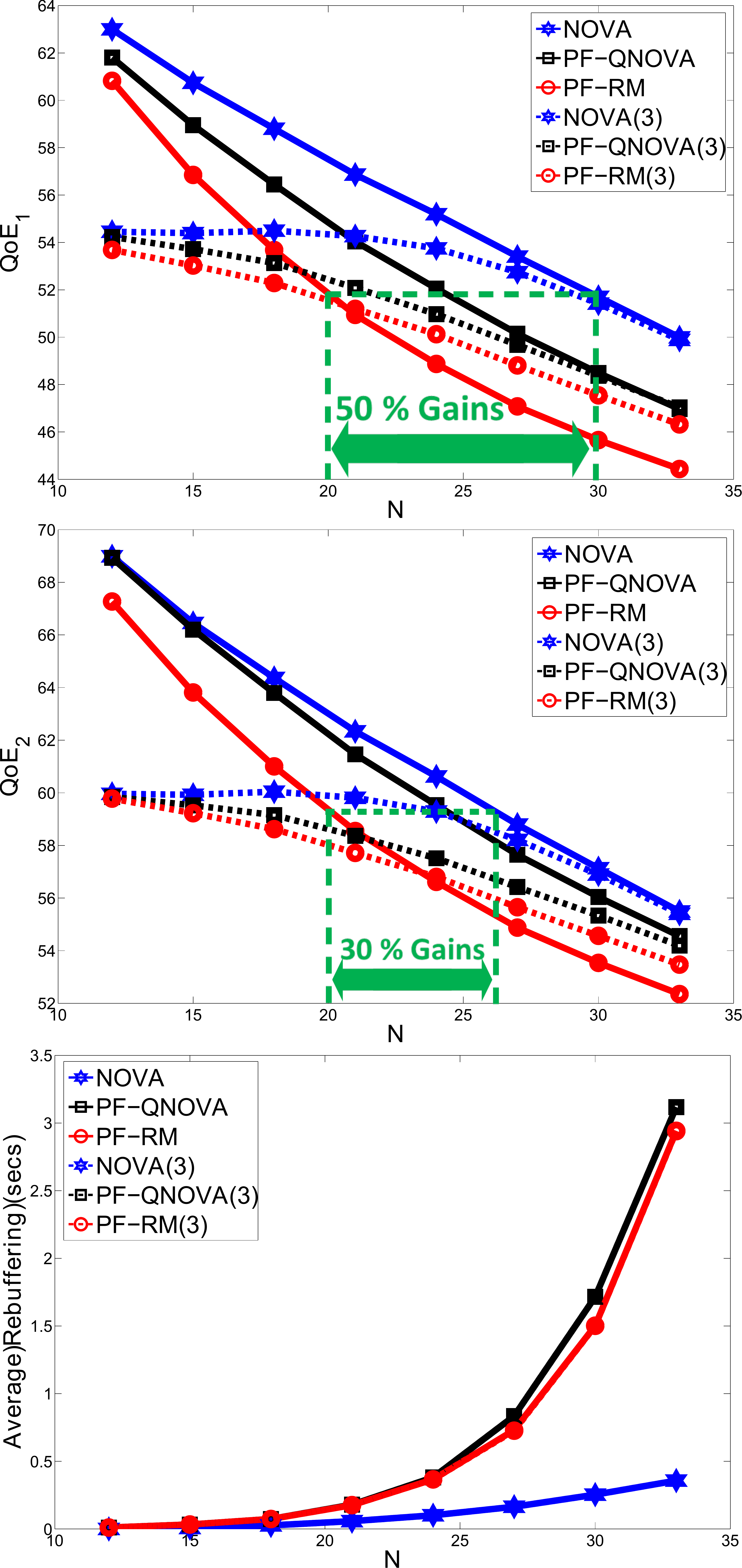}
\else
\includegraphics[scale=.29]{../../../papers/video_varaiability_buffer_and_cost_aware/simulations/thesis_simlns/main_het_network_R/combinedR.pdf}
\fi
	\caption{Performance gains using NOVA: Streaming movie Route 66.}
	\label{perf_gains_route}
\end{figure}

\begin{figure}[ht]
	\centering
		\ifarxivmode
\includegraphics[scale=.27]{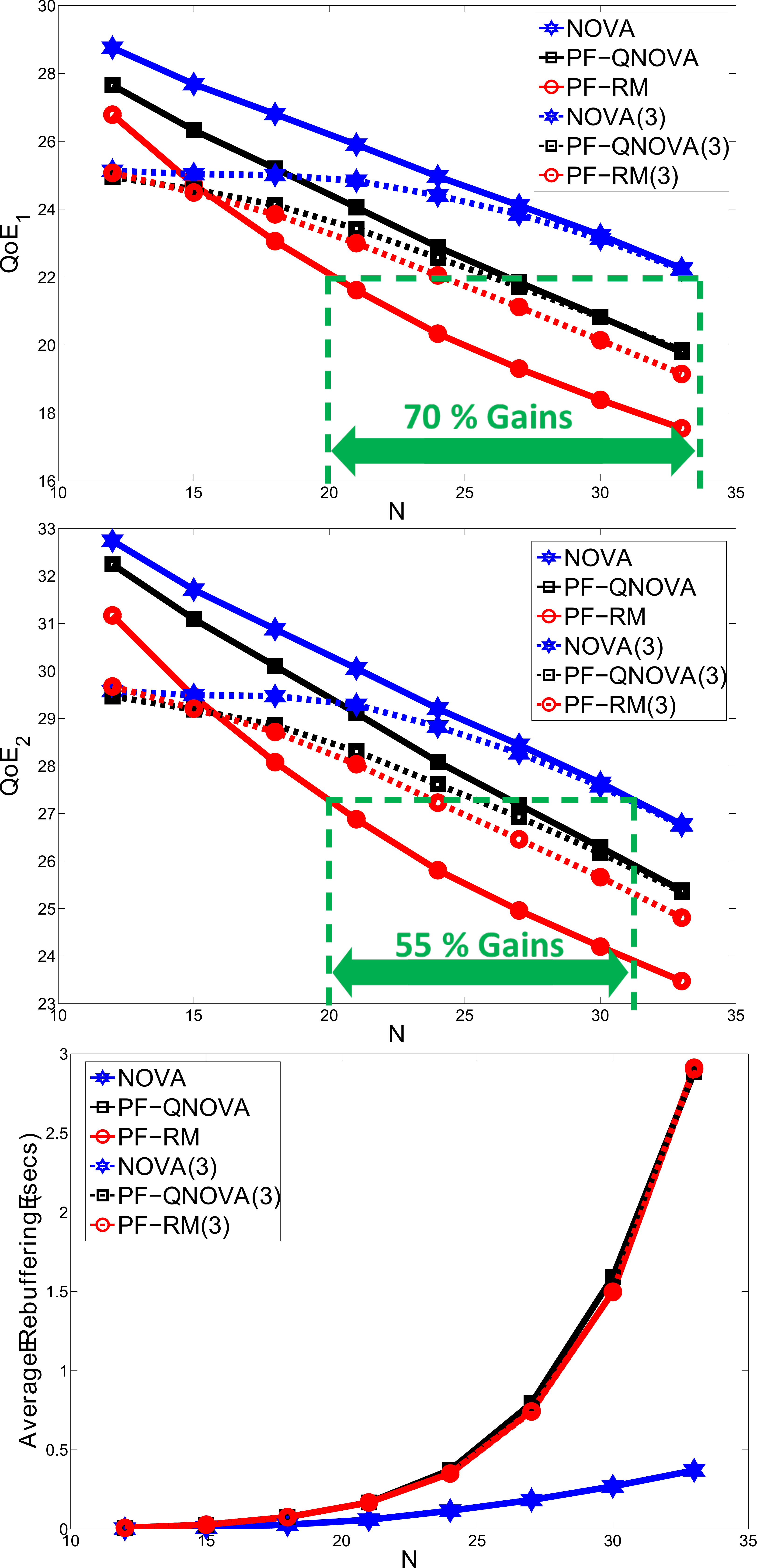}
\else
\includegraphics[scale=.27]{../../../papers/video_varaiability_buffer_and_cost_aware/simulations/thesis_simlns/main_het_network_V/combinedV.pdf}
\fi
	\caption{Performance gains using NOVA: Streaming movie Valkaama.}
	\label{perf_gains_valkaama}
\end{figure}

We also compared the performance of NOVA, PF-QNOVA and PF-RM 
for the movies Oceania, Route 66 and Valkaama separately, and these results are depicted in Figs.~\ref{perf_gains_oceania}-\ref{perf_gains_valkaama}.

\section{Extensions}
\label{nova_qoe_ch_extensions}

In this section, we discuss following important extensions of NOVA:
\begin{itemize}
\item NOVA for general QoE models;
\item NOVA for general channel models;
	\item NOVA under other, e.g. legacy, resource allocation policies;
	\item the performance of quality adaptation in NOVA, referred to as QNOVA, when used for a standalone video client;
	\item the presence of, and sharing with, other traffic;
	\item discrete network resources;
	\item video client implementation considerations for NOVA such as 
\begin{itemize}
	\item discrete levels of quality adaptation, 
	\item choice of $\epsilon$, $(h^B_i(.))_{\iiN}$ and $(h^D_i(.))_{\iiN}$,
	\item reduction of startup delay and frequency of rebuffering, 
	\item playback buffer limits,
	\item and, video playback pauses;
\end{itemize}
	\item and, the performance of NOVA in stochastic networks, i.e.,
	networks with dynamically varying number of video clients.
\end{itemize}

\subsection{More general QoE models}
\label{more_general_QoE_models}
NOVA can be used for a larger class of QoE models,
and still retain its optimality characteristics.
For instance, we can consider QoE models such as
\begin{eqnarray*}
e^S_i \left(q_i\right)=
   m^{U^Q_i,S}_i \left(q_i\right)	- U^V_i \left(\mbox{Var}^S \left(q_i\right)\right),
\end{eqnarray*}
where $ m^{U^Q_i,S}_i \left(q_i\right)$ is a generalized mean defined as
\begin{eqnarray*}
m^{U^Q_i,S}_i \left(q_i\right)\defeq\frac{\sum_{s=1}^{S} l_{i,s} U^Q_i \left(q_{i,s}\right)}{\sum_{s'=1}^{S} l_{i,s'} },
\end{eqnarray*}
and $U^Q_i$ is a twice differentiable concave increasing function.
We only need two simple modifications to the algorithm NOVA in order to allow for the above QoE model:
\begin{enumerate}
	\item Modify objective function \eqref{defn_phiQ} of the 
optimization problem QNOVA$_i(\bm{\theta}_i,f_i)$ associated with the quality adaptation of video client $i$
as given below:
\begin{eqnarray*}
\phi^{U^Q_i,Q}\left(q_i,\bm{\theta}_i,f_i\right) 
&=&\left(U^E_i\right)^{'}\left( \mu_i - U^V_i \left(v_i\right)\right) \left( U^Q_i \left(q_i\right)
-  \left(U^V_i\right)^{'} \left(v_i\right) \left(  q_i - m_i\right)^2\right)
 \\\nonumber&& - \frac{h^B_i\left(b_i\right)}{\left(1+\overline{\beta}_i\right)} f_{i} \left(q_i\right)
 - \frac{p^d_i h^D_i\left(d_i\right)}{\overline{p}_i } f_{i} \left(q_i\right),
\end{eqnarray*}
where we have only replaced a $q_i$ term appearing in \eqref{defn_phiQ} with $U^Q_i \left(q_i\right)$.
\item Modify  the update rule \eqref{mu_update_NOVA} for $\mu_{i,s_i} $ as follows:
\begin{eqnarray}
 \mu_{i,s_i+1} &=& \mu_{i,s_i} 
  + \epsilon  \left( \frac{l_{i,s_i+1}}{\lambda_{i,s_i}} U^Q_i \left(q^*_{i,s_i+1}\right) - \mu_{i,s_i}   \right),
  \end{eqnarray}
  so that $\mu_{i,s_i}$ keeps track of the generalized mean.
\end{enumerate}
We can show that under the new QoE model, NOVA with the above modifications is still asymptotically optimal,
i.e., we can obtain a result similar to Theorem \ref{main_optimality_theorem}.
This generalization allows us to accommodate QoE models involving generalized mean 
such as those proposed in \cite{chao_paper_on_metrics_F1_F2}.

\subsection{More general channel models}
\label{more_general_channel_models}
For notational simplicity, we assumed that the network allocation constraint in each slot is a real valued function.
However, we can consider more general channel models prevalent in many practical networks like wireless networks using OFDM,
where the resource allocation to a video client is the sum of the resource allocation over 
several sub-resources $w\in{\cal W}$ (for e.g., orthogonal subcarriers in OFDM)
where ${\cal W}$ is a finite set of sub-resources in the network.
It is easy to extend the preceding discussion to consider such networks.
In particular, we can extend the resource allocation algorithm RNOVA$\left(\mathbf{b},c\right)$ proposed in NOVA 
to obtain RNOVA-GC$\left(\mathbf{b},\left(c_w\right)_{w\in{\cal W}}\right)$ given below:
	\begin{eqnarray*}
  \max_{\mathbf{r}}	&& \sum_{i\in {\cal N}}   h^B_i\left(b_i\right)r_i
\\\mbox{ subject to }& & c_w\left( \left(r_{i,w}\right)_{\iiN} \right)\le 0, \ \forall \ w\in{\cal W},
\\& & r_i=\sum_{w\in{\cal W}} r_{i,w} \ge r_{i,\min} \ \forall \ i\in{\cal N},
\\& & r_{i,w} \ge 0, \ \forall \ w\in{\cal W}, \ \forall \ i\in{\cal N}.
\end{eqnarray*}
 where the optimization variable $r_{i,w}$ represents the resource allocation to video client $i$ over sub-resource $w$,
 and $r_i$ represents the cumulative resource allocation to video client $i$.
If the natural generalization of assumptions on network allocation constraints (e.g., stationary ergodic, Assumption-SF etc) discussed earlier hold,
then we can show that the above extension of NOVA (which uses RNOVA-GC$\left(\mathbf{b},\left(c_w\right)_{w\in{\cal W}}\right)$
for network resource allocation) is also asymptotically optimal.
Similar extensions will typically be possible in general settings where the capacity region can be described using a finite number of convex functions.

	\subsection{NOVA under other resource allocation policies, and QNOVA for a standalone video client}	
	\label{nova_fixed_resource_allocation_and_for_standalone_client}
	
	Let us consider the problem of optimizing video delivery in scenarios where we cannot modify or optimize
	the resource allocation policies, e.g., legacy systems or networks where resource allocation is driven by other considerations.
	Note in such scenarios, we can still control the quality adaptation at the video clients.
	In this section, we show that the quality adaptation component in NOVA is still optimal for such scenarios.
	
	Further, we show that QNOVA, i.e., quality adaptation component of NOVA,
	carries out optimal quality adaptation for a standalone video client.
	
	\subsubsection{NOVA under other resource allocation policies}	
	\label{nova_fixed_resource_allocation}
	Consider a network where the resource allocation policy cannot be 
	modified or optimized for video delivery.
	This would be the case in legacy networks where the	video clients 
	have to operate under a predetermined resource allocation policy like proportional fair allocation policy (see \cite{kushner_prop_fair_conv}),
	or other proprietary (unknown) allocation policies.
	The following result says that QNOVA is asymptotically optimal for any
	feasible stationary resource allocation policy $\left(\mathbf{r}(c)\right)_{c\in{\cal C}} $
	(see Definition \ref{defn_feasiible_stationary_resource_allocation_policy} 
	of a feasible stationary resource allocation policy)

	\begin{corollary}
	QNOVA is asymptotically optimal for any feasible stationary resource allocation policy $\left(\mathbf{r}(c)\right)_{c\in{\cal C}} $.
	\end{corollary}
	\begin{IEEEproof}
		This result follows once we note that under the given network resource allocation policy $\left(\mathbf{r}(c)\right)_{c\in{\cal C}} $,
	the offline optimization problem formulation OPT$_{\cal N}(S)$,
	discussed in Section \ref{section_Optimal_finite_horizon_policy},
	breaks into $N$ single video client offline problem formulations $\left(\mbox{OPT}_{  \left\{i\right\}}(S)\right)_{\iiN}$.
	Recall that ${\cal N}$ is the set of video clients considered,
	and we have added the subscript ${\cal N}$ in OPT$_{\cal N}(S)$ to emphasize the dependence of the 
	problem formulation on the set of video clients.
	Thus, for $\iiN$, 
	the offline optimization problem $\mbox{OPT}_{  \left\{i\right\}}(S)$ is
  obtained by only considering the terms in the objective function and constraints of	OPT$_{\cal N}(S)$
  that involve video client $i$, and ensuring that the allocation constraints correspond to the fixed resource allocation 
	$\left( r_i(c) \right)_{\ciC}$ associated with the video client.
	\end{IEEEproof}
	This result sheds light on an important feature of NOVA that the optimality properties of 
  the adaptation component QNOVA are \emph{insensitive} to the resource allocation in the network
  as long as the resource allocation policy is stationary.	
	The gains from QNOVA are also explored using simulations in Section \ref{section_simulations} for a scenario with
	the legacy proportionally fair resource allocation schemes.

	\subsubsection{QNOVA for optimizing a standalone video client}	
	\label{nova_for_standalone_client}
	
	We now move away from the network setting, and consider a video client $i^*$ with associated resource allocations
	$\left(R_{i^*,k}\right)_{k\ge 1}$ modeled as an exogenous stationary ergodic process
	(i.e., $R_{i^*,k}$ is the random variable modeling the resource allocation to the client in slot $k$).
	This is a reasonable model for a standalone video client accessing video servers in a wide range of scenarios 
	involving wired networks and wireless networks.
	We have the following important optimality property for QNOVA when used for a standalone video client.
	\begin{corollary}\label{adaptation_in_NOVA_is_optimal_for_single_client}
	QNOVA is asymptotically optimal for a standalone video client 
	if the associated resource allocation $\left(R_{i^*,k}\right)_{k\ge 1}$ is an exogenous (i.e., independent of quality adaptation decisions) 
	stationary ergodic process.
	\end{corollary}
	\begin{IEEEproof}
	This result directly follows by setting ${\cal N}=\left\{i^*\right\}$
	and defining the capacity regions using $\left(R_{i^*,k}\right)_{k\ge 1}$.
	\end{IEEEproof}
	The above result is significant since the optimization of adaptation in standalone video clients is an important problem in practice,
	and the result provides optimality guarantees for the solution QNOVA to this problem.
  Further, this result also reinforces the insensitivity of optimality of the adaptation component in NOVA.
  
\begin{figure}[ht]
	\centering
	\ifarxivmode
\includegraphics[scale=.43]{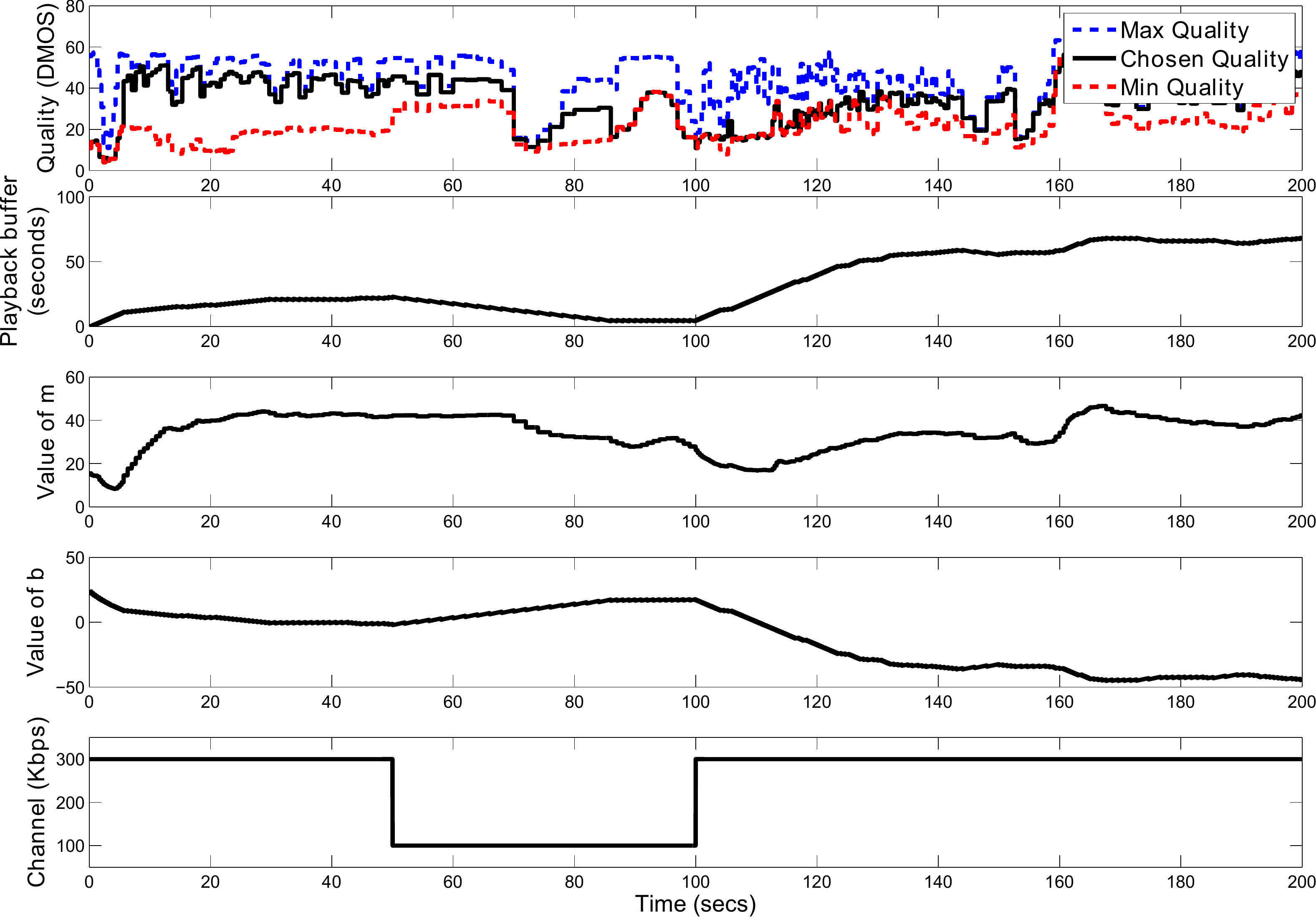}
\else
\includegraphics[scale=.43]{figs/single_user_cap_drop.pdf}
\fi
	\caption{Quality adaptation responding to a capacity drop}
	\label{single_video client_trace}
\end{figure}

  An evolution of the parameters of QNOVA and associated quality adaptation is depicted in Figure \ref{single_video client_trace},
  which also illustrates the response of the quality adaptation in QNOVA to an abrupt capacity drop between time instants 50 secs and 100 secs (see the last subplot).
  We see that the value of the parameter $b_{i^*,k}$ starts to increase (see the fourth subplot) following the drop,
  and eventually becomes large enough to force the selection of representations with the least size (see the first subplot).

	\subsection{NOVA and sharing network resources with other traffic}
	\label{nova_with_other_traffic}
	
	Video traffic due to stored video is typically carried in networks 
	supporting other types of traffic such as voice traffic, real time video traffic, data traffic due to file downloads etc.
	If the network capacity consumed by other sources of traffic can be modeled as 
	an exogenous stationary ergodic process,
	then we can extend the optimality result for NOVA given in Theorem \ref{main_optimality_theorem} to this scenario.
	This can be shown by incorporating the consumption of capacity
	by other traffic into the stationary ergodic random process 
	$\left(C_k\right)_k$ (i.e., the allocation constraints associated with the video clients).
	This observation about optimality of NOVA is useful
	as it covers scenarios where the stored video traffic has to compete with other sources of 
	traffic that have higher priority, for e.g., voice traffic
	and video traffic generated by real time interactive applications like phone calls, video conferencing etc.
	However, note that in these scenarios, the traffic due to stored video 
	(is low priority to the network and) is only supported using residual capacity in the network.
	
	A more fair approach is to set aside a fixed part of the available network capacity
	for carrying stored video traffic. This could however lead to inefficient use of network resources if there are 
	too few or too many video clients streaming stored video.
	A better approach can be obtained by using a more flexible division of network capacity.
	For instance, we can use the approach developed below which extends NOVA 
	for use in the presence of other data traffic, e.g., associated with users downloading long files.
	Let ${\cal N}_D$ denote the set of data users.
	The following modification of the resource allocation component in NOVA can be used to this end:
	\begin{eqnarray*} 
\nonumber \max_{  \left( (r_i)_{\iiN \cup{\cal N}_D}\right) } 
	&& \hspace{0cm}p_V \sum_{i\in {\cal N}}  h^B_i (b_{i,k}) r_i 
	+ \sum_{j\in {\cal N}_D} \left(U^D_j\right)' \left(\rho_{j,k}\right) r_j
\\\mbox{s.t. }& & c_k\left( (r_i)_{i\in{\cal N}\cup{\cal N}_D}\right)\le 0,
\ r_i \ge r_{i,\min} \ \forall \ i\in{\cal N} \cup {\cal N}_D.
\end{eqnarray*}
The resource allocation in slot $k$ is carried out by obtaining an optimal solution
$\left( (r^*_{i,k})_{\iiN \cup{\cal N}_D}\right)$ 
to the above optimization problem,
where $(r^*_{i,k})_{\iiN}$ and $(r^*_{j,k})_{j\in{\cal N}_D}$
are respectively the resource allocation to the video clients and data users.
Here, $U^D_j$ is a twice differentiable concave function for each data user $j\in{\cal N}_D$, 
and $c_k$ is a convex function characterizing the capacity region in slot $k$. 
The parameter $\rho_{j,k}$ tracks the mean resource allocation to 
data user $j\in{\cal N}_D$, and is updated at the beginning of each slot $k$ as follows:
$$\rho_{j,k+1} = \rho_{j,k} + \epsilon \left(r^*_{j,k} - \rho_{j,k} \right).$$
The constant $p_V$ determines the priority given to video clients, 
i.e., the higher the value of $p_V$, the higher is the priority given to video over data users.

We can show that NOVA with the above modification in resource allocation is once again asymptotically optimal,
i.e., we can establish an optimality result like that given in Theorem \ref{main_optimality_theorem},
where we compare the performance of NOVA against that of the optimal offline algorithm solving
OPT-D$(S)$, which has objective function
\begin{eqnarray*}
p_V \sum_{i\in {\cal N}}  U^E_i \left(  e^S_i \left(q_i\right)  \right)
+ \sum_{j\in {\cal N}_D} U^D_j \left(  e^{K_S,data}_j \left(r_j\right) \right),
\end{eqnarray*}
and same constraints as in OPT$(S)$ and additional constraints, $r_j \ge r_{j,\min}$ for each $j\in{\cal N}_D$,
on the resource allocation to the data users.
Here, for each data user $j\in{\cal N}_D$,
\begin{eqnarray*}
e^{K_S,data}_j \left(r_j\right)=
   m^{K_S}_j \left(r_j\right)	,
\end{eqnarray*}
represents the QoE for a data user $j\in{\cal N}_D$
and is equal to the mean resource allocation to the video client, i.e.,
this QoE model assumes that the data users care only about their long term time average resource allocation.
The objective function of OPT-D$(S)$ also indicates that the constant $p_V$ 
controls the tradeoff between QoE delivered to video clients versus that delivered to the data users
under an optimal solution.
Also, note that if there are no video clients (i.e., ${\cal N}$ is empty), the above objective corresponds to that of a network 
carrying out fair resource allocation to the data users
where the fairness is implicitly decided by the choice of functions $\left(U^D_j\right)_{j\in{\cal N}_D}$.
In particular, if $U^D_j$ is $\log(.)$ for each $j\in{\cal N}_D$, 
then the resulting allocation is proportionally fair.
The above discussion also sheds light 
on an important property of the resource allocation component in NOVA 
that it requires only a simple modification of legacy schedulers like proportionally fair schedulers,
and hence is well suited for use in current networks.
	
	\subsection{NOVA implementation considerations}
	In this subsection, we discuss implementation considerations related to NOVA
	focusing on the resource allocation component of NOVA in Subsection \ref{NOVA_with_discrete_resource_blocks},
	and the quality adaptation component of NOVA in rest of the subsection.
	
	\subsubsection{Discrete network resources}
	\label{NOVA_with_discrete_resource_blocks}	
	
	In many practical settings, the set of feasible resource allocations in a slot is \emph{discrete}.
	For instance, the basic unit of resource allocation in LTE is a Resource Block (RB)
	which comprises several OFDM sub-carriers for a given time slot,
	and an RB can be assigned to at most one video client.
	In such cases, we can use a discrete approximation RNOVA-DISCRETE$\left(\mathbf{b}_k,c_k\right)$
	of RNOVA$\left(\mathbf{b}_k,c_k\right)$
given below to obtain the resource allocation in slot $k$:
	\begin{eqnarray*} 
  \max_{\mathbf{r}}	&& \sum_{i\in {\cal N}}   h^B_i\left(b_{i,k}\right)r_i
\\\mbox{ subject to }& & \mathbf{r} \in {\cal R}^{discrete}_{c_k},
\end{eqnarray*}
where ${\cal R}^{discrete}_{c_k}$ is the discrete set of permissible (i.e., permitted by the practical constraints) 
resource allocation vectors satisfying the allocation constraint $c_k$ for slot $k$.
In many practical settings, allocation constraints are essentially linear,
and hence, we can obtain computationally efficient approaches to solve the
discrete convex optimization problem RNOVA-DISCRETE$\left(\mathbf{b}_k,c_k\right)$
by exploiting the linearity of the objective and constraint functions.
We consider an example of this in Section \ref{section_PNOVA_practical_implementation} where we obtain 
the \emph{optimal} solution (and not just a good solution) to the above problem by just finding the maximum of $N$ scalars.

	\subsubsection{Discrete quality adaptation}
	\label{NOVA_with_finite_number_of_representations}	
		
	Till now, we assumed that we have a continuous set of quality choices
	for each segment.
	However, in practice, video segments are only available in a \emph{finite} number of representations.
	Thus, we modify the optimization problem QNOVA$_i(\bm{\theta}_i,f_i)$ 
	associated with the adaptation in NOVA as follows: 
	the representation chosen for segment $s$ of video client $i$
	is the one with quality equal to the optimal solution to QNOVA$_i$-FINITE$(\bm{\theta}_{i,s},f_{i,s},{\cal Q}_{i,s})$
	given below
	\begin{eqnarray*} 
 \nonumber \max_{q_i} &&  \phi^Q\left(q_i,\bm{\theta}_{i,s},f_{i,s}\right)
\\\mbox{ subject to }& & 0 \le q_i \le q_{\max},
\\& & q_i\in{\cal Q}_{i,s},
\end{eqnarray*}
where we have modified QNOVA$_i(\bm{\theta}_{i,s},f_{i,s})$ by
imposing an additional restriction that the quality should be picked from the set
${\cal Q}_{i,s}$ of available quality choices for segment $s$ of video client $i$.
For instance, if segment 10 of video client 1 has 4 representations of sizes 
200, 300, 500 and 1000 kb with the corresponding 
quality measurements being DMOS values equal to 38, 48, 62 and 83 respectively,
then ${\cal Q}_{1,10}=\left\{38, \  48,\   62,\   83\right\}$, and $f_{1,10}(38)=200$ kbps, $f_{1,10}(48)=300$ kbps etc.

When using QNOVA$_i$-FINITE$(\bm{\theta}_{i,s},f_{i,s},{\cal Q}_{i,s})$ for adaptation, 
extra care is needed while choosing the function $U^V_i$ (which in turn decides the penalty for variability) due to the 
structure of the objective function
\begin{eqnarray*}
\phi^Q\left(q_i,\bm{\theta}_i,f_i\right) 
&=&\left(U^E_i\right)^{'}\left( \mu_i - U^V_i \left(v_i\right)\right) \left( q_i
-  \left(U^V_i\right)^{'} \left(v_i\right) \left(q_i - m_i\right)^2\right)
 \\\nonumber&& - \frac{h^B_i\left(b_i\right)}{\left(1+\overline{\beta}_i\right)} f_{i} \left(q_i\right)
 - \frac{p^d_i h^D_i\left(d_i\right)}{\overline{p}_i } f_{i} \left(q_i\right),
\end{eqnarray*}
This is because a very high value of $(U^V_i)'$ could potentially inhibit 
the above discrete approximation of NOVA's adaptation from 
selecting representations that correspond to  quality choices greater than $m_i$.
This is especially the case when the number of quality choices is small.
This happens due to the fact that the term $\left(q_i - m_i\right)^2$ could be large
for quality choices in the discrete set ${\cal Q}_{i,s}$ 
that are larger than $m_{i,s}$.

	Quality adaptation using 
	QNOVA$_i$-FINITE$(\bm{\theta}_{i,s},f_{i,s},{\cal Q}_{i,s})$
	can be efficiently carried out as it involves a simple task of evaluating the objective function
	for a few quality choices.
	
	\subsubsection{Choice of $\epsilon$, $(h^B_i(.))_{\iiN}$ and $(h^D_i(.))_{\iiN}$}
	\label{choiceof_epsilon_etc}
  The choice of the constant $\epsilon$, and the functions $(h^B_i(.))_{\iiN}$ and $(h^D_i(.))_{\iiN}$
  used in NOVA can have a significant impact on the convergence and tracking ability of NOVA operating in non-stationary regimes involving short duration videos,
  discrete quality adaptation (discussed above) etc. 
	Although choosing small $\epsilon$ is beneficial for long videos, it can significantly affect the performance (initial transient and tracking ability) of NOVA
	for short videos. 
	We have observed that choices of $\epsilon$ in the range 0.05 to 0.1 typically work well,
	and often a good choice can be made using trial and error for the system under consideration.

	Setting $(h^D_i(.))_{\iiN}$ as linear functions, with the scaling obtained using trial and error,
	worked well in our simulations. 
	Good choices of the scaling depend on certain features of QR tradeoffs (of the videos) 
	like their first and second order derivatives 
	(or first and second order differences in the case of discrete quality adaptation).
	This dependence follows from the dependence of the quality adaptation on the first order derivatives 
	of the QR tradeoffs (see \eqref{KKT_QNOVA_gradient_zero}).
	
	Unlike in the case of the choice of $(h^D_i(.))_{\iiN}$ where (simple) linear functions were enough,
	we used $h^B_i(.)$ that have the following structure in 
	Section \ref{section_simulations} (see \eqref{hb_used_for_simulations}):
	\begin{eqnarray*}
		h^B_i(b_i)= h_{i,0} \left( \frac{b_i}{0.05} +  \max \left( \frac{b_i-h_{i,1}}{0.05},0\right)^2\right),
		\end{eqnarray*}
	with carefully chosen constant $h_{i,0}$. Also, note that the constant 0.05 corresponds to $\epsilon$ associated with NOVA updates.
	The linear structure of $h^D_i(d_{i,k})$ was enough to meet the \emph{average} cost constraints in NOVA,
	whereas the above structure of $h^B_i(b_{i,k})$ 
	allows NOVA to meet average rebuffering constraints (i.e., \eqref{OPTS_bound_rebuf}
	which requires that rebuffering be asymptotically negligible)
	and a \emph{stronger per-slot requirement} (unlike the `average requirement') that there is no rebuffering; we explain this below.
	The constant $h_{i,1}$ is picked to be equal to or (a bit) greater than $b_{i,0}-20$ (where $b_{i,0}$ is the initialization of the parameter $b_{i,k}$)
	so that $h^B_i(b_i)$ increases more quickly (i.e., quadratically)
	when $\frac{b_i}{\epsilon}$ is close to $\frac{b_{i,0}-20}{\epsilon}$,
	and is very large when  $\frac{b_i}{\epsilon}$ is close to $\frac{b_{i,0}-5}{\epsilon}$
	so as to force QNOVA to select lower quality representations.
	This feature of QNOVA  is desirable since (we have argued the following in Section \ref{section_NOVA} after presenting the algorithm NOVA)
	$$\frac{b_{i,k}}{\epsilon} - \frac{b_{i,0}}{\epsilon}\approx  \left(  \frac{k \tau_{slot}}{ \left(1+ \overline{\beta}_i\right)  } - \mbox{Duration of video downloaded till now} \right). $$
	For instance, if  $\overline{\beta}_i=0$ (i.e., video client $i$ prefers not to see any rebuffering),
	then 
	our choice of $h_{i,1}$ ensures that
	$h^B_i(b_i)$ is large when the playback buffer has video content of duration less than 5 seconds (i.e.,  $\frac{b_i}{\epsilon}$ is close to $\frac{b_{i,0}-5}{\epsilon}$)
	so that QNOVA starts to pick lower (if not lowest) quality representations.
	This was our main motivation for using $h^B_i(b_{i,k})$ with the above structure,
	and setting $h_{i,1}\approx b_{i,0}-20$.

	Good choices of the constant $h_{i,0}$  depend again on the characteristics of the QR tradeoffs like the first and second order derivatives 
	(as in the case of good scaling constants for $(h^D_i(.))_{\iiN}$),
	and we obtained them via trial and error for the system under consideration.

	\subsubsection{Reducing startup delay and the frequency of rebufffering}
Video client optimization also requires attention to issues like reduction of 
  frequency of rebuffering events and playback startup delay.
  Frequency of rebuffering events can be reduced by forcing the video client 
  to delay the resumption of playback after a rebuffering event until there is sufficient amounts of video content in the playback buffer.
  
  We can reduce the start up delay by appropriately choosing the initial conditions.
  For instance, we can pick large values for $b_{i^*,0}$,
  and small values for $m_{i^*,0}$ (recall that $b_{i^*,0}$ and $m_{i^*,0}$ denote the initialization of 
  parameters $b_{i^*,k}$ and $m_{i^*,s}$ used in NOVA) to encourage selection of representations with smaller size
  so that they are downloaded quickly at the beginning.	
  An evolution of NOVA's parameters using such an initialization is depicted in Figure \ref{single_video client_trace}
  where we see that, in the beginning, large $b_{i^*,0}$ and small $m_{i^*,0}$  encourages 
  the selection of representations with lower STQ and thus, smaller size.

  \subsubsection{Playback buffer limits}
  In practical systems, we might have to operate NOVA under an additional constraint on
  the size of the playback buffer.
  Hardware limitations on memory could be a reason for this constraint although
  the latest smartphones, tablets, laptops etc
  have plenty of memory and memory limitations are no longer a major concern in the design of video clients.
  However, it is interesting to note that even  data delivery cost considerations 
  can force us to impose this constraint
  especially 
  when the chances of video client abandonment are high, i.e.,
  the possibility of a video client terminating the video playback without viewing the entire video is high.
  For instance, suppose that we do not impose any constraint on the size of playback buffer, 
  and hence there are no constraints on the amount of video data 
  downloaded by the video client that has not been viewed yet.
  Then, if the video client receives very high resource allocation
  (that is well above even the largest compression rates of available representations), 
  then QNOVA will aggressively download the video segments exploiting the good resource allocation.
  But, all the downloaded data would be wasted if there is a video client abandonment.
  In such a scenario, the video client and/or content provider might have to pay for the delivery of these wasted (i.e., not viewed) segments.
  Thus, high data delivery costs and video client abandonment concerns would motivate the use of playback buffer limits
  for some types of content.
  
  Note that there is an interesting tradeoff between the size of the playback buffer 
  (small buffers might reduce wastage of video data in the event of a video client abandonment) 
  and the ability of QNOVA (and other adaptation algorithms) to exploit periods of good resource allocation.
  For instance, consider a user running a video client on a mobile that is moving away from a base station in  a cellular network
  so that the user initially sees better wireless channels but they are declining.
  Imposing a playback buffer limit will adversely impact the ability of the video client
  to exploit high initial resource availability.
  Therefore, we conclude that
  it may be useful to impose limits on playback buffer size,
  and this has to be carefully chosen after taking into account the data delivery costs, 
  and the possibility of video client abandonment etc.

  In the presence of limits on the size of playback buffer, we can modify
  QNOVA to slow down the rate of segment download when capacity is abundant. 
  A simple modification would be to stop segment download requests once a playback buffer limit is reached.
  A better modification would use a `smoother' approach where we keep reducing the segment download rate
  as we approach the playback buffer limit, for e.g., we could 
  force QNOVA to delay the next segment download request by a duration proportional to 
  $$
  \max \left( \frac{1}{PB_{lim}-PB_{cur}}
  -\frac{1}{0.5PB_{lim}}
  , 0\right)
  $$
  where $PB_{lim}$ is the playback buffer limit and $PB_{cur}$ is the current state of the playback buffer.
  This would ensure that QNOVA slows  segment download rate once the playback buffer is large enough, 
  (i.e., greater than $0.5PB_{lim}$),
  and stops once it reaches the limit.
  
  Note that under NOVA, the issue of wastage of data under video client abandonment is mainly relevant 
  in scenarios where the network resource allocation is exceptionally high.
  This is due to the fact that QNOVA will switch to higher quality representations 
  under high resource allocation, 
  and hence, excessive buffering of video data can occur only when 
  average resource allocation is consistently higher than average compression rate of the largest representations
  (or if the video client has set $\overline{\beta}_i$ in NOVA less than zero).

\subsubsection{Video playback pauses}
  If a video client $\iiN$ pauses the playback of a video, 
  then we stop the use of update rule \eqref{b_allocate_update}
  which increments the value of the variable $b_{i,k}$,
  and resume the use of the update rule when the video client resumes its playback.
  Recall that (see Section \ref{section_NOVA}) $b_{i,k}$ serves an indicator of risk of violation of rebuffering constraints
  of video client $i$,
  and a large value of $b_{i,k}$ would force the selection of low quality representations by the video client and 
  ensure higher priority in resource allocation to the video client.
  Hence, by temporarily pausing the use of update rule \eqref{b_allocate_update},
  we ensure that we are not unnecessarily forcing the video client to lower its quality
  or forcing the network controller to give higher priority to a paused video client.
  
  We can use the same idea of temporarily pausing the use of update rule \eqref{b_allocate_update}
  when the content provider inserts ads during the playback of a video.

	\subsection{NOVA in stochastic networks}
	\label{nova_in_stochastic_networks}
	
	Till now, we focused on networks with a static number of video clients (since the set ${\cal N}$ is a fixed set) and data users.
	An important feature of real world networks will be dynamics in the number of video clients.
	Motivated by this, in this section, we study stochastic networks where video clients arrive into the network,
	utilize network resources to stream video content, and depart.	
	
	We start by exploring some of the new challenges associated with video delivery optimization problem 
	in stochastic networks.
	Recall that we formulated the video delivery optimization problem as an optimization problem OPT-BASIC
	(and more formally as OPT$(S)$).
	Similarly, we could formulate video delivery optimization problem in stochastic networks 
	as the following `stochastic' extension OPT-BASIC-STOCH associated with a time window of $K$ slots:
	\begin{eqnarray*} 
 &&\hspace{-1.3cm}\max \quad \frac{1}{\left|{\cal N} [0,K]\right|}\sum_{\iiN [0,K]}  U^E_i  \left(  \mbox{Mean Quality}_i  - \mbox{Quality Variability}_i \right)
\\ \mbox{subject to} && \mbox{Rebuffering$_i$, Cost$_i$, and Stochastic Network constraints,}
\end{eqnarray*}
	where ${\cal N}[0,K]$ is the set of video clients who arrive into the network during the $K$ slots.
	Note that the new objective is the average of $U^E_i$ of the QoE of video clients 
	utilizing the network resources during the $K$ slots,
	where the functions $U^E_i$ implicitly decide fairness in the delivery of QoE to the video clients.
	The \emph{stochastic} network constraints reflect the network resource allocation constraints associated with
	the time varying number of video clients in different slots,
	and this video client dynamics introduces another potential source of variability in network resource allocation.
	
	Solving OPT-BASIC-STOCH presents new challenges due to 
	the video client dynamics considered in the formulation.
	Firstly, note that the time duration (i.e., the number of slots) that a video client spends in the network
	can depend on the resource allocation to it. 
	For instance, a video client receiving high resource allocation over many slots could leave the network early
	after downloading all the segments of its video before the completion of video playback.
	However, even if we ignore this dependence on resource allocation 
	(this dependence will be negligible if
	there are sufficient choices of representations for segments and the quality adaptation algorithm responds to very high or very low
	resource allocation by appropriately adjusting the effective rate of download of segments),
	there are issues related to video client dynamics that need to be carefully tackled.
	For instance, consider a video client $i^*$ 
	that arrives into the network at time slot $\underline{k}$ and leaves the network
	at time slot $\overline{k}$.
	Now, let ${\cal N}_1$ denote the set of video clients in the network after the arrival of video client $i^*$ in slot $\underline{k}$,
	and we refer to this set as the state of the network in slot $\underline{k}$.
	Suppose that the state of the network does not change until slot $0.5\left(\overline{k}+\underline{k}\right)$,
	and (the arrival/departure of another video client causes) the network state changes to ${\cal N}_2$ in slot $0.5\left(\overline{k}+\underline{k}\right)$
	(for simplicity, assume that this is an integer),
	and suppose that the network state remains the same until slot $\overline{k}$ (when video client $i^*$ leaves).
	Suppose the video client downloads $S_1$ segments until slot $0.5\left(\overline{k}+\underline{k}\right)$, 
	and $S_2$ segments in the remaining slots.
	Then, 
	\begin{eqnarray*}\label{variance_calculation_under_dynamics}
	\mbox{Var} \left(\left(q_{i^*,s}\right)_{1:(S_1+S_2)}\right)
	&=&
	\frac{S_1\mbox{Var} \left(\left(q_{i^*,s}\right)_{1:(S_1)}\right)}{S_1+S_2} 
	+\frac{S_2\mbox{Var} \left(\left(q_{i^*,s}\right)_{(S_1+1):(S_1+S_2)}\right)}{S_1+S_2} 
	\\\nonumber&&\hspace{-1cm}+ \frac{S_1 \left(\mbox{Mean} \left(\left(q_{i^*,s}\right)_{1:(S_1)}\right)-\mbox{Mean} \left(\left(q_{i^*,s}\right)_{1:(S_1+S_2)}\right)\right)^2}{S_1+S_2} 
	\\\nonumber&&\hspace{-1cm}+\frac{S_2\left(\mbox{Mean} \left(\left(q_{i^*,s}\right)_{(S_1+1):(S_1+S_2)}\right)-\mbox{Mean} \left(\left(q_{i^*,s}\right)_{1:(S_1+S_2)}\right)\right)^2}{S_1+S_2},
	\end{eqnarray*}
	where 
	$$\mbox{Mean} \left(\left(q_{i^*,s}\right)_{1:(S)}\right)= \frac{1}{S}\sum_{s=1}^S q_{i^*,s},$$
	$$\mbox{Var} \left(\left(q_{i^*,s}\right)_{1:(S)}\right)= \frac{1}{S}\sum_{s=1}^S \left(q_{i^*,s} - \mbox{Mean} \left(\left(q_{i^*,s}\right)_{1:(S)}\right)\right)^2.$$
	From the above expression, we see that
	$\mbox{Var} \left(\left(q_{i^*,s}\right)_{1:(S_1+S_2)}\right)$ (and hence the QoE)
	of the video client $i^*$ 
	is a complex function of resource allocation and quality adaptation associated with the various network states
	seen during its stay in the network,
	for e.g., note that $S_1$ and $S_2$ depend on the resource allocation to the video client $i^*$ during 
	$\left[\underline{k},0.5\left(\overline{k}+\underline{k}\right)\right]$
	and
	$\left[0.5\left(\overline{k}+\underline{k}\right),\overline{k}\right]$
	respectively
	which in turn depend on the network states ${\cal N}_1$ and ${\cal N}_2$ respectively
	associated with these time windows.
	This observation 
	suggests that a direct extension of the approach in the preceding discussion,
	which is based on considering a static network (or a \emph{single} network state), is difficult.
	However, this can be done under additional assumptions on the nature of video client dynamics and its impact on the stochastic network constraints,
	and this is the focus of most of the rest of this section.
	
	Although a theoretical analysis of the performance of NOVA in stochastic networks is difficult without 
	additional simplifying assumptions, the following three features of NOVA suggest that we can expect it to perform well in stochastic networks also.
\begin{enumerate}
\item \textbf{Adaptation in NOVA is optimal and insensitive to resource allocation:} 
This property, explored in Section \ref{nova_fixed_resource_allocation_and_for_standalone_client},
essentially guarantees that the adaptation in NOVA is optimal as long the resource allocation can be modeled as a stationary ergodic process.
Thus, if the video client dynamics results in a resource allocation to the video clients which is stationary ergodic, 
the \emph{quality adaptation} in NOVA will perform well.
However, this argument cannot be extended to argue optimality of \emph{resource allocation} under video client dynamics.

\item \textbf{The tracking ability of NOVA:} 
Although NOVA was studied for scenarios where a fixed set of video clients see stationary variations in capacity and video QR tradeoffs, 
NOVA has tracking ability built into it which allows it to perform well in non-stationary regimes.
Such non-stationary regimes could include networks with video client dynamics,
settings where the video and/or channels exhibit non-stationary behavior etc.
The tracking ability follows from the structure of the update rules for the parameters used in NOVA
where the current decision is weighted by $\epsilon$.
For instance, consider the update rule \eqref{m_update_NOVA} for NOVA parameter $m_{i,s_i} $ repeated below:
$$m_{i,s_i+1} = m_{i,s_i} 
  + \epsilon \left(U^E_i\right)^{'}\left( \mu_i - U^V_i \left(v_i\right)\right)\left(U^V_i\right)^{'} \left(v_i\right)
  \left( \frac{l_{i,s_i+1}}{\lambda_{i,s_i}} q^*_{i,s_i+1}- m_{i,s_i}   \right).$$
  Recall that the parameter $m_{i,s_i} $ is responsible for tracking the mean quality of video client $i$.
  This rule keeps updating the value of the estimate of mean based on the quality $q^*_{i,s_i+1}$ of the current segment under consideration.
	The choice of $\epsilon$ decides the impact of the current quality $q^*_{i,s_i+1}$.
	Similarly, we can identify the impact of current decisions on other NOVA parameters.
	This influence of current information allows NOVA to track and adapt in non-stationary settings.
	Further, we can control the tracking ability by controlling $\epsilon$ since we can 
	increase the impact of current decisions on the update, and thus on the tracking ability, by increasing $\epsilon$. 
	However, note that an excessively high value of $\epsilon$ can degrade the performance of the algorithm,
	as NOVA parameters will not be able to converge due to
	their evolution being swayed by even small changes in the network.

	\item \textbf{Optimal for static setting:} NOVA comes with strong optimality guarantees for a static setting with a fixed set of video clients.
	This, along with the tracking ability discussed above, suggests that we can expect good performance under certain assumptions on the video client dynamics.
	For instance, this would be the case if the video client dynamics	were `slow'.
\end{enumerate}

\section{Implementing NOVA: An example}
\label{section_PNOVA_practical_implementation}

In this section, we discuss an example for an implementation of NOVA for a network shared
by video clients and data users.
To simplify the exposition, we will make simplifying assumptions.
We also discuss the issues related to signaling requirements, information exchange and complexity 
towards the end.

\subsection{Setting}
\label{section_setting}
We consider a base station in a cellular network supporting a dynamic number of video clients and data users,
i.e., the base station is the network controller responsible for network resource allocation.
Let ${\cal N}(k)$ denote the set of video clients and ${\cal N}_D(k)$ denote the set of data users in the network in slot $k$.
The priority given to video clients is determined by the parameter $p_V>0$ (discussed in Section \ref{nova_with_other_traffic}).
Let the duration of each slot be equal to $\tau_{slot}$ seconds and that of each segment be $l_{seg}$ seconds.

In each slot $k$, let 
$c_k\left(\left(r_{i,k}\right)_{i\in{\cal N}(k)\cup{\cal N}_D(k)}\right)\le 0$ 
defined below
\begin{eqnarray}\label{example_lin_cap_constraints}
c_k\left(\left(r_{i,k}\right)_{i\in{\cal N}(k)\cup{\cal N}_D(k)}\right)
=  \sum_{i\in{\cal N}(k)} \frac{r_i}{p_{i,k}} +  \sum_{i\in{\cal N}_D(k)} \frac{r_i}{p_{i,k}}-1,
\end{eqnarray}
describe the capacity region in slot $k$,
where $p_{i,k}$ and $p_{j,k}$ denotes the peak rate seen by video client $i\in{\cal N}(k)$
and data user $j\in{\cal N}_D(k)$ respectively in slot $k$, i.e., 
$p_{i,k}$ is the maximum rate that can be allocated to video client $i\in{\cal N}(k)$ in slot $k$ when we allocate all the resources 
to this video client (and none to others).
We assume that the video clients have no cost constraints (so that we can ignore the variables $d_{i,s}$) and that
the QoE model is given by
\begin{eqnarray*}
e^S_i \left(q_i\right)=
   m^S_i \left(q_i\right)	- c_v \mbox{Var}^S \left(q_i\right),
\end{eqnarray*}
for some positive constant $c_v$. Further, we set $U^E_i(e)=e$ for each $\iiN (k)$.
Also, let $\overline{\beta}_i=0$ for each $\iiN (k)$.
We also set $r_{i,min}=0$ for each $\iiN$ (and ignore the requirement that it should be positive).
For each data user $j\in{\cal N}_D(k)$, we use the following QoE model (see Section \ref{nova_with_other_traffic} for a discussion of this QoE model)
\begin{eqnarray*}
e^{K_S,data}_j \left(r_j\right)=
   m^{K_S}_j \left(r_j\right).
\end{eqnarray*}

For the $s$th segment of video client $i\in{\cal N}(k)$, $ {\cal Q}_{i,s}$ is the (finite) set of available quality choices
for the segment and $f_{i,s}(q)$ denotes the compression rate of the representation associated with 
a quality choice $q_i \in {\cal Q}_{i,s}$.

\subsection{Detailed algorithm}

As for the proportional fair scheduler, the base station uses $\rho_{j,k}$ to track the mean rate allocation to data user $j\in{\cal N}_D(k)$.
Since the update of variable $b_{i,k}$ requires the knowledge of segment download completions
(see the update rule \eqref{b_adapt_update} of NOVA),
the base station either has to be able look at the data stream of video clients to infer segment download completions,
or rely on signaling from the video clients that indicate segment download completions.
We focus on the latter setting in this section.
Let the base station store an current estimate $b^B_{i,k}$ of $b_{i,k}$ for each video client $i\in{\cal N}(k)$,
and we discuss the update rule for this estimate in the algorithm presented later in the section.
Each video client $\iiN(k)$  uses index $s_i$ to track the number of segments downloaded by video client $i\in{\cal N}(k)$,
the parameter $m_{i,s}$ to track mean quality, and
uses the parameter $b_{i,k}$ to obtain a proxy for the risk of rebuffering.
Thus, each video client $i\in{\cal N}(k)$ stores the current value of $s_i$, $m_{i,.}$ and $b_{i,.}$.
The base station stores the current value of variable $\rho_{j,k}$ for each data user $j\in{\cal N}_D(k)$
in addition to the current value of variable $b^B_{i,k}$ for each video client $i\in{\cal N}(k)$.
The detailed algorithm is given below.
\\\line(1,0){514}\vspace{-.5cm}
\\\line(1,0){514}
\begin{flushleft}
\textbf{INIT}:
The base station initializes $b^B_{i,0}$ for each video client $i\in{\cal N}(k)$,
and $\rho_{j,0}$ for each data user $j\in{\cal N}_D(k)$. Each video client $i\in{\cal N}(k)$ initializes $m_{i,0}$ and $b_{i,0}$.
Let $s_{i,0}=1$ for each video client $i\in{\cal N}(k)$.
\\
\end{flushleft}
In each slot $k\ge 0$, carry out the following steps:
\\ \textbf{RNOVA-BS}: At the beginning of slot $k$, base station \emph{estimates} current capacity region $c_k$, and allocates rate
$r^*_{i,k}$ to each video client $i\in{\cal N}(k)$ and $r^*_{j,k}$ to each data user $j\in{\cal N}_D(k)$,
where
$\left(r^*_{i,k}\right)_{i\in{\cal N}(k)\cup{\cal N}_D(k)}$  is an optimal solution to
	\begin{eqnarray} 
\max_{\mathbf{r} \ge 0}	&& p_V \sum_{i\in {\cal N}(k)} h ( b^B_{i,k} ) r_i +  \sum_{j\in {\cal N}_D(k)}  \frac{ r_j}{\rho_{j,k}}
\label{RPNOVA_objective}
\\\mbox{ subject to }& & 
c_k\left( \left(r_{i,k}\right)_{i\in{\cal N}(k)\cup{\cal N}_D(k)}\right) \le 0,
\label{RPNOVA_channel_constraints}
\end{eqnarray}
and update 
\begin{eqnarray*}
  b^B_{i,k+1} &=&  b^B_{i,k} + \epsilon  \left( \tau_{slot}  \right), \ \forall \ i\in {\cal N}(k)
 \\ \rho_{j,k+1} &=&  \rho_{j,k} +  \epsilon     \left(r^*_{j,k} - \rho_{j,k}\right), \ j\in {\cal N}_D(k).
\end{eqnarray*}
\\ \textbf{RNOVA-VC}: At the beginning of each slot $k$, each video client $i\in{\cal N}(k)$ updates
$b_{i,k}$ as follows
\begin{eqnarray}
  b_{i,k+1} &=&  b_{i,k} + \epsilon  \left(  \tau_{slot}  \right),
\end{eqnarray}
\\\textbf{QNOVA-VC}: During slot $k$, if any video client $i\in{\cal N}(k)_v$ finishes transmission of $s_i$ th segment, 
the video client $i$ sends an END-OF-SEG$_i$ to the BS, retrieves (this should be locally available at this time) ${\cal Q}_{i,s_i+1}$ and $f_{i,s_i+1}$ for the $(s+1)$th (i.e., next) segment, 
and picks segment corresponding to quality $q^*_{i,s_i+1}$ obtained using
	\begin{eqnarray} 
 q^*_{i,s_i+1} =  \mbox{argmax}_{q_i \in {\cal Q}_{i,s_i+1}}  
  \left(q_i-  c_v \left(q_i - m_{i,s_i}\right)^2  -h(b_{i,k}) f_{i,s_i+1} \left(q_i\right)\right)
\label{QPNOVA_objective}
\end{eqnarray}
and update $m_{i,s_i+1}$, $b_{i,k+1}$ and $s_i$ as follows:
\begin{eqnarray}
  m_{i,s_i+1} &=& m_{i,s_i} 
  + \epsilon   \left( q^*_{i,s_i+1}- m_{i,s_i}   \right),
\\b_{i,k+1} &=&  b_{i,k+1} -  \epsilon \left( l_{seg}\right),
\\\nonumber s_i &=& s_i+1.
\end{eqnarray}
\\\textbf{BS-REC-SIG}:
On receiving END-OF-SEG$_i$ from video client $i\in{\cal N}(k)$, the base station updates (overwrites)
\begin{eqnarray}
b^B_{i,k+1} &=&  b^B_{i,k+1} - \epsilon  \left( l_{seg}\right),\vspace{-.5cm}
\label{update_BS_b_on_rec_signal}
\end{eqnarray}
\line(1,0){514}

\subsubsection{Signaling required between the base station and video clients}

Each video client $i\in{\cal N}(k)$ sends an END-OF-SEG$_i$ signal to the base station when it completes downloading a segment.
Here, END-OF-SEG$_i$ can be viewed as a control signal carrying ID of the video client sending it and data to indicate
that this is an END-OF-SEG control signal.
When the base station receives the END-OF-SEG$_i$ signal from video client $i\in{\cal N}(k)$, it updates $b^B_{i,k}$ using \eqref{update_BS_b_on_rec_signal}.
It can be verified that the update mechanism along with the signaling ensures that
$b^B_{i,k}$, which is base stations's estimate for $b_{i,k}$, is equal to $b_{i,k}$ most of the time (except for the time duration between sending END-OF-SEG$_i$ and the reception of the signal at the basestaion).
Note that if the segment lengths are variable,
the video clients will also have to send the length of the next segment being downloaded 
as it is required (see \ref{b_adapt_update}) by the base station to update $b^B_{i,k+1}$.

\subsubsection{Information flow}
\label{about_data_required}
NOVA uses two types of `external' data- (a) channel capacity data $c_{.}$,
and (b) $ {\cal Q}_{i,.}$ and $f_{i,.}$ for each video client $i\in{\cal N}(k)$.
They are described in more detail below.

We assume that the base station  knows (or knows with reasonable accuracy) the current value of $c_k$ for each slot $k$,
e.g., the current value of peak rates $\left(\left(p_{i,k}\right)_{i\in{\cal N}(k)},\left(p_{j,k}\right)_{j\in{\cal N}_D(k)}\right)$ for each video client in each slot $k$.
The video clients could measure this and inform the base station.

For each video client $i\in{\cal N}(k)$, on completion of download of segment $s$, we assume that
video client knows $ {\cal Q}_{i,s+1}$ and $f_{i,s+1}(.)$ for the $s+1$th segment.
One simple way to ensure this is to make sure that 
when a video client starts downloading a certain segment, 
the video client has requested the video server to ensure that 
video server has sent $ {\cal Q}_{i,.}$ and $f_{i,.}(.)$ for the next few segments to the video client.
Note that this is not a difficult requirement to meet even for live videos (and clearly not for stored videos).
In the worst case, if this information is not available, then we could use 
a concave function (e.g., $\log(.)$) of the size of the segment as a proxy for quality.

\begin{figure}[ht]
	\centering
	\ifarxivmode
\includegraphics[scale=.55]{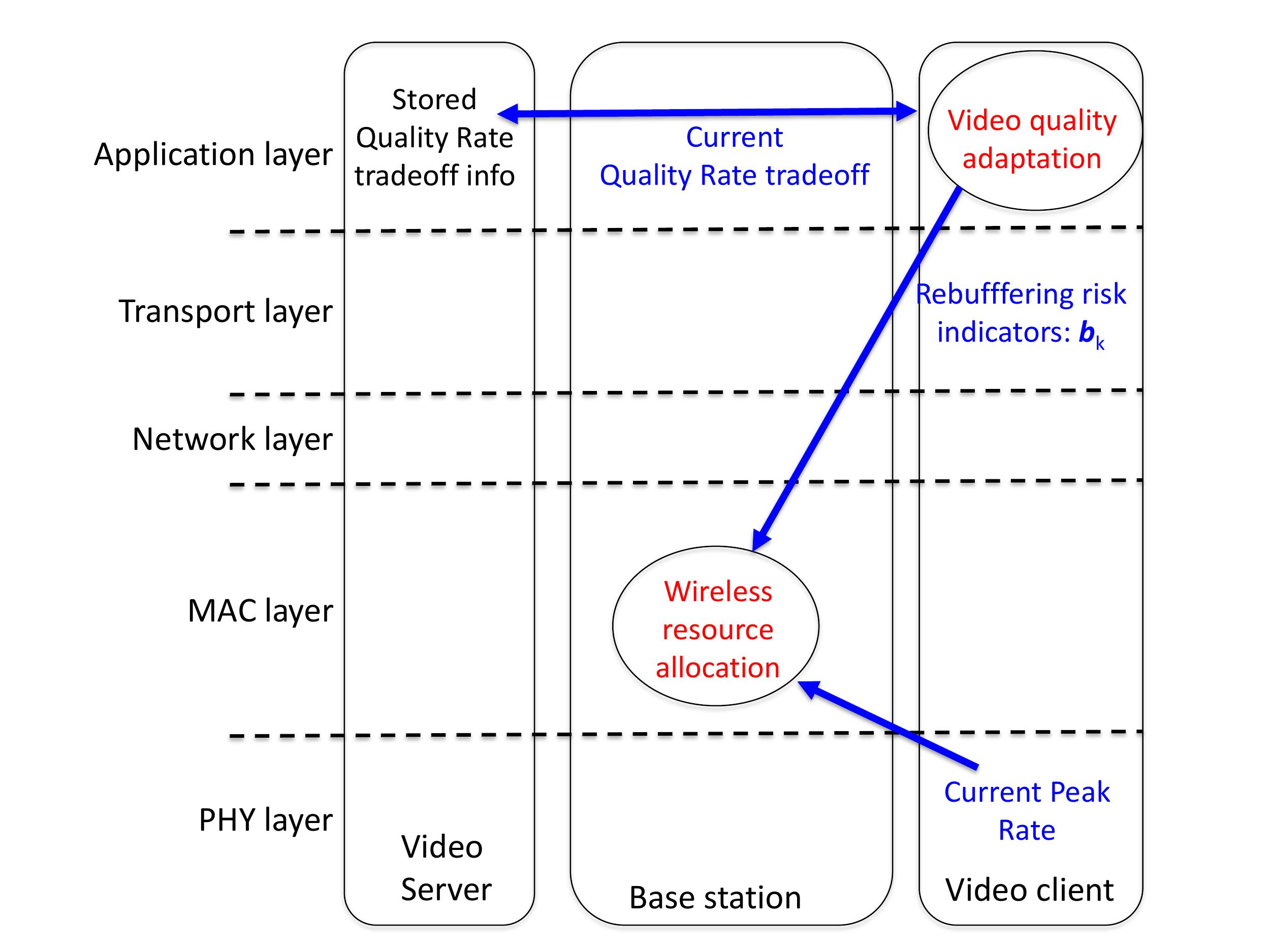}	
	\else
\includegraphics[scale=.55]{figs/cross_layer_fig.pdf}	
	\fi
	\caption{Cross Layer Information Flow}
	\label{cross_layer_fig}
\end{figure}

The flow of information across various layers of the network for this implementation of NOVA is depicted in
Fig.~\ref{cross_layer_fig}.

\subsubsection{Complexity}
The optimization problem used in resource allocation (described in \eqref{RPNOVA_objective}-
\eqref{RPNOVA_channel_constraints}) is a linear program. 
Further, we can exploit the structure of allocation constraints given in \eqref{example_lin_cap_constraints} and linearity of objective,
to show that will be carrying out optimal resource allocation if we pick 
a video client or data user that has the highest value a metric,
and assign the peak rate to that video client or data user.
The metric for this setting is equal to $p_V h ( b^B_{i,k} )p_{i,k}$ for video clients $\iiN(k)$, and $\frac{p_{j,k}}{\rho_{j,k}}$ for data users $j\in{\cal N}_D(k)$).

\section{Proof of optimality of NOVA} 
\label{section_proof_optimality_NOVA} 
This section is devoted to the proof of the previously stated Theorem \ref{main_optimality_theorem} related to optimality of NOVA.
In Subsection \ref{section_optstat}, 
we study an auxiliary optimization problem OPTSTAT and 
obtain Theorem \ref{main_NOVA_with_theta_pi_is_optimal}
which suggests that we can prove the main optimality result Theorem \ref{main_optimality_theorem} for NOVA 
if we establish an appropriate convergence result for NOVA's parameters.
In Subsection \ref{relate_NOVA_to_diff_inclusion}, we study 
an auxiliary differential inclusion (given in \eqref{mhat_ode_rule}-\eqref{rhohat_ode_rule})
which evolves according to average dynamics of NOVA,
and obtain a convergence result for the differential inclusion.
In Subsection \ref{convergence_of_diff_inclusion_and_theorem}, 
we view NOVA's update equations (\eqref{m_update_NOVA}-\eqref{lambda_update_NOVA} and \eqref{sigma_NOVA_update}-\eqref{rho_NOVA_update})
as an asynchronous stochastic approximation update (see, e.g., \cite{kushner_text} for reference),
and relate this stochastic approximation update to 
the auxiliary differential inclusion (in \eqref{mhat_ode_rule}-\eqref{rhohat_ode_rule}),
and use this relationship
to establish desired convergence of NOVA's parameters
using the convergence result for the auxiliary differential inclusion
established in Subsection \ref{relate_NOVA_to_diff_inclusion}.

\begin{figure}[ht]
	\centering
	\ifarxivmode
	\includegraphics[scale=.5]{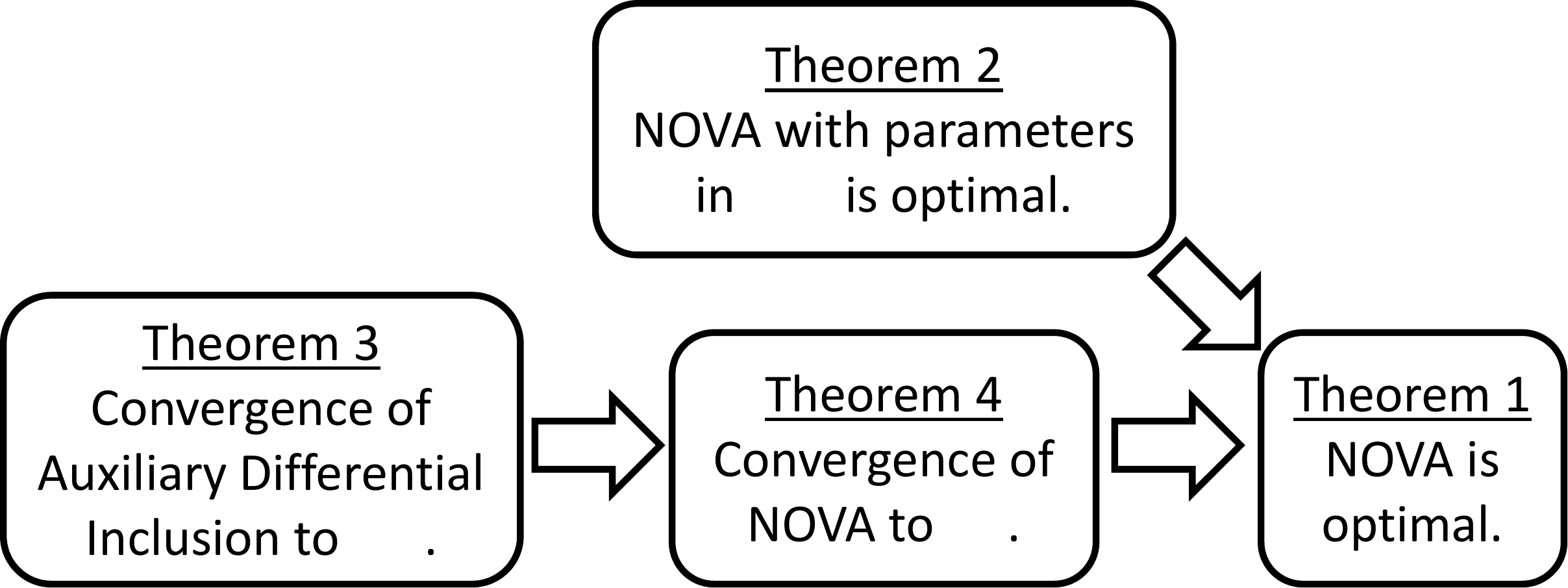}
	\else
	\includegraphics[scale=.5]{figs/nova_proof_diagram.pdf}
	\fi
	\caption{An outline of proof of optimality of NOVA}	
	\label{nova_proof_diagram}
\end{figure}

\subsection{OPTSTAT: An auxiliary optimization problem related to the offline optimization formulation}
\label{section_optstat}
The offline optimization formulation OPT$(S)$ mainly involves time and segment averages of various quantities.
By contrast, the formulation of OPTSTAT discussed in this section is based on the expected value of the corresponding quantities evaluated under the stationary distribution of $\left(C_k\right)_k$ and $\left(F_{i,s},L_{i,s}\right)_{s\ge 0}$ for each $i\in{\cal N}$.

Recall (see Section \ref{section_System_model}) that $\left(C_k\right)_k$ is stationary ergodic random process 
with marginal distribution $\left(\pi^{\cal C}(c)\right)_{c\in{\cal C}}$, and
let $C^{\pi}$ denote a random variable with distribution $\left(\pi^{\cal C}(c):c\in{\cal C}\right)$.
Also, recall that for each $i\in{\cal N}$, $\left(F_{i,s},L_{i,s}\right)_{s\ge 0}$ is a stationary ergodic process 
with marginal distribution 
$\left(\pi^{{\cal F},{\cal L}}_i\left( f_i,l_i  \right)\right)_{\left(f_{i},l_i\right)\in{\cal F L}_i} $.
We let $\left(F^{\pi}_{i},L^{\pi}_{i}\right)$ denote random variables with distribution
$\left(\pi^{{\cal F},{\cal L}}_i\left( f_i,l_i  \right)\right)_{\left(f_{i},l_i\right)\in{\cal F L}_i} $.

Let $\left(\mathbf{r} \left(c\right)\right)_{c\in{\cal C}} $ be a vector (of vectors) representing 
the reward allocation $\mathbf{r} \left(c\right)$ ($\in{\mathbb{R}^N}$) to the video clients for each $c\in{\cal C}$.
Although we are abusing the notation introduced earlier where $\mathbf{r}(t)$ denoted the allocation to the video clients in slot $t$,
one can differentiate between the functions based on the context in which they are being discussed.
Similarly, we let $q_i\left(f,l\right)$ denote the quality associated with a segment of video client $i$ 
with $(f,l)\in{\cal F L}_i$.
Mimicking the definition of $\phi_S\left(\left(\mathbf{q}\right)_{1:S}\right)$,
$m^S_i \left(q_i\right)$
and $\mbox{Var}^S_i \left(q_i\right)$ in Section \ref{section_Optimal_finite_horizon_policy}, 
we let
\begin{eqnarray}\label{OPTSTAT_objective_expression}
\phi_{\pi}\left(  \left(\left(q_i \left( f_i,l_i  \right) \right)_{\left(f_{i},l_i\right)\in{\cal F L}_i}\right)_{i\in{\cal N}}\right)
=
\sum_{i\in {\cal N}} U^E_i \left(  \mbox{Mean} \left( q_i\left( F^{\pi}_{i},L^{\pi}_i \right)  \right)
	- U^V_i \left( \mbox{Var} \left( q_i\left( F^{\pi}_{i},L^{\pi}_i \right)  \right) \right)\right),
\end{eqnarray}
where  
\begin{eqnarray*} 
\mbox{Mean} \left( q_i\left( F^{\pi}_{i},L^{\pi}_i \right)  \right)&=&
 \frac{\exn \left[ L^{\pi}_i q_i\left( F^{\pi}_{i},L^{\pi}_i \right) \right]}{\exn \left[ L^{\pi}_i\right]},
 \\\mbox{Var} \left( q_i\left( F^{\pi}_{i},L^{\pi}_i \right)  \right) &=&
 \frac{\exn \left[ L^{\pi}_i 
 \left(q_i\left( F^{\pi}_{i},L^{\pi}_i \right) - \mbox{Mean} \left( q_i\left( F^{\pi}_{i},L^{\pi}_i \right)  \right)\right)^2
  \right]}{\exn \left[ L^{\pi}_i\right]}.
 \end{eqnarray*} 
Now, consider the optimization problem OPTSTAT given below:
\begin{eqnarray} 
\label{OPTSTAT_eta_obj}
 \max_{  \left(\left(q_i \left( f_i,l_i  \right) \right)_{\left(f_{i},l_i\right)\in{\cal F L}_i}\right)_{i\in{\cal N}},
 \left(\mathbf{r} \left(c\right)\right)_{c\in{\cal C}}} & & \ \phi_{\pi}\left(  \left(\left(q_i \left( f_i,l_i  \right) \right)_{\left(f_{i},l_i\right)\in{\cal F L}_i}\right)_{i\in{\cal N}}\right)
\\ \mbox{ subject to }  c\left(\mathbf{r}\left(c\right))\right)& \le &0, \ \forall \ c\in {\cal C},
\label{OPTSTAT_eta_feasibility}
\\  q_i \left( f_i,l_i  \right) & \ge & 0, \  \forall \ \left(f_{i},l_i\right)\in{\cal F L}_i,\ \forall \ {i\in {\cal N}},
\label{OPTSTAT_eta_q_non_neg}
\\   q_i \left( f_i,l_i  \right) & \le &q_{\max}, \  \forall \ \left(f_{i},l_i\right)\in{\cal F L}_i,\ \forall \ {i\in {\cal N}},
\label{OPTSTAT_eta_q_upper_bound}
\\   r_{i}\left(c\right) & \ge & r_{i,\min}, \ \forall \ c\in {\cal C}, \ \forall \ {i\in {\cal N}},
\label{OPTSTAT_eta_r_non_neg}
\\   p^d_i \frac{ \exn \left[L^{\pi}_i F^{\pi}_{i} \left( q_i\left( F^{\pi}_{i},L^{\pi}_i \right) \right) \right] } 
{  \overline{p}_i \exn \left[L^{\pi}_i   \right]}
  & \le &1,\ \forall \ {i\in {\cal N}},
\label{OPTSTAT_eta_cost_bound}
\\  \frac{ \exn \left[L^{\pi}_i F^{\pi}_{i} \left( q_i\left( F^{\pi}_{i},L^{\pi}_i \right) \right) \right] } {\left(1+\overline{\beta}_i\right)  \exn \left[L^{\pi}_i   \right]}
  & \le &\frac{ \exn \left[  r_i  \left(C^{\pi}\right)  \right] }{ \tau_{slot} } ,\ \forall \ {i\in {\cal N}}.
\label{OPTSTAT_eta_bound_rebuf}
\end{eqnarray}
We obtained the above formulation by replacing the time and segment averages of various quantities in OPT$(S)$ (see \eqref{OPTS_obj}-\eqref{OPTS_bound_rebuf})
with the expected value of the corresponding quantities.
Note that in the constraint $c\left(\mathbf{r}\left(c\right))\right)\le 0$ given in \eqref{OPTSTAT_eta_feasibility}, 
$c$ appearing as argument of $\mathbf{r}(c)$ is an index (for the corresponding element in ${\cal C}$)
whereas the other $c$ is the associated function with argument $\mathbf{r}(c)$.
Similarly, in the term $F^{\pi}_{i} \left( q_i\left( F^{\pi}_{i},L^{\pi}_i \right) \right) $, 
the argument $\left( F^{\pi}_{i},L^{\pi}_i \right)$ serves as an index
whereas $F^{\pi}_{i}\left( . \right)$ is the (random) function.

For $\delta\ge 0$, let OPTSTAT$_{\delta}$ denote a modification of optimization problem OPTSTAT
with objective \eqref{OPTSTAT_eta_obj}, constraints \eqref{OPTSTAT_eta_feasibility}-\eqref{OPTSTAT_eta_r_non_neg}, and the following constraints
\begin{eqnarray} 
   p^d_i \frac{ \exn \left[L^{\pi}_i F^{\pi}_{i} \left( q_i\left( F^{\pi}_{i},L^{\pi}_i \right) \right) \right] } 
{  \overline{p}_i \exn \left[L^{\pi}_i   \right]}
  & \le & 1 + \delta,\ \forall \ {i\in {\cal N}},
\label{delta_modified_OPTSTAT_eta_cost_bound}
\\  \frac{ \exn \left[L^{\pi}_i F^{\pi}_{i} \left( q_i\left( F^{\pi}_{i},L^{\pi}_i \right) \right) \right] } {\left(1+\overline{\beta}_i\right)  \exn \left[L^{\pi}_i   \right]}
  & \le &\frac{ \exn \left[  r_i  \left(C^{\pi}\right)  \right] }{ \tau_{slot} }  + \delta ,\ \forall \ {i\in {\cal N}}.
\label{delta_modified_OPTSTAT_eta_bound_rebuf}
\end{eqnarray}
Hence, OPTSTAT$_{\delta}$ is obtained by relaxing constraints \eqref{OPTSTAT_eta_cost_bound} and \eqref{OPTSTAT_eta_bound_rebuf}
of OPTSTAT by $\delta$.
Let OPTSTATVAL and OPTSTATVAL$_{\delta}$ denote the optimal value of OPTSTAT and OPTSTAT$_{\delta}$ respectively.
Clearly, for any $\delta\ge 0$, OPTSTATVAL $\le$ OPTSTATVAL$_{\delta}$  and we have equality when $\delta=0$.

The next result presents properties related to the optimal solution of OPTSTAT, and optimal values of OPTSTAT and OPTSTAT$_{\delta}$.
Part (a) states says that OPTSTAT is a nice convex optimization problem, 
and part (b) states that the optimal quality choices obtained by solving OPTSTAT are unique.
Part (c) shows that for any video client $\iiN$, the optimal quality choices for any two segments with the same QR tradeoff are the same
irrespective of their lengths. In part (d), we establish continuity of optimal value OPTSTATVAL$_{\delta}$ of OPTSTAT$_{\delta}$ at $\delta=0$.

\begin{lemma}\label{results_about_OPTSTAT}
(a) OPTSTAT is a convex optimization problem satisfying Slater's condition.
\\(b) Let 
 $\left(\left(\left(q^{\pi,1}_i \left( f_i,l_i  \right) \right)_{\left(f_{i},l_i\right)\in{\cal F L}_i}\right)_{i\in{\cal N}},
 \left(\mathbf{r}^{\pi,1} \left(c\right)\right)_{c\in{\cal C}}\right)$ 
 and
 \\$\left(\left(\left(q^{\pi,2}_i \left( f_i,l_i  \right) \right)_{\left(f_{i},l_i\right)\in{\cal F L}_i}\right)_{i\in{\cal N}},
 \left(\mathbf{r}^{\pi,2} \left(c\right)\right)_{c\in{\cal C}}\right)$ 
 denote two optimal solutions to OPTSTAT. 
 Then, $q^{\pi,1}_i \left( f_i,l_i  \right)=q^{\pi,2}_i \left( f_i,l_i  \right)$
 for each $\left( f_i,l_i  \right)\in {\cal F L}_i$ for each $\iiN$.
 \\(c) Let 
   $\left(\left(\left(q^{\pi,1}_i \left( f_i,l_i  \right) \right)_{\left(f_{i},l_i\right)\in{\cal F L}_i}\right)_{i\in{\cal N}},
 \left(\mathbf{r}^{\pi,1} \left(c\right)\right)_{c\in{\cal C}}\right)$ 
 denote an optimal solution to OPTSTAT. 
 Then, given $\left( f_i,l_i  \right)\in {\cal F L}_i$,
 $q^{\pi,1}_i \left( f_i,l_i  \right)=q^{\pi,1}_i \left( f_i,l'_i  \right)$
 for each $l'_i$ such that $\left( f_i,l'_i  \right)\in {\cal F L}_i$ and for each $\iiN$, i.e.,
 quality choices for any two segments with the same QR tradeoff are the same
irrespective of their lengths.
\\(d) $\lim_{\delta\rightarrow 0}$ OPTSTATVAL$_{\delta}$=OPTSTATVAL.
\end{lemma}
\begin{IEEEproof}
Convexity properties of the objective and constraint functions of OPTSTAT are easy to establish using 
the convexity of the functions in ${\cal C}$ and $\cup_{\iiN}{\cal F}_i$ once 
we establish convexity of $\mbox{Var} \left( q_i\left( F^{\pi}_{i},L^{\pi}_i \right)  \right)$.
This can be done using arguments similar to those in Lemma 3 (a) of \cite{VJGdVAA_TAC2012},
and we can show that OPTSTAT is a convex optimization problem.
Using Assumption-SF, 
we can show that it also satisfies Slater's condition.

Proofs for part (b) is similar to that for Lemma 3 (b) in \cite{VJGdVAA_TAC2012}.

From (a), we can conclude that the KKT conditions are necessary and sufficient for optimality. 
Thus, there exist non-negative constants $\left(b^{\pi,1}_i\right)_{\iiN}$, $\left(d^{\pi,1}_i\right)_{\iiN}$, 
$\left(\left({\gamma}_i^{\pi,1}\left(f_{i},l_i\right)\right)_{\left(f_{i},l_i\right)\in {\cal FL}_i}\right)_{i\in{\cal N}}$
and
$\left(\left({\overline{\gamma}^{\pi,1}_i\left(f_{i},l_i\right)}\right)_{\left(f_{i},l_i\right)\in {\cal FL}_i}\right)_{i\in{\cal N}}$,
such that
\begin{eqnarray}
\nonumber \left(U^E_i\right)^{'} \left(  m^{\pi,1}_i	- U^V_i \left(  v^{\pi,1}_i \right)\right)
\left(1- 2 \left(U^V_i \right)^{'}\left( v^{\pi,1}_i \right) 
\left(q^{\pi,1}_i\left(f_{i},l_i\right) -    m^{\pi,1}_i  \right)\right) \hspace{-5.7cm} &&
\\\nonumber	 + \gamma^{\pi,1}_i\left(f_{i},l_i\right) - \overline{\gamma}^{\pi,1}_i\left(f_{i},l_i\right)
-  \frac{b^{\pi,1}_i}{\left(1+\overline{\beta}_i\right)} \left(f_{i}\right)^{'} \left(q^{\pi,1}_i\left(f_{i},l_i\right)\right)
 \hspace{-5.7cm}&&
\\	 -  p^d_i \frac{d^{\pi,1}_i}{ \overline{p}_i} \left( f_{i}\right)^{'} \left(q^{\pi,1}_i\left(f_{i},l_i\right)\right) &=&0
	\  \forall \ \left(f_{i},l_i\right)\in {\cal FL}_i,\ \forall \ {i\in {\cal N}},
	\label{pi_1_KKT_OPTSTAT_gradient_zero_quality_condition}
\\\label{pi_1_complimentary_slackness_cond1}
 \gamma^{\pi,1}_i\left(f_{i},l_i\right) q^{\pi,1}_i\left(f_{i},l_i\right)	& = & 0, 		\  \forall \ \left(f_{i},l_i\right)\in {\cal FL}_i,\ \forall \ {i\in {\cal N}},
	\\\label{pi_1_complimentary_slackness_cond2}
	 \overline{\gamma}^{\pi,1}_i\left(f_{i},l_i\right) \left(q_{\max} - q^{\pi,1}_i\left(f_{i},l_i\right)\right)  & = & 0 , 		
	\  \forall \ \left(f_{i},l_i\right)\in {\cal FL}_i,\ \forall \ {i\in {\cal N}}.
\end{eqnarray}
where for each $i\in{\cal N}$,
\begin{eqnarray*}
m^{\pi,1}_i &=& \frac{\exn \left[ L^{\pi}_i q^{\pi,1}_i\left( F^{\pi}_{i},L^{\pi}_i \right) \right]}{\exn \left[ L^{\pi}_i\right]} ,
\\v^{\pi,1}_i &=&   \mbox{Var} \left( q^{\pi,1}_i\left( F^{\pi}_{i},L^{\pi}_i \right)  \right).
\end{eqnarray*}
Using 
\eqref{pi_1_KKT_OPTSTAT_gradient_zero_quality_condition},
\eqref{pi_1_complimentary_slackness_cond1} and
\eqref{pi_1_complimentary_slackness_cond2}, 
we can conclude that,
given $\left( f_i,l_i  \right)\in {\cal F L}_i$,
 $q^{\pi,1}_i \left( f_i,l_i  \right)$ is an optimal solution to
 \begin{eqnarray}
\nonumber \max_{0\le q \le q_{\max}} \left(U^E_i\right)^{'} \left(  m^{\pi,1}_i	- U^V_i \left(  v^{\pi,1}_i \right)\right)
\left(q-  \left(U^V_i \right)^{'}\left( v^{\pi,1}_i \right) 
\left(q -    m^{\pi,1}_i  \right)^2\right)
	-  \frac{b^{\pi,1}_i}{\left(1+\overline{\beta}_i\right)} f_{i} \left(q\right)
 -  p^d_i \frac{d^{\pi,1}_i}{ \overline{p}_i}  f_{i} \left(q\right),
	\label{pi_1_KKT_OPTSTAT_sub_optmzn_problem}
\end{eqnarray} 
Using the above observation, we can conclude that $q^{\pi,1}_i \left( f_i,l_i  \right)$ is the unique optimal solution to
\\QNOVA$_i \left(\left(m^{\pi,1}_i,m^{\pi,1}_i,v^{\pi,1}_i,\left(h^B_i\right)^{-1}\left(b^{\pi,1}_i\right),\left(h^D_i\right)^{-1}\left(d^{\pi,1}_i\right)\right), 
f_i\right)$
(where the uniqueness is due to the strict concavity of the objective)
which is independent of $l_i$ and part (c) follows.
 
 Part (d) is a result related to the continuity of optimal value of OPTSTAT$_{\delta}$
 and this follows from Theorem 2.1 in \cite{fiacco_sensitivity_analysis}
	which provides sufficient conditions for verifying continuity of the optimal value.
 We can verify that the conditions given in Theorem 2.1 by noting that the objective 
 function $\phi_{\pi}\left( . \right)$ (defined in \eqref{OPTSTAT_objective_expression})
 of OPTSTAT
 is a continuous function (which follows from the continuity of the functions $\mbox{Mean} \left( . \right)$,
  $\mbox{Var} \left( . \right)$ and $U^V_i \left(. \right)$),
 and by establishing  the upper semicontinuity, lower semicontinuity and compactness at $\delta=0$
 of the feasible region ${\cal QR}_{\delta}$ of OPTSTAT$_{\delta}$ defined below
 \begin{eqnarray*}
 {\cal QR}_{\delta}&=&   \left\{ \left(\left(\left(q_i \left( f_i,l_i  \right) \right)_{\left(f_{i},l_i\right)\in{\cal F L}_i}\right)_{i\in{\cal N}},
 \left(\mathbf{r} \left(c\right)\right)_{c\in{\cal C}}\right): \left(\left(\left(q_i \left( f_i,l_i  \right) \right)_{\left(f_{i},l_i\right)\in{\cal F L}_i}\right)_{i\in{\cal N}}, \left(\mathbf{r} \left(c\right)\right)_{c\in{\cal C}}\right)
 \right.
 \\&&\left. \ \mbox{ satisfies the constraints
 \eqref{OPTSTAT_eta_feasibility}, \eqref{OPTSTAT_eta_q_non_neg}, \eqref{OPTSTAT_eta_q_upper_bound}, 
\eqref{OPTSTAT_eta_r_non_neg}, \eqref{delta_modified_OPTSTAT_eta_cost_bound} and \eqref{delta_modified_OPTSTAT_eta_bound_rebuf}}
 \right\}.
  \end{eqnarray*}
  The compactness of ${\cal QR}_{\delta}$ follows from the boundedness of the set,
  and the continuity of the functions associated with the constraints 
  \eqref{OPTSTAT_eta_feasibility}, \eqref{OPTSTAT_eta_q_non_neg}, \eqref{OPTSTAT_eta_q_upper_bound}, 
\eqref{OPTSTAT_eta_r_non_neg}, \eqref{delta_modified_OPTSTAT_eta_cost_bound} and \eqref{delta_modified_OPTSTAT_eta_bound_rebuf}.
Proofs of upper and lower semicontinuity of ${\cal QR}_{\delta}$, although not difficult, are long and have been omitted for brevity.

\end{IEEEproof}

We let $  \left(\left(q^{\pi}_i\left(f\right) \right)_{f\in{\cal F}_i}\right)_{i\in{\cal N}}$
denote the optimal quality choices associated with different quality rate tradeoffs.
Note that we have dropped the dependence of the optimal quality choices on segment length based 
on the observation in Lemma \ref{results_about_OPTSTAT} (c).
Let $ \left( \left(\left(q^{\pi}_i\left(f\right) \right)_{f\in{\cal F}_i}\right)_{i\in{\cal N}},
\left(\mathbf{r}^{\pi} \left(c\right)\right)_{c\in{\cal C}}\right)$
be an optimal solution to OPTSTAT, and let $\mathbf{b}^{\pi}$ and $\mathbf{d}^{\pi}$ denote the associated Lagrange multipliers for the 
constraints \eqref{OPTSTAT_eta_cost_bound} and  \eqref{OPTSTAT_eta_bound_rebuf} respectively.
Using the above result, we can conclude that the KKT conditions are necessary and sufficient for optimality. 
Hence, there exist non-negative constants (referred to in the sequel as Lagrange multipliers associated with 
the optimal solution)
$\left(\chi^{\pi}(c)\right)_{c\in{\cal C}}$, 
$\left(\left({\gamma}_i^{\pi}(f)\right)_{f\in{\cal F}_i}\right)_{i\in{\cal N}}$,
$\left(\left({\overline{\gamma}_i(f)}^{\pi}\right)_{f\in{\cal F}_i}\right)_{i\in{\cal N}}$,
$\left(\bm{\omega}^{\pi}(c)\right)_{c\in{\cal C}}$,
$\mathbf{d}^{\pi}$ and $\mathbf{b}^{\pi}$
such that
\begin{eqnarray}
\nonumber \left(U^E_i\right)^{'} \left(  m^{\pi}_i	- U^V_i \left(  v^{\pi}_i \right)\right)
\left(1- 2 \left(U^V_i \right)^{'}\left( v^{\pi}_i \right) 
\left(q^{\pi}_i(f) -    m^{\pi}_i  \right)\right)+ \gamma^{\pi}_i(f) - \overline{\gamma}^{\pi}_i(f)\hspace{-4.75cm}&&
	\\  	-  \frac{b^{\pi}_i}{\left(1+\overline{\beta}_i\right)} \left(f\right)^{'} \left(q^{\pi}_i(f)\right)
 -  p^d_i \frac{d^{\pi}_i}{ \overline{p}_i} \left(f\right)^{'} \left(q^{\pi}_i(f)\right) &=&0,
	\  \forall \ f\in {\cal F}_i, \forall \ {i\in {\cal N}},
	\label{KKT_OPTSTAT_gradient_zero_quality_condition}
\\ -  \chi^{\pi}(c) \frac{ \partial c \left(\mathbf{r}^{\pi}(c)\right) }{\partial r_i}
	  + \frac{b^{\pi}_i }{\tau_{slot}}
	  + \omega^{\pi}_{i}(c) &=&0, \  \forall \ c\in{\cal C}, \forall \ {i\in {\cal N}},
	  \label{KKT_OPTSTAT_gradient_zero_rate_condition}
\end{eqnarray}
\begin{eqnarray}\label{complimentary_slackness_cond1}
 \gamma^{\pi}_i(f) q^{\pi}_i(f)	& = & 0, 		\  \forall \ f\in {\cal F}_i, \forall \ {i\in {\cal N}},
	\\\label{complimentary_slackness_cond2}
	 \overline{\gamma}^{\pi}_i(f) \left(q_{\max} - q^{\pi}_i(f)\right)  & = & 0 , 		
	\  \forall \ f\in {\cal F}_i, \forall \ {i\in {\cal N}},
	\\\label{complimentary_slackness_cond3}
	\chi^{\pi}(c)   c \left(\mathbf{r}^{\pi}(c)\right)&=&0,
	\\\label{complimentary_slackness_cond4}  
	\frac{\omega^{\pi}_{i}(c)}{K_S} \left(r^{\pi}_{i}(c)- r_{i,\min}\right)	 & = & 0, 	\  \forall \ c\in{\cal C}, \forall \ {i\in {\cal N}}, 
	 \\\label{complimentary_slackness_cond5} d^{\pi}_i\left(  1
	 -   \frac{p^d_i \sigma^{\pi}_i  } 
{  \overline{p}_i } \right) & = & 0 \ \forall \ i\in {\cal N},
	 \\\label{complimentary_slackness_cond6} b^{\pi}_i\left(  \frac{\sigma^{\pi}_i } {\left(1+\overline{\beta}_i\right)} - \frac{ \rho^{\pi}_i }{ \tau_{slot} } \right) & = & 0   \ \forall \ i\in {\cal N}.
\end{eqnarray}
where for each $i\in{\cal N}$,
\begin{eqnarray}
\label{defn_mpi}
m^{\pi}_i &=& \frac{\exn \left[ L^{\pi}_i q^{\pi}_i\left( F^{\pi}_{i} \right) \right]}{\exn \left[ L^{\pi}_i\right]} ,
\\\label{defn_vpi}
v^{\pi}_i &=&   \mbox{Var} \left( q^{\pi}_i\left( F^{\pi}_{i} \right)  \right),
\\\label{defn_sigmapi}
\sigma^{\pi}_i &=& \frac{ \exn \left[L^{\pi}_i F^{\pi}_{i} \left( q^{\pi}_i\left( F^{\pi}_{i} \right) \right) \right] } 
{  \exn \left[L^{\pi}_i   \right]},
\\\label{defn_lambdapi}
\lambda^{\pi}_i &=& \exn \left[ L^{\pi}_i\right].
\end{eqnarray}
Thus $m^{\pi}_i$, $v^{\pi}_i$ and $\sigma^{\pi}_i$ are the (statistical) mean quality, variance in quality and mean segment size for video client $i$
associated with optimal solution to OPTSTAT.
Also, let
\begin{eqnarray}\label{defn_xpi}
{\cal X}^{\pi} &=& \left\{\left(\bm{\rho}^{\pi},\mathbf{b}^{\pi},\mathbf{d}^{\pi}\right):
\mbox{there is an optimal solution }\right.
\\&&\nonumber\left.
\left(\left(\left(q^{\pi}_i\left(f\right) \right)_{f\in{\cal F}_i}\right)_{i\in{\cal N}},
\left(\mathbf{r}^{\pi} \left(c\right)\right)_{c\in{\cal C}}\right)
\mbox{to OPTSTAT with }\rho^{\pi}_i = \exn \left[  r^{\pi}_i \left(C^{\pi}\right)\right] \mbox{ for each } i\in{\cal N}, \mbox{ and with}\right.
\\&&\nonumber\left.
\mbox{$\mathbf{b}^{\pi}$ and $\mathbf{d}^{\pi}$ as the associated optimal Lagrange multipliers }
\mbox{for constraints \eqref{OPTSTAT_eta_cost_bound} and  \eqref{OPTSTAT_eta_bound_rebuf} respectively}
\right\}.
\end{eqnarray}

In the next result, we present three useful properties of the optimal solution to OPTSTAT.
The result in part (a) below provides a video client level optimality result 
which essentially suggests that we can decouple the quality adaptation of the video clients.
It states that the component $\left(q^{\pi}_i\left(f\right) \right)_{f\in{\cal F}_i}$ 
of the optimal solution to OPTSTAT
associated with video client $\iiN$ is itself an optimal solution to an optimization problem 
which can be solved by the video client $i$.
This result hints at the possibility of distributing the task of quality adaptation across the video clients
so that each video client manages its own adaptation.
The result in part (b) points out that we only need to know a few parameters (specifically, the optimal Lagrange multipliers associated with the rebuffering constraints)
associated with the quality adaptation to carry out optimal resource allocation.
This suggests that we could potentially decouple the task of optimal resource allocation
from quality adaptation.
Part (c) states that that when  NOVA parameter $\bm{\theta}_{i,s}$
of video client $i$ is in the set ${\cal H}^*_i$ defined below
\begin{eqnarray}\label{defn_Hstar_i}
{\cal H}^*_i \defeq \left\{\left(m^{\pi}_i,m^{\pi}_i,v^{\pi}_i,\left(h^B_i\right)^{-1}\left(b^{\pi}_i\right),
\left(h^D_i\right)^{-1}\left(d^{\pi}_i\right)\right):
 \left(\bm{\rho}^{\pi},\mathbf{b}^{\pi},\mathbf{d}^{\pi}\right)\in{\cal X}^{\pi}\right\},
\end{eqnarray}
we can obtain optimal quality choices for OPTSTAT by using NOVA.

\begin{lemma}\label{optimality_of_q_pi}
For parts  (a) and (b) of this result, suppose 
 $\left(\bm{\rho}^{\pi},\mathbf{b}^{\pi},\mathbf{d}^{\pi}\right)\in{\cal X}^{\pi}$
 and let the associated optimal solution be
$\left(\left(\left(q^{\pi}_i\left(f\right) \right)_{f\in{\cal F}_i}\right)_{i\in{\cal N}},
\left(\mathbf{r}^{\pi} \left(c\right)\right)_{c\in{\cal C}} \right)$.
\\(a) For each $\iiN$,   $\left(q^{\pi}_i\left(f\right) \right)_{f\in{\cal F}_i}$ 
is the unique optimal solution to the following optimization problem
\begin{eqnarray*}
&& \hspace{-2cm} \max_{  \left(\left(q_i\left(f\right) \right)_{f\in{\cal F}_i}\right)} U^E_i \left(   \frac{\exn \left[ L^{\pi}_i q_i\left( F^{\pi}_{i} \right) \right]}{\exn \left[ L^{\pi}_i\right]}	- U^V_i \left( \mbox{Var} \left( q_i\left( F^{\pi}_{i} \right)  \right) \right)\right)
\\&& \hspace{-1cm}\nonumber  - \sum_{i\in{\cal N}}  d^{\pi}_i \left( \frac{ p^d_i   }{  \overline{p}_i }  \right)
\left( \frac{ \exn \left[L^{\pi}_i F^{\pi}_{i} \left( q_i\left( F^{\pi}_{i} \right) \right) \right] } 
{  \exn \left[L^{\pi}_i   \right]}    \right)
 -  \sum_{i\in{\cal N}}  \frac{ b^{\pi}_i}{ \left(1+ \overline{\beta}_i\right)} 
 \left(  \frac{ \exn \left[L^{\pi}_i F^{\pi}_{i} \left( q_i\left( F^{\pi}_{i} \right) \right) \right] } 
{  \exn \left[L^{\pi}_i   \right]}  \right),
\\  q_i(f) & \ge & 0, \  \forall \ f\in{\cal F}_i,
\\   q_i(f) & \le &q_{\max}, \  \forall \ f\in{\cal F}_i.
 \end{eqnarray*}
 \\(b)  $\left(\mathbf{r}^{\pi} \left(c\right)\right)_{c\in{\cal C}}$ is an optimal solution to the following optimization problem
\begin{eqnarray*}
 && \hspace{-3cm}\exn \left[\sum_{i\in{\cal N}} b^{\pi}_i r_i \left(C^{\pi}\right)  \right] ,
\\ \mbox{ s.t. }  c\left(\mathbf{r}\left(c\right))\right) &\le& 0, \ \forall \ c\in {\cal C},
\\   r_{i}\left(c\right)  &\ge& r_{i,\min}, \ \forall \ c\in {\cal C}, \forall \ {i\in {\cal N}}.
\end{eqnarray*}
\\(c) The following holds for each $\iiN$: If $\bm{\theta}^{\pi}_i\in {\cal H}^*_i$, then
$q^*_i \left(\bm{\theta}^{\pi}_i,f \right) = q^{\pi}_i\left(f\right)$
for each $f\in{\cal F}_i$.
\end{lemma}
\begin{IEEEproof}
As with the case of OPTSTAT, we can show that KKT conditions are necessary and sufficient for optimality 
for the optimization problem considered in part (a).
Now, the result follows by using \eqref{KKT_OPTSTAT_gradient_zero_quality_condition}, \eqref{complimentary_slackness_cond1} and \eqref{complimentary_slackness_cond2}
to conclude that $\left(q^{\pi}_i\left(f\right) \right)_{f\in{\cal F}_i}$ 
satisfies these conditions.
Proof of part (b) is similar to that of (a), and can be completed by using
the fact that $\left(\mathbf{r}^{\pi} \left(c\right)\right)_{c\in{\cal C}}$ satisfies \eqref{KKT_OPTSTAT_gradient_zero_rate_condition},
\eqref{complimentary_slackness_cond3} and \eqref{complimentary_slackness_cond4}.

Using the necessary optimality conditions for OPTSTAT given in \eqref{KKT_OPTSTAT_gradient_zero_quality_condition}, \eqref{complimentary_slackness_cond1} and \eqref{complimentary_slackness_cond2},
 we can show that $q^{\pi}_i\left(f\right)$ satisfies the 
 sufficient optimality conditions  \eqref{KKT_QNOVA_gradient_zero}-\eqref{KKTQRNOVA_bounded_above_constraints} for QNOVA$_i\left(\bm{\theta}^{\pi}_i,f \right) $
 (following an approach similar to that used in the proof of part (c) of Lemma \ref{results_about_OPTSTAT}).
 Then part (c) follows from the fact that QNOVA$_i\left(\bm{\theta}^{\pi}_i,f \right) $
 has a unique optimal solution.

\end{IEEEproof}

We use the observation in part (c) and properties of OPTSTAT to prove the next result which is 
the main result for this subsection and is an important intermediate result used in the proof of 
main optimality of NOVA given in Theorem \ref{main_optimality_theorem}.
The result states that the performance of NOVA with its parameters $\bm{\theta}_{i,s}$ picked from the set ${\cal H}^*_i$ for each $\iiN$
is asymptotically optimal.
Further, this result suggests that we can prove Theorem \ref{main_optimality_theorem}
if we can show that the updates \eqref{m_update_NOVA}-\eqref{lambda_update_NOVA} of NOVA guide 
the 
parameters $\left(\bm{\theta}_{i,s}\right)_{s\ge 1}$ of video client $i$ to ${\cal H}^*_i$
for each video client $\iiN$.
This motivates the study of convergence behavior of NOVA which is the main focus of the rest of this section.

\begin{theorem}\label{main_NOVA_with_theta_pi_is_optimal}
Suppose $\bm{\theta}^{\pi}_i \in {\cal H}^*_i$ for each $\iiN$.
 Then, for almost all sample paths
\begin{eqnarray*}
\lim_{S\rightarrow \infty} 
\left(\phi_S\left(\left( \left( q^*_i \left(\bm{\theta}^{\pi}_i , f_{i,s} \right)\right)_{\iiN} \right)_{1\le s \le S}  \right)
-\phi^{opt}_S\right)
=0.
\end{eqnarray*}
\end{theorem}
\begin{IEEEproof}
For a fixed $S$, 
consider an optimal solution $\left(\left(\mathbf{q}^S\right)_{1:S},\left(\mathbf{r}^S\right)_{1:K_S}\right)$ to OPT$(S)$.
Without loss of generality (we prove this below), we assume that the optimal solution $\left(\left(\mathbf{q}^S\right)_{1:S},\left(\mathbf{r}^S\right)_{1:K_S}\right)$ 
satisfies the following two conditions:
\\(a) $q^{S}_{i,s_1}=q^{S}_{i,s_2}$ for any two segments $s_1$ and $s_2$ such that
$f_{i,s_1}=f_{i,s_2}$ and $l_{i,s_1}=l_{i,s_2}$. 
\\(b) $\mathbf{r}^S_{k_1}=\mathbf{r}^S_{k_2}$ for any two slots $k_1$ and $k_2$ such that
$c_{k_1}=c_{k_2}$.
\\We show that we can always find an optimal solution satisfying these conditions.
Consider an optimal solution $\left(\left(\mathbf{q}^{0,S}\right)_{1:S},\left(\mathbf{r}^{0,S}\right)_{1:K_S}\right)$ to OPT$(S)$
that does not satisfy the conditions.
We can obtain another optimal solution $\left(\left(\mathbf{q}^{1,S}\right)_{1:S},\left(\mathbf{r}^{1,S}\right)_{1:K_S}\right)$ to OPT$(S)$
satisfying this condition by letting
\begin{eqnarray*}
q^{1,S}_{i,s'} &=& \frac{ \sum_{s=1}^{S} I \left(f_{i,s}=f_{i,s'},l_{i,s}=l_{i,s'}\right) q^{0,S}_{i,s}}
{ \sum_{s=1}^{S} I \left(f_{i,s}=f_{i,s'},l_{i,s}=l_{i,s'}\right) },
\ \forall \ 1\le s' \le S,
\\r^{1,S}_{i,k'}&=& \frac{ \sum_{k=1}^{K_S} I \left( c_{k}=c_{k'}\right) r^{0,S}_{i,k}}
{ \sum_{k=1}^{K_S}I \left( c_{k}=c_{k'}\right) },
\ \forall \ 1\le k' \le K_S.
\end{eqnarray*}
It is clear that $\left(\left(\mathbf{q}^{1,S}\right)_{1:S},\left(\mathbf{r}^{1,S}\right)_{1:K_S}\right)$ satisfies the conditions (a) and (b).
Further, using the structure of the optimization problem OPT$(S)$,
we can show that $\left(\left(\mathbf{q}^{1,S}\right)_{1:S},\left(\mathbf{r}^{1,S}\right)_{1:K_S}\right)$
is also an optimal solution to OPT$(S)$.

Now, we return to the proof of the main result and
consider an optimal solution $\left(\left(\mathbf{q}^S\right)_{1:S},\left(\mathbf{r}^S\right)_{1:K_S}\right)$ to OPT$(S)$
satisfying the conditions (a) and (b) so that the component $q^S_{i,s}$ in the optimal solution assocaited with quality adaptation
for segment $s$ of video client $i$ depends only on $(f,l)$.
Hence, we can obtain a function $q^S_i(f,l)$ for $(f,l)\in {\cal FL}_i$, 
such that $q^S_i(f,l)$ denotes the quality associated with this optimal solution for a segment $s$ with QR tradeoff $f_{i,s}=f$ and length $l_{i,s}=l$.
Similarly, we can obtain a function $r^S_i(c)$ for $\ciC$
such that $r^S_i(c)$ denotes the resource allocation associated with the optimal solution for a slot $k$ with allocation constraint $c_k=c$.
Let  $\pi^{{\cal F},{\cal L},S}_i\left( f_{i},l_{i}  \right)
= \frac{\sum_{s=1}^{S} I( f_{i,s}=f_{i},\ l_{i,s}=l_{i}  )}{S}$ for  $(f,l)\in {\cal FL}_i$
be the empirical distribution for the occurrence of segments with QR tradeoff $f$ and length $l$,
and $\pi^{{\cal C},S}\left( c \right)
= \frac{\sum_{k=1}^{K_S} I( c_k=c  )}{K_S}$ for  $\ciC$
be the empirical distribution for the occurrence of allocation constraint $c$.

The mean quality for video client $i$ corresponding to the  optimal solution $\left(\left(\mathbf{q}^S\right)_{1:S},\left(\mathbf{r}^S\right)_{1:K_S}\right)$ 
is given by
\begin{eqnarray*}
m^S_i \left(q^S_i\right) &=&\frac{\sum_{s=1}^{S} l_{i,s} q^S_{i,s}}{\sum_{s=1}^{S} l_{i,s} }
= \frac{
\sum_{ \left( f_i,l_i  \right) \in {\cal FL}_i }\pi^{{\cal F},{\cal L},S}_i\left( f_i,l_i  \right) 
l_{i} q^S_i\left( f_i,l_i  \right)  }
{ \sum_{ \left( f_i,l_i  \right) \in {\cal FL}_i }\pi^{{\cal F},{\cal L},S}_i\left( f_i,l_i  \right) 
l_{i}}
\\&=& \frac{
\sum_{ \left( f_i,l_i  \right) \in {\cal FL}_i }\pi^{{\cal F},{\cal L}}_i\left( f_i,l_i  \right) 
l_{i} q^S_i\left( f_i,l_i  \right)  }
{ \sum_{ \left( f_i,l_i  \right) \in {\cal FL}_i } \pi^{{\cal F},{\cal L}}_i\left( f_i,l_i  \right) 
l_{i}}
 + \delta_m(S)
\end{eqnarray*}
where $\delta_m(S)$ is a function such that $\lim_{S\rightarrow \infty}\delta_m(S)=0$ a.s. (in this proof, `a.s.' stands for `for almost all sample paths').
This limiting behavior of $\delta_m(S)$ follows from the boundedness of the terms involved, and the
fact that 
$\left(L_{i,s} ,F_{i,s} \right)_{s\ge 0}$ is a stationary ergodic process
as a result of which 
$\lim_{S\rightarrow \infty}\left|\pi^{{\cal F},{\cal L},S}_i\left( f_i,l_i  \right)- \pi^{{\cal F},{\cal L}}_i\left( f_i,l_i  \right) \right|=0$ a.s.
for each  $(f,l)\in {\cal FL}_i$, i.e.,
the empirical distribution converges to the stationary distribution.
Recall that $\left(\pi^{{\cal F},{\cal L}}_i\left( f_i,l_i  \right)\right)_{\left(f_{i},l_i\right)\in{\cal F L}_i} $
is the marginal distribution associated with the stationary ergodic process $\left(F_{i,s},L_{i,s}\right)_{s\ge 0}$.
Using similar calculations, we can obtain a function $\delta_{e,1}(S)$
satisfying $\lim_{S\rightarrow \infty}\delta_{e,1}(S)=0$ a.s.
such that the optimal value for OPT$(S)$, i.e. $\phi^{opt}_S = \phi_S\left(\left(\mathbf{q}^S\right)_{1:S}\right)$
can be expressed as
\begin{eqnarray*}
  \phi^{opt}_S
 =\sum_{i\in {\cal N}} U^E_i \left(  \mbox{Mean} \left( q^S_i\left( F^{\pi}_{i},L^{\pi}_i \right)  \right)
	- U^V_i \left( \mbox{Var} \left( q^S_i\left( F^{\pi}_{i},L^{\pi}_i \right)  \right) \right)\right)
 	+ \delta_{e,1}(S).
 	\end{eqnarray*}
 	where the first term on the right hand side is equal to the objective function of OPTSTAT (given in \eqref{OPTSTAT_objective_expression})
 	evaluated at $  \left(\left(q^S_i \left( f_i,l_i  \right) \right)_{\left(f_{i},l_i\right)\in{\cal F L}_i}\right)_{i\in{\cal N}}$.

Again using the fact that $\lim_{S\rightarrow \infty}\left|\pi^{{\cal F},{\cal L},S}_i\left( f_i,l_i  \right)_i- \pi^{{\cal F},{\cal L}}\left( f_i,l_i  \right) \right| = 0$  a.s.
and
$\lim_{S\rightarrow \infty} \left|\pi^{{\cal C},S}\left( c  \right)- \pi^{{\cal C}}\left( c  \right) \right| = 0$  a.s.,
together with arguments similar to those above,
we can show that
  $\left(\left(\left(q^S_i \left( f_i,l_i  \right) \right)_{\left(f_{i},l_i\right)\in{\cal F L}_i}\right)_{i\in{\cal N}},
 \left(\mathbf{r}'_{S} \left(c\right)\right)_{c\in{\cal C}}\right)$ 
is a feasible solution to the optimization problem OPTSTAT$_{\delta_{e,2}(S)}$ (i.e., optimization problem OPTSTAT with constraints loosened by $\delta_{e,2}(S)$)
for an appropriately chosen function $\delta_{e,2}(S)$ satisfying $\lim_{S\rightarrow \infty}\delta_{e,2}(S)=0$ a.s.. 
Hence,
\begin{eqnarray}\label{ineq_realating_optstat_and_opt_values}
  \phi^{opt}_S \le \mbox{OPTSTATVAL}_{\delta_{e,2}(S)}	+ \delta_{e,1}(S)
 \end{eqnarray}
From Lemma \ref{results_about_OPTSTAT} (d), 
we have that $\lim_{\delta\rightarrow 0}$ OPTSTATVAL$_{\delta}$=OPTSTATVAL.
Using Lemma \ref{optimality_of_q_pi} (c) and the fact that $\left(F_{i,s},L_{i,s}\right)_{s\ge 0}$ is a stationary ergodic process for each $\iiN$,
we have 
\begin{eqnarray*}
\lim_{S\rightarrow \infty} 
\left( \phi_S\left(\left( \left( q^*_i \left(\bm{\theta}^{\pi}_i , f_{i,s} \right)\right)_{\iiN} \right)_{1\le s \le S}  \right)
-\mbox{OPTSTATVAL}\right)
=0 \ \mbox{a.s.}.
\end{eqnarray*}
Now, the result follows by using these two observations, \eqref{ineq_realating_optstat_and_opt_values} and the fact that 
$$\phi_S\left(\left( \left( q^*_i \left(\bm{\theta}^{\pi}_i , f_{i,s} \right)\right)_{\iiN} \right)_{1\le s \le S}  \right)
\le\phi^{opt}_S.$$
 
\end{IEEEproof}

\subsection{An auxiliary differential inclusion related to NOVA}
\label{relate_NOVA_to_diff_inclusion}
In the previous subsection, we stated Theorem \ref{main_NOVA_with_theta_pi_is_optimal}
which suggests that we can prove the main optimality result for NOVA 
if we establish an appropriate convergence result for NOVA.
In this subsection, we study 
an auxiliary differential inclusion 
which evolves according to average dynamics of NOVA.
The main goal of this subsection is to study the convergence of the differential inclusion 
which in turn will help us obtain the desired convergence of parameters of NOVA in the next subsection.

For the rest of this section, 
we additionally consider the evolution of auxiliary parameters
$ \left(\sigma_{i,s_i}\right)_{s_i\ge 1} $ and $ \left(\rho_{i,k}\right)_{k\ge 1} $
associated with NOVA
which evolve according to update rules discussed next.
We update the auxiliary parameter $\sigma_{i,s_i}$ based on the quality $q^*_{i,s_i+1}$
(shorthand for $q^*_i(\bm{\theta}_{i,s_i}, f_{i,s_i+1})$, $\bm{\theta}_{i,s_i}=(m_{i,s_i},\mu_{i,s_i},v_{i,s_i},b_{Q,i,s_i},d_{i,s_i})$)
chosen by NOVA for $(s_i+1)$th segment of video client $\iiN$
as follows:
\begin{eqnarray}
\sigma_{i,s_i+1} &=&  \sigma_{i,s_i} + \epsilon 
\left( \frac{ l_{i,s_i+1} f_{i,s_i}\left(  q^*_{i,s_i+1}\right)  }{\lambda_{i,s_i} } - \sigma_{i,s_i} \right),
\label{sigma_NOVA_update}
\end{eqnarray}
Thus, the auxiliary parameter $\sigma_{i,s_i}$ tracks the mean segment size of the segments downloaded by video client $\iiN$.
We update the parameter $\bm{\rho}_k$ based on the resource allocation $\mathbf{r}^*_k\in{\cal R}^*\left(\mathbf{b}_k,c_k\right)$
in slot $k$ as described below
\begin{eqnarray}\label{rho_NOVA_update}
  \rho_{i,k+1} &=&  \rho_{i,k} + \epsilon  \left(r^*_{i,k} - \rho_{i,k}\right) \ \forall \ \iiN.
\end{eqnarray}
Thus, the auxiliary parameter $\bm{\rho}_k$  tracks the mean resource allocation to video clients.
Note that the auxiliary parameters $\sigma_{i,.}$ and $\bm{\rho}_.$ 
do not affect the allocation or adaptation in NOVA.

Next, let
\begin{eqnarray}\label{defn_H}
{\cal H}&=& \left\{ \left(\mathbf{m},\bm{\mu},\mathbf{v},\mathbf{b},\mathbf{d},\bm{\lambda},\bm{\sigma},\bm{\rho}\right)\in \mathbb{R}^{8N}:
\mbox{ for each }\iiN, \right.
 \\\nonumber&&\left.0\le m_i,\mu_i\le q_{\max},
0\le v_i\le q^2_{\max},\ \underline{b}\le b_i\le \overline{b},\  \underline{d}\le d_i\le \overline{d}, \right.
 \\\nonumber&&\left.l_{\min}\le  \lambda_i \le l_{\max}, \ l_{\min}f_{\min}\le  \sigma_i \le l_{\max}f_{\max},\ r_{i,\min} \le \rho_i \le r_{\max} \right\}.
\end{eqnarray}
Note that the parameters $\left( \mathbf{m}_s,\bm{\mu}_s,\mathbf{v}_s,\mathbf{b}_k,\mathbf{d}_s,\bm{\lambda}_s,\bm{\sigma}_s,\bm{\rho}_k\right)_{s,k}$
associated with NOVA remain in ${\cal H}$ (see Lemma \ref{parameters_are_bounded}).
For each video client $\iiN$, we use the variables
$\widehat{m}_i(t)$, $\widehat{\mu}_i(t)$, $\widehat{v}_i(t)$, $\widehat{b}_i(t)$, $\widehat{d}_i(t)$, 
$ \widehat{\lambda}_i(t)$, $\widehat{\sigma}_i(t)$ and $\widehat{\rho}_i(t)$
to track the average dynamics of the parameters 
$m_{i,s_i}$, $\mu_{i,s_i}$, $v_{i,s_i}$, $b_{i,k}$, $d_{i,s_i}$, $\lambda_{i,s_i}$, $\sigma_{i,s_i}$ and $\rho_{i,k}$
respectively associated with NOVA (explained in detail in the sequel before Lemma \ref{differential_inclusion_nice}).
Let $\widehat{\bm{\Theta}}(t)=\left(\widehat{\mathbf{m}}(t),\widehat{\bm{\mu}}(t),\widehat{\mathbf{v}}(t),\widehat{\mathbf{b}}(t),
\widehat{\mathbf{d}}(t),\widehat{\bm{\lambda}}(t),\widehat{\bm{\sigma}}(t),\widehat{\bm{\rho}}(t)\right)\in {\cal H}$
and
$\widehat{\bm{\theta}}_i(t)=(\widehat{m}_i(t),\widehat{\mu}_i(t),\widehat{v}_i(t),\widehat{b}_i(t),\widehat{d}_i(t))$ for each $\iiN$,
i.e., $\widehat{\bm{\theta}}_i(t)$ includes the components in $\widehat{\bm{\Theta}}(t)$ that affect the quality adaptation of video client $\iiN$.

The main focus of this subsection is the following differential inclusion which describes 
the evolution of $\left(\widehat{\bm{\Theta}}(t)\right)_{t\ge 0}$:
\\\line(1,0){514}\vspace{-.5cm}
\begin{center}
Auxiliary differential inclusion related to NOVA\vspace{-.7cm}
\end{center}
\line(1,0){514}
\\$\widehat{\bm{\Theta}}(0) \in {\cal H}$
and for almost all $t\ge 0$ and each $i\in{\cal N}$,
\begin{eqnarray}\label{mhat_ode_rule}
 \bdt{\widehat{m}}_i(t) &=&
 \frac{ \left(U^E_i\right)^{'}\left( \widehat{\mu}_i(t) - U^V_i \left(\widehat{v}_i(t)\right)\right)
 \left(U^V_i\right)^{'} \left(\widehat{v}_i(t)\right)}
 {u_i \left(\widehat{\bm{\Theta}}(t) \right)}
\\\nonumber &&\left( \frac{ E \left[ L^{\pi}_i q^*_i\left( \widehat{\bm{\theta}}_i(t) ,F^{\pi}_{i} \right)\right]}{\widehat{\lambda}_i(t)} 
 - \widehat{m}_i(t)   \right),
\\ \bdt{\widehat{\mu}}_i(t) &=&
 \frac{1} {u_i \left(\widehat{\bm{\Theta}}(t) \right)}\left( \frac{ E \left[ L^{\pi}_i q^*_i\left( \widehat{\bm{\theta}}_i(t) ,F^{\pi}_{i} \right)\right]}{\widehat{\lambda}_i(t)} 
 - \widehat{\mu}_i(t)   \right),
 \label{muhat_ode_rule}
 \\\bdt{\widehat{v}}_i(t) &=& \frac{1}{u_i \left(\widehat{\bm{\Theta}}(t) \right)}\left( \frac{E \left[  L^{\pi}_i\left(  q^*_i\left( \widehat{\bm{\theta}}_i(t) ,F^{\pi}_{i} \right) -\widehat{m}_i(t) \right)^2\right] }{\widehat{\lambda}_i(t)}  
  - \widehat{v}_i(t)   \right),
\label{vhat_ode_rule}
\\ \bdt{\widehat{b}}_i(t) &=&  \frac{1}{ \left(1+ \overline{\beta}_i\right)  } 
-    \frac{ E \left[  L^{\pi}_i\right] }{u_i \left(\widehat{\bm{\Theta}}(t) \right)} + \widehat{z}^b_i \left(\widehat{\bm{\Theta}}(t)\right),
\label{bhat_ode_rule}
\\\bdt{\widehat{d}}_i(t) &=& \frac{1}{u_i \left(\widehat{\bm{\Theta}}(t) \right)}
\left( \frac{p^d_i E \left[  L^{\pi}_i F^{\pi}_{i}\left(  q^*_i\left( \widehat{\bm{\theta}}_i(t) ,F^{\pi}_{i} \right)\right)\right]  }{  \overline{p}_i } - \widehat{\lambda}_i(t)   \right)
\\&&\nonumber+ \widehat{z}^d_i \left(\widehat{\bm{\Theta}}(t)\right),
\label{dhat_ode_rule}
\\ \bdt{\widehat{\lambda}}_i(t) &=& \frac{1}{u_i \left(\widehat{\bm{\Theta}}(t) \right)} \left(E \left[  L^{\pi}_i\right] - \widehat{\lambda}_i(t)\right),
\label{lambdahat_ode_rule}
\\\bdt{\widehat{\sigma}}_i(t) &=& \frac{1}{u_i \left(\widehat{\bm{\Theta}}(t) \right)}
\left(\frac{ E \left[  L^{\pi}_i F^{\pi}_{i}\left(  q^*_i\left( \widehat{\bm{\theta}}_i(t) ,F^{\pi}_{i} \right)\right)\right]  }{ \widehat{\lambda}_i(t) } - 
\widehat{\sigma}_i(t) \right),
\label{sigma_ode_rule}
\\\bdt{\widehat{\rho}}_i(t) &=& \frac{1}{\tau_{slot}}\left( \frac{  \overline{r}^*_i\left(\widehat{\mathbf{b}}(t)\right)  }{\tau_{slot}}- \widehat{\rho}_i(t) \right),
\label{rhohat_ode_rule}
\end{eqnarray}
where 
\begin{eqnarray} 
u_i \left(\widehat{\bm{\Theta}}(t) \right)= \tau_{slot} \frac{ E \left[  L^{\pi}_i F^{\pi}_{i}\left(  q^*_i\left( \widehat{\bm{\theta}}_i(t) ,F^{\pi}_{i} \right)\right)\right]  }{ E\left[ r^*_i\left(\widehat{\mathbf{b}}(t),C^{\pi}\right)\right] },
\label{u_i_t_def}
\end{eqnarray}
and $\mathbf{r}^*\left(\widehat{\mathbf{b}}(t),c\right)\in {\cal R}^*\left(\widehat{\mathbf{b}}(t),c\right)$
for each $\ciC$.
Here, 
\begin{eqnarray}\label{defn_zb_and_zd}
\left(\widehat{\mathbf{z}}^b \left(\widehat{\bm{\Theta}}(t)\right),\widehat{\mathbf{z}}^d \left(\widehat{\bm{\Theta}}(t)\right)\right)
\in -{\cal Z}_{\cal H} \left(\bm{\Theta}\right).
\end{eqnarray}
\line(1,0){514}
\\Here $ \widehat{z}^b_i \left(\widehat{\bm{\Theta}}(t)\right)$
and  $\widehat{z}^d_i \left(\widehat{\bm{\Theta}}(t)\right)$
are terms mimicking the role of the operators
$\left[  . \right]_{\underline{b}}$ and $\left[ .\right]_{\underline{d}}$
in \eqref{b_adapt_update} and \eqref{d_adapt_update},
and ensure that $\left(\widehat{\bm{\Theta}}(t)\right)_{t\ge 0}$ stays in ${\cal H}$
(see Section 4.3 of \cite{kushner_text} for a discussion about projected stochastic approximation).
For $\boldsymbol{\Theta}=\left(\mathbf{m},\bm{\mu},\mathbf{v},\mathbf{b},\mathbf{d},\bm{\lambda},\bm{\sigma},\bm{\rho}\right)\in{\cal H}$, 
${\cal Z}_{\cal H} \left(\boldsymbol{\Theta} \right)\subset \mathbb{R}^{2N}$ is the set containing only the zero element 
when $\left(\mathbf{b},\mathbf{d}\right)$ is in the interior of the set
$$
{\cal H}_{\cal BD}=\left\{\left(\mathbf{b},\mathbf{d}\right)\in \mathbb{R}^{2N}: \mbox{ for each }\iiN, \ b_i \ge \underline{b},\ d_i \ge \underline{d} \right\},
$$
and for $\left(\mathbf{b},\mathbf{d}\right)$ on the boundary of the set ${\cal H}_{\cal BD}$, 
${\cal Z}_{\cal H} \left(\boldsymbol{\Theta} \right)$ is the convex cone generated by the 
outer normals at $\left(\mathbf{b},\mathbf{d}\right)$ of the faces of ${\cal H}_{\cal BD}$ on which $\left(\mathbf{b},\mathbf{d}\right)$ lies.
Thus, for a given $\boldsymbol{\Theta} $, $-{\cal Z}_{\cal H} \left(\boldsymbol{\Theta} \right)$
contains reflection terms pointing in the right directions to keep $\left(\widehat{\bm{\Theta}}(t)\right)_{t\ge 0}$  in ${\cal H}$.
For $\bm{\Theta}$ with $\left(\mathbf{b},\mathbf{d}\right)$ in the interior of ${\cal H}_{\cal BD}$,
\begin{eqnarray}\label{reflection_terms_are_zero_for_theta_in_interior}
\widehat{z}^b_i = 0, \ \widehat{z}^d_i = 0, \ \forall \ \iiN, 
\ \forall \ \left(\widehat{\mathbf{z}}^b ,\widehat{\mathbf{z}}^d\right) \in -{\cal Z}_{\cal H}\left(\bm{\Theta}\right),
\end{eqnarray}
Also note that for all $\bm{\Theta}\in{\cal H}$
\begin{eqnarray}\label{reflection_terms_are_non_negative}
\widehat{z}^b_i \ge 0, \ \widehat{z}^d_i \ge 0, \ \forall \ \iiN, 
\ \forall \ \left(\widehat{\mathbf{z}}^b ,\widehat{\mathbf{z}}^d\right) \in -{\cal Z}_{\cal H}\left(\bm{\Theta}\right),
\end{eqnarray}
i.e., the components of all the terms in $-{\cal Z}_{\cal H}\left(\bm{\Theta}\right)$ are non-negative
and this is clear from the definition of the set ${\cal H}_{\cal BD}$ which indicates that the reflection terms are needed only 
needed when the parameters hit a lower bound.
Thus, $ \widehat{z}^b_i \left(\widehat{\bm{\Theta}}(t)\right)$
and  $\widehat{z}^d_i \left(\widehat{\bm{\Theta}}(t)\right)$
are terms mimicking the role of the operators
$\left[  . \right]_{\underline{b}}$ and $\left[ .\right]_{\underline{d}}$.
Also, note that $u_i \left(. \right)$ is a set valued map (and hence \eqref{mhat_ode_rule}-\eqref{rhohat_ode_rule} describes 
a differential inclusion) since the denominator  
 $E\left[ r^*_i\left(\widehat{\mathbf{b}}(t),C^{\pi}\right)\right]$ in  \eqref{u_i_t_def}
 is a set valued map.
Finally, note that the above definition only requires that $\left(\widehat{\bm{\Theta}}(t)\right)_{t\ge 0}$
is differentiable for \emph{almost} all $t\ge 0$, i.e., 
we are considering the class of absolutely continuous functions $\left(\widehat{\bm{\Theta}}(t)\right)_{t\ge 0}$
that satisfy \eqref{mhat_ode_rule}-\eqref{rhohat_ode_rule}.

Although we will rigorously establish the relationship between the evolution of parameters of NOVA
and \eqref{mhat_ode_rule}-\eqref{rhohat_ode_rule} in the next subsection,
we can see that the differential inclusion \eqref{mhat_ode_rule}-\eqref{rhohat_ode_rule}
reflects the average dynamics of the evolution of parameters in NOVA
by comparing \eqref{mhat_ode_rule}-\eqref{rhohat_ode_rule} against the update rules 
\eqref{m_update_NOVA}-\eqref{lambda_update_NOVA} and \eqref{sigma_NOVA_update}-\eqref{rho_NOVA_update}
in NOVA.
For instance, this is apparent when we compare the update rule 
$$\mu_{i,s_i+1}-  \mu_{i,s_i} = 
  \epsilon  \left( \frac{l_{i,s_i+1}}{\lambda_{i,s_i}} q^*_{i,s_i+1}- \mu_{i,s_i}   \right)$$
  for NOVA parameter $\mu_{i,s_i+1}$ given in \eqref{mu_update_NOVA},
  against \eqref{muhat_ode_rule} describing the evolution of
the parameter $ \widehat{\mu}_i(t)$.
Note that the rate of change of $ \widehat{\mu}_i(t)$ given in \eqref{muhat_ode_rule}
has a scaling term $\frac{1}{u_i \left(\widehat{\bm{\Theta}}(t) \right)} $
which corresponds to the segment download rate of video client $i$ at time $t$
(and $u_i \left(\widehat{\bm{\Theta}}(t) \right)$ defined in \eqref{u_i_t_def} corresponds to expected segment download time of video client $i$ at time $t$).
This scaling by segment download rate is naturally expected 
for the rate of change of parameters
$\widehat{m}_i(t)$, $\widehat{\mu}_i(t)$, $\widehat{v}_i(t)$, $\widehat{d}_i(t)$, $ \widehat{\lambda}_i(t)$,  and $\widehat{\sigma}_i(t)$
which correspond to NOVA parameters of video client $i$ that are updated when a segment download is completed,
and thus we can view $\frac{1}{u_i \left(\widehat{\bm{\Theta}}(t) \right)} $
as the update rate associated with these parameters.
Similarly, we can view the constant scaling term  $\frac{1}{\tau_{slot}}$ in \eqref{rhohat_ode_rule} describing the evolution of $ \widehat{\rho}_i(t) $
as the corresponding update rate
by noting that the associated (auxiliary) NOVA parameter $\rho_{i,k}$ is updated 
at the beginning of every slot, i.e., once every $\tau_{slot}$ seconds.
Finally, note that \eqref{bhat_ode_rule} describing the evolution of $ \widehat{b}_i(t) $
can be rewritten as
$$
 \bdt{\widehat{b}}_i(t) =\frac{1}{\tau_{slot}}\left( \frac{\tau_{slot}}{ \left(1+ \overline{\beta}_i\right)  } \right)
-    \frac{ 1 }{u_i \left(\widehat{\bm{\Theta}}(t) \right)} \left(E \left[  L^{\pi}_i\right]\right) + \widehat{z}^b_i \left(\widehat{\bm{\Theta}}(t)\right),
$$
and presence of the two scaling terms $\frac{1}{\tau_{slot}}$ and $\frac{1}{u_i \left(\widehat{\bm{\Theta}}(t) \right)} $ reflects the fact that
the corresponding NOVA parameter $b_{i,k}$ is updated 
at the beginning of every slot (using \eqref{b_allocate_update}) and when a segment download of video client $i$ is completed (using \eqref{b_adapt_update}).
Thus, we can expect that \eqref{mhat_ode_rule}-\eqref{rhohat_ode_rule}
captures the average dynamics of NOVA,
and the presence of the two video client dependent update rates $\frac{1}{\tau_{slot}}$ and $\frac{1}{u_i \left(\widehat{\bm{\Theta}}(t) \right)} $
reflects the \emph{asynchronous} nature of the evolution of parameters in NOVA where
different video clients are updating their parameters at their own (possibly time varying) rates.

Now, we study the differential inclusion \eqref{mhat_ode_rule}-\eqref{rhohat_ode_rule}
to identify properties that will help us to study convergence behavior.
The next result shows that the differential inclusion is `well behaved'.

\begin{lemma}\label{differential_inclusion_nice}
The differential inclusion \eqref{mhat_ode_rule}-\eqref{rhohat_ode_rule}
is well defined, i.e., 
there exists an absolutely continuous function that solves \eqref{mhat_ode_rule}-\eqref{rhohat_ode_rule}
for any $\widehat{\bm{\Theta}}(0) \in {\cal H}$.
Further, these solutions are Lipschitz continuous and stay in ${\cal H}$ and hence are bounded.
\end{lemma}
\begin{IEEEproof}
The existence of solution follows from 
the proof of Theorem \ref{main_stoch_approx_result} (in Subsection \ref{convergence_of_diff_inclusion_and_theorem})
where we obtain a solution 
satisfying \eqref{mhat_ode_rule}-\eqref{rhohat_ode_rule}.
The boundedness follows from
\eqref{defn_zb_and_zd} and arguments similar to those in Lemma \ref{parameters_are_bounded}.
The Lipschitz continuity of the paths follows from the fact that all the terms on the right hand side
of \eqref{mhat_ode_rule}-\eqref{rhohat_ode_rule} 
can be bounded above for $\widehat{\bm{\Theta}}(t) \in {\cal H}$.
\end{IEEEproof}

\begin{definition} \emph{Stationary resource allocation policy:}
Let $\left(\mathbf{r}(c)\right)_{c\in{\cal C}} $ be a $|{\cal C}|$ length vector (of vectors)
where $\mathbf{r}(c)\in{\mathbb R}^N_+$. 
We refer to $\left(\mathbf{r}(c)\right)_{c\in{\cal C}} $
as a stationary resource allocation policy 
as we can associate $\left(\mathbf{r}(c)\right)_{c\in{\cal C}} $ with a resource allocation policy
that allocates resource $\mathbf{r}(c)$ in a slot $k$ when $C_k=c$,
and thus the policy carries out the resource allocation in a slot only based on the allocation constraint in the slot.
\end{definition}
\begin{definition}\label{defn_feasiible_stationary_resource_allocation_policy}
\emph{Feasible stationary resource allocation policy:}
We say that a stationary resource allocation policy
$\left(\left(\mathbf{r} \left(c\right)\right)_{c\in{\cal C}}\right)$  is feasible if
\begin{eqnarray*}
\mathbf{r} \left(c\right) \ge \mathbf{r}_{\min} \mbox{ and } c \left(\mathbf{r} \left(c\right)\right) \le 0, \ \forall \ c\in{\cal C}.
\end{eqnarray*}
\end{definition}
\begin{definition}\emph{Stationary quality adaptation policy for video client $i$:}
\\Let $\left(q_i\left( f_i \right)\right)_{f_i \in {\cal F}_i} \in {\mathbb R}^{{\cal F}_i}_+$.
We refer to $\left(q_i\left( f_i \right)\right)_{f_i \in {\cal F}_i}$
as a stationary quality adaptation policy for video client $\iiN$ as we can associate 
$\left(q_i\left( f_i \right)\right)_{f_i \in {\cal F}_i}$
with a quality adaptation policy for video client $i$ that chooses quality $q_i\left( f_i \right)$
for each segment $s$ with QR tradeoff $F_{i,s}=f_i$,
and thus the policy carries out quality adaptation for a segment based only on the QR tradeoff of that segment.
\end{definition}
\begin{definition}\emph{Feasible stationary quality adaptation policy for video client $i$:}
We say that a stationary quality adaptation policy $\left(q_i\left( f_i \right)\right)_{f_i \in {\cal F}_i}$ for video client $i$
is feasible if
$0 \le q_i\left( f_i \right) \le q_{\max}$ for each $f_i \in {\cal F}_i.$
\end{definition}
Next, we define the set $\widetilde{\cal H}\subset {\mathbb{R}}^{8N}$ as 
\begin{eqnarray}\label{defn_calH}
\widetilde{{\cal H}} &=& \bigg\{ \left(\mathbf{m},\bm{\mu},\mathbf{v},\mathbf{b},\mathbf{d},\bm{\lambda},\bm{\sigma},\bm{\rho}\right)\in{\cal H}:
\ \lambda_i=E [ L^{\pi}_i ] \ \forall \ \iiN;
 \\\nonumber&&\exists \mbox{ a feasible stationary resource allocation policy} \left(\mathbf{r} \left(c\right)\right)_{c\in{\cal C}} \mbox{ s.t. }
  \frac{\exn \left[  r_i \left(C^{\pi} \right)  \right]}{\tau_{slot}}=\rho_{i} \ \forall \ \iiN;
 \\\nonumber&&\mbox{for each }\iiN,\  \exists \  \mbox{ there is a feasible stationary quality adaptation scheme}
  \left(\left(q_i\left( f_i \right)\right)_{f_i \in {\cal F}_i}\right)
 \\\nonumber&& 
 \mbox{such that }
 \frac{ E \left[ L^{\pi}_i q_i\left( F^{\pi}_i \right)\right]}{ E \left[ L^{\pi}_i \right]}=\mu_{i},
 \ \mbox{Var} \left( q_i\left( F^{\pi}_{i} \right)  \right)
 \le v_{i}\le q^2_{\max}, \ \frac{\exn\left[L^{\pi}_i F^{\pi}_{i} \left( q_i\left( F^{\pi}_{i} \right) \right) \right]}{ \exn \left[ L^{\pi}_i\right]}
 \le\sigma_{i}\le f_{\max}  \bigg\}.
\end{eqnarray}
We can view $\widetilde{\cal H}$ as the set of `achievable' 
parameters in ${\cal H}$,
i.e., for any element $\left(\mathbf{m},\bm{\mu},\mathbf{v},\mathbf{b},\mathbf{d},\bm{\lambda},\bm{\sigma},\bm{\rho}\right)\in{\cal H}$
there is some feasible stationary resource allocation policy with mean resource allocation per unit time $\bm{\rho}$, 
and there is some feasible stationary quality adaptation policy for each $i$ that has a mean quality $\mu_i$, 
variance in quality which is at least $v_i$ and mean segment size which is at least $\sigma_i$ 
(and satisfies $\lambda_i=E [ L^{\pi}_i ] \ \forall \ \iiN$).

It can be verified that  $\widetilde{\cal H}$ is a bounded, closed and convex set (using an approach similar to that in Lemma 5 (b) in \cite{VJGdVAA_TAC2012}).
Hence, we conclude that for any $\bm{\Theta}\in{\cal H}$, 
there exists a unique projection of  $\widetilde{\bm{\Theta}}\in{\cal H}$
on the set $\widetilde{{\cal H}}$. Let $\proj{.}$ denote this projection operator.
Hence, for any $\bm{\Theta}\in{\cal H}$,
$d_{8N} \left(\bm{\Theta},\widetilde{{\cal H}} \right)
=d_{8N} \left(\bm{\Theta},\proj{\bm{\Theta}} \right).$
The next result states that, irrespective of the initialization, the differential inclusion converges to the 
bounded, closed and convex set $\widetilde{{\cal H}}$ of achievable parameters.
\begin{lemma}\label{results_about_eventually_reaching_H_tilda}
There exist finite constants $\chi_0>0$ and $\chi_1$ such that for any initialization $\widehat{\bm{\Theta}}(0) \in{\cal H}$, 
$$ \frac{d }{d t}d_{8N} \left( \widehat{\bm{\Theta}}(t) ,{\cal H}\right)
\le -\chi_0 d_{8N} \left( \widehat{\bm{\Theta}}(t) ,{\cal H}\right) 
+ \chi_1 d_N \left( \widehat{\bm{\lambda}}(t),\bm{\lambda}^{\pi}  \right).$$
Hence,
$$\lim_{t\rightarrow \infty} d_{8N}\left( \widehat{\bm{\Theta}}(t) ,{\cal H}\right)=0.$$
\end{lemma}
\begin{IEEEproof}
The result is an application of a generalization of Lemma 3 in \cite{Stolyar_gradient_descent}.
It can be proved by modifying the arguments in \cite{Stolyar_gradient_descent}, 
and using the update rule \eqref{lambdahat_ode_rule} for $\widehat{\bm{\lambda}}(t)$
which can be also written as
$$\bdt{\widehat{\lambda}}_i(t) = \frac{1}{u_i \left(\widehat{\bm{\Theta}}(t) \right)} d_{1}\left( \widehat{\lambda}_i(t), E \left[  L^{\pi}_i\right] \right).$$

\end{IEEEproof}

In the next result, we provide the main convergence result for the differential inclusion \eqref{mhat_ode_rule}-\eqref{rhohat_ode_rule}
which states that $ \widehat{\bm{\Theta}}(t)$ converges to the following set
\begin{eqnarray}\nonumber
{\cal H}^*&=& \left\{ \left(\mathbf{m},\bm{\mu},\mathbf{v},\mathbf{b},\mathbf{d},\bm{\lambda},\bm{\sigma},\bm{\rho}\right)\in{\cal H}:
\left(\bm{\rho},\left(h^B_i\left(b_i\right)\right)_{\iiN},\left(h^D_i\left(d_i\right)\right)_{\iiN}\right)\in{\cal X}^{\pi}, \right.
 \\&&\left. \mbox{and for each }\iiN,\ m_i=\mu_i=m^{\pi}_i,\ v_i=v^{\pi}_i \right\}\label{defn_Hstar}
\end{eqnarray}
Recall that Theorem \ref{main_NOVA_with_theta_pi_is_optimal}
suggested that we can prove Theorem \ref{main_optimality_theorem},
if we can show that the updates \eqref{m_update_NOVA}-\eqref{lambda_update_NOVA} guide NOVA
parameters $\left(\bm{\theta}_{i,s}\right)_{s\ge 1}$ of video client $i$ to the set ${\cal H}^*_i$ (defined in \eqref{defn_Hstar_i})
for each video client $\iiN$.
Note that for each $\iiN$, ${\cal H}^*_i$ is a set obtained by projecting ${\cal H}^*$ on a lower dimensional space 
(by considering only video client $i$'s components and `dropping' the components $\left(\bm{\lambda},\bm{\sigma},\bm{\rho}\right)$).
Hence, the following result along with Theorem \ref{main_stoch_approx_result} 
(which relates evolution of NOVA parameters to the differential inclusion)
help us to establish the desired convergence property for NOVA parameters.
\begin{theorem}\label{convergence_of_differential_inclusion}
(a) For $\widehat{\bm{\Theta}}=\left(\widehat{\mathbf{m}},\widehat{\bm{\mu}},\widehat{\mathbf{v}},\widehat{\mathbf{b}},
\widehat{\mathbf{d}},\widehat{\bm{\lambda}},\widehat{\bm{\sigma}},\widehat{\bm{\rho}}\right)\in {\cal H}$,
and some $\left(\bm{\rho}^{\pi},\mathbf{b}^{\pi},\mathbf{d}^{\pi}\right)\in{\cal X}^{\pi}$,
let 
\begin{eqnarray}\nonumber
L \left( \widehat{\bm{\Theta}}\right)
&\defeq& - \sum_{i\in{\cal N}}  \left(1+ \overline{\beta}_i\right)   \widehat{\lambda}_i  
U^E_i \left( \widehat{\mu}_i -  U^V_i \left( \widehat{v}_i  \right)\right) 
+ \sum_{i\in{\cal N}}  \left(1+ \overline{\beta}_i\right)   \widehat{\lambda}_i  \left(  \widehat{m}_i  -  m^{\pi}_i\right)^2
\\\nonumber&& + \sum_{i\in{\cal N}} \left(1+ \overline{\beta}_i\right)  \widehat{\lambda}_i   d^{\pi}_i 
\left( \frac{ p^d_i \widehat{\sigma}_i }{  \overline{p}_i } - 1 \right)
 + \sum_{i\in{\cal N}} \left(1+ \overline{\beta}_i\right)   \int_{\underline{d} }^{ \widehat{d}_i } 
\left(h^D_i\left(e\right) - d^{\pi}_i \right)de
\\\nonumber&& +  \sum_{i\in{\cal N}} \left( \widehat{\lambda}_i b^{\pi}_i \widehat{\sigma}_i 
- \tau_{slot} b^{\pi}_i \widehat{\rho}_i \right)
 + \sum_{i\in{\cal N}} \sigma^{\pi}_i \int_{ \underline{b}}^{ \widehat{b}_i } \left(h^B_i\left(e\right) -  b^{\pi}_i\right)de
\\&& + \frac{\chi_2}{\chi_0} d_{8N} \left( \widehat{\bm{\Theta}},\widetilde{{\cal H}} \right),\label{defn_Lyp}
\end{eqnarray}
where $\chi_0$ is the positive constant from Lemma \ref{results_about_eventually_reaching_H_tilda}, 
and $\chi_2$ is a large positive constant (the value is given in the proof).
If $\widehat{\bm{\Theta}}(0) \in{\cal H}$, then for almost all $t$
\begin{eqnarray*}
\frac{d L \left(\widehat{\bm{\Theta}}(t)\right)}{dt}
\begin{cases}
\le 0, \ \forall \ \widehat{\bm{\Theta}}(t)\in{\cal H},
\\< 0, \ \forall \ \widehat{\bm{\Theta}}(t)\notin{\cal H}^*.
\end{cases}
\end{eqnarray*}
\\(b) If $\widehat{\bm{\Theta}}(0) \in{\cal H}$, 
then 
$$\lim_{t\rightarrow \infty} d_{8N}\left( \widehat{\bm{\Theta}}(t) ,{\cal H}^*\right)=0.$$
\end{theorem}
\begin{IEEEproof}
The proof of part (b) relies on the analysis of the drift of the Lyapunov function $L(.)$ defined in \eqref{defn_Lyp} of part (a)
for
$\widehat{\bm{\Theta}}=\left(\widehat{\mathbf{m}},\widehat{\bm{\mu}},\widehat{\mathbf{v}},\widehat{\mathbf{b}},
\widehat{\mathbf{d}},\widehat{\bm{\lambda}},\widehat{\bm{\sigma}},\widehat{\bm{\rho}}\right)\in {\cal H}$
where $\left(\bm{\rho}^{\pi},\mathbf{b}^{\pi},\mathbf{d}^{\pi}\right)\in{\cal X}^{\pi}$ (defined in \eqref{defn_xpi}).
Here
$\chi_0$ is the positive constant from Lemma \ref{results_about_eventually_reaching_H_tilda}, 
and $\chi_2$ is a positive constant whose value is chosen to be large enough to satisfy certain conditions,
and is specified towards the end of the proof (above \eqref{define_Delta}).
The choice of several terms in the Lyapunov function $L(.)$
given above are motivated by the choice of Lyapunov functions in 
\cite{joseph_gustavo_varaiance_aware_video_tech_report}
and \cite{Stolyar_primal_dual}.
The first term $- \sum_{i\in{\cal N}}  \left(1+ \overline{\beta}_i\right)   \widehat{\lambda}_i  
U^E_i \left( \widehat{\mu}_i -  U^V_i \left( \widehat{v}_i  \right)\right) $
is similar to terms in the Lyapunov function in 
\cite{joseph_gustavo_varaiance_aware_video_tech_report},
and resembles an estimate for
a scaled version of the objective of OPT$(S)$ (see \eqref{OPTBASIC_objective})
since $\widehat{\mu}_i(.) $ tracks mean quality and $\widehat{v}_i(.)$ tracks variance in quality
so that a negative drift would suggest that the estimate is decreasing and we are moving in the right direction.
The terms in the second line 
$\sum_{i\in{\cal N}} \left(1+ \overline{\beta}_i\right)  \widehat{\lambda}_i   d^{\pi}_i 
\left( \frac{ p^d_i \widehat{\sigma}_i }{  \overline{p}_i } - 1 \right)
 + \sum_{i\in{\cal N}} \left(1+ \overline{\beta}_i\right)   \int_{\underline{d} }^{ \widehat{d}_i } 
\left(h^D_i\left(e\right) - d^{\pi}_i \right)de$
are similar to the terms chosen in the Lyapunov function in 
\cite{Stolyar_primal_dual}.
However,  note that our choice and analysis of the Lyapunov function has many novel elements.
For instance, the term $\sum_{i\in{\cal N}}  \left(1+ \overline{\beta}_i\right)   \widehat{\lambda}_i  \left(  \widehat{m}_i  -  m^{\pi}_i\right)^2$
(which can be viewed as weighted distance of `$\widehat{\mathbf{m}}$' component in $\widehat{\bm{\Theta}}(t)$
to `$\widehat{\mathbf{m}}$' component of elements in ${\cal H}^*$)
allows us to accommodate objectives involving variability terms (i.e., non-zero $U^V_i \left( \widehat{v}_i  \right)$)
and the
terms in $\sum_{i\in{\cal N}}b^{\pi}_i \left( \widehat{\lambda}_i  \widehat{\sigma}_i 
- \tau_{slot}  \widehat{\rho}_i \right)
 + \sum_{i\in{\cal N}} \sigma^{\pi}_i \int_{ \underline{b}}^{ \widehat{b}_i } \left(h^B_i\left(e\right) -  b^{\pi}_i\right)de$
 which allow us to accommodate the rebuffering constraints 
 (which involves comparing averages over time time scales).
 Further, note that our analysis of $L(.)$ will establish a convergence result for a differential inclusion
 associated with an algorithm NOVA which, unlike those in \cite{joseph_gustavo_varaiance_aware_video_tech_report}
and \cite{Stolyar_primal_dual}, uses \emph{asynchronous} updates.

\begin{figure}[ht]
	\centering
	\ifarxivmode
	\includegraphics[scale=.35]{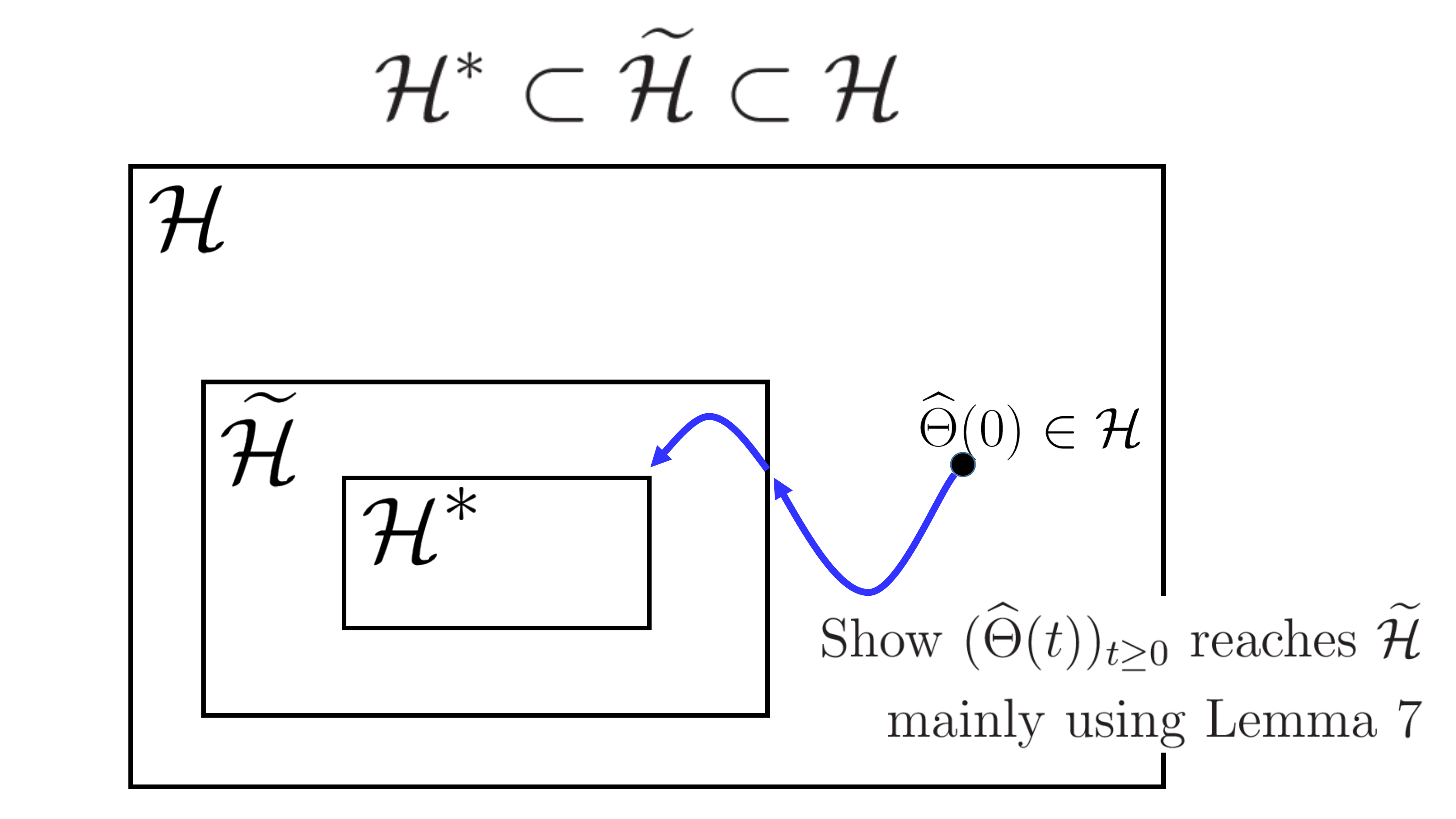}
	\else
	\includegraphics[scale=.35]{figs/conv_proof_fig.pdf}
	\fi
	\caption{${\cal H}^* \subset 
\widetilde{\cal H} \subset 
{\cal H}$, and we show that $\widehat{\bm{\Theta}}(t))_{t\ge 0}$ reaches $\widetilde{\cal H}$
	mainly using Lemma \ref{results_about_eventually_reaching_H_tilda}}	
	\label{conv_proof_fig}
\end{figure}

In the first part of the proof, we establish that 
the Lyapunov function $L$ has a non-positive drift,
and that the drift is strictly negative outside ${\cal H}^*$.
Note that since $\left(\bm{\rho}^{\pi},\mathbf{b}^{\pi},\mathbf{d}^{\pi}\right)\in{\cal X}^{\pi}$, there is some optimal solution
$\left(\left(\left(q^{\pi}_i\left(f\right) \right)_{f\in{\cal F}_i}\right)_{i\in{\cal N}},
\left(\mathbf{r}^{\pi} \left(c\right)\right)_{c\in{\cal C}}\right)$
to OPTSTAT with $\rho^{\pi}_i = \exn \left[  r^{\pi}_i \left(C^{\pi}\right)\right]$ for each $i\in{\cal N}$
and 
$\mathbf{b}^{\pi}$ and $\mathbf{d}^{\pi}$ as the associated optimal Lagrange multipliers
for the constraints \eqref{OPTSTAT_eta_cost_bound} and  \eqref{OPTSTAT_eta_bound_rebuf} respectively.

The proof is a bit lengthy, and the following is rough outline of the initial steps in the proof
(RHS is shorthand for right hand side):
\begin{eqnarray*}
\frac{d L \left(\widehat{\bm{\Theta}}(t)\right)}{dt}
&\le& \mbox{RHS of \eqref{line_1_der_L}} 
\\&\le& \mbox{RHS of \eqref{line_1half_der_L}} 
\\&\le& \mbox{RHS of \eqref{line_2_der_L}} 
\\&\le& \mbox{RHS of \eqref{line_3_der_L}} 
\\&\le& \mbox{RHS of \eqref{line_last_der_L}} 
\end{eqnarray*}
We use definition of $L(.)$ and \eqref{mhat_ode_rule}-\eqref{rhohat_ode_rule}
to obtain \eqref{line_1_der_L}.
We obtain
\eqref{line_1half_der_L}
mainly counting on (optimality) properties of optimal solution to
QNOVA$_i \left(\widehat{\bm{\theta}}_i(t),f_i\right)$ and Lemma \ref{concavity_and_derivative_of_obj_of_QNOVA} (c),
and we obtain
\eqref{line_2_der_L}
mainly counting on (optimality) properties of optimal solutions to RNOVA$\left(\widehat{\mathbf{b}} (t),c\right)$
and QNOVA$_i \left(\widehat{\bm{\theta}}^{(m^{\pi}_i)}_i(t),f_i\right)$,
where $\widehat{\bm{\theta}}^{(m^{\pi}_i)}_i(t)=(m^{\pi}_i,\widehat{\mu}_i(t),\widehat{v}_i(t),\widehat{b}_i(t),\widehat{d}_i(t))$.
In step
\eqref{line_3_der_L},
we mainly collect projection (projection of $\widehat{\bm{\Theta}}(t)$ on $\widetilde{{\cal H}}$) error terms 
and terms containing $\left| \widehat{\lambda}_i(t)- \lambda^{\pi}_i  \right|$, 
bound them 
and nullify the role of these terms by using \eqref{lambdahat_ode_rule} and Lemma \ref{results_about_eventually_reaching_H_tilda}
and picking large enough $\chi_2$.
Finally, we obtain
\eqref{line_last_der_L}
mainly counting on properties of optimal solution and optimal Lagrange multipliers of OPTSTAT.
In \eqref{line_last_der_L}, we have an upper bound for $\frac{d L \left(\widehat{\bm{\Theta}}(t)\right)}{dt}$
as a sum of several functions which are shown to be non-positive using aforementioned properties,
and we use additional arguments to
establish strict negativity of $\frac{d L \left(\widehat{\bm{\Theta}}(t)\right)}{dt}$
outside ${\cal H}^*$ and conclude proof of part (a).

Using the definition of $L(.)$ and \eqref{mhat_ode_rule}-\eqref{rhohat_ode_rule}, we have that
\begin{eqnarray}
 \label{line_1_der_L}
&&\hspace{-.5cm}\frac{d L \left(\widehat{\bm{\Theta}}(t)\right)}{dt}
\le - \sum_{i\in{\cal N}}\left(1+ \overline{\beta}_i\right)  \widehat{\lambda}_i(t)  \left(U^E_i\right)^{'} \left( \widehat{\mu}_i(t) -  U^V_i \left( \widehat{v}_i(t)  \right)\right)
\\\nonumber&&\hspace{3cm}\left(  
\frac{1}{u_i(t)}\left( \frac{ E \left[ L^{\pi}_i q^*_i\left(t\right)\right]}{\widehat{\lambda}_i(t)} 
 - \widehat{\mu}_i(t)   \right)- 
  \frac{\left(U^V_i\right)^{'} \left( \widehat{v}_i(t)  \right) }{u_i(t)}\left( \frac{E \left[  L^{\pi}_i\left(  q^*_i\left(t\right) -\widehat{m}_i(t) \right)^2\right] }{\widehat{\lambda}_i(t)} 
  - \widehat{v}_i(t)   \right)               \right)
    \\\nonumber &&+ \sum_{i\in{\cal N}}  2 \left(1+ \overline{\beta}_i\right)   \widehat{\lambda}_i (t)
 \frac{ \left(U^E_i\right)^{'}\left( \widehat{\mu}_i(t) - U^V_i \left(\widehat{v}_i(t)\right)\right)\left(U^V_i\right)^{'} \left(\widehat{v}_i(t)\right)}
 {u_i(t)}
 \left(  \widehat{m}_i(t)  -  m^{\pi}_i\right)
\left( \frac{ E \left[ L^{\pi}_i q^*_i\left(t\right)\right]}{\widehat{\lambda}_i(t)} 
 - \widehat{m}_i(t)   \right)
 \\\nonumber && + \sum_{i\in{\cal N}} \left(1+ \overline{\beta}_i\right)   \widehat{\lambda}_i(t)  d^{\pi}_i 
\left( \frac{ p^d_i   }{  \overline{p}_i }  \right)
 \frac{1}{u_i(t)}
\left(\frac{ E \left[  L^{\pi}_i F^{\pi}_{i}\left(  q^*_i\left(t\right)\right)\right]  }{ \widehat{\lambda}_i(t) } - 
\widehat{\sigma}_i(t) \right)
\\\nonumber && + \sum_{i\in{\cal N}} \left(1+ \overline{\beta}_i\right)   \widehat{\lambda}_i(t) 
\left(h^D_i\left(\widehat{d}_i(t)\right) - d^{\pi}_i \right) 
\frac{1}{u_i(t)}
\left( p^d_i\frac{ E \left[  L^{\pi}_i F^{\pi}_{i}\left(  q^*_i\left(t\right)\right)\right]  }{ \widehat{\lambda}_i(t) \overline{p}_i } - 1 \right)
\\\nonumber &&
\\\nonumber && +  \sum_{i\in{\cal N}}  \widehat{\lambda}_i(t) b^{\pi}_i 
\frac{1}{u_i(t)}
\left(\frac{ E \left[  L^{\pi}_i F^{\pi}_{i}\left(  q^*_i\left(t\right)\right)\right]  }{ \widehat{\lambda}_i(t) } - 
\widehat{\sigma}_i(t) \right)
 -  \sum_{i\in{\cal N}}   b^{\pi}_i 
\left( \frac{  E\left[ r^*_i(t) \right] }{\tau_{slot}}- \widehat{\rho}_i(t) \right) 
\\\nonumber && + \sum_{i\in{\cal N}} \sigma^{\pi}_i  \left(h^B_i\left(\widehat{b}_i(t)\right) -  b^{\pi}_i\right) 
\left(\frac{1}{ \left(1+ \overline{\beta}_i\right)  } 
-    \frac{ E \left[  L^{\pi}_i\right] }{u_i(t)}\right)
\\\nonumber&&- \chi_2 d_{8N} \left( \widehat{\bm{\Theta}}(t) ,\widetilde{{\cal H}} \right)
+\frac{\chi_2\chi_1}{\chi_0} d_N \left( \widehat{\bm{\lambda}}(t),\bm{\lambda}^{\pi}  \right)
+\sum_{i\in{\cal N}}\bdt{ \widehat{\lambda}}_i(t) \widetilde{l}_{i1} \left(\widehat{\bm{\Theta}}(t)\right)
\\\nonumber&&- +\sum_{i\in{\cal N}}\left(h^B_i\left(\widehat{b}_i(t)\right) -  b^{\pi}_i\right)
\left(\frac{E \left[  L^{\pi}_i F^{\pi}_{i}\left(  q^*_i\left(t\right)\right)\right]}{u_i(t)}
 -  \frac{ E\left[ r^*_i(t) \right] } { \tau_{slot}  }\right)
 ,
\end{eqnarray} 
where we have collected terms involving $\bdt{ \widehat{\lambda}}_i(t) $ and grouped them together in the term $\bdt{ \widehat{\lambda}}_i(t) \widetilde{l}_{i1} \left(\widehat{\bm{\Theta}}(t)\right)$ so that for each $\iiN$,
\begin{eqnarray*} 
\widetilde{l}_{1i} \left(\widehat{\bm{\Theta}}\right)
&=&
-   \left(1+ \overline{\beta}_i\right)   
U^E_i \left( \widehat{\mu}_i -  U^V_i \left( \widehat{v}_i  \right)\right) 
+  \left(1+ \overline{\beta}_i\right)   \left(  \widehat{m}_i  -  m^{\pi}_i\right)^2
\\\nonumber&& + \sum_{i\in{\cal N}} \left(1+ \overline{\beta}_i\right)   d^{\pi}_i 
\left( \frac{ p^d_i \widehat{\sigma}_i }{  \overline{p}_i } - 1 \right)
+     b^{\pi}_i \widehat{\sigma}_i.
\end{eqnarray*} 
For brevity, we have not explicitly indicated the dependence of many terms above on $ \widehat{\bm{\Theta}}(t)$. For instance, 
$u_i \left(\widehat{\bm{\Theta}}(t) \right)$ is shorthand for $u_i(t)$,
and $ E \left[ L^{\pi}_i q^*_i\left(t\right)\right]$ 
is shorthand for $ E \left[ L^{\pi}_i q^*_i\left( \widehat{\bm{\theta}}_i(t) ,F^{\pi}_{i} \right)\right]$
where 
\\$\widehat{\bm{\theta}}_i(t)=(\widehat{m}_i(t),\widehat{\mu}_i(t),\widehat{v}_i(t),\widehat{b}_i(t),\widehat{d}_i(t))$.
Also, note that we also added 
 $$\sum_{i\in{\cal N}}\left(h^B_i\left(\widehat{b}_i(t)\right) -  b^{\pi}_i\right)
\left( \frac{E \left[  L^{\pi}_i F^{\pi}_{i}\left(  q^*_i\left(t\right)\right)\right]}{u_i(t)}
 -  \frac{ E\left[ r^*_i(t) \right] } { \tau_{slot}  }\right)
 $$
to the right hand side of \eqref{line_1_der_L} where
$E\left[ r^*_i(t) \right]$ is shorthand for $E\left[ r^*_i\left(\widehat{\mathbf{b}}(t),C^{\pi}\right)\right]$
where $\mathbf{r}^*\left(\widehat{\mathbf{b}}(t),c\right)\in {\cal R}^*\left(\widehat{\mathbf{b}}(t),c\right)$
for each $\ciC$.
This does not change the inequality since this expression evaluates to zero from the definition of $u_i(.)$ in \eqref{u_i_t_def}.
The terms 
$\sum_{i\in{\cal N}} \left(1+ \overline{\beta}_i\right)   \widehat{\lambda}_i(t) 
\left(h^D_i\left(\widehat{d}_i(t)\right) - d^{\pi}_i \right) \widehat{z}^d_i \left(\widehat{\bm{\Theta}}(t)\right)$
and $ \sum_{i\in{\cal N}} \sigma^{\pi}_i  \left(h^B_i\left(\widehat{b}_i(t)\right) -  b^{\pi}_i\right) \widehat{z}^b_i \left(\widehat{\bm{\Theta}}(t)\right) $
have also been dropped from right hand side of \eqref{line_1_der_L} noting that they are less than or equal to zero.
To see this, note that $ \widehat{z}^b_i \left(\widehat{\bm{\Theta}}(t)\right)\ge 0$
(from \eqref{reflection_terms_are_non_negative})
which is equal to zero unless $\widehat{b}_i(t)=\underline{b}$ 
(from \eqref{reflection_terms_are_zero_for_theta_in_interior})
for which $h^B_i\left(\widehat{b}_i(t)\right) =0\le b^{\pi}_i$.
Hence, $ \sum_{i\in{\cal N}} \sigma^{\pi}_i  \left(h^B_i\left(\widehat{b}_i(t)\right) -  b^{\pi}_i\right) \widehat{z}^b_i \left(\widehat{\bm{\Theta}}(t)\right) \le 0$,
and we can similarly show that
$\sum_{i\in{\cal N}} \left(1+ \overline{\beta}_i\right)   \widehat{\lambda}_i(t) 
\left(h^D_i\left(\widehat{d}_i(t)\right) - d^{\pi}_i \right) \widehat{z}^d_i \left(\widehat{\bm{\Theta}}(t)\right)\le 0$.

Consider the right hand side of \eqref{line_1_der_L}, and group the terms containing $q^*_i(t)$ except those in the fifth line
(i.e., the line with the term $\left(  \widehat{m}_i(t)  -  m^{\pi}_i\right)$)
to note that we have negative of a scaled (by  $\left(1+ \overline{\beta}_i\right)/ u_i(t)$) version  of the expectation of the objective of  QNOVA$_i \left(\widehat{\bm{\theta}}_i(t),f_i\right)$, i.e., 
$E \left[  L^{\pi}_i\phi^Q\left(q^*_i(t),\widehat{\bm{\theta}}_i(t),F^{\pi}_{i}\right)\right]$.
Recall that $q^*_i(t)$ in the above calculations is a shorthand for $q^*_i\left(\widehat{\bm{\theta}}_i(t),F^{\pi}_{i}  \right)$.
Now, let $q^{*,m^{\pi}_i}_i(t)$ denote the shorthand for $q^*_i\left(\widehat{\bm{\theta}}^{(m^{\pi}_i)}_i(t),F^{\pi}_{i}  \right)$
where $\widehat{\bm{\theta}}^{(m^{\pi}_i)}_i(t)=(m^{\pi}_i,\widehat{\mu}_i(t),\widehat{v}_i(t),\widehat{b}_i(t),\widehat{d}_i(t))$, i.e., 
 $\widehat{\bm{\theta}}_i(t)$ with the first component set to $m^{\pi}_i$ (defined in \eqref{defn_mpi}).
 Next, we replace $q^*_i(t)$ appearing in the above inequality with $q^{*,m^{\pi}_i}_i(t)$,
 incorporate the correction term associated with this replacement into a function $\Delta_1 \left(\widehat{\bm{\Theta}}(t)\right)$,
 and rewrite \eqref{line_1_der_L} as
\begin{eqnarray}
 \label{line_1half_der_L}
&&\hspace{-.5cm}\frac{d L \left(\widehat{\bm{\Theta}}(t)\right)}{dt}
\le \Delta_1 \left(\widehat{\bm{\Theta}}(t)\right) - \sum_{i\in{\cal N}}\left(1+ \overline{\beta}_i\right)  \widehat{\lambda}_i(t)  \left(U^E_i\right)^{'} \left( \widehat{\mu}_i(t) -  U^V_i \left( \widehat{v}_i(t)  \right)\right)
\\\nonumber&&\hspace{2.1cm}\left(  
\frac{1}{u_i(t)}\left( \frac{ E \left[ L^{\pi}_i q^{*,m^{\pi}_i}_i\left(t\right)\right]}{\widehat{\lambda}_i(t)} 
 - \widehat{\mu}_i(t)   \right) - 
  \frac{\left(U^V_i\right)^{'} \left( \widehat{v}_i(t)  \right) }{u_i(t)}\left( \frac{E \left[  L^{\pi}_i\left(  q^{*,m^{\pi}_i}_i\left(t\right) -m^{\pi}_i \right)^2\right] }{\widehat{\lambda}_i(t)} 
  - \widehat{v}_i(t)   \right)               \right)
 \\\nonumber && + \sum_{i\in{\cal N}} \left(1+ \overline{\beta}_i\right)   \widehat{\lambda}_i(t)  d^{\pi}_i 
\left( \frac{ p^d_i   }{  \overline{p}_i }  \right)
 \frac{1}{u_i(t)}
\left(\frac{ E \left[  L^{\pi}_i F^{\pi}_{i}\left(  q^{*,m^{\pi}_i}_i\left(t\right)\right)\right]  }{ \widehat{\lambda}_i(t) } - 
\widehat{\sigma}_i(t) \right)
\\\nonumber && + \sum_{i\in{\cal N}} \left(1+ \overline{\beta}_i\right)   \widehat{\lambda}_i(t) 
\left(h^D_i\left(\widehat{d}_i(t)\right) - d^{\pi}_i \right) 
\frac{1}{u_i(t)}
\left( p^d_i\frac{ E \left[  L^{\pi}_i F^{\pi}_{i}\left(  q^{*,m^{\pi}_i}_i\left(t\right)\right)\right]  }{ \widehat{\lambda}_i(t) \overline{p}_i } - 1 \right)
\\\nonumber && +  \sum_{i\in{\cal N}}  \widehat{\lambda}_i(t) b^{\pi}_i 
\frac{1}{u_i(t)}
\left(\frac{ E \left[  L^{\pi}_i F^{\pi}_{i}\left(  q^{*,m^{\pi}_i}_i\left(t\right)\right)\right]  }{ \widehat{\lambda}_i(t) } - 
\widehat{\sigma}_i(t) \right)
-  \sum_{i\in{\cal N}}   b^{\pi}_i 
\left( \frac{  E\left[ r^*_i(t) \right] }{\tau_{slot}}- \widehat{\rho}_i(t) \right) 
\\\nonumber && + \sum_{i\in{\cal N}} \sigma^{\pi}_i  \left(h^B_i\left(\widehat{b}_i(t)\right) -  b^{\pi}_i\right) 
\left(\frac{1}{ \left(1+ \overline{\beta}_i\right)  } 
-    \frac{ E \left[  L^{\pi}_i\right] }{u_i(t)}\right)
\\\nonumber&&+\sum_{i\in{\cal N}}\left(h^B_i\left(\widehat{b}_i(t)\right) -  b^{\pi}_i\right)
\left(\frac{E \left[  L^{\pi}_i F^{\pi}_{i}\left(  q^{*,m^{\pi}_i}_i\left(t\right)\right)\right]}{u_i(t)}
 -  \frac{ E\left[ r^*_i(t) \right] } { \tau_{slot}  }\right)
 \\\nonumber&&- \chi_2 d_{8N} \left( \widehat{\bm{\Theta}}(t) ,\widetilde{{\cal H}} \right)
 +\frac{\chi_2\chi_1}{\chi_0} d_N \left( \widehat{\bm{\lambda}}(t),\bm{\lambda}^{\pi}  \right)
 +\sum_{i\in{\cal N}}\bdt{ \widehat{\lambda}}_i(t) l_{1i} \left(\widehat{\bm{\Theta}}(t)\right),
\end{eqnarray}
where
\begin{eqnarray}\nonumber
 \Delta_1 \left(\widehat{\bm{\Theta}}(t)\right)= - \sum_{i\in{\cal N}} \frac{\left(1+ \overline{\beta}_i\right)}{ u_i(t) }
 E \left[  L^{\pi}_i  
 \left(  \phi^Q\left(q^*_i \left(\widehat{\bm{\theta}}_i(t),F^{\pi}_{i} \right),\widehat{\bm{\theta}}_i(t),F^{\pi}_{i} \right)- \phi^Q\left(q^*_i \left(\widehat{\bm{\theta}}^{(m^{\pi}_i)}_i(t),F^{\pi}_{i} \right),\widehat{\bm{\theta}}^{(m^{\pi}_i)}_i(t),F^{\pi}_{i} \right) \hspace{-1cm}   \right.      \right.
\\\left.\left.-  \left(U^E_i\right)^{'}\left( \widehat{\mu}_i(t) - U^V_i \left(\widehat{v}_i(t)\right)\right)\left(U^V_i\right)^{'} \left(\widehat{v}_i(t)\right)
  2 \left(  \widehat{m}_i(t)  -  m^{\pi}_i\right)
\left(  q^*_i\left(t\right)  - \widehat{m}_i(t)  \right)\right) \right]\label{expression_for_DELTA_1}
 \end{eqnarray}
 and for each $\iiN$,
 \begin{eqnarray*}
 l_{1i} \left(\widehat{\bm{\Theta}}(t)\right)=
 \tilde{l}_{1i} \left(\widehat{\bm{\Theta}}(t)\right) 
+2  \frac{\left(1+ \overline{\beta}_i\right)}{ u_i(t) }
   \left(U^E_i\right)^{'}\left( \widehat{\mu}_i(t) - U^V_i \left(\widehat{v}_i(t)\right)\right)\left(U^V_i\right)^{'} \left(\widehat{v}_i(t)\right)
   \left(  \widehat{m}_i(t)  -  m^{\pi}_i\right)\widehat{m}_i(t).
  \end{eqnarray*}
  Note that we have included the terms in the third line of  \eqref{line_1_der_L}
  in \eqref{expression_for_DELTA_1} after replacing $\widehat{\lambda}_i(t)$ 
  with $E \left[  L^{\pi}_i\right]$,
  and the correction for the modification is included as 
  the second term of $ l_{1i} \left(\widehat{\bm{\Theta}}(t)\right)$ defined above.
   From the definition \eqref{u_i_t_def} of $u_i(t)$, we have that
   \begin{eqnarray}\label{defn_umin_umax}
u_{\min}\defeq\frac{\tau_{slot}l_{\min}f_{\min}}{r_{\max}}, \ u_{\max}\defeq\frac{\tau_{slot}l_{\max}f_{\max}}{r_{\min}}
\end{eqnarray}
are lower and upper bounds respectively on $u_i(t)$ for each $\iiN$.
  Note that, in the term $\sum_{i\in{\cal N}}\bdt{ \widehat{\lambda}}_i(t) l_{1i} \left(\widehat{\bm{\Theta}}(t)\right)$,
  we are collecting the terms scaled by $\bdt{ \widehat{\lambda}}_i(t)$ 
  together so as to bound it by picking a large enough $\chi_2$ since 
  $ \bdt{ \widehat{\lambda}}_i(t)
  \le \frac{1}{u_{\min}} d_{N} \left( \widehat{\bm{\lambda}}(t),\bm{\lambda}^{\pi}  \right)
 \le \frac{1}{u_{\min}} d_{8N} \left( \widehat{\bm{\Theta}} (t) ,\widetilde{{\cal H}} \right)$.  
 Using the bounded nature of the terms involved in $ l_{1i} \left(\widehat{\bm{\Theta}}\right)$ and $\tilde{l}_{1i} \left(\widehat{\bm{\Theta}}\right) $, 
 we can show that there is some finite $\chi_3$ such that
  \begin{eqnarray}
 \label{can_bound_l1}
 \frac{\chi_2\chi_1}{\chi_0} d_N \left( \widehat{\bm{\lambda}}(t),\bm{\lambda}^{\pi}  \right)+
 \sum_{\iiN}l_{1i} \left(\widehat{\bm{\Theta}}\right)  \bdt{ \widehat{\lambda}}_i(t)
 \le \chi_3 d_{8N} \left( \widehat{\bm{\Theta}}(t) ,\widetilde{{\cal H}} \right) 
 \end{eqnarray}
 holds for any $\widehat{\bm{\Theta}}(t)\in {\cal H}$.

If we group the terms containing $q^{*,m^{\pi}_i}_i(t)$ and $\mathbf{r}^*(t)$,
we find that the right hand side of \eqref{line_1half_der_L}
contains negative of scaled versions of optimal value of objective functions of QNOVA$_i \left(\widehat{\bm{\theta}}^{(m^{\pi}_i)}_i(t),f_i\right)$ (i.e., $\phi^Q\left(q^*_i(t),\widehat{\bm{\theta}}^{(m^{\pi}_i)}_i(t),f_i\right)$)
and those of RNOVA$\left(\widehat{\mathbf{b}} (t),c\right)$ (i.e., 
$\phi^R\left(\mathbf{r}^*(t),\widehat{\mathbf{b}} (t),c\right)$).
Now using the optimality of $q^{*,m^{\pi}_i}_i(t)$ and $\mathbf{r}^*(t)$ with respect to 
QNOVA$_i \left(\widehat{\bm{\theta}}^{(m^{\pi}_i)}_i(t),f_i\right)$ and RNOVA$\left(\widehat{\mathbf{b}} (t),c\right)$,
and using the fact that
$  q^{\pi}_i\left(f\right)$ and $\mathbf{r}^{\pi} \left(c\right)$
are feasible solutions for these optimization problems, 
we obtain the following inequality from \eqref{line_1half_der_L}
(obtained by replacing $q^{*,m^{\pi}_i}_i(t)$ and $\mathbf{r}^*(t)$ with
$  q^{\pi}_i\left(f\right)$ and $\mathbf{r}^{\pi} \left(c\right)$ in \eqref{line_1half_der_L}
and adding the correction term $\Delta_2 \left(\widehat{\bm{\Theta}}(t)\right)$ associated with this replacement)
\begin{eqnarray}
 \label{line_2_der_L}
&&\hspace{-1.3cm}\frac{d L \left(\widehat{\bm{\Theta}}(t)\right)}{dt}
\le \Delta_1 \left(\widehat{\bm{\Theta}}(t)\right) + \Delta_2 \left(\widehat{\bm{\Theta}}(t)\right)
\\\nonumber&& - \sum_{i\in{\cal N}}\left(1+ \overline{\beta}_i\right)  \widehat{\lambda}_i(t)  \left(U^E_i\right)^{'} \left( \proj{\widehat{\mu}_i(t)} -  U^V_i \left( \proj{\widehat{v}_i(t)}  \right)\right)
\left(  
\frac{1}{u_i(t)}\left( \frac{m^{\pi}_i E \left[  L^{\pi}_i\right]}{ \widehat{\lambda}_i(t)}
 - \proj{\widehat{\mu}_i(t)}   \right) \right.
 \\&&\left.\nonumber \hspace{4cm}-   \frac{\left(U^V_i\right)^{'} \left( \proj{\widehat{v}_i(t)}  \right) }{u_i(t)}\left( 
  \frac{E \left[  L^{\pi}_i\left( q^{\pi}_i\left(F^{\pi}_{i}\right) - m^{\pi}_i \right)^2\right] }{\widehat{\lambda}_i(t)} 
  -  \proj{\widehat{v}_i(t) }  \right)               \right)
\\\nonumber && + \sum_{i\in{\cal N}} \left(1+ \overline{\beta}_i\right)   \widehat{\lambda}_i(t)  d^{\pi}_i 
\left( \frac{ p^d_i   }{  \overline{p}_i }  \right)
 \frac{1}{u_i(t)}
\left(\frac{\sigma^{\pi}_i E \left[  L^{\pi}_i\right]  }{ \widehat{\lambda}_i(t) } - 
\proj{\widehat{\sigma}_i(t)} \right)
\\\nonumber && + \sum_{i\in{\cal N}} \left(1+ \overline{\beta}_i\right)   \widehat{\lambda}_i(t) 
\left(h^D_i\left(\widehat{d}_i(t)\right) - d^{\pi}_i \right) \frac{1}{u_i(t)}
\left( p^d_i\frac{ \sigma^{\pi}_i E \left[  L^{\pi}_i\right]   }{ \widehat{\lambda}_i(t) \overline{p}_i } - 1 \right)
\\\nonumber && +  \sum_{i\in{\cal N}}  \widehat{\lambda}_i(t) b^{\pi}_i 
\frac{1}{u_i(t)}
\left(\frac{ \sigma^{\pi}_i E \left[  L^{\pi}_i\right]   }{ \widehat{\lambda}_i(t) } - 
\proj{\widehat{\sigma}_i(t)} \right)
-  \sum_{i\in{\cal N}}   b^{\pi}_i 
\left( \frac{\rho^{\pi}_i}{\tau_{slot}}- \proj{ \widehat{\rho}_i(t)} \right) 
\\\nonumber && + \sum_{i\in{\cal N}} \sigma^{\pi}_i  \left(h^B_i\left(\widehat{b}_i(t)\right) -  b^{\pi}_i\right) 
\left(\frac{1}{ \left(1+ \overline{\beta}_i\right)  } 
-    \frac{ E \left[  L^{\pi}_i\right] }{u_i(t)}\right)
+\sum_{i\in{\cal N}}\left(h^B_i\left(\widehat{b}_i(t)\right) -  b^{\pi}_i\right)
\left(\frac{ \sigma^{\pi}_i E \left[  L^{\pi}_i\right]  }{u_i(t)}
 -  \frac{\rho^{\pi}_i}{\tau_{slot}} \right)
\\\nonumber&&++\frac{\chi_2\chi_1}{\chi_0} d_N \left( \widehat{\bm{\lambda}}(t),\bm{\lambda}^{\pi}  \right)  +\sum_{i\in{\cal N}}\bdt{ \widehat{\lambda}}_i(t) l_{1i} \left(\widehat{\bm{\Theta}}(t)\right)
- \chi_2 d_{8N} \left( \widehat{\bm{\Theta}}(t) ,\widetilde{{\cal H}} \right) 
+l_2 \left(\widehat{\bm{\Theta}}(t)\right),
\end{eqnarray}
  where
 $m^{\pi}_i$, $v^{\pi}_i$ and $\sigma^{\pi}_i$ are defined in
\eqref{defn_mpi}-\eqref{defn_sigmapi}, (and $\bm{\rho}^{\pi}$ was chosen at the beginning of the proof- see below \eqref{defn_Lyp})
  \begin{eqnarray}
 \Delta_2 \left(\widehat{\bm{\Theta}}(t)\right)&=& - \frac{1}{\tau_{slot}}   E\left[  \phi^R\left(\mathbf{r}^*(t),\widehat{\mathbf{b}} (t),C^{\pi}\right) - \phi^R\left(\mathbf{r}^{\pi}\left(C^{\pi}\right),\widehat{\mathbf{b}} (t),C^{\pi}\right)  \right] \hspace{0cm}
 \\\nonumber && \hspace{-.7cm}- \sum_{i\in{\cal N}} \frac{\left(1+ \overline{\beta}_i\right)}{ u_i(t) }
 E \left[  L^{\pi}_i  
 \left(  \phi^Q\left(q^*_i \left(  \proj{\widehat{\bm{\theta}}^{(m^{\pi}_i)}_i}(t),F^{\pi}_{i} \right),\proj{\widehat{\bm{\theta}}^{(m^{\pi}_i)}_i}(t),F^{\pi}_{i} \right) 
- \phi^Q\left(q^{\pi}_i \left(F^{\pi}_{i} \right),\proj{\widehat{\bm{\theta}}^{(m^{\pi}_i)}_i}(t),F^{\pi}_{i} \right) \right)  \right],
 \end{eqnarray}
and $\widetilde{.}$ is the projection operator mapping elements in ${\cal H}$ to the set $\widetilde{{\cal H}}$. 
Here
 \begin{eqnarray*}
\left(\proj{\widehat{\mathbf{m}}(t)},\proj{\widehat{\bm{\mu}}(t)},\proj{\widehat{\mathbf{v}}(t)},\proj{\mathbf{b}(t)},
\proj{\mathbf{d}(t)},\proj{\widehat{\bm{\lambda}}(t)},\proj{\widehat{\bm{\sigma}}(t)},\proj{\widehat{\bm{\rho}}(t)}\right)
 &\defeq& \proj{\widehat{\bm{\Theta}}(t)}.
  \end{eqnarray*}
 Due to the definition of $\widetilde{{\cal H}}$ (see \eqref{defn_calH}),
  $\proj{\widehat{\mathbf{m}}(t)}=\widehat{\mathbf{m}}(t)$, $\proj{\widehat{\mathbf{b}}(t)}=\widehat{\mathbf{b}}(t)$
  and $\proj{\widehat{\mathbf{d}}(t)}=\widehat{\mathbf{d}}(t)$
  as \eqref{defn_calH} does not impose any additional restrictions on these components. Also, for each $\iiN$,
  \begin{eqnarray*}
 \proj{\widehat{\bm{\theta}}_i(t)}&\defeq& 
 \left(\widehat{m}_i(t),\proj{\widehat{\mu}_i(t)},\proj{\widehat{v}_i(t)},\widehat{b}_i(t),\widehat{d}_i(t)\right),
 \\\proj{\widehat{\bm{\theta}}^{(m^{\pi}_i)}_i} (t)&\defeq&
 \left(m^{\pi}_i,\proj{\widehat{\mu}_i(t)},\proj{\widehat{v}_i(t)},\widehat{b}_i(t),\widehat{d}_i(t)\right).
 \end{eqnarray*}
 Note that we have replaced 
 components of 
 $\widehat{\bm{\mu}}(t)$, $\widehat{\mathbf{v}}(t)$, $\widehat{\bm{\lambda}}(t)$, $\widehat{\bm{\sigma}}(t)$
 and $\widehat{\bm{\rho}}(t)$
appearing in \eqref{line_1half_der_L}
with those of $\proj{\widehat{\bm{\mu}}(t)}$, $\proj{\widehat{\mathbf{v}}(t)}$, $\proj{\widehat{\bm{\lambda}}(t)}$, $\proj{\widehat{\bm{\sigma}}(t)}$ and $\proj{\widehat{\bm{\rho}}(t)}$ respectively,
and in \eqref{line_2_der_L}, we have added the function $l_2 \left(.\right) $ defined below to account for 
these replacements:
 \begin{eqnarray}
&&\hspace{-.5cm}l_2 \left(\widehat{\bm{\Theta}}(t)\right) 
=  
- \sum_{i\in{\cal N}}\left(1+ \overline{\beta}_i\right)  \widehat{\lambda}_i(t)  
\left(\left(U^E_i\right)^{'} \left( \widehat{\mu}_i(t) -  U^V_i \left( \widehat{v}_i(t)  \right)\right)
-\left(U^E_i\right)^{'} \left( \proj{\widehat{\mu}_i(t)} -  U^V_i \left( \proj{\widehat{v}_i(t)}  \right)\right)\right)
\\\nonumber&&\hspace{2.8cm}\left(  
\frac{1}{u_i(t)}\left( \frac{m^{\pi}_i E \left[  L^{\pi}_i\right]}{ \widehat{\lambda}_i(t)}
 - \proj{\widehat{\mu}_i(t)}   \right)
 -  \frac{\left(U^V_i\right)^{'} \left( \proj{\widehat{v}_i(t)}  \right) }{u_i(t)}\left( 
  \frac{E \left[  L^{\pi}_i\left( q^{\pi}_i\left(F^{\pi}_{i}\right) - m^{\pi}_i \right)^2\right] }{\widehat{\lambda}_i(t)} 
  -  \proj{\widehat{v}_i(t) }  \right)               \right)
 \\\nonumber&&  - \sum_{i\in{\cal N}}\left(1+ \overline{\beta}_i\right)  \widehat{\lambda}_i(t)  \left(U^E_i\right)^{'} \left( \widehat{\mu}_i(t) -  U^V_i \left( \widehat{v}_i(t)  \right)\right)
\\\nonumber&&\hspace{5cm}\left(
  \frac{\left(U^V_i\right)^{'} \left( \proj{\widehat{v}_i(t)}  \right)  - \left(U^V_i\right)^{'} \left( \widehat{v}_i(t)  \right) }{u_i(t)}\left( 
  \frac{E \left[  L^{\pi}_i\left( q^{\pi}_i\left(F^{\pi}_{i}\right) - m^{\pi}_i \right)^2\right] }{\widehat{\lambda}_i(t)} 
  -  \proj{\widehat{v}_i(t) }  \right)               \right)
  \\\nonumber&&
- \sum_{i\in{\cal N}}\left(1+ \overline{\beta}_i\right)  \widehat{\lambda}_i(t)  \left(U^E_i\right)^{'} \left( \widehat{\mu}_i(t) -  U^V_i \left( \widehat{v}_i(t)  \right)\right)
\left(  
\left(\proj{\widehat{\mu}_i(t) }
 - \widehat{\mu}_i(t)   \right) - 
  \frac{\left(U^V_i\right)^{'} \left( \widehat{v}_i(t)  \right) }{u_i(t)}\left( 
  \proj{\widehat{v}_i(t) }  - \widehat{v}_i(t)   \right)               \right)
    \\\nonumber && + \sum_{i\in{\cal N}} \left(1+ \overline{\beta}_i\right)   \widehat{\lambda}_i(t)  d^{\pi}_i 
\left( \frac{ p^d_i   }{  \overline{p}_i }  \right)
 \frac{1}{u_i(t)}
\left(\proj{\widehat{\sigma}_i(t)}  - 
\widehat{\sigma}_i(t) \right)
\\\nonumber && -  \sum_{i\in{\cal N}}  \widehat{\lambda}_i(t) b^{\pi}_i 
\frac{1}{u_i(t)}
\left( \proj{\widehat{\sigma}_i(t)} - 
\widehat{\sigma}_i(t) \right)
 -  \sum_{i\in{\cal N}}   b^{\pi}_i 
\left( \proj{\widehat{\rho}_i(t)} - \widehat{\rho}_i(t) \right) .
\end{eqnarray}
 Using the bounded nature of the  terms involved in the above expression for $l_2 \left(\widehat{\bm{\Theta}}(t)\right)$ 
 (note that here we are using the fact that the functions $\left(U^V_i,U^E_i\right)_{\iiN}$ have Lipschitz continuous derivatives
 due to Assumptions U.V and U.E),
 we can show that there exists some large enough finite constant $\chi_4$ such that 
 \begin{eqnarray}
 \label{can_bound_l2}
 l_2 \left(\widehat{\bm{\Theta}}(t)\right) \le \chi_4 d_{8N} \left( \widehat{\bm{\Theta}}(t) ,\widetilde{{\cal H}} \right) 
 \end{eqnarray}
  holds for any $\widehat{\bm{\Theta}}(t)\in {\cal H}$.
 We can replace the term
 $\widehat{\lambda}_i(t)$ 
 appearing in the denominators of terms in \eqref{line_2_der_L}
 with $E \left[  L^{\pi}_i\right]$,
 and add a term $l_{3i} \left(\widehat{\bm{\Theta}}(t)\right)\bdt{ \widehat{\lambda}}_i(t) $ to
 account for the change in the expression due to the replacement such that all the terms in 
 $l_{3i} \left(\widehat{\bm{\Theta}}\right)$ are bounded.
 Using the boundedness of terms involved in $l_{3i} \left(\widehat{\bm{\Theta}}\right)$, 
  and then using arguments similar to that used in obtaining \eqref{can_bound_l1},
 we can show that there is some $\chi_5$ such that
  \begin{eqnarray}
 \label{can_bound_l3}
 \sum_{\iiN} l_{3i} \left(\widehat{\bm{\Theta}}(t)\right)\bdt{ \widehat{\lambda}}_i(t)  \le \chi_5 d_{8N} \left( \widehat{\bm{\Theta}}(t) ,\widetilde{{\cal H}} \right) 
 \end{eqnarray}
  holds for any $\widehat{\bm{\Theta}}(t)\in {\cal H}$.
 Thus, we can use the observations in \eqref{can_bound_l1}, \eqref{can_bound_l2} and \eqref{can_bound_l3} 
 along with \eqref{line_2_der_L} to conclude that
 \begin{eqnarray}
 \label{line_3_der_L}
&&\hspace{-.5cm}\frac{d L \left(\widehat{\bm{\Theta}}(t)\right)}{dt}
\le \Delta_1 \left(\widehat{\bm{\Theta}}(t)\right) + \Delta_2 \left(\widehat{\bm{\Theta}}(t)\right) 
- \sum_{i\in{\cal N}}\left(1+ \overline{\beta}_i\right)  \widehat{\lambda}_i(t)  \left(U^E_i\right)^{'} \left( \proj{\widehat{\mu}_i(t)} -  U^V_i \left( \proj{\widehat{v}_i(t)}  \right)\right)
\\\nonumber&&\hspace{4.5cm}\left(  
\frac{1}{u_i(t)}\left( m^{\pi}_i 
 - \proj{\widehat{\mu}_i(t)}   \right)- 
  \frac{\left(U^V_i\right)^{'} \left( \proj{\widehat{v}_i(t)}  \right) }{u_i(t)}\left( 
  \frac{E \left[  L^{\pi}_i\left( q^{\pi}_i\left(F^{\pi}_{i}\right) - m^{\pi}_i \right)^2\right] }
  {E \left[  L^{\pi}_i\right]} 
  -  \proj{\widehat{v}_i(t) }  \right)               \right)
\\\nonumber && + \sum_{i\in{\cal N}} \left(1+ \overline{\beta}_i\right)   \widehat{\lambda}_i(t)  d^{\pi}_i 
\left( \frac{ p^d_i   }{  \overline{p}_i }  \right)
 \frac{1}{u_i(t)}
\left(\sigma^{\pi}_i    - 
\proj{\widehat{\sigma}_i(t)} \right)
 + \sum_{i\in{\cal N}} \left(1+ \overline{\beta}_i\right)   \widehat{\lambda}_i(t) 
\left(h^D_i\left(\widehat{d}_i(t)\right) - d^{\pi}_i \right) \frac{1}{u_i(t)}
\left( p^d_i\frac{ \sigma^{\pi}_i    }{ \overline{p}_i } - 1 \right)
\\\nonumber && +  \sum_{i\in{\cal N}}  \widehat{\lambda}_i(t) b^{\pi}_i 
\frac{1}{u_i(t)}
\left( \sigma^{\pi}_i  - \proj{\widehat{\sigma}_i(t)} \right)
 -  \sum_{i\in{\cal N}}   b^{\pi}_i 
\left( \rho^{\pi}_i- \proj{ \widehat{\rho}_i(t)} \right) 
\\\nonumber&& + \sum_{i\in{\cal N}}   \left(h^B_i\left(\widehat{b}_i(t)\right) -  b^{\pi}_i\right) 
\left(\frac{\sigma^{\pi}_i}{ \left(1+ \overline{\beta}_i\right)  } 
-    \frac{ E \left[  L^{\pi}_i\right] \sigma^{\pi}_i }{u_i(t)}\right)
+\sum_{i\in{\cal N}}\left(h^B_i\left(\widehat{b}_i(t)\right) -  b^{\pi}_i\right)
\left(\frac{ \sigma^{\pi}_i E \left[  L^{\pi}_i\right]  }{u_i(t)}
 -  \frac{\rho^{\pi}_i}{\tau_{slot}} \right)
 \\\nonumber&& +
\left(\chi_3+\chi_4+\chi_5- \chi_2\right) d_{8N} \left( \widehat{\bm{\Theta}}(t) ,\widetilde{{\cal H}} \right). 
\end{eqnarray}
Let $\chi_2 = \chi_3+\chi_4+\chi_5+1$, and let
 \begin{eqnarray}
 \label{define_Delta}
\Delta \left(\bm{\Theta}\right)
&=& \Delta_1 \left(\bm{\Theta}\right) + \Delta_2 \left(\bm{\Theta}\right) + \Delta^{(d)} \left(\bm{\Theta}\right) + \Delta^{(b)} \left(\bm{\Theta}\right)  
+  \Delta^{(\pi,q)} \left(\bm{\Theta}\right)   +  \Delta^{(\pi,r)} \left(\bm{\Theta}\right) + \Delta_3 \left(\bm{\Theta}\right),
\end{eqnarray}
where 
\begin{eqnarray*}
\Delta^{(\pi,q)} \left(\widehat{\bm{\Theta}}(t)\right)  &=& - \sum_{i\in{\cal N}}\frac{\left(1+ \overline{\beta}_i\right)   \widehat{\lambda}_i(t)}{u_i(t)} \left( \left(U^E_i\right)^{'} \left( \proj{\widehat{\mu}_i(t)} -  U^V_i \left( \proj{\widehat{v}_i(t)}  \right)\right)\right. 
\\&&\left.\left(  \left( m^{\pi}_i  - \proj{\widehat{\mu}_i(t)}   \right) -    \left(U^V_i\right)^{'} \left( \proj{\widehat{v}_i(t)}  \right)\left( 
  v^{\pi}_i  -  \proj{\widehat{v}_i(t) }  \right)               \right) \right.
\\&&\left.\nonumber \hspace{-.2cm} + \sum_{i\in{\cal N}}  d^{\pi}_i \left( \frac{ p^d_i   }{  \overline{p}_i }  \right)
\left(\sigma^{\pi}_i    -  \proj{\widehat{\sigma}_i(t)} \right)
 +  \sum_{i\in{\cal N}}   b^{\pi}_i \left( \frac{ \sigma^{\pi}_i }{\left(1+ \overline{\beta}_i\right)} -  \frac{\proj{\widehat{\sigma}_i(t)}}{\left(1+ \overline{\beta}_i\right)} \right)\right),
 \\\Delta^{(\pi,r)} \left(\widehat{\bm{\Theta}}(t)\right)  &=&  -  \sum_{i\in{\cal N}}   b^{\pi}_i  \left( \rho^{\pi}_i- \proj{ \widehat{\rho}_i(t)} \right) ,
\\ \Delta^{(d)} \left(\widehat{\bm{\Theta}}(t)\right)  &=& \sum_{i\in{\cal N}} \left(1+ \overline{\beta}_i\right)   \widehat{\lambda}_i(t) 
\left(h^D_i\left(\widehat{d}_i(t)\right) - d^{\pi}_i \right) \frac{1}{u_i(t)}
\left( p^d_i\frac{ \sigma^{\pi}_i    }{ \overline{p}_i } - 1 \right),
 \\  \Delta^{(b)} \left(\widehat{\bm{\Theta}}(t)\right) &=&
 \sum_{i\in{\cal N}}   \left(h^B_i\left(\widehat{b}_i(t)\right) -  b^{\pi}_i\right) 
\left(\frac{\sigma^{\pi}_i}{ \left(1+ \overline{\beta}_i\right)  } 
-   \rho^{\pi}_i \right),
\\\nonumber  \Delta_3 \left(\widehat{\bm{\Theta}}(t)\right) &=& - d_{8N} \left( \widehat{\bm{\Theta}}(t) ,\widetilde{{\cal H}} \right),
\end{eqnarray*}
and recall that 
\begin{eqnarray*}
 \Delta_1 \left(\widehat{\bm{\Theta}}(t)\right) &=&\hspace{-.3cm} - \sum_{i\in{\cal N}} \frac{\left(1+ \overline{\beta}_i\right)}{ u_i(t) }
 E \left[  L^{\pi}_i  
 \left(  \phi^Q\left(q^*_i \left(\widehat{\bm{\theta}}_i(t),F^{\pi}_{i} \right),\widehat{\bm{\theta}}_i(t),F^{\pi}_{i} \right) - \phi^Q\left(q^*_i \left(\widehat{\bm{\theta}}^{(m^{\pi}_i)}_i(t),F^{\pi}_{i} \right),\widehat{\bm{\theta}}^{(m^{\pi}_i)}_i(t),F^{\pi}_{i} \right)   \right.      \right.
\\&&\nonumber\left.\left. \hspace{2.4cm} -  \left(U^E_i\right)^{'}\left( \widehat{\mu}_i(t) - U^V_i \left(\widehat{v}_i(t)\right)\right)\left(U^V_i\right)^{'} \left(\widehat{v}_i(t)\right)
  2 \left(  \widehat{m}_i(t)  -  m^{\pi}_i\right)
\left(  q^*_i\left(t\right)  - \widehat{m}_i(t)  \right)\right) \right],
\\ \Delta_2 \left(\widehat{\bm{\Theta}}(t)\right) &=& - \frac{1}{\tau_{slot}}   E\left[  \phi^R\left(\mathbf{r}^*(t),\widehat{\mathbf{b}} (t),C^{\pi}\right) - \phi^R\left(\mathbf{r}^{\pi}\left(C^{\pi}\right),\widehat{\mathbf{b}} (t),C^{\pi}\right)  \right] \hspace{0cm}
 \\&&\nonumber \hspace{-1cm}- \sum_{i\in{\cal N}} \frac{\left(1+ \overline{\beta}_i\right)}{ u_i(t) }
 E \left[  L^{\pi}_i  
 \left(  \phi^Q\left(q^*_i \left(\proj{\widehat{\bm{\theta}}^{(m^{\pi}_i)}_i}(t),F^{\pi}_{i} \right),\proj{\widehat{\bm{\theta}}^{(m^{\pi}_i)}_i}(t),F^{\pi}_{i} \right) 
  - \phi^Q\left(q^{\pi}_i \left(F^{\pi}_{i} \right),\proj{\widehat{\bm{\theta}}^{(m^{\pi}_i)}_i}(t),F^{\pi}_{i} \right) \right)  \right].
\end{eqnarray*}
Hence, we can rewrite \eqref{line_3_der_L} as follows:
 \begin{eqnarray}
 \label{line_last_der_L}
\frac{d L \left(\widehat{\bm{\Theta}}(t)\right)}{dt}
&\le& \Delta \left(\widehat{\bm{\Theta}}(t)\right).
\end{eqnarray}
Next, we show that all the functions
$\Delta_1 \left(\bm{\Theta}\right) $, $ \Delta_2 \left(\bm{\Theta}\right) $, $ \Delta^{(d)} \left(\bm{\Theta}\right) $, $ \Delta^{(b)} \left(\bm{\Theta}\right)  
$, $  \Delta^{(\pi,q)} \left(\bm{\Theta}\right)   $, $  \Delta^{(\pi,r)} \left(\bm{\Theta}\right) $
and $ \Delta_3 \left(\bm{\Theta}\right)$,
from the definition \eqref{define_Delta} of $\Delta \left(\bm{\Theta}\right)$ 
are non-positive for $\bm{\Theta}\in{\cal H}^*$
so that $\Delta \left(\bm{\Theta}\right)$ is non-positive
for all $\bm{\Theta}\in{\cal H}^*$, 
and that $\Delta \left(\bm{\Theta}\right)<0$ for $\bm{\Theta}\notin{\cal H}^*$.

The non-positivity of the functions $\Delta^{(d)} \left(\widehat{\bm{\Theta}}(t)\right)$ and $\Delta^{(b)} \left(\widehat{\bm{\Theta}}(t)\right)$ 
follows from complementary slackness conditions for OPTSTAT given in \eqref{complimentary_slackness_cond5}-\eqref{complimentary_slackness_cond6},
and the following observations about the optimal solution to OPTSTAT which follows from 
the feasibility of the optimal solution (specifically that it satisfies
\eqref{OPTSTAT_eta_cost_bound} and \eqref{OPTSTAT_eta_bound_rebuf})
$$ p^d_i\frac{ \sigma^{\pi}_i    }{ \overline{p}_i } \le 1,\ \frac{\sigma^{\pi}_i}{ \left(1+ \overline{\beta}_i\right)  } \le   \rho^{\pi}_i \ \iiN.$$
For instance,
if we consider the term 
$ \left(h^B_i\left(\widehat{b}_i(t)\right) -  b^{\pi}_i\right) 
\left(\frac{\sigma^{\pi}_i}{ \left(1+ \overline{\beta}_i\right)  } 
-   \rho^{\pi}_i \right)$
in the definition of $\Delta^{(b)} \left(\widehat{\bm{\Theta}}(t)\right)$,
we see that if  $b^{\pi}_i>0$,
$\ \frac{\sigma^{\pi}_i}{ \left(1+ \overline{\beta}_i\right)  } =   \rho^{\pi}_i$ due to \eqref{complimentary_slackness_cond6},
and thus the term is zero.
The case for $b^{\pi}_i=0$ follows from the feasibility condition \eqref{OPTSTAT_eta_bound_rebuf} and the fact that $h^B_i\left(\widehat{b}_i(t)\right)\ge 0$.

Next, we show that $ \Delta^{(\pi,q)} \left(\widehat{\bm{\Theta}}(t)\right)$
is non-positive and that it is negative unless
 $ \proj{\widehat{\bm{\mu}}(t)} = \mathbf{m}^{\pi}$ and 
 $  \proj{\widehat{\mathbf{v}}(t) } = \mathbf{v}^{\pi}$.
 Since $\proj{\widehat{\bm{\Theta}}(t)}$ is an element of $\widetilde{\cal H}$, for each $\iiN$,
 there is some feasible quality adaptation policy $\left(\left(q_i\left(f_i\right) \right)_{f_i\in{\cal F}_i}\right)$ satisfying
 (see definition of $\widetilde{\cal H}$ in \eqref{defn_calH})
  \begin{eqnarray}
  \label{conditions_on_achievable_parameters_for_adaptation}
  \proj{\widehat{\mu}_i(t)}= \frac{ E \left[ L^{\pi}_i q_i\left( F^{\pi}_i \right)\right]}{ E \left[ L^{\pi}_i \right]},\
 \proj{\widehat{v}_i(t) } \ge \mbox{Var} \left( q_i\left( F^{\pi}_{i} \right)  \right),
  \  \proj{\widehat{\sigma}_i(t) } \ge \frac{\exn\left[L^{\pi}_i F^{\pi}_{i} \left( q_i\left( F^{\pi}_{i} \right) \right) \right]}{ \exn \left[ L^{\pi}_i\right]},
 \end{eqnarray}
 Using Lemma \ref{optimality_of_q_pi} (a),
 noting that $\left(\left(q^{\pi}_i\left(f_i\right) \right)_{f_i\in{\cal F}_i}\right)$
 is the unique optimal solution and 
 $\left(\left(q_i\left(f_i\right) \right)_{f_i\in{\cal F}_i}\right)$ is a feasible solution
 to the optimization problem considered in Lemma \ref{optimality_of_q_pi} (a),
 we have
 \begin{eqnarray}
  \nonumber&&\hspace{-1cm}U^E_i \left(  m^{\pi}_i- U^V_i \left( v^{\pi}_i \right)\right)
 - \sum_{i\in{\cal N}}  d^{\pi}_i \left( \frac{ p^d_i   }{  \overline{p}_i }  \right)
\sigma^{\pi}_i -  \sum_{i\in{\cal N}}  \frac{ b^{\pi}_i}{ \left(1+ \overline{\beta}_i\right)} 
\sigma^{\pi}_i 
\\\label{user_i_optimality_gap_1}&\ge& U^E_i \left(   \frac{\exn \left[ L^{\pi}_i q_i\left( F^{\pi}_{i} \right) \right]}{\exn \left[ L^{\pi}_i\right]}	- U^V_i \left( \mbox{Var} \left( q_i\left( F^{\pi}_{i} \right)  \right) \right)\right)
 \\\nonumber&&- \sum_{i\in{\cal N}}  d^{\pi}_i \left( \frac{ p^d_i   }{  \overline{p}_i }  \right)
\left( \frac{ \exn \left[L^{\pi}_i F^{\pi}_{i} \left( q_i\left( F^{\pi}_{i} \right) \right) \right] } 
{  \exn \left[L^{\pi}_i   \right]}    \right)
 -  \sum_{i\in{\cal N}}  \frac{ b^{\pi}_i}{ \left(1+ \overline{\beta}_i\right)} 
 \left(  \frac{ \exn \left[L^{\pi}_i F^{\pi}_{i} \left( q_i\left( F^{\pi}_{i} \right) \right) \right] } 
{  \exn \left[L^{\pi}_i   \right]}  \right)
\\\label{user_i_optimality_gap_2}&\ge& U^E_i \left(  \proj{\widehat{\mu}_i(t)}
- U^V_i \left( \proj{\widehat{\mathbf{v}}(t) } \right)\right)
- \sum_{i\in{\cal N}}  d^{\pi}_i \left( \frac{ p^d_i   }{  \overline{p}_i }  \right) \proj{\widehat{\sigma}_i(t) }
 -  \sum_{i\in{\cal N}}  \frac{ b^{\pi}_i}{ \left(1+ \overline{\beta}_i\right)} \proj{\widehat{\sigma}_i(t) },
 \end{eqnarray}
 where the second inequality follows from \eqref{conditions_on_achievable_parameters_for_adaptation}.
 Since 
 $\left(\left(q^{\pi}_i\left(f_i\right) \right)_{f_i\in{\cal F}_i}\right)$
 is the unique optimal solution,
 the inequality in \eqref{user_i_optimality_gap_1} is strict unless 
 $\left(\left(q^{\pi}_i\left(f_i\right) \right)_{f_i\in{\cal F}_i}\right)=\left(\left(q_i\left(f_i\right) \right)_{f_i\in{\cal F}_i}\right)$.
 Also, the inequality in \eqref{user_i_optimality_gap_2} is strict unless 
 $\proj{\widehat{\mu}_i(t)}= \frac{ E \left[ L^{\pi}_i q_i\left( F^{\pi}_i \right)\right]}{ E \left[ L^{\pi}_i \right]}$
 and $ \proj{\widehat{v}_i(t) } = \mbox{Var} \left( q_i\left( F^{\pi}_{i} \right)  \right)$ for each $\iiN$.
 Now, since $U^E_i(.)$ and $-U^V_i(.)$ are concave functions of their arguments, and since $\left(U^E_i\right)'(.)$ is non-negative, we have
  \begin{eqnarray*}
  &&\hspace{-.7cm}U^E_i \left(  m^{\pi}_i- U^V_i \left( v^{\pi}_i \right)\right)
\le U^E_i \left(   \proj{\widehat{\mu}_i(t)}  - U^V_i \left( \proj{\widehat{v}_i(t)}  \right) \right)
\\\nonumber &&+\left(U^E_i\right)^{'} \left( \proj{\widehat{\mu}_i(t)} -  U^V_i \left( \proj{\widehat{v}_i(t)}  \right)\right)
\left( \proj{\widehat{\mu}_i(t)}-  m^{\pi}_i
-  \left(U^V_i\right)^{'} \left( \proj{\widehat{v}_i(t)}  \right)\left( 
  v^{\pi}_i  -  \proj{\widehat{v}_i(t) }  \right)
\right).
 \end{eqnarray*}
 By combining the above inequality and \eqref{user_i_optimality_gap_2},
 we have that $ \Delta^{(\pi,q)} \left(\widehat{\bm{\Theta}}(t)\right)$
is non-positive.
Further, since the inequality in \eqref{user_i_optimality_gap_1} is strict unless 
 $\left(\left(q^{\pi}_i\left(f_i\right) \right)_{f_i\in{\cal F}_i}\right)=\left(\left(q_i\left(f_i\right) \right)_{f_i\in{\cal F}_i}\right)$,
 and
 the inequality in \eqref{user_i_optimality_gap_2} is strict unless 
 $\proj{\widehat{\mu}_i(t)}= \frac{ E \left[ L^{\pi}_i q_i\left( F^{\pi}_i \right)\right]}{ E \left[ L^{\pi}_i \right]}$
 and $ \proj{\widehat{v}_i(t) } = \mbox{Var} \left( q_i\left( F^{\pi}_{i} \right)  \right)$ for each $\iiN$,
 we can conclude
 $ \proj{\widehat{\bm{\mu}}(t)} = \mathbf{m}^{\pi}$ and 
 $  \proj{\widehat{\mathbf{v}}(t) } = \mathbf{v}^{\pi}$.
\begin{eqnarray}\label{delta_pi_q_stictly_negative_if_mu_not_mpi_v_not_vpi}
\Delta^{(\pi,q)} \left(\widehat{\bm{\Theta}}(t)\right)
=0 \mbox{ only if }\ \proj{\widehat{\bm{\mu}}(t)} = \mathbf{m}^{\pi} \mbox{ and }  \proj{\widehat{\mathbf{v}}(t) } = \mathbf{v}^{\pi}.
\end{eqnarray}
Using similar arguments along with Lemma \ref{optimality_of_q_pi} (b), 
 we can show that 
 \begin{eqnarray}\label{delta_pi_r_stictly_negative_if_rho_not_rhopi}
\Delta^{(\pi,r)} \left(\widehat{\bm{\Theta}}(t)\right)
=0 \mbox{ only if }\ \proj{ \widehat{\bm{\rho}}_i(t)}=\bm{\rho}^{\pi} .
\end{eqnarray}

Next we consider the term $\Delta_1 \left(\widehat{\bm{\Theta}}(t)\right)$.
Using Lemma \ref{concavity_and_derivative_of_obj_of_QNOVA} (c), 
we can show that $\Delta_1 \left(\widehat{\bm{\Theta}}(t)\right)\le 0$
and that
\begin{eqnarray}\label{delta2_stictly_negative_if_m_not_mpi}
\Delta_1 \left(\widehat{\bm{\Theta}}(t)\right)=0 \mbox{ only if } \widehat{\mathbf{m}}(t) =  \mathbf{m}^{\pi}.
\end{eqnarray}

Next we consider the term $\Delta_2 \left(\widehat{\bm{\Theta}}(t)\right)$.
Using the fact that $q^*_i \left(\proj{\widehat{\bm{\theta}}^{(m^{\pi}_i)}_i}(t),f_{i} \right)$ and $\mathbf{r}^*(t)$
are optimal solutions to optimization problems QNOVA$\left(\proj{\widehat{\bm{\theta}}^{(m^{\pi}_i)}_i}(t),f_{i} \right)$
and RNOVA$\left(\widehat{\mathbf{b}} (t),c\right)$ respectively, 
we can conclude that $\Delta_2 \left(\widehat{\bm{\Theta}}(t)\right)\le 0$.

Also, note that $ \Delta_3 \left(\bm{\Theta}\right)=- d_{8N} \left( \widehat{\bm{\Theta}} ,\widetilde{{\cal H}} \right)$ is non-positive, and 
 \begin{eqnarray}\label{delta_3_unless_theta_in_calH}
\Delta_3 \left(\widehat{\bm{\Theta}}(t)\right) = 0 
\mbox{ only if }\ \proj{\widehat{\bm{\mu}}(t)} = \widehat{\bm{\mu}}(t), 
\ \proj{\widehat{\mathbf{v}}(t) } = \widehat{\mathbf{v}}(t) \mbox{ and }  \proj{\widehat{\bm{\rho}}(t)} = \widehat{\bm{\rho}}(t).
\end{eqnarray}

 Next, we argue that
 $\Delta \left(\widehat{\bm{\Theta}}(t)\right)=0 $
only if 
$$\left(\widehat{\bm{\rho}}(t),\left(h^B_i\left(\widehat{b}_i (t)\right)\right)_{\iiN}
,\left(h^D_i\left(\widehat{d}_i (t)\right)\right)_{\iiN}\right)\in {\cal X}^{\pi}.$$
Suppose that $\Delta \left(\widehat{\bm{\Theta}}(t)\right)=0 $.
Then,
$ \Delta^{(\pi,q)} \left(\widehat{\bm{\Theta}}(t)\right)   +  \Delta^{(\pi,r)} \left(\widehat{\bm{\Theta}}(t)\right) + \Delta_3 \left(\widehat{\bm{\Theta}}(t)\right)=0 $,
and from \eqref{delta_pi_q_stictly_negative_if_mu_not_mpi_v_not_vpi}, \eqref{delta_pi_r_stictly_negative_if_rho_not_rhopi}
and \eqref{delta_3_unless_theta_in_calH}, we can conclude that
 $ \widehat{\bm{\mu}}(t) = \mathbf{m}^{\pi}$, 
 $  \widehat{\mathbf{v}}(t) = \mathbf{v}^{\pi}$
 and $ \widehat{\bm{\rho}}_i(t)=\bm{\rho}^{\pi}$.
 We also have that $\Delta_2 \left(\widehat{\bm{\Theta}}(t)\right)=0$, and hence
  \begin{eqnarray*}
      \phi^R\left(  \mathbf{r}^*\left(\widehat{\mathbf{b}} (t),c\right),\widehat{\mathbf{b}} (t),c\right) 
      &= &\phi^R\left(\mathbf{r}^{\pi}\left(c\right),\widehat{\mathbf{b}} (t),c\right),
 \ \forall  \ \ciC,
 \\   \phi^Q\left(q^*_i \left(\proj{\widehat{\bm{\theta}}^{(m^{\pi}_i)}_i}(t),f_i \right),\proj{\widehat{\bm{\theta}}^{(m^{\pi}_i)}_i}(t),f_i \right) 
 &=& \phi^Q\left(q^{\pi}_i \left(f_i \right),\proj{\widehat{\bm{\theta}}^{(m^{\pi}_i)}_i}(t),f_i \right),
 \\&& \hspace{3cm}\ \forall \ f_i\in{\cal F}_i,\ \forall \ \iiN,
\end{eqnarray*}
where recall that 
$\proj{\widehat{\bm{\theta}}^{(m^{\pi}_i)}_i}(t)=
 \left(m^{\pi}_i,\proj{\widehat{\mu}_i(t)},\proj{\widehat{v}_i(t)},\widehat{b}_i(t),\widehat{d}_i(t)\right)$.
 Since (from earlier observations in this paragraph) $ \widehat{\bm{\mu}}(t) = \mathbf{m}^{\pi}$
 and $  \widehat{\mathbf{v}}(t) = \mathbf{v}^{\pi}$,
 we have $\proj{\widehat{\bm{\theta}}^{(m^{\pi}_i)}_i}(t)=
 \left(m^{\pi}_i,m^{\pi}_i,v^{\pi}_i,\widehat{b}_i(t),\widehat{d}_i(t)\right)$.
Hence, $\mathbf{r}^{\pi}\left(c\right)$ is an optimal solution to
RNOVA$\left(\widehat{\mathbf{b}} (t),c\right)$ for each $\ciC$,
and 
$q^{\pi}_i \left(f_i \right)$ is an optimal solution to
QNOVA$\left(\proj{\widehat{\bm{\theta}}^{(m^{\pi}_i)}_i}(t),f_i\right)$ for each $f_i\in{\cal F}_i$ and $\iiN$.
 Hence for each $\ciC$,
 $\mathbf{r}^{\pi}\left(c\right) $
 satisfies the optimality conditions 
 \eqref{rvbar_gradient_zero_condtion}-\eqref{rvbar_complimentary_slackness_for_resource_allocation_lower_bound}
 for RNOVA$\left(\widehat{\mathbf{b}} (t),c\right)$.
 Denote the associated optimal Lagrange multipliers in these conditions by
$\left(\chi^{'}(c)\right)_{c\in{\cal C}}$ and $\left(\bm{\omega}^{'}(c)\right)_{c\in{\cal C}}$.
 Similarly, $q^{\pi}_i\left( f_i\right) $
 satisfies the optimality conditions \eqref{KKT_QNOVA_gradient_zero}-\eqref{KKTQRNOVA_bounded_above_constraints}
 for QNOVA$\left(\proj{\widehat{\bm{\theta}}^{(m^{\pi}_i)}_i}(t),f_i\right)$.
Let $\left(\left({\gamma}_i^{'}(f)\right)_{f\in{\cal F}_i}\right)_{i\in{\cal N}}$,
$\left(\left({\overline{\gamma}_i(f)}^{'}\right)_{f\in{\cal F}_i}\right)_{i\in{\cal N}}$,
denote the associated optimal Lagrange multipliers.

Thus, $\left(\left(\left(q^{\pi}_i\left(f\right) \right)_{f\in{\cal F}_i}\right)_{i\in{\cal N}},
\left(\mathbf{r}^{\pi} \left(c\right)\right)_{c\in{\cal C}}\right)$
together with the non-negative constants
$\left(\chi^{'}(c)\right)_{c\in{\cal C}}$, $\left(\bm{\omega}^{'}(c)\right)_{c\in{\cal C}}$,
\\$\left(\left({\gamma}_i^{'}(f)\right)_{f\in{\cal F}_i}\right)_{i\in{\cal N}}$
and
$\left(\left({\overline{\gamma}_i(f)}^{'}\right)_{f\in{\cal F}_i}\right)_{i\in{\cal N}}$
satisfy the optimality conditions for OPTSTAT given in \eqref{KKT_OPTSTAT_gradient_zero_quality_condition}-\eqref{complimentary_slackness_cond6}
with $b^{\pi}_i$ replaced by $\widehat{b}_i(t)$ and $d^{\pi}_i$ replaced by $\widehat{d}_i(t)$ for each video client $\iiN$.
Note that \eqref{complimentary_slackness_cond5}-\eqref{complimentary_slackness_cond6}
are satisfied since $\Delta^{(d)} \left(\widehat{\bm{\Theta}}(t)\right) + \Delta^{(b)} \left(\widehat{\bm{\Theta}}(t)\right)=0$ 
(since $\Delta \left(\widehat{\bm{\Theta}}(t)\right)=0 $).
Hence, we have shown that
\begin{eqnarray}\label{delta_stictly_negative_if_rhobd_not_in_calX}
\Delta \left(\widehat{\bm{\Theta}}(t)\right)=0
\mbox{ only if }
\left(\widehat{\bm{\rho}}(t),\left(h^B_i\left(\widehat{b}_i (t)\right)\right)_{\iiN},\left(h^D_i\left(\widehat{d}_i (t)\right)\right)_{\iiN}\right)\in {\cal X}^{\pi}
\end{eqnarray}
 
Now, the above discussion along with \eqref{delta_pi_q_stictly_negative_if_mu_not_mpi_v_not_vpi}, 
\eqref{delta_pi_r_stictly_negative_if_rho_not_rhopi}
\eqref{delta_3_unless_theta_in_calH},
and \eqref{delta_stictly_negative_if_rhobd_not_in_calX} allow us to conclude that for almost all $t$
 \begin{eqnarray}\label{main_negative_drift_claim}\hspace{-.5cm}
\frac{d L \left(\widehat{\bm{\Theta}}(t)\right)}{dt}
\le \Delta \left(\widehat{\bm{\Theta}}(t)\right) \mbox{ where }
\Delta \left(\widehat{\bm{\Theta}}\right)\le 0
\ \forall \ \bm{\Theta}\in{\cal H}, \
\Delta \left(\bm{\Theta}\right) < 0 \ \forall \ \bm{\Theta}\notin{\cal H}^*.\end{eqnarray}
This completes proof of part (a) of the theorem.
Proof of part (b) can be completed using \eqref{main_negative_drift_claim}
and an appropriate choice of a continuous function bounding $\Delta \left(\bm{\Theta}\right) $ from above
(obtained by replacing $\frac{1}{u_i(t)}$ with $\frac{1}{u_{\max}}$, where $u_{\max}$ is defined in \eqref{defn_umin_umax},
in the constituent functions of $\Delta \left(\bm{\Theta}\right) $)
so that the new function also satisfies \eqref{main_negative_drift_claim} after replacement of $\Delta \left(\bm{\Theta}\right) $ 
in \eqref{defn_umin_umax}.
The details are left to the reader.

\end{IEEEproof}

\subsection{Convergence of NOVA and proof of Theorem \ref{main_optimality_theorem}}
\label{convergence_of_diff_inclusion_and_theorem}

In Subsection \ref{section_optstat}, we obtained Theorem \ref{main_NOVA_with_theta_pi_is_optimal}
which says that
for almost all sample paths
\begin{eqnarray*}
\lim_{S\rightarrow \infty} 
\left(\phi_S\left(\left( \left( q^*_i \left(\bm{\theta}^{\pi}_i , f_{i,s} \right)\right)_{\iiN} \right)_{1\le s \le S}  \right)
-\phi^{opt}_S\right)
=0,
\end{eqnarray*}
for each $\bm{\theta}^{\pi}_i \in {\cal H}^*_i$ and each $\iiN$.
This suggests that we can prove the main optimality result for NOVA 
if we establish convergence of NOVA's parameters to the set ${\cal H}^*_i$.
The main focus of this subsection is Theorem \ref{main_stoch_approx_result} which relates 
NOVA to the auxiliary differential inclusion \eqref{mhat_ode_rule}-\eqref{rhohat_ode_rule},
and thus obtains the desired convergence result for NOVA by using 
the convergence result obtained in Theorem \ref{convergence_of_differential_inclusion} for the differential inclusion.
Our approach here relies on viewing the update equations (\eqref{m_update_NOVA}-\eqref{lambda_update_NOVA} and \eqref{sigma_NOVA_update}-\eqref{rho_NOVA_update})
of NOVA as an asynchronous stochastic approximation update equation (see Chapter 12 of \cite{kushner_text} for a
detailed discussion on asynchronous stochastic approximation)
to relate NOVA to the differential inclusion.
After obtaining the convergence result for NOVA in Theorem \ref{main_stoch_approx_result},
we conclude this section with the proof of Theorem \ref{main_optimality_theorem}.

In this subsection, we use the superscript $\epsilon$ on 
NOVA parameters $(m^{\epsilon}_{i,s})_{\iiN}$, $(\mu^{\epsilon}_{i,s})_{\iiN}$, $(v^{\epsilon}_{i,s})_{\iiN}$,
$(b^{\epsilon}_{Q,i,s})_{\iiN}$, 
$(b^{\epsilon}_{R,i,k})_{\iiN}$, 
$(b^{\epsilon}_{i,k})_{\iiN}$, 
$(d^{\epsilon}_{i,s})_{\iiN}$, $(\lambda^{\epsilon}_{i,s})_{\iiN}$, $(\sigma^{\epsilon}_{i,s})_{\iiN}$ and $(\rho^{\epsilon}_{i,k})_{\iiN}$
to emphasize their dependence on $\epsilon$ (see NOVA updates in \eqref{b_allocate_update}-\eqref{lambda_update_NOVA}).
We refer to the update 
of NOVA parameters $\left(m_{i,s_i},\mu_{i,s_i},v_{i,s_i},b_{i,k},d_{i,s_i},\lambda_{i,s_i}\right)$
in
\eqref{m_update_NOVA}-\eqref{lambda_update_NOVA}
carried out after the selection of segment quality for video client $i$ (following a segment download) as a Q$_i$-update,
and we refer to the update \eqref{b_allocate_update} on $\mathbf{b}_k$ carried out at the beginning of each slot $k$ as an R-update.
Let $\delta \tau^{\epsilon}_{Q,i,s}$ denote the time (in seconds) between the $s$th and $(s+1)$th Q$_i$-updates.
Let $\delta \tau^{\epsilon}_{R,k}$ denote the time between the $k$th and $(k+1)$th R-updates,
i.e., $\delta \tau^{\epsilon}_{R,k}=\tau_{slot}$ for each $k$.
Let 
$$ \tau^{\epsilon}_{R,k} = \epsilon \sum_{j=0}^{k-1} \delta \tau^{\epsilon}_{R,j},
 \ \tau^{\epsilon}_{Q,i,s} = \epsilon \sum_{j=0}^{s-1} \delta \tau^{\epsilon}_{Q,i,j} $$
denote $\epsilon$ times the cumulative time for the first $k$ R-updates and $s$ Q$_i$-updates respectively.

Next, we define time interpolated processes \\$\left(\widehat{\mathbf{m}}^{\epsilon}(t),\widehat{\bm{\mu}}^{\epsilon}(t),\widehat{\mathbf{v}}^{\epsilon}(t),\widehat{\mathbf{b}}^{\epsilon}(t),
\widehat{\mathbf{d}}^{\epsilon}(t),\widehat{\bm{\lambda}}^{\epsilon}(t),\widehat{\bm{\sigma}}^{\epsilon}(t),\widehat{\bm{\rho}}^{\epsilon}(t)\right)$
associated with NOVA's parameters.
For each $\iiN$ and for $t\in \left[ \tau^{\epsilon}_{Q,i,s},\tau^{\epsilon}_{Q,i,s+1} \right)$, let
$\widehat{m}^{\epsilon}_i(t)=m^{\epsilon}_{i,s}$,
$\widehat{\mu}^{\epsilon}_i(t)=\mu^{\epsilon}_{i,s}$,
$\widehat{v}^{\epsilon}_i(t)=v^{\epsilon}_{i,s}$,
$\widehat{b}^{\epsilon}_{Q,i}(t)=b^{\epsilon}_{Q,i,s}$,
$\widehat{d}^{\epsilon}_i(t)=d^{\epsilon}_{i,s}$,
$\widehat{\lambda}^{\epsilon}_i(t)=\lambda^{\epsilon}_{i,s}$ and
$\widehat{\sigma}^{\epsilon}_i(t)=\sigma^{\epsilon}_{i,s}$.
Also, for $t\in \left[k\epsilon,(k+1)\epsilon\right)$,
let $\widehat{b}^{\epsilon}_{R,i}(t)=b^{\epsilon}_{R,i,k}$ and $\widehat{\rho}^{\epsilon}_i(t)=\rho^{\epsilon}_{i,k}$.
Recall that $b^{\epsilon}_{Q,i,s}$ and $b^{\epsilon}_{R,i,k}$ 
are auxiliary variables used in the description of NOVA (in Section \ref{section_NOVA}).

For each $t$, let 
\begin{eqnarray*}
\widehat{\bm{\Theta}}^{\epsilon}_Q(t)&=&
\left(\widehat{\mathbf{m}}^{\epsilon}(t),\widehat{\bm{\mu}}^{\epsilon}(t),\widehat{\mathbf{v}}^{\epsilon}(t),
\widehat{\mathbf{b}}^{\epsilon}_Q(t),
\widehat{\mathbf{d}}^{\epsilon}(t),\widehat{\bm{\lambda}}^{\epsilon}(t),\widehat{\bm{\sigma}}^{\epsilon}(t),\widehat{\bm{\rho}}^{\epsilon}(t)\right),
\\\widehat{\bm{\Theta}}^{\epsilon}_R(t)&=&\left(\widehat{\mathbf{m}}^{\epsilon}(t),\widehat{\bm{\mu}}^{\epsilon}(t),\widehat{\mathbf{v}}^{\epsilon}(t)
,\widehat{\mathbf{b}}^{\epsilon}_R(t),
\widehat{\mathbf{d}}^{\epsilon}(t),\widehat{\bm{\lambda}}^{\epsilon}(t),\widehat{\bm{\sigma}}^{\epsilon}(t),\widehat{\bm{\rho}}^{\epsilon}(t)\right),
\end{eqnarray*}
Note that definitions $\widehat{\bm{\Theta}}^{\epsilon}_Q(.)$
and                   $\widehat{\bm{\Theta}}^{\epsilon}_R(.)$
are different only for components $3N+1$ to $4N$.
The next result states that for small enough $\epsilon$,
the time interpolated versions of NOVA parameters 
$\widehat{\bm{\Theta}}^{\epsilon}_Q(.)$
and $\widehat{\bm{\Theta}}^{\epsilon}_R(.)$
stay close to the set ${\cal H}^*$ (defined in \eqref{defn_Hstar})
most of the time over long time windows.
The proof relies on relating $\widehat{\bm{\Theta}}^{\epsilon}_Q(.)$
and $\widehat{\bm{\Theta}}^{\epsilon}_R(.)$ 
associated with NOVA to the auxiliary differential inclusion \eqref{mhat_ode_rule}-\eqref{rhohat_ode_rule}
by viewing the update equations \eqref{m_update_NOVA}-\eqref{lambda_update_NOVA}
of NOVA as an asynchronous stochastic approximation update equation,
and using Theorem \ref{convergence_of_differential_inclusion}
which states that the differential inclusion converge to the set ${\cal H}^*$.
\begin{theorem}\label{main_stoch_approx_result}
Let $\widehat{\bm{\Theta}}^{\epsilon}_Q(0)=\widehat{\bm{\Theta}}^{\epsilon}(0)\in{\cal H}$.
Then, the fraction of time in the time interval $[0,T]$ that $\widehat{\bm{\Theta}}^{\epsilon}_Q(.)$
and $\widehat{\bm{\Theta}}^{\epsilon}_R(.)$
spend in a small neighborhood of ${\cal H}^*$
converges to one in probability as $\epsilon\rightarrow 0$
and $T\rightarrow \infty$.
\end{theorem} 
\begin{IEEEproof}
This result follows from an extension of Theorem 3.4 in Chapter 12 of \cite{kushner_text}
which relates asynchronous stochastic approximation \eqref{m_update_NOVA}-\eqref{lambda_update_NOVA}
to its associated differential inclusion \eqref{mhat_ode_rule}-\eqref{rhohat_ode_rule}.
Theorem 3.4 cannot be directly applied mainly because 
condition (A3.8) (given in Section 12.3.3, page 418 of \cite{kushner_text}) concerning the time between the (asynchronous) updates is not be satisfied in our problem setting (discussed later in the proof).
Below, we discuss why Theorem 3.4 can not be directly applied,
and an appropriate extension to prove our result.
To explain this in more detail, we introduce some notation.

In order to simplify our discussion, we consider the special case of NOVA, NOVA-L1 (described below)
which exhibits the key ideas involved in the extension.
NOVA-L1 corresponds to a setting with a single video client with $U^V_1$ and $U^E_1$ equal to (linear) identity functions, 
and no cost constraints.
We also assume the allocation in slot $k$ is $r^*_{1}\left(b^{\epsilon}_{1,k},c_k\right)$
where $r^*_{1}\left(b,c\right)$ is a continuous (single valued) function of $b$ for each $c$
(instead of the set valued mapping ${\cal R}^*\left(\mathbf{b}_k,c_k\right)$ associated with NOVA).
In the single video client case, $r^*_{1}\left(b,c\right)$ actually does not depend on $b$.
However, below we will not use this property, and will only rely on the continuity of $r^*_{1}\left(b,c\right)$ (with respect to $b$) 
so as to facilitate the extension of the proof to more general settings.
To further simplify the notation, we assume that $\overline{\beta}_1=0$ and all the segments have the same length $l_1$.
Hence, we need only track $m_{1,.}$ and $b_{1,.}$ since resource allocation and quality adaptation only depend on
these parameters (see \eqref{defn_phiR} and \eqref{defn_phiQ}).
For this special case, the algorithm NOVA-L1 works as follows:
\\\line(1,0){514}
\begin{flushleft}
\textbf{NOVA-L1.0}: Initialize: $m^{\epsilon}_{1,0},\ b^{\epsilon}_{1,0}$.
Let $s_1=0$.
\end{flushleft}
In each slot $k\ge 0$, carry out the following steps:
\\ \textbf{RNOVA-L1}: At the beginning of slot $k$, 
let $b^{\epsilon}_{R,1,k} =  b^{\epsilon}_{1,k}$,
allocate rate $r^*_{1}\left(b^{\epsilon}_{1,k},c_k\right)$ to video client 1,
and update $b^{\epsilon}_{1,k}$ as follows:
\begin{eqnarray}\label{L1_slot_b_update}
  b^{\epsilon}_{1,k+1} &=&  b^{\epsilon}_{1,k} + \epsilon \tau_{slot}.
\end{eqnarray}\\
 \textbf{QNOVA-L1}: In slot $k$, if video client 1 finishes transmission of the $s_1$ th segment, 
 let $b^{\epsilon}_{Q,1,s_1+1} = b^{\epsilon}_{1,k+1}$,
choose quality $q^*_1 \left(\left(m^{\epsilon}_{1,s_1},b^{\epsilon}_{Q,1,s_1+1}\right), f_{1,s_1+1}\right)$ denoted as $q^*_{1,s_1+1}$ for brevity,
and update $m^{\epsilon}_{1,s_1+1}$, $b^{\epsilon}_{1,k+1}$ and $s_1$ as follows:
\begin{eqnarray}\label{L1_seg_m_update}
  m^{\epsilon}_{1,s_1+1} &=& m^{\epsilon}_{1,s_1} 
  + \epsilon   \left(q^*_{1,s_1+1}- m^{\epsilon}_{1,s_1}   \right),
\\b^{\epsilon}_{1,k+1} &=& \left[  b^{\epsilon}_{1,k+1}   - \epsilon l_{1} \right]_{\underline{b}},
\label{L1_seg_b_update}
\\\nonumber s_1 &=& s_1+1.
\end{eqnarray}
\line(1,0){514}
\\Note that
$b^{\epsilon}_{Q,1,s}$ is the value of $b^{\epsilon}_{1,.}$ used in choosing quality for $s$th segment,
and $b^{\epsilon}_{R,1,k}$ is the value of $b$ used in choosing the allocation in $k$th slot.
These are book-keeping variables, and do not affect the evolution of the algorithm over time.
In this proof, we refer to the updates \eqref{L1_seg_m_update}-\eqref{L1_seg_b_update} in QNOVA-L1 as a Q-update (dropping the subscript `1' used in Section
\ref{section_NOVA} since there is just one video client),
and the update \eqref{L1_slot_b_update} as an R-update.

Let $\delta \tau^{\epsilon}_{Q,s}$ denote the time (in seconds) between the $s$th and $(s+1)$th Q-updates,
i.e., time required to download the $s$th segment.
Let $\delta \tau^{\epsilon}_{R,k}$ denote the time between the $k$th and $(k+1)$th R-updates.
Note that $\delta \tau^{\epsilon}_{R,k}=\tau_{slot}$ for each $k$ whereas 
$\delta \tau^{\epsilon}_{Q,s}$ can potentially be different for different $s$.
Let 
$$ \tau^{\epsilon}_{R,k} = \epsilon \sum_{j=0}^{k-1} \delta \tau^{\epsilon}_{R,j},
 \ \tau^{\epsilon}_{Q,s} = \epsilon \sum_{j=0}^{s-1} \delta \tau^{\epsilon}_{Q,j} $$
denote $\epsilon$ times the cumulative time for the first $k$ R-updates and $s$ Q-updates respectively.
We let 
\begin{eqnarray*}
\tau^{\epsilon}_{R}(t)&=&\tau^{\epsilon}_{R,k}, \ t\in \left[k\epsilon,(k+1)\epsilon\right),
\\\tau^{\epsilon}_{Q}(t)&=&\tau^{\epsilon}_{Q,s},\  t\in \left[s\epsilon,(s+1)\epsilon\right),
\\\widehat{m}^{\epsilon}_1(t)&=&m^{\epsilon}_{1,s},  \ t\in \left[ \tau^{\epsilon}_{Q,s},\tau^{\epsilon}_{Q,s+1} \right)
\\\widehat{b}^{\epsilon}_Q(t)&=&b^{\epsilon}_{Q,1,s}, \ t\in \left[ \tau^{\epsilon}_{Q,s},\tau^{\epsilon}_{Q,s+1} \right)
\\\widehat{b}^{\epsilon}_R(t)&=&b^{\epsilon}_{R,1,k}, \ t\in \left[ \tau^{\epsilon}_{R,k},\tau^{\epsilon}_{R,k+1} \right)
\end{eqnarray*}
Let
$\widehat{\bm{\Theta}}^{\epsilon}_Q(t)=  \left(\widehat{m}^{\epsilon}_1(t),\widehat{b}^{\epsilon}_Q(t)\right)$. 
Let ${\cal F}_{Q,s}$ denote a sigma-algebra that measures at least 
$m^{\epsilon}_{1,i},\ b^{\epsilon}_{Q,1,i},\ \tau^{\epsilon}_{Q,1,i}$ and $F_{1,i}$
for each $i\le s+1$,
and $b^{\epsilon}_{R,1,k}$ and $C_k$
for each $k\le \frac{\tau^{\epsilon}_{Q,1,s+1}}{\epsilon \tau_{slot}}$.

	\begin{figure}[ht]
	\centering
		\ifarxivmode
\includegraphics[scale=.41]{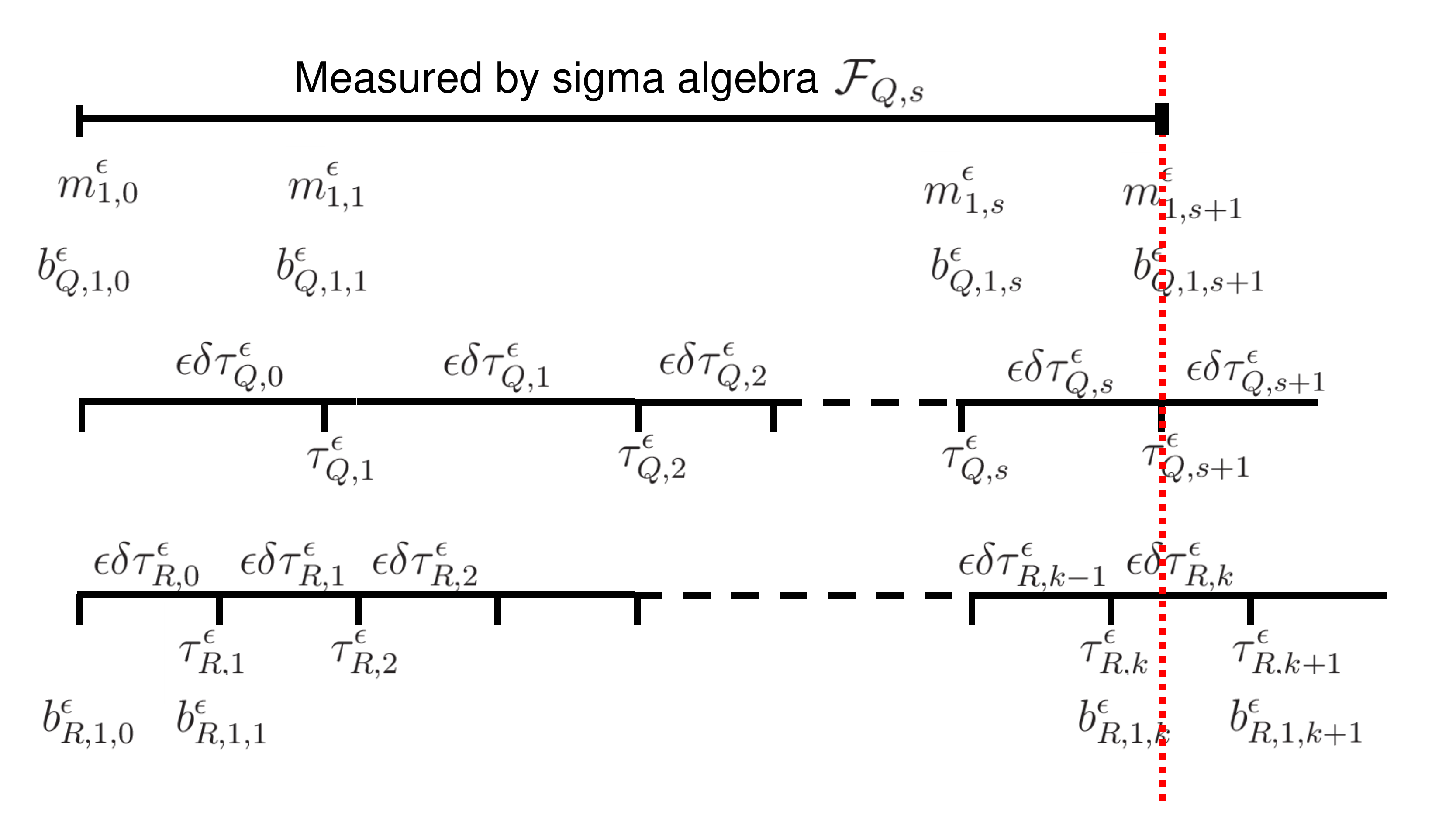}
\else
\includegraphics[scale=.41]{figs/stoch_approx_proof_fig.pdf}
\fi
	\caption{NOVA-L1: Asynchronous updates and associated variables}
	\label{stoch_approx_proof_fig client_trace}
\end{figure}

The stochastic approximation algorithm in our setting is different in two aspects from 
the one studied in Theorem 3.4 of \cite{kushner_text}.
First, since the time $\delta \tau^{\epsilon}_{Q,s+1}$ required to download segment $s+1$ depends on the exact instant during the slot in which the segment download begins
\footnote{This requirement may be met in real systems if practical constraints force segment downloads to begin at slot boundaries.},
we do not satisfy the necessary condition (A3.8) (and related assumptions (A3.10) and (A3.14) given in Section 12.3.3, page 418 of \cite{kushner_text})) given in Theorem 3.4.
More precisely, condition (A3.8) 
requires that the conditional expectation of  $\delta \tau^{\epsilon}_{Q,s}$ with respect to the sigma algebra ${\cal F}_{Q,s}$
depends on $\tau^{\epsilon}_{Q,s+1}$ only through the value of $\widehat{\bm{\Theta}}^{\epsilon}_Q(t)$
at $t=\tau^{\epsilon}_{Q,s+1}$ (and conditions (A3.10) and (A3.14) are related to the continuity properties and averaging behavior of this
conditional expectation function).
The approach in \cite{kushner_text} relies on this assumption to prove 
the following result (which in turn is used to prove the main result)
associated with the component $\tau_{Q}(t)$
of the weak limits (associated with $\tau^{\epsilon}_{Q}(t)$):
\begin{eqnarray}
\tau_Q(t) &= & \int_{0}^{t} \overline{u}_Q \left( \widehat{\bm{\Theta}}_Q( \tau_{Q}(s)  )  \right)ds.
\label{NOVA_L1_result_on_evolution_of_tau_hat_Q}
\end{eqnarray}
where for $ \bm{\Theta}=(m,b)$
\begin{eqnarray*} 
\overline{u}_{Q} \left(  \bm{\Theta}\right) &=&
 \tau_{slot} l_1 \frac{ E \left[   F^{\pi}_{1}\left(  
q^*_{1}\left( \bm{\Theta}  ,F^{\pi}_{1} \right)
\right)\right]  }{ E\left[r^*_1\left( b  ,C^{\pi}  \right)\right] }.
\end{eqnarray*}
Note that $\overline{u}_Q \left( \widehat{\bm{\Theta}}_Q( \tau_{Q}(t)  )  \right)$
corresponds to  $u_i\left( \widehat{\bm{\Theta}}_Q( \tau_{Q}(t)  )  \right)$ 
with $u_i$ as defined in \eqref{u_i_t_def} for the setting in NOVA-L1.
Further, we can intuitively see why \eqref{NOVA_L1_result_on_evolution_of_tau_hat_Q} should hold
by noting that 
$\tau_Q(t)$ is roughly the time required to download the first $t/\epsilon$ segments (for small $\epsilon$)
and viewing
$\overline{u}_Q \left( \widehat{\bm{\Theta}}_Q( \tau_{Q}(s)  )  \right)$ (appearing in the right hand side)
as the expected instantaneous segment download time for segment $s/\epsilon$,
so that the expression in the right hand side can be viewed 
as an integral (or roughly the sum) of segment download durations of the first $t/\epsilon$ segments.

The first goal of the discussion below is to argue that
\eqref{NOVA_L1_result_on_evolution_of_tau_hat_Q}
can be established for NOVA-L1.
Our argument relies on the fact that when considering time required for the download of a large number of segments,
the starting time (and the capacity of the slot associated with that instant) of the download of the first segment makes a negligible impact
(unlike the conditional expectation of $\delta \tau^{\epsilon}_{Q,s}$ with respect to ${\cal F}_{Q,s}$,
considered in condition (A3.8) of \cite{kushner_text}, which depends on $\tau^{\epsilon}_{Q,s+1}$).

The second aspect which in our setting is different from that in \cite{kushner_text} is the fact that the update rules for the
$b^{\epsilon}_{Q,1,s}$ and $b^{\epsilon}_{R,1,k}$ 
are different from those considered in Theorem 3.4 in that
they are determined by the evolution of parameter
$b^{\epsilon}_{1,k}$ which is \emph{updated on two time scales}.
Thus, we must also argue that the weak limits associated with $b^{\epsilon}_{Q,1,s}$ and $b^{\epsilon}_{R,1,k}$
are the same, and that the common limit $\widehat{b}_1(t)$ satisfies (special case of \eqref{bhat_ode_rule} for NOVA-L1)
given below
\begin{eqnarray}\label{nova_l1_b_weak_limit_property}
\bdt{\widehat{b}}_1(t) = 1
-    \frac{ l_1 }{u_1 \left(\widehat{\bm{\Theta}}_Q(t) \right)} + \widehat{z}^b_1 \left(\widehat{\bm{\Theta}}_Q(t)\right),
\end{eqnarray}
where $\widehat{z}^b_1 \left(\widehat{\bm{\Theta}}_Q(t)\right)\in  -{\cal Z}_{\cal H} \left( \widehat{\bm{\Theta}}_Q(t) \right)$.
Later in the proof, we show that \eqref{nova_l1_b_weak_limit_property} follows once we establish \eqref{NOVA_L1_result_on_evolution_of_tau_hat_Q}.

First, let us focus on the proof of \eqref{NOVA_L1_result_on_evolution_of_tau_hat_Q}.
If we let
\begin{eqnarray}\nonumber
\overline{{U}}^{\epsilon}_{Q}(t) 
&=& \epsilon \sum_{i=0}^{\frac{t}{\epsilon}-1}
 \overline{u}_{Q}\left(  \widehat{\bm{\Theta}}^{\epsilon}_Q(\tau^{\epsilon}_{Q,i})\right),\mbox{ and}
\\\label{interp_ubar_defn}
\widetilde{{U}}^{\epsilon}_{Q}(t) 
&=& \epsilon \sum_{i=0}^{\frac{t}{\epsilon}-1}
 \left( \delta \tau^{\epsilon}_{Q}(i)-\overline{u}_{Q}\left(  \widehat{\bm{\Theta}}^{\epsilon}_Q(\tau^{\epsilon}_{Q,i}) \right)\right),
\end{eqnarray}
then
\begin{eqnarray*}
\tau^{\epsilon}_{Q}(t)=\epsilon \sum_{i=0}^{\frac{t}{\epsilon}-1} \delta \tau^{\epsilon}_{Q}(i)
= \overline{{U}}^{\epsilon}_{Q}(t)
+\widetilde{{U}}^{\epsilon}_{Q}(t).
\end{eqnarray*}
Next, let 
\begin{eqnarray}
W^{\epsilon}_{Q} \left(t\right)  
= \tau^{\epsilon}_{Q}(t) - \overline{{U}}^{\epsilon}_{Q}(t)
= \widetilde{{U}}^{\epsilon}_{Q}(t).
\label{defining_W}
\end{eqnarray} 
For ease of reference, we are using notation similar to that used in Chapters 8 and 12 of \cite{kushner_text}
(and hence we are using redundant notation, for e.g., $W^{\epsilon}_{Q} = \widetilde{{U}}^{\epsilon}_{Q}$).

Now, fix $t$ and $\tau$. For any integer $p$, let $t_i\le t$, $i\le p$,
Let $h(.)$ be an arbitrary bounded, continuous and real valued function of its arguments.
Hence,
\begin{eqnarray}
\label{main_equation_for_checking_martingale_with_h_and_W}
 &&E\left[ h \left( \tau^{\epsilon}_{Q}(t_i),\widehat{\bm{\Theta}}^{\epsilon}_Q(\tau^{\epsilon}_{Q} (t_i) ),i\le p \right)
\left(W^{\epsilon}_{Q} \left(t+\tau\right)  - W^{\epsilon}_{Q} \left(t\right) \right) \right]
\\\nonumber &&-E\left[ h \left( \tau^{\epsilon}_{Q}(t_i),\widehat{\bm{\Theta}}^{\epsilon}_Q(\tau^{\epsilon}_{Q} (t_i) ),i\le p \right)
\left(\widetilde{{U}}^{\epsilon}_{Q} \left(t+\tau\right)  - \widetilde{{U}}^{\epsilon}_{Q} \left(t\right) \right) \right]=0.
\end{eqnarray}
If we show that the expression 
\begin{eqnarray}
\label{stoch_approx_proof_noise_averages_out_claim}
E\left[ h \left( \tau^{\epsilon}_{Q}(t_i),\widehat{\bm{\Theta}}^{\epsilon}_Q(\tau^{\epsilon}_{Q} (t_i) ),i\le p \right)
\left(\widetilde{{U}}^{\epsilon}_{Q} \left(t+\tau\right)  - \widetilde{{U}}^{\epsilon}_{Q} \left(t\right) \right) \right]
\end{eqnarray}
appearing in the above equation goes to zero as $\epsilon\rightarrow 0$,
then we can use \eqref{main_equation_for_checking_martingale_with_h_and_W}
along with an approach similar to that in proof of Theorem 2.1 in Chapter 8 of \cite{kushner_text}
to show that \eqref{NOVA_L1_result_on_evolution_of_tau_hat_Q} holds.

Hence, we next focus on showing that the expression in \eqref{stoch_approx_proof_noise_averages_out_claim} goes to zero as $\epsilon\rightarrow 0$.
For some fixed $\Delta>0$, let 
$$I_j^{\epsilon,\Delta}= \left\{i: \frac{j\Delta}{\epsilon}\le i \le \frac{\left(j+1\right)\Delta}{\epsilon}\right\}.$$
Then,
\begin{eqnarray}\label{stoch_approx_diff_in_tilde_U}
&& \hspace{-1cm} \mbox{limsup}_{\epsilon\rightarrow 0} E\left[ h \left( \tau^{\epsilon}_{Q}(t_i),\widehat{\bm{\Theta}}^{\epsilon}_Q(\tau^{\epsilon}_{Q} (t_i) ),i\le p \right)
\left(\widetilde{{U}}^{\epsilon}_{Q} \left(t+\tau\right)  - \widetilde{{U}}^{\epsilon}_{Q} \left(t\right) \right) \right]
\\\nonumber&=& \mbox{limsup}_{\epsilon\rightarrow 0} 
E  \Bigg[h \left( \tau^{\epsilon}_{Q}(t_i),\widehat{\bm{\Theta}}^{\epsilon}_Q(\tau^{\epsilon}_{Q} (t_i) ),i\le p \right) 
\left(\epsilon\sum_{i=\frac{t}{\epsilon}}^{\frac{t+\tau}{\epsilon}-1}
 \left(\delta \tau^{\epsilon}_{Q}(i)-\overline{u}_{Q}\left(  \widehat{\bm{\Theta}}^{\epsilon}_Q(\tau^{\epsilon}_{Q,i}) \right)\right) \right) \Bigg]
 \\\nonumber&=& \lim_{\Delta\rightarrow 0}  \mbox{limsup}_{\epsilon\rightarrow 0}  
 E\Bigg[ h \left( \tau^{\epsilon}_{Q}(t_i),\widehat{\bm{\Theta}}^{\epsilon}_Q(\tau^{\epsilon}_{Q} (t_i) ),i\le p \right)
 \left( \sum_{j=\frac{t}{\Delta}}^{\frac{t+\tau}{\Delta}-1} \epsilon 
 \sum_{i\in I_j^{\epsilon,\Delta}}\left(\delta \tau^{\epsilon}_{Q}(i)-\overline{u}_{Q}\left(  \widehat{\bm{\Theta}}^{\epsilon}_Q(\tau^{\epsilon}_{Q,i}) \right)\right) \right) \Bigg]
  \\\label{stoch_approx_proof_noise_averages_out_claim_2}&=& \lim_{\Delta\rightarrow 0}  \mbox{limsup}_{\epsilon\rightarrow 0}  
 E\Bigg[ h \left( \tau^{\epsilon}_{Q}(t_i),\widehat{\bm{\Theta}}^{\epsilon}_Q(\tau^{\epsilon}_{Q} (t_i) ),i\le p \right)
 \\\nonumber&&\hspace{7cm}
 \left( \sum_{j=\frac{t}{\Delta}}^{\frac{t+\tau}{\Delta}-1} \epsilon 
 \expn_{{\cal F}_{Q,\frac{ j \Delta}{\epsilon}}}\left[\sum_{i\in I_j^{\epsilon,\Delta}}
 \left(  \delta \tau^{\epsilon}_{Q}(i)
 -\overline{u}_{Q}\left(  \widehat{\bm{\Theta}}^{\epsilon}_Q(\tau^{\epsilon}_{Q,i}) \right)  \right)\right] \right) \Bigg]
\end{eqnarray}
where the third equality holds since ${\cal F}_{Q,\frac{ j \Delta}{\epsilon}}$ measures 
$\left( \tau^{\epsilon}_{Q}(t_i),\widehat{\bm{\Theta}}^{\epsilon}_Q(\tau^{\epsilon}_{Q} (t_i) ),i\le p \right)$
due to the fact that 
$t_i\le p$ for each $i\le p$ and $j\ge \frac{t}{\Delta}$.
Next, we show that \eqref{stoch_approx_proof_noise_averages_out_claim_2}
goes to zero by picking small enough $\epsilon$ and $\delta$.

Due to the bounded nature of the quantities involved in the update rules 
\eqref{L1_slot_b_update}-\eqref{L1_seg_b_update} for $\widehat{\bm{\Theta}}^{\epsilon}$, 
we have that
$\max_{i\in I_j^{\epsilon,\Delta}}
 \left|  \widehat{\bm{\Theta}}^{\epsilon}_Q(\tau^{\epsilon}_{Q,i+1})   
 -      \widehat{\bm{\Theta}}^{\epsilon}_Q(\tau^{\epsilon}_{Q,i}) \right| = O(\epsilon)$
 and hence
 \begin{eqnarray}\label{theta_does_not_change_much_in_I}
 \max_{i\in I_j^{\epsilon,\Delta}}
 \left|  \widehat{\bm{\Theta}}^{\epsilon}_Q(\tau^{\epsilon}_{Q,j})   
 -      \widehat{\bm{\Theta}}^{\epsilon}_Q(\tau^{\epsilon}_{Q,i}) \right| = O(\Delta).
\end{eqnarray}
Thus, using continuity of $\overline{u}_{Q}\left( . \right)$,
we have
$$
\lim_{\Delta\rightarrow 0}  \mbox{limsup}_{\epsilon\rightarrow 0}  
 E\left[ 
   \sum_{j=\frac{t}{\Delta}}^{\frac{t+\tau}{\Delta}-1} \epsilon
  \left|\sum_{i\in I_j^{\epsilon,\Delta}}
\left( \overline{u}_{Q}\left(  \widehat{\bm{\Theta}}^{\epsilon}_Q(\tau^{\epsilon}_{Q,i}) \right)  
 -  \overline{u}_{Q}\left(  \widehat{\bm{\Theta}}^{\epsilon}_Q(\tau^{\epsilon}_{Q,j}) \right)\right)
 \right|  \right]
 =0.
$$
Hence, to show that \eqref{stoch_approx_diff_in_tilde_U} is zero (and hence \eqref{stoch_approx_proof_noise_averages_out_claim} goes to zero), 
it is enough to show that the following term appearing in \eqref{stoch_approx_proof_noise_averages_out_claim_2} satisfies
\begin{eqnarray*}
 &&\lim_{\Delta\rightarrow 0}  \mbox{limsup}_{\epsilon\rightarrow 0}  
 E\Bigg[ h \left( \tau^{\epsilon}_{Q}(t_i),\widehat{\bm{\Theta}}^{\epsilon}_Q(\tau^{\epsilon}_{Q} (t_i) ),i\le p \right)
 \\&&\hspace{1cm}\left( \sum_{j=\frac{t}{\Delta}}^{\frac{t+\tau}{\Delta}-1} \epsilon 
 \expn_{{\cal F}_{Q,\frac{ j \Delta}{\epsilon}}}\left[\sum_{i\in I_j^{\epsilon,\Delta}}
 \left(  \delta \tau^{\epsilon}_{Q}(i)
 - \overline{u}_{Q}\left(  \widehat{\bm{\Theta}}^{\epsilon}_Q(\tau^{\epsilon}_{Q,j}) \right)
    \right) \right]\right) \Bigg]=0.
\end{eqnarray*}

Let 
\begin{eqnarray*}
 \varsigma^{\epsilon,\Delta}_{Q,j}
 = \frac{\epsilon}{\Delta} \expn_{{\cal F}_{Q,\frac{ j \Delta}{\epsilon}}} \left[\sum_{i\in I_j^{\epsilon,\Delta}}  \delta \tau^{\epsilon}_{Q}(i)\right]
 \end{eqnarray*}
denote the conditional expectation of the average time (in seconds) for the $\frac{\Delta}{\epsilon}$ updates indexed by the set $I_j^{\epsilon,\Delta}$.
Since, for each segment index $i$ and channel slot index $k$
\begin{eqnarray*}
\widehat{\bm{\Theta}}^{\epsilon}_Q  \left(\tau^{\epsilon}_{Q} \left(i \epsilon \right) \right)
&=& \left(\widehat{m}^{\epsilon}_1(\tau^{\epsilon}_{Q} \left(i \epsilon \right) ),\widehat{b}^{\epsilon}_Q(\tau^{\epsilon}_{Q} \left(i \epsilon \right) )\right)  
= \left(m^{\epsilon}_{1,i},b^{\epsilon}_{Q,1,i}\right),
\\\widehat{b}^{\epsilon}_R  \left(\tau^{\epsilon}_{R} \left(k \epsilon \right) \right)
&=&\widehat{b}^{\epsilon}_R  \left( \tau^{\epsilon}_{R,k} \right)
=b^{\epsilon}_{R,1,k}.
 \end{eqnarray*}
we have
\begin{eqnarray}
\label{upper_bnd_varsigma}
\sum_{i\in I_j^{\epsilon,\Delta}}  
l_1 F_{Q,i}\left( q^*_{Q} \left(\widehat{\bm{\Theta}}^{\epsilon}_Q  \left(\tau^{\epsilon}_{Q} \left(i \epsilon \right) \right), F_{Q,i}\right)\right)
 \ge  \sum_{{k}= \left\lceil \frac{\tau^{\epsilon}_{Q} \left(j \Delta\right)}{\epsilon \tau_{slot}}\right\rceil }
 ^{\left(\frac{ \Delta \varsigma^{\epsilon,\Delta}_{Q,j}}{\epsilon \tau_{slot}}-2\right)+\left\lceil \frac{\tau^{\epsilon}_{Q} \left(j \Delta\right)}{\epsilon \tau_{slot}}\right\rceil} 
 r^* \left(\widehat{b}^{\epsilon}_R  \left(\tau^{\epsilon}_{R} \left(k \epsilon \right) \right), C_{k}  \right),
\end{eqnarray}
and
\begin{eqnarray}
\label{lower_bnd_varsigma}
\sum_{i\in I_j^{\epsilon,\Delta}}  
l_1 F_{Q,i}\left( q^*_{Q} \left(\widehat{\bm{\Theta}}^{\epsilon}_Q  \left(\tau^{\epsilon}_{Q} \left(i \epsilon \right) \right), F_{Q,i}\right)\right) 
 \le   \sum_{{k}= \left\lfloor \frac{\tau^{\epsilon}_{Q} \left(j \Delta\right)}{\epsilon \tau_{slot}}\right\rfloor }
 ^{\left(\frac{ \Delta \varsigma^{\epsilon,\Delta}_{Q,j}}{\epsilon \tau_{slot}}+2\right)
 +\left\lfloor \frac{\tau^{\epsilon}_{Q} \left(j \Delta\right)}{\epsilon \tau_{slot}}\right\rfloor} 
 r^* \left(\widehat{b}^{\epsilon}_R  \left(\tau^{\epsilon}_{R} \left(k \epsilon \right) \right), C_{k}  \right).
\end{eqnarray}
In the above inequalities, the left hand side is equal to the total size of the segments 
indexed by the set $ I_j^{\epsilon,\Delta}$,
and the right hand side is roughly equal to the total allocation over the slots during which these segments are downloaded.
The term in the left hand side of \eqref{upper_bnd_varsigma} satisfies
\begin{eqnarray}
\label{F_LLN_and_cont_stoch_approx_proof}
&&\hspace{-2cm} \sum_{i\in I_j^{\epsilon,\Delta}}  
l_1 F_{Q,i}\left( q^*_{Q} \left(\widehat{\bm{\Theta}}^{\epsilon}_Q  \left(\tau^{\epsilon}_{Q} \left(i \epsilon \right) \right), F_{Q,i}\right)\right)\hspace{-0cm}
= \frac{\Delta}{\epsilon} E\left[  l_1 F^{\pi}_{Q}\left(
   q^*_{Q} \left(\widehat{\bm{\Theta}}^{\epsilon}_Q  \left(\tau^{\epsilon}_{Q} \left(j \Delta \right) \right) , F^{\pi}_{Q}\right)\right)
    \right] \hspace{0cm}
\\\nonumber&&\hspace{-.2cm}  + \sum_{i\in I_j^{\epsilon,\Delta}}  
\left(l_1 F_{Q,i}\left( q^*_{Q} \left(\widehat{\bm{\Theta}}^{\epsilon}_Q  \left(\tau^{\epsilon}_{Q} \left(i \epsilon \right) \right), F_{Q,i}\right)\right)
-  l_1 F_{Q,i}  
\left(q^*_{Q}\left( \widehat{\bm{\Theta}}^{\epsilon}_Q \left(\tau^{\epsilon}_{Q} \left(j \Delta \right) \right), F_{Q,i}\right)\right)\right)
 \\\nonumber&&\hspace{-.2cm}+\sum_{i\in I_j^{\epsilon,\Delta}}  \left(   
  l_1 F_{Q,i}  \left(q^*_{Q}\left( \widehat{\bm{\Theta}}^{\epsilon}_Q 
 \left(\tau^{\epsilon}_{Q} \left(j \Delta \right) \right), F_{Q,i}\right)\right)-  E\left[  l_1 F^{\pi}_{Q}\left(
   q^*_{Q} \left(\widehat{\bm{\Theta}}^{\epsilon}_Q  \left(\tau^{\epsilon}_{Q} \left(j \Delta \right) \right) , F^{\pi}_{Q}\right)\right)
    \right]\right).
\end{eqnarray}
Similarly, the term in the right hand side of \eqref{upper_bnd_varsigma} satisfies
\begin{eqnarray}\label{C_LLN_and_cont_stoch_approx_proof}
 &&\hspace{-1.5cm}\sum_{{k}= \left\lceil \frac{\tau^{\epsilon}_{Q} \left(j \Delta\right)}{\epsilon \tau_{slot}}\right\rceil }
 ^{\left(\frac{ \Delta \varsigma^{\epsilon,\Delta}_{Q,j}}{\epsilon \tau_{slot}}-2\right)
 +\left\lceil \frac{\tau^{\epsilon}_{Q} \left(j \Delta\right)}{\epsilon \tau_{slot}}\right\rceil} 
 r^* \left(\widehat{b}^{\epsilon}_R  \left(\tau^{\epsilon}_{R} \left(k \epsilon \right) \right), C_{k}  \right) 
\ge \left(\frac{ \Delta \varsigma^{\epsilon,\Delta}_{Q,j}}{\epsilon \tau_{slot}}-2\right) 
 E \left[r^* \left( \widehat{b}^{\epsilon}_Q 
 \left(\tau^{\epsilon}_{Q} \left(j \Delta\right)\right), C^{\pi}  \right)\right]
 \\ \nonumber && 
 -\left(\frac{ \Delta \varsigma^{\epsilon,\Delta}_{Q,j}}{\epsilon \tau_{slot}}\right) 
 \max_{  \left\lceil \frac{\tau^{\epsilon}_{Q} \left(j \Delta\right)}{\epsilon \tau_{slot}}\right\rceil \le k \le
 \left(\frac{ \Delta \varsigma^{\epsilon,\Delta}_{Q,j}}{\epsilon \tau_{slot}}-2\right)+\left\lceil \frac{\tau^{\epsilon}_{Q} \left(j \Delta\right)}{\epsilon \tau_{slot}}\right\rceil}
\left| r^* \left(\widehat{b}^{\epsilon}_R  \left(\tau^{\epsilon}_{R} \left(k \epsilon \right) \right), C_{k}  \right) 
  -r^* \left(\widehat{b}^{\epsilon}_Q  \left(\tau^{\epsilon}_{Q} \left(j \Delta \right) \right), C_{k}  \right) \right|
 \\\nonumber&&
 +  \sum_{{k}= \left\lceil \frac{\tau^{\epsilon}_{Q} \left(j \Delta\right)}{\epsilon \tau_{slot}}\right\rceil }
 ^{\left(\frac{ \Delta \varsigma^{\epsilon,\Delta}_{Q,j}}{\epsilon \tau_{slot}}-2\right)+\left\lceil \frac{\tau^{\epsilon}_{Q} \left(j \Delta\right)}{\epsilon \tau_{slot}}\right\rceil} 
 \left(
 r^* \left(\widehat{b}^{\epsilon}_Q  \left(\tau^{\epsilon}_{Q} \left(j \Delta \right) \right) , C_{k}  \right)
  -  E \left[r^* \left(\widehat{b}^{\epsilon}_Q  \left(\tau^{\epsilon}_{Q} \left(j \Delta \right) \right), C^{\pi}  \right)\right] \right)
\end{eqnarray}

Using \eqref{upper_bnd_varsigma}, \eqref{F_LLN_and_cont_stoch_approx_proof} and \eqref{C_LLN_and_cont_stoch_approx_proof},
we have
\begin{eqnarray}\label{sum_times_upper_bound_stoch_approx_proof}
  \Delta \varsigma^{\epsilon,\Delta}_{Q,j} 
&\le& \Delta \overline{u}_{Q} \left( \widehat{\bm{\Theta}}^{\epsilon}_Q \left(\tau^{\epsilon}_{Q} \left(j \Delta\right)\right)\right)
    + \Delta E \left[r^* \left( \widehat{b}^{\epsilon}_Q  \left(\tau^{\epsilon}_{Q} \left(j \Delta \right) \right) , C^{\pi}  \right)\right]
 \\\nonumber&&\left( 2 O \left( \frac{\epsilon}{\Delta} \right)+  \Bigg|
\frac{ \sum_{i\in I_j^{\epsilon,\Delta}}     
  l_1 F_{Q,i}  \left(q^*_{Q}\left( \widehat{\bm{\Theta}}^{\epsilon}_Q \left(\tau^{\epsilon}_{Q} \left(j \Delta \right) \right) , F_{Q,i}\right)\right) 
    }{ \left( \frac{\Delta}{\epsilon} \right)}\right.
   \\\nonumber&&\hspace{4cm}\left.
  -  E\left[  l_1 F^{\pi}_{Q}\left(    
  q^*_{Q} \left( \widehat{\bm{\Theta}}^{\epsilon}_Q \left(\tau^{\epsilon}_{Q} \left(j \Delta \right) \right) , F^{\pi}_{Q}\right)\right)\right]
   \Bigg|\right.
    \\\nonumber&&\left.+  \left(\frac{  \varsigma^{\epsilon,\Delta}_{Q,j}}{ \tau_{slot}}\right)
\Bigg|\frac{
   \sum_{{k}= \left\lceil \frac{\tau^{\epsilon}_{Q} \left(j \Delta\right)}{\epsilon \tau_{slot}}\right\rceil }
 ^{\left(\frac{ \Delta \varsigma^{\epsilon,\Delta}_{Q,j}}{\epsilon \tau_{slot}}-2\right)+\left\lceil \frac{\tau^{\epsilon}_{Q} \left(j \Delta\right)}{\epsilon \tau_{slot}}\right\rceil} 
 \left(
 r^* \left(\widehat{b}^{\epsilon}_Q  \left(\tau^{\epsilon}_{Q} \left(j \Delta \right) \right) , C_{k}  \right)
\right) }
  {\left(\frac{ \Delta \varsigma^{\epsilon,\Delta}_{Q,j}}{\epsilon \tau_{slot}}-2\right)} 
   -  E \left[r^* \left( \widehat{b}^{\epsilon}_Q  \left(\tau^{\epsilon}_{Q} \left(j \Delta \right) \right) , C^{\pi}  \right)\right] \Bigg|\right.
    \\\nonumber&&\left.
    +\left(\frac{  \varsigma^{\epsilon,\Delta}_{Q,j}}{ \tau_{slot}}\right) 
 \max_{  \left\lceil \frac{\tau^{\epsilon}_{Q} \left(j \Delta\right)}{\epsilon \tau_{slot}}\right\rceil \le k \le
 \left(\frac{ \Delta \varsigma^{\epsilon,\Delta}_{Q,j}}{\epsilon \tau_{slot}}-2\right)+\left\lceil \frac{\tau^{\epsilon}_{Q} \left(j \Delta\right)}{\epsilon \tau_{slot}}\right\rceil}
 \left| r^* \left(\widehat{b}^{\epsilon}_R  \left(\tau^{\epsilon}_{R} \left(k \epsilon \right) \right), C_{k}  \right) 
  -r^* \left(\widehat{b}^{\epsilon}_Q  \left(\tau^{\epsilon}_{Q} \left(j \Delta \right) \right), C_{k}  \right) \right|      \right.
 \\\nonumber&&\left.
     + \max_{i\in I_j^{\epsilon,\Delta}}  
\left|l_1 F_{Q,i}\left( q^*_{Q} \left(\widehat{\bm{\Theta}}^{\epsilon}_Q  \left(\tau^{\epsilon}_{Q} \left(i \epsilon \right) \right), F_{Q,i}\right)\right)
-  l_1 F_{Q,i}  \left(q^*_{Q}\left( \widehat{\bm{\Theta}}^{\epsilon}_Q 
 \left(\tau^{\epsilon}_{Q} \left(j \Delta \right) \right), F_{Q,i}\right)\right)\right|\right).
\end{eqnarray}

Then, using \eqref{lower_bnd_varsigma} and arguments similar to those above, we have
\begin{eqnarray}\label{sum_times_lower_bound_stoch_approx_proof}
\Delta \varsigma^{\epsilon,\Delta}_{Q,j} 
&\ge& \Delta \overline{u}_{Q} \left( \widehat{\bm{\Theta}}^{\epsilon}_Q \left(\tau^{\epsilon}_{Q} \left(j \Delta\right)\right)\right)
    - \Delta E \left[r^* \left( \widehat{b}^{\epsilon}_Q  \left(\tau^{\epsilon}_{Q} \left(j \Delta \right) \right) , C^{\pi}  \right)\right]
  \\\nonumber&& \left( 2 O \left( \frac{\epsilon}{\Delta} \right) +  \Bigg|
\frac{ \sum_{i\in I_j^{\epsilon,\Delta}}     
  l_1 F_{Q,i}  \left(q^*_{Q}\left( \widehat{\bm{\Theta}}^{\epsilon}_Q \left(\tau^{\epsilon}_{Q} \left(j \Delta \right) \right) , F_{Q,i}\right)\right) 
    }{ \left( \frac{\Delta}{\epsilon} \right)}\right.
   \\\nonumber&&\hspace{4cm}\left.
  -  E\left[  l_1 F^{\pi}_{Q}\left(    
  q^*_{Q} \left( \widehat{\bm{\Theta}}^{\epsilon}_Q \left(\tau^{\epsilon}_{Q} \left(j \Delta \right) \right) , F^{\pi}_{Q}\right)\right)\right]
   \Bigg|\right.
    \\\nonumber&&\left.+  \left(\frac{  \varsigma^{\epsilon,\Delta}_{Q,j}}{ \tau_{slot}}\right)
\Bigg|\frac{
   \sum_{{k}= \left\lfloor \frac{\tau^{\epsilon}_{Q} \left(j \Delta\right)}{\epsilon \tau_{slot}}\right\rfloor }
 ^{\left(\frac{ \Delta \varsigma^{\epsilon,\Delta}_{Q,j}}{\epsilon \tau_{slot}}+2\right)+\left\lfloor \frac{\tau^{\epsilon}_{Q} \left(j \Delta\right)}{\epsilon \tau_{slot}}\right\rfloor} 
 \left(
 r^* \left( \widehat{b}^{\epsilon}_Q  \left(\tau^{\epsilon}_{Q} \left(j \Delta \right) \right) , C_{k}  \right)
\right) }
  {\left(\frac{ \Delta \varsigma^{\epsilon,\Delta}_{Q,j}}{\epsilon \tau_{slot}}+2\right)}
    -  E \left[r^* \left( \widehat{b}^{\epsilon}_Q  \left(\tau^{\epsilon}_{Q} \left(j \Delta \right) \right) , C^{\pi}  \right)\right] \Bigg|\right.
    \\\nonumber&&\left.
    +\left(\frac{  \varsigma^{\epsilon,\Delta}_{Q,j}}{ \tau_{slot}}\right) 
 \max_{  \left\lfloor \frac{\tau^{\epsilon}_{Q} \left(j \Delta\right)}{\epsilon \tau_{slot}}\right\rfloor \le k \le
 \left(\frac{ \Delta \varsigma^{\epsilon,\Delta}_{Q,j}}{\epsilon \tau_{slot}}+2\right)+\left\lfloor \frac{\tau^{\epsilon}_{Q} \left(j \Delta\right)}{\epsilon \tau_{slot}}\right\rfloor}
 \left| r^* \left(\widehat{b}^{\epsilon}_R  \left(\tau^{\epsilon}_{R} \left(k \epsilon \right) \right), C_{k}  \right)
  -r^* \left(\widehat{b}^{\epsilon}_Q  \left(\tau^{\epsilon}_{Q} \left(j \Delta \right) \right), C_{k}  \right) \right|      \right.
 \\\nonumber&&\left.
     + \max_{i\in I_j^{\epsilon,\Delta}}  
\left|l_1 F_{Q,i}\left( q^*_{Q} \left(\widehat{\bm{\Theta}}^{\epsilon}_Q  \left(\tau^{\epsilon}_{Q} \left(i \epsilon \right) \right), F_{Q,i}\right)\right)
-  l_1 F_{Q,i}  \left(q^*_{Q}\left( \widehat{\bm{\Theta}}^{\epsilon}_Q 
 \left(\tau^{\epsilon}_{Q} \left(j \Delta \right) \right), F_{Q,i}\right)\right)\right|\right)
 \end{eqnarray}
Using the boundedness of the terms, and the fact that $\left(F_{Q,i}\right)_{i\ge 1}$ and $(C_k)_{k\ge 1}$ are stationary ergodic, 
the terms appearing in lines 2-5 of \eqref{sum_times_upper_bound_stoch_approx_proof} and \eqref{sum_times_lower_bound_stoch_approx_proof}
converge in mean to zero (i.e., we have $L^1$ convergence to 0).
Also, the terms in the last four lines of \eqref{sum_times_upper_bound_stoch_approx_proof} and \eqref{sum_times_lower_bound_stoch_approx_proof} can be made as small as needed by picking a small enough $\Delta$ 
due to the absolute continuity of $q^*_1(.,f)$ and $r^*_1(,c) $ for each $f$ and $c$.
and since we have \eqref{theta_does_not_change_much_in_I}
and
\begin{eqnarray} \label{b_does_not_change_much_in_Ij}
 \max_{  \left\lfloor \frac{\tau^{\epsilon}_{Q} \left(j \Delta\right)}{\epsilon \tau_{slot}}\right\rfloor \le k \le
 \left(\frac{ \Delta \varsigma^{\epsilon,\Delta}_{Q,j}}{\epsilon \tau_{slot}}+2\right)+\left\lfloor \frac{\tau^{\epsilon}_{Q} \left(j \Delta\right)}{\epsilon \tau_{slot}}\right\rfloor} \left| \widehat{b}^{\epsilon}_R  \left(\tau^{\epsilon}_{R} \left(k \epsilon \right) \right) 
  -\widehat{b}^{\epsilon}_Q  \left(\tau^{\epsilon}_{Q} \left(j \Delta \right) \right) \right| = O(\Delta).
  \end{eqnarray}
  where the argument for the above property is similar to \eqref{theta_does_not_change_much_in_I}
  and using the fact that $\varsigma^{\epsilon,\Delta}_{Q,j}$ is bounded by the constant $\delta \tau_{\max}$.
Thus, the expression in \eqref{stoch_approx_proof_noise_averages_out_claim_2} is equal to zero,
and consequently \eqref{stoch_approx_proof_noise_averages_out_claim} is also zero.
The rest of the proof is similar to that in \cite{kushner_text}.

In the above arguments, 
we used the property that the resource allocation $r^*_{1}\left(b_{1,k},c_k\right)$ in slot $k$  is a continuous function of $b_{1,k}$ for each $c_k$.
These arguments can be extended if the resource
 allocation in slot $k$ is picked from 
 ${\cal R}^*\left(b^{\epsilon}_{1,k},c_k\right)$
 where ${\cal R}^*\left(b,c\right)$ is an upper semi-continuous
 set valued map of $b$ taking compact convex values (i.e.,
 ${\cal R}^*\left(b,c\right)$ is a convex compact set for each $b$ and $\ciC$
 ) for each $\ciC$.
 For instance, we used \eqref{theta_does_not_change_much_in_I}, \eqref{b_does_not_change_much_in_Ij}
and the continuity of $\overline{u}_{Q}\left( . \right)$ to argue earlier (see below \eqref{theta_does_not_change_much_in_I}) that
$$
\lim_{\Delta\rightarrow 0}  \mbox{limsup}_{\epsilon\rightarrow 0}  
 E\left[ 
   \sum_{j=\frac{t}{\Delta}}^{\frac{t+\tau}{\Delta}-1} \epsilon
  \left|\sum_{i\in I_j^{\epsilon,\Delta}}
\left( \overline{u}_{Q}\left(  \widehat{\bm{\Theta}}^{\epsilon}_Q(\tau^{\epsilon}_{Q,i}) \right)  
 -  \overline{u}_{Q}\left(  \widehat{\bm{\Theta}}^{\epsilon}_Q(\tau^{\epsilon}_{Q,j}) \right)\right)
 \right|  \right]
 =0.$$
 Note that (after relaxing the continuity assumption)
\begin{eqnarray*} 
\overline{u}_{Q} \left(  \bm{\Theta}\right) &=&
 \tau_{slot} l_1 \frac{ E \left[   F^{\pi}_{1}\left(  
q^*_{1}\left( \bm{\Theta}  ,F^{\pi}_{1} \right)
\right)\right]  }{ E\left[ r^*_1\left(b,C^{\pi}\right)\right] }.
\end{eqnarray*}
for $ \bm{\Theta}=(m_1,b_1)$, $r^*_1\left(b_1,c\right)\in {\cal R}^*\left(b_1,c\right)$
for each $\ciC$.
Now we can pick
$r^*_1\left(\widehat{b}^{\epsilon}_Q  \left( \tau^{\epsilon}_{Q,j} \right),c \right)
\in{\cal R}^*\left( \widehat{b}^{\epsilon}_Q  \left( \tau^{\epsilon}_{Q,j} \right),c\right)$
for each $\ciC$ and for each $j$ 
such that 
\begin{eqnarray}\label{intermediate_result_to_be_proved_for_set_valued_maps_in_SA_proof}
\lim_{\Delta\rightarrow 0}  \mbox{limsup}_{\epsilon\rightarrow 0}  
 E\left[ 
   \sum_{j=\frac{t}{\Delta}}^{\frac{t+\tau}{\Delta}-1} \epsilon
 \left|  \sum_{i\in I_j^{\epsilon,\Delta}}
 \left(\tau_{slot} l_1 \frac{ E \left[   F^{\pi}_{1}\left(  
q^*_{1}\left(   \widehat{\bm{\Theta}}^{\epsilon}_Q(\tau^{\epsilon}_{Q,i})   ,F^{\pi}_{1} \right)
\right)\right]  }{ E\left[ r^*_1\left( \widehat{b}^{\epsilon}_Q  \left( \tau^{\epsilon}_{Q,i} \right) ,C^{\pi}\right)\right]   }\right.\right.\right.
 \\\nonumber&&\left.\left.\left.\hspace{-6.5cm}- 
 \tau_{slot} l_1 \frac{ E \left[   F^{\pi}_{1}\left(  
q^*_{1}\left(   \widehat{\bm{\Theta}}^{\epsilon}_Q(\tau^{\epsilon}_{Q,j})   ,F^{\pi}_{1} \right)
\right)\right]  }{ E\left[ r^*_1\left( \widehat{b}^{\epsilon}_Q  \left( \tau^{\epsilon}_{Q,j} \right) ,C^{\pi}\right)\right]   }\right)
 \right|  \right]
 =0.
\end{eqnarray}
This follows from the following two properties:
\\(i) \ the continuity of QR tradeoffs
and $q^*_{i}\left(.\right)$,
and 
\\(ii) the fact that for small enough $\Delta$,
we can always pick 
$r^*_1\left(\widehat{b}^{\epsilon}_Q  \left( \tau^{\epsilon}_{Q,j} \right),c \right)
\in {\cal R}^*\left( \widehat{b}^{\epsilon}_Q  \left( \tau^{\epsilon}_{Q,j} \right),c\right)$
for each $\ciC$ and for each $j$ 
such that
$$   \left|\frac{\epsilon}{\Delta}\sum_{i\in I_j^{\epsilon,\Delta}}
 \frac{1}{ E\left[ r^*_1\left( \widehat{b}^{\epsilon}_Q  \left( \tau^{\epsilon}_{Q,i} \right) ,C^{\pi}\right)\right]  }
 - \frac{1}{ E\left[ r^*_1\left( \widehat{b}^{\epsilon}_Q  \left( \tau^{\epsilon}_{Q,j} \right) ,C^{\pi}\right) \right]  } 
  \right| 
$$
is small enough.
\\Property (ii) follows from \eqref{b_does_not_change_much_in_Ij} and the fact that
$$
\overline{\cal R}^*_{inv}\left( b \right)=
\left\{
\frac{1}{ E\left[ r^*_1\left( b ,C^{\pi}\right) \right]  }:
r^*_1\left( b ,c\right) \in {\cal R}^*\left(b,c\right) 
\ \forall \ \ciC
\right\},
$$
is an upper semicontinuous set valued map of $b$ taking compact convex values.
These properties of $\overline{\cal R}^*_{inv}\left( b \right)$ are essentially consequences
of the fact that ${\cal R}^*\left(b,c\right) $ is an upper semicontinuous set valued map of $b$ taking compact convex values for each $c$ 
(from Lemma \ref{continuity_of_solutions_to_QNOVA_and_RNOVA} (b)).
The compactness of $\overline{\cal R}^*_{inv}\left( b \right)$ follows from the compactness of ${\cal R}^*\left(b,c\right) $ for each $c$,
and the fact that $\overline{\cal R}^*_{inv}\left( b \right)$ is obtained using a continuous map on 
the elements of $\left({\cal R}^*\left(b,c\right)\right)_{\ciC}$.
To show convexity, let $\alpha\in[0,1]$ and consider $x,y \in \overline{\cal R}^*_{inv}\left( b \right)$,
i.e., there exists $r^*_x\left( b ,c\right),\ r^*_y\left( b ,c\right) \in {\cal R}^*\left(b,c\right) $
for each $\ciC$ such that
$x= \frac{1}{ E\left[ r^*_x\left( b ,C^{\pi}\right) \right]  }, 
\ y=\frac{1}{ E\left[ r^*_y\left( b ,C^{\pi}\right) \right]  }$
. Then,
$\left(\alpha x + \left(1-\alpha\right)y\right)\in \overline{\cal R}^*_{inv}\left( b \right) $
since
$\left(\alpha x + \left(1-\alpha\right) y\right)
= \frac{1}{  E\left[ r^*_{xy}\left( b ,C^{\pi}\right) \right]  }$
where $r^*_{xy}\left( b ,c\right)= \alpha' r^*_x\left( b ,c\right) + \left(1-\alpha'\right)  r^*_y\left( b ,c\right) \in  {\cal R}^*\left( b ,c\right)$
(due to convexity of $ {\cal R}^*\left( b ,c\right)$), and
$\alpha'=\frac{\alpha x }{\alpha x + \left(1-\alpha\right) y}\in[0,1]$.
Next, we show that $\overline{\cal R}^*_{inv}\left( . \right)$ is an upper semicontinuous set valued map.
Note that since $\overline{\cal R}^*_{inv}\left( . \right)$ is uniformly compact, 
$\overline{\cal R}^*_{inv}\left( . \right)$ is upper semicontinuous if it is closed
(see \cite{fiacco_sensitivity_analysis} or \cite{smirnov_differential_inclusions}
for a discussion about upper semicontinuous, uniformly compact and closed set valued maps).
Consider any sequence $\left(b_n\right)_{n\ge 1}$ converging to $b$,
and consider any sequence 
$\left(x_n\right)_{n\ge 1}$ such that $x_n \in \overline{\cal R}^*_{inv}\left( b_n \right)$ for each $n$,
and $x_n$ converges to some $x$.
Then, $\overline{\cal R}^*_{inv}\left( b \right)$ is closed at $b$ if $x\in \overline{\cal R}^*_{inv}\left( b \right)$.
Since, $x_n \in \overline{\cal R}^*_{inv}\left( b_n \right)$,
there exists $r_{(n)}\left( c \right) \in {\cal R}^*\left(b_n,c\right) $
for each $\ciC$ such that
$x_n= \frac{1}{ E\left[ r_{(n)}\left( C^{\pi}\right) \right]  }$.
For any $c\in{\cal C}$, we can obtain a convergent subsequence $\left(r_{(n_{k_c})}\left( c \right)\right)_{k_c\ge 1}$
that converges to some $r\left( c \right) \in {\cal R}^*\left(b_n,c\right) $ (due to upper semicontinuity of $ {\cal R}^*\left(b,c\right) $).
Since ${\cal C}$ is finite, we can keep picking subsequences (of subsequences) to obtain a 
sequence of indices $(n_k)_{k\ge 1}$
such that for each $c\in{\cal C}$,
$\left(r_{(n_k)}\left( c \right)\right)_{k\ge 1}$
converges to some $r\left( c \right) \in {\cal R}^*\left(b,c\right) $.
Using this convergence property for each $\ciC$, and noting that $x_n$ converges to $x$, 
we can conclude that $x= \frac{1}{ E\left[ r\left( C^{\pi}\right) \right]  }\in \overline{\cal R}^*_{inv}\left( b \right)$.
Thus, $\overline{\cal R}^*_{inv}\left( . \right)$ is a closed map,
and hence is upper semicontinuous.

Now that we have shown that $\overline{\cal R}^*_{inv}\left( b \right)$
is an upper semicontinuous set valued map of $b$ taking compact convex values,
we use this observation to show that property (ii) holds.
Due to upper semicontinuity of $\overline{\cal R}^*_{inv}\left( b \right)$ and \eqref{b_does_not_change_much_in_Ij},
for each $\delta>0$ and $i\in I_j^{\epsilon,\Delta}$, 
we can find $  \overline{r}_{inv,i} \left(  \widehat{b}^{\epsilon}_Q  \left( \tau^{\epsilon}_{Q,i} \right)  \right)
\in  \overline{\cal R}^*_{inv}\left(  \widehat{b}^{\epsilon}_Q  \left( \tau^{\epsilon}_{Q,i} \right)  \right)$
such that
\\$\left| \frac{1}{ E\left[ r^*_1\left( \widehat{b}^{\epsilon}_Q  \left( \tau^{\epsilon}_{Q,i} \right) ,C^{\pi}\right)\right]  }
- \overline{r}_{inv,i}\left(  \widehat{b}^{\epsilon}_Q  \left( \tau^{\epsilon}_{Q,i} \right)  \right) \right|<\delta$
by picking $\Delta$ small enough. 
Due to convexity of $\overline{\cal R}^*_{inv}\left( b \right)$, 
there are $r^*_1\left(\widehat{b}^{\epsilon}_Q  \left( \tau^{\epsilon}_{Q,j} \right),c \right)
\in {\cal R}^*\left( \widehat{b}^{\epsilon}_Q  \left( \tau^{\epsilon}_{Q,j} \right),c\right)$
for each $\ciC$ and for each $j$ 
that satisfies
$$   \frac{1}{ E\left[ r^*_1\left( \widehat{b}^{\epsilon}_Q  \left( \tau^{\epsilon}_{Q,j} \right) ,C^{\pi}\right) \right]  }  
= \frac{\epsilon}{\Delta}\sum_{i\in I_j^{\epsilon,\Delta}}
 \overline{r}_{inv,i}   \left(  \widehat{b}^{\epsilon}_Q  \left( \tau^{\epsilon}_{Q,i} \right)  \right).
$$

Then, we see that property (ii) holds since
\begin{eqnarray*}    \left|\frac{\epsilon}{\Delta}\sum_{i\in I_j^{\epsilon,\Delta}}
 \frac{1}{ E\left[ r^*_1\left( \widehat{b}^{\epsilon}_Q  \left( \tau^{\epsilon}_{Q,i} \right) ,C^{\pi}\right)\right]  }
 - \frac{1}{ E\left[ r^*_1\left( \widehat{b}^{\epsilon}_Q  \left( \tau^{\epsilon}_{Q,j} \right) ,C^{\pi}\right) \right]  } 
  \right|<\delta
\end{eqnarray*} 
for $\Delta$ small enough.
Since properties (i) and (ii) hold, \eqref{intermediate_result_to_be_proved_for_set_valued_maps_in_SA_proof} follows.
We can similarly extend other arguments in our proof (that relied to continuity 
$r^*_{1}\left(.,c\right)$) to the case when the resource
 allocation in slot $k$ is picked from 
 ${\cal R}^*\left(b^{\epsilon}_{1,k},c_k\right)$
 by using the fact that ${\cal R}^*\left(b,c\right)$ is an upper semi-continuous
 set valued map of $b$ taking compact convex values for each $\ciC$.

Now, we focus on proving the result in \eqref{nova_l1_b_weak_limit_property} 
for the weak limit component $\widehat{b}^{\epsilon}_Q(.)$
associated with time interpolated version
$\widehat{b}^{\epsilon}_Q(.)$ of the parameter $b^{\epsilon}_{Q,1,s}$.
Similar to \eqref{defining_W},
we start by rewriting
$\widehat{b}^{\epsilon}_Q(\tau^{\epsilon}_{Q}(t))$ as given below:
\begin{eqnarray} \label{expand_b_Q_L1}
\hspace{-.43cm}\widehat{b}^{\epsilon}_Q(\tau^{\epsilon}_{Q}(t)) = 
b^{\epsilon}_{1,0} + \widetilde{G}^{\epsilon}_b(t) 
+\epsilon \sum_{i=0}^{\frac{t}{\epsilon} -1} 
  \left( \overline{u}_{Q}\left(  \widehat{\bm{\Theta}}^{\epsilon}_Q(\tau^{\epsilon}_{Q,i}) \right) - l_1  \right)
   + \epsilon \sum_{i=0}^{\frac{t}{\epsilon} -1}  Z^{\epsilon}_{1,i}
   +   E^{\epsilon}_b(t) ,
\end{eqnarray}
where
\begin{eqnarray*} 
 \widetilde{G}^{\epsilon}_b(t) &= &\epsilon \sum_{i=0}^{\frac{t}{\epsilon}-1}
 \left( \delta \tau^{\epsilon}_{Q}(i)-\overline{u}_{Q}\left(  \widehat{\bm{\Theta}}^{\epsilon}_Q(\tau^{\epsilon}_{Q,i}) \right)\right),
 \\E^{\epsilon}_b(t)  &= & \epsilon\tau_{slot} \left( \frac{ \sum_{i=0}^{\frac{t}{\epsilon}-1} \delta \tau^{\epsilon}_{Q}(i)}{ \tau_{slot}  }
  - \left\lfloor \frac{ \sum_{i=0}^{\frac{t}{\epsilon}-1} \delta \tau^{\epsilon}_{Q}(i)}{ \tau_{slot}  }\right\rfloor\right)
\end{eqnarray*}
and 
for each $i$, $Z^{\epsilon}_{1,i}$ is the term that accounts for the reflection term 
associated with the update in \eqref{L1_seg_b_update} for the $s$th segment
due to the operator $\left[  . \right]_{\underline{b}}$.
Note that $E^{\epsilon}_b(t)$ accounts for the fact that R-updates
(that increment $b^{\epsilon}_{1,.}$ by $\epsilon \tau_{slot}$)
only occur at slot boundaries which are separated by $\tau_{slot}$ seconds.
Note that $Z^{\epsilon}_{1,i} \in -{\cal Z}_{\cal H} \left(   \widehat{\bm{\Theta}}^{\epsilon}_Q(\tau^{\epsilon}_{Q,i}) \right)$
and $|E^{\epsilon}_b(t) |\le \epsilon\tau_{slot}$ so that the last term in \eqref{expand_b_Q_L1} is $O(\epsilon)$.

Now, fix $t$ and $\tau$. For any integer $p$, let $t_i\le t$, $i\le p$,
Let $h(.)$ be an arbitrary bounded, continuous and real valued function of its arguments.
The proof of \eqref{nova_l1_b_weak_limit_property}  can be completed just like in \cite{kushner_text} (see pages 414-415 and pages 251-257) once we prove that
\begin{eqnarray}\label{stoch_approx_diff_in_tilde_G_b}
 \mbox{limsup}_{\epsilon\rightarrow 0} E\left[ 
 h \left( \tau^{\epsilon}_{Q}(t_i),\widehat{\bm{\Theta}}^{\epsilon}_Q(\tau^{\epsilon}_{Q} (t_i) ),i\le p \right)
\left(\widetilde{{ G}}^{\epsilon}_{b} \left(t+\tau\right)  - \widetilde{{ G}}^{\epsilon}_{b} \left(t\right) \right) \right]
=0.
\end{eqnarray}
But, note that $ \widetilde{G}^{\epsilon}_b(t)=\overline{{U}}^{\epsilon}_{Q}(t) $ for each $t$ (defined in \eqref{interp_ubar_defn}),
and hence, \eqref{stoch_approx_diff_in_tilde_G_b} holds since 
we have established this property earlier for the same expression in \eqref{stoch_approx_proof_noise_averages_out_claim}.

We have now established that the weak limit $\widehat{b}_Q(.)$ satisfies \eqref{nova_l1_b_weak_limit_property}.
The weak limit components $\widehat{b}_R(.)$ (associated with $b^{\epsilon}_{R,1,k}$)
also satisfies \eqref{nova_l1_b_weak_limit_property} since
$$\widehat{b}^{\epsilon}_Q(t) - \widehat{b}^{\epsilon}_R(t) =O( \epsilon)$$
which follows from the bounded nature of the increase of $b^{\epsilon}_{R,1,k}$ due to \eqref{b_allocate_update}, and 
the fact that the number of slots between two segment download completion instants is bounded.

Thus, we have argued that for the special case NOVA-L1,
we can extend Theorem 3.4 of \cite{kushner_text}
as described in the above discussion.
These arguments can be extended to the general setting considered in NOVA also.
In particular, we can study NOVA given in \eqref{m_update_NOVA}-\eqref{lambda_update_NOVA} and \eqref{sigma_NOVA_update}-\eqref{rho_NOVA_update}
and relate it to the auxiliary differential inclusion \eqref{mhat_ode_rule}-\eqref{rhohat_ode_rule}.
We can also show that the following claim in Theorem 3.4 of \cite{kushner_text} holds:
the fraction of time in the time interval $[0,T]$ that $\widehat{\bm{\Theta}}^{\epsilon}(.)$
and $\widehat{\bm{\Theta}}^{\epsilon}_R(.)$
spends in a small neighborhood of set of limit points of the differential inclusion in \eqref{mhat_ode_rule}-\eqref{rhohat_ode_rule}
converges to one in probability as $\epsilon\rightarrow 0$
and $T\rightarrow \infty$.
Now, the main claim of this result now follows from this observation and Theorem \ref{convergence_of_differential_inclusion} 
where we have shown that the 
set of limit points of the differential inclusion in \eqref{mhat_ode_rule}-\eqref{rhohat_ode_rule}
is contained in ${\cal H}^*$.

We conclude the proof by verifying that the remaining conditions given in Theorem 3.4 of \cite{kushner_text} are satisfied.
Due to Lemma \ref{parameters_are_bounded}, we can view NOVA given in \eqref{m_update_NOVA}-\eqref{lambda_update_NOVA} and \eqref{sigma_NOVA_update}-\eqref{rho_NOVA_update}
as a constrained stochastic approximation which satisfies condition (3.1) in Chapter 12 of \cite{kushner_text},
and hence the set $H$ in the discussion of Theorem 3.4
corresponds to ${\cal H}$ in our problem setting.
Although the initialization in NOVA, specifically $\mathbf{b}_0$ and $\mathbf{d}_0$, does not ensure that we start in ${\cal H}$,
we enter and stay in ${\cal H}$ in a finite number of slots (as shown in Lemma \ref{parameters_are_bounded}),
and thus we can view NOVA as a constrained stochastic approximation.
Note that the random variables $\xi^{\epsilon}_{s,i}$ considered in the discussion of Theorem 3.4
corresponds to $\left(F_{i,s},L_{i,s}\right)$ in our setting,
and the condition given in (A3.11) concerning these random variables is clearly satisfied as they 
take values in a finite set.
Condition (A3.1) and (A3.12) can be verified using the boundedness of the quantities associated with these conditions.
In particular, note that
$0<\delta \tau_{\min} \le \delta \tau^{\epsilon}_{Q,s}\le \delta \tau_{\max}<\infty$
where 
$\delta \tau_{\min}=\frac{f_{\min}l_{min}}{r_{\max}}$ and 
$\delta \tau_{\max}=\frac{f_{\max}l_{max}}{\min_{\iiN}r_{i,\min}}$.
The condition (A3.13) is satisfied
since, for each $i\in{\cal N}$, $\left(F_{i,s},L_{i,s}\right)_{s\ge 0}$ is a stationary ergodic process.
Conditions (A3.6), (A3.7) and (A3.9) can be verified 
by letting
$\beta^{\epsilon}_{n,\alpha}$ and $\Delta^{\epsilon}_{n,\alpha}$ to be identically zero,
and $g^{\epsilon}_{n,\alpha}=\overline{g}_{\alpha}$ 
where $g^{\epsilon}_{n,\alpha}=\overline{g}_{\alpha}$ are defined
based on the right hand side of the differential inclusion in \eqref{mhat_ode_rule}-\eqref{rhohat_ode_rule}
taking appropriate care of the terms associated with the update rates.
For instance, 
let $\bm{\Theta}=\left(\mathbf{m},\bm{\mu},\mathbf{v},\mathbf{b},\mathbf{d},\bm{\lambda},\bm{\sigma},\bm{\rho}\right)\in {\cal H}$
and 
let $\bm{\theta}_i=(m_i,\mu_i,v_i,b_i,d_i)$ for each $\iiN$,
and for each $i\in{\cal N}$, we define $\overline{g}_{i + N}$ based on \eqref{muhat_ode_rule} as given below:
\begin{eqnarray*}
\overline{g}_{i + N}   \left(\bm{\Theta}\right) &=&
  \frac{ E \left[ L^{\pi}_i 
 q^*_{i}\left(\bm{\theta}_i  ,F^{\pi}_{i} \right)
 \right]}{\lambda_i} 
 - \mu_i   ,
 \label{gbar_for_muhat_ode_rule}
 \end{eqnarray*}
 Note that we have not verified conditions (A3.8), (A3.10) and (A3.14)
 since our discussion at the beginning of the proof allows us to avoid using these assumptions.
\end{IEEEproof}
We have the following corollary of Theorem \ref{main_stoch_approx_result}
which says that for small enough $\epsilon$ and after running NOVA for long enough,
video client $i$'s NOVA parameter stays close to ${\cal H}^*_i$(defined in \eqref{defn_Hstar_i}) most of the time
with high probability.
\begin{corollary}\label{corollary_of_stoch_approx}
Let $\widehat{\bm{\Theta}}^{\epsilon}(0)\in{\cal H}$ and $S_{\epsilon}=\frac{ S}{\epsilon}$.
Then for each $\iiN$, the following holds:
for any $\delta>0$, the fraction of segment indices for which $\left(\bm{\theta}_{i,s}\right)_{1 \le s\le S_{\epsilon}}$
is in a $\delta$-neighborhood of ${\cal H}^*_i$
converges to one in probability as $\epsilon$ goes to zero and $S$ goes to infinity.
\end{corollary}
\begin{IEEEproof}
The corollary follows by using Theorem \ref{main_stoch_approx_result}
to conclude that
the fraction of time in the time interval $[0,T]$ that $\widehat{\bm{\theta}}^{\epsilon}_i(.)$
spends in a small neighborhood of ${\cal H}^*_i$ 
converges to one in probability as $\epsilon\rightarrow 0$
and $T\rightarrow \infty$.
Recall that here $\widehat{\bm{\theta}}^{\epsilon}_i(t)=\left(\widehat{m}^{\epsilon}_i(t),\widehat{\mu}^{\epsilon}_i(t),\widehat{v}^{\epsilon}_i(t),
\widehat{b}^{\epsilon}_{Q,i}(t),\widehat{d}^{\epsilon}_i(t)\right)$.
Note that here we are also using the fact that for each video client $\iiN$,
the amount of time between updates is bounded below.
\end{IEEEproof}

We have now obtained all the intermediate results required to prove Theorem \ref{main_optimality_theorem} which is given below.
\begin{IEEEproof}[\textbf{Proof of Theorem \ref{main_optimality_theorem}}]
Part (a) of Theorem \ref{main_optimality_theorem}
states that 
NOVA satisfies the constraints on rebuffering and cost asymptotically, i.e., for each $\iiN$
\begin{eqnarray}
\label{repeat_NOVA_satifies_rebuf_constraints}
\mbox{limsup}_{S\rightarrow \infty}\beta_{i,S}\left(\left(q^*_i\right)_{1:S},\left(r^*_i\right)_{1:K_S}\right)  &\le& \overline{\beta}_i,
\\\label{repeat_NOVA_satifies_cost_constraints}
\mbox{limsup}_{S\rightarrow \infty}p_{i,S}\left( \left(q^*_i\right)_{1:S} \right) &\le& \overline{p}_i.
\end{eqnarray}
We first prove that NOVA satisfies rebuffering constraints, i.e., \eqref{repeat_NOVA_satifies_rebuf_constraints}.
Note that 
\begin{eqnarray}\label{nova_satisfies_rebuf_constraint_asymptotically}
\beta_{i,S}\left(\left(q^*_i\right)_{1:S},\left(r^*_i\right)_{1:K_S}\right)
= \frac{ \frac{\sum_{s=1}^{S} l_{i,s} f_{i,s} \left(q^*_{i,s}\right)}
{ \frac{1}{\tau_{slot}K_S} \sum_{k=1}^{K_S} r^*_{i,k}  }}
{\sum_{s=1}^{S} l_{i,s}}
- 1.
\end{eqnarray}
Let $T_i(S)$ (measured in seconds) denote the time at which the download of the first $S$ segments of video client $i$ completes.
Then, based on NOVA update rules \eqref{b_allocate_update} and \eqref{b_adapt_update}, 
and removing the projection operator, we get the following lower bound on
the value $b_{Q,i,S}$:
 \begin{eqnarray*} 
 b_{Q,i,S} &\ge&  b_{Q,i,0} + \epsilon \left( \frac{\tau_{slot} \left\lfloor  \frac{T_i(S)}{\tau_{slot} }\right\rfloor }{ \left(1+ \overline{\beta}_i\right)  }  - \sum_{s=1}^{S} l_{i,s}\right)
 \\&\ge&  b_{Q,i,0} - \frac {\epsilon \tau_{slot} }{\left(1+ \overline{\beta}_i\right) } 
 + \epsilon \left( \frac{ T_i(S) }{ \left(1+ \overline{\beta}_i\right)  }  - \sum_{s=1}^{S} l_{i,s}\right).
\end{eqnarray*}
Hence,
 \begin{eqnarray} \label{bound_on_cum_time_to_download} \frac{ T_i(S) }{ \sum_{s=1}^{S} l_{i,s}    }  \le \left(1+ \overline{\beta}_i\right) +
 \left(1+ \overline{\beta}_i\right) \left(\frac{b_{Q,i,S}- b_{Q,i,0} + \frac {\epsilon \tau_{slot} }{\left(1+ \overline{\beta}_i\right) }  } 
 {\epsilon \sum_{s=1}^{S} l_{i,s} }\right).
\end{eqnarray}

Now, if we let $K_i(S)$ denote the (random variable associated with) the number of slots which video client $i$ takes to download $S$ segments, 
then we can express the term appearing in the left hand side of above inequality as
 \begin{eqnarray} \label{another_exprsn_for_fraction_of_time_spent_rebuf}
\frac{ T_i(S) }{ \sum_{s=1}^{S} l_{i,s+1}    } =
\frac{\tau_{slot} \frac{\sum_{s=1}^{S} l_{i,s} f_{i,s} \left(q^*_{i,s}\right)}
{ \frac{1}{K_i(S)} \sum_{k=1}^{K_i(S)} r^*_{i,k}  }}
{\sum_{s'=1}^{S} l_{i,s'}}
 + o(S).
 \end{eqnarray} 
 Now note that any limit point of the sequence $\frac{1}{K_S} \sum_{k=1}^{K_S} r^*_{i,k}  $
 is also a limit point of the sequence $\frac{1}{K_i(S)} \sum_{k=1}^{K_i(S)} r^*_{i,k}  $
 since we can uniformly bound $K_i(S)-K_i(S-1)$.
 Thus, using \eqref{nova_satisfies_rebuf_constraint_asymptotically}, \eqref{another_exprsn_for_fraction_of_time_spent_rebuf}, \eqref{bound_on_cum_time_to_download}
 and the fact that $b_{Q,i,S}$ is bounded (see Lemma \ref{parameters_are_bounded}), 
 we can conclude that
 \eqref{repeat_NOVA_satifies_rebuf_constraints} also holds.
 
Next, we prove that NOVA asymptotically satisfies the cost constraints, i.e., \eqref{repeat_NOVA_satifies_cost_constraints}.
Note that the cost per unit video duration associated with the first $S$ segments under NOVA  for video client $i$ is 
 \begin{eqnarray} \label{cost_under_nova}
p_{i,S}\left( \left(q^*_i\right)_{1:S} \right) 
=
 p^d_i \frac{\sum_{s_i=1}^{S} l_{i,s_i} f_{i,s_i} \left(q^*_{i,s_i}\right)} { \sum_{s_i=1}^{S} l_{i,s_i} }.
  \end{eqnarray} 
Now, using the NOVA update rule \eqref{d_adapt_update} for parameter $d_{i,s_i}$, we have
$$ d_{i,s_i+1} \ge d_{i,s_i} + \epsilon 
\left(p^d_i \frac{l_{i,s_i+1}  f_{i,s_i+1}\left(  q^*_{i,s_i+1}\right)  }{  \overline{p}_i   } - \lambda_{i,s_i} \right).$$
Summing both sides of the above inequality from $s_i=1$ to $S$, we have
\begin{eqnarray} \label{int_step_proving_NOVA_satisfies_cost_constraint}
\frac{ p^d_i}{  \overline{p}_i  } \sum_{s_i=1}^S l_{i,s_i+1}  f_{i,s_i+1}\left(  q^*_{i,s_i+1}\right)   
\le  \sum_{s_i=1}^S  \lambda_{i,s_i}  + \frac{ d_{i,S+1} - d_{i,1}}{\epsilon }.
\end{eqnarray}
Next, note that by summing both sides of the NOVA update rule \eqref{lambda_update_NOVA} for the parameter $\lambda_{i,s_i}$ from $s_i=1$ to $S$, 
and rearranging the terms, we have
$$\sum_{s_i=1}^S  \lambda_{i,s_i} =  \sum_{s_i=1}^S  l_{i,s_i+1} -  \frac{\lambda_{i,S + 1} - \lambda_{i,1}}{\epsilon}  .$$
Combining this with \eqref{int_step_proving_NOVA_satisfies_cost_constraint}
and dividing by $\sum_{s_i=1}^S  l_{i,s_i+1} $,
we have
\begin{eqnarray*} 
\frac{ p^d_i}{  \overline{p}_i  } \frac{ \sum_{s_i=1}^S l_{i,s_i+1}  f_{i,s_i+1}\left(  q^*_{i,s_i+1}\right)  }{\sum_{s_i=1}^S  l_{i,s_i+1} } 
\le 1
-  \frac{\lambda_{i,S + 1} - \lambda_{i,1}}{\epsilon\sum_{s_i=1}^S  l_{i,s_i+1} }  
+ \frac{ d_{i,S+1} - d_{i,1}}{\epsilon \sum_{s_i=1}^S  l_{i,s_i+1} }.
\end{eqnarray*}
Now, the result in \eqref{repeat_NOVA_satifies_cost_constraints}
follows from \eqref{cost_under_nova} and the above inequality 
by noting that the terms $\lambda_{i,S + 1}$, $\lambda_{i,1}$, $d_{i,S+1}$ and $d_{i,1}$ are bounded (from Lemma \ref{parameters_are_bounded}),
and that $l_{\min}S\le \sum_{s_i=1}^S  l_{i,s_i+1}\le l_{\max}S$.

Next, we prove part (b) of Theorem \ref{main_optimality_theorem} regarding the optimality of NOVA.
Using Corollary  \ref{corollary_of_stoch_approx} (which says that $\left(\bm{\theta}_{i,s}\right)_{1 \le s\le S_{\epsilon}}$
essentially converges to ${\cal H}^*_i$) 
and Lemma \ref{continuity_of_solutions_to_QNOVA_and_RNOVA} (a)
(which says that $q^*_i \left(\bm{\theta}_i,f_i\right) $ is a continuous function of $\bm{\theta}_i$),
we can conclude that for $\bm{\theta}^{\pi}_i\in {\cal H}^*_i$
 \begin{eqnarray*}
\lim_{S\rightarrow \infty} \lim_{\epsilon \rightarrow 0}
\left(
\phi_{S_{\epsilon}}\left(\left( \left( q^*_i \left(\bm{\theta}_{i,s} , f_{i,s} \right)\right)_{\iiN} \right)_{1\le s \le S_{\epsilon}}  \right)
-\phi_{S_{\epsilon}}\left(\left( \left( q^*_i \left(\bm{\theta}^{\pi}_i , f_{i,s} \right)\right)_{\iiN} \right)_{1\le s \le S_{\epsilon}}  \right)\right)
\end{eqnarray*}
goes to zero in probability.
Now, part (b) of Theorem \ref{main_optimality_theorem} follows from the above observation and 
Theorem \ref{main_NOVA_with_theta_pi_is_optimal}
which states that for each $\iiN$ and for almost all sample paths
\begin{eqnarray*}
\lim_{S\rightarrow \infty} 
\left(\phi_S\left(\left( \left( q^*_i \left(\bm{\theta}^{\pi}_i , f_{i,s} \right)\right)_{\iiN} \right)_{1\le s \le S}  \right)
-\phi^{opt}_S\right)
=0.
\end{eqnarray*}

\end{IEEEproof}

\section{Conclusions}
\label{section_conclusion}
We developed a simple online algorithm NOVA for optimizing video delivery well suited for today's networks supporting DASH-based clients.
Interesting future directions include
an exploration of the potential of \emph{learning} user preferences,
and developing (and analyzing) `NOVA-like' algorithms for networks contention based medium access
by modulating the back-off timers using information about parameters like $b_{i,k}$.

\bibliographystyle{ieeetr}
\bibliography{proposal_refs}
\end{document}